# Construction of planar multilayer dyadic Green's functions by Fourier expansion method — Part I: Isotropic media

Huai-Yi Xie*（謝懷毅）

(Final Modified Date: 2017/10/20)

*Email: damoxie156@gmail.com

*Website: https://scholar.google.com.tw/citations?user=XxULS7cAAAAJ&hl=zh-TW&oi=ao

* Corresponding author

## Abstract

In this paper, we have derived planar multilayer dyadic Green's functions by Fourier expansion method and have checked its correctness by comparing results for reflected electric fields from dipole emissions near such structures available in previous literature. Furthermore, we show how these dyadic Green's functions can be applied to calculate reflected fields from a dipole source with arbitrary orientations. We believe our formulation will be powerful in the modeling of molecular fluorescence near these structures.

## 1. Introduction

Green's function method plays an important role in many applications in mathematics and physics. This has become a useful technique for obtaining the solutions to a linear and inhomogeneous partial differential equation with some appropriate boundary conditions [1]. *From a mathematical viewpoint, the Green function can be regarded as a kernel which converts a partial differential equation to an integral equation. In physical viewpoint, the Green function is closely related to the field produced by a point source, specifically determined by the boundary conditions of the problem* [2]. Once we have the solutions for the Green's function, we can easily obtain the solutions with arbitrary distribution of source.

In Electrodynamics, starting from Maxwell's equations, we obtain vector wave equations for both electric and magnetic fields. Let us focus on electric vector wave equation, which corresponds to dyadic Green's function partial differential equation but with a point source [2]. In addition, in general, we have the optical reciprocity property that dyadic Green's function is symmetric in the source and observer coordinates unless there is asymmetry in the permittivity or permeability tensor [3-4]. With its application for certain specific geometry, the construction of dyadic Green's function becomes an important topic and has been investigated in many literatures. There are two popular approaches to this: *One is based on the expansion of the dyadic Green's function in terms of eigenfunctions which can be obtained by solving the Sturm-Liouville problem, and then obtain its coefficients by matching the specific boundary conditions. The other is based on transforming the dyadic Green's function partial differential equation to reciprocal space, and then obtain its coefficients by matching the specific boundary conditions.* These two constructing methods each has its advantages and disadvantages.

The former is "customized", having obvious rules of how to construct the dyadic Green's function but eigenfunctions are more complicated-computed—to solve Sturm-Liouville problem is challenging under complicated boundary conditions. To our knowledge, this fundamental theory has been adopted with details provided in Tai's book [5] where multi-planar dyadic Green's functions have been used to calculate the decay rates of molecules in a vicinity of an interface [6]. However, there are some explicit theoretical errors in the calculation of the double mirror and have been corrected by extending and symmetrizing the solution forms [7]. Furthermore, how to construct the dyadic Green's function for inhomogeneous media has also been introduced [8]. Moreover, some scientists have constructed the dyadic Green's functions with many different boundary shapes such as planar multilayers [9-10], concentric spheres [11] and concentric ellipses [12], furthermore extending the concentric spheres to the eccentric spheres [13] and the cluster spheres [14] according to the addition theory. By the way, with anisotropic media of some specific forms of both permittivity and permeability (e.g. uniaxial [15] and gyrotropic media [16]), the dyadic Green's function has been constructed previously in the literature.

The latter is "intuitive", convenient to include some physical effects such as anisotropic and nonlocal effect of materials but less customized. To our knowledge, dyadic Green's function by Fourier expansion method for a single interface has been constructed and then used to study optical scattering in the presence of surface roughness [17] and furthermore applied to study corrugated films [18]. Moreover, the decay rate of molecules in the vicinity of corrugated films has been calculated as well [19]. In addition, multi-planar dyadic Green's functions by Fourier expansion method has also been developed [20]. Some scientists try to add nonlocal optical effect to dyadic Green's function by Fourier expansion method for a single interface [21-22] which can be reduced to the local case [17]. Above all, with anisotropic media of some specific forms of both permittivity and permeability like gyrotropic media, the multi-planar dyadic Green's functions by Fourier expansion method has also been discussed [23].

Once the dyadic Green's function is constructed, we can obtain electromagnetic fields in a straightforward way. However, for more complicated boundary shapes, like a combination of all boundaries of nanoparticles and multi-planar layers, constructing dyadic Green's function to obtain electromagnetic field has become difficult. One way to solve this problem is using the so-called Lippmann-Schwinger equation in which we apply both the "unperturbed" dyadic Green's function and the "unperturbed" incident field to obtain the total (unperturbed + perturbed) field everywhere. In our previous works, we make a lot of effort to construct more effective and accuracy theoretical models to fit the experimental data obtained by scanning electron microscopy (SEM), and then determine the critical features to the distribution of nanoparticles on a substrate [24-26].

Here, in this paper, we will re-construct the multi-planar Green's functions by Fourier expansion method, writing down all explicit analytic forms in each case with arbitrary locations of both the source and the field. *So in the following context, we can easily check if computational reflected electric fields are correct "analytically"— not just numerically such as Ref. [20].*

## 2. Green's function representation by Fourier expansion method

In the beginning, according to Fourier expansion in both real ($\boldsymbol{\rho} - \boldsymbol{\rho}'$) and reciprocal space ($\mathbf{k}_n$), we have:

$$\begin{cases} \mathbf{g}_n(z, z') = \int d^2(\boldsymbol{\rho} - \boldsymbol{\rho}') e^{-i\mathbf{k}_n \bullet (\boldsymbol{\rho} - \boldsymbol{\rho}')} \mathbf{G}(\mathbf{r}, \mathbf{r}') \\ \mathbf{G}(\mathbf{r}, \mathbf{r}') = \frac{1}{(2\pi)^2} \int d^2 \mathbf{k}_n e^{i\mathbf{k}_n \bullet (\boldsymbol{\rho} - \boldsymbol{\rho}')} \mathbf{g}_n(z, z') \end{cases}, \qquad (1)$$

where $\mathbf{G}(\mathbf{r}, \mathbf{r}')$ is in real space and $\mathbf{g}_n(z, z')$ is in reciprocal space, respectively. Also, both are in dyadic forms and in general, both satisfy the reciprocal properties that are $\mathbf{G}(\mathbf{r}, \mathbf{r}') = \mathbf{G}(\mathbf{r}', \mathbf{r})$ and $\mathbf{g}_n(z, z') = \mathbf{g}_n(z', z)$.

Next, the second-order differential wave equation for dyadic Green's function $\mathbf{G}(\mathbf{r}, \mathbf{r}')$ satisfies:

$$\nabla \times \nabla \times \mathbf{G} - k^2 \mathbf{G} = \mathbf{I} \delta(\mathbf{r} - \mathbf{r}'), \qquad (2)$$

where $k$ is the wavenumber of the medium. In component forms of Eq. (2), we

have:

$$\left[\nabla \times \nabla \times \mathbf{G} - k^2 \mathbf{G}\right]_{ij} = \delta_{ij}\delta(\mathbf{r}-\mathbf{r}')$$
$$\Rightarrow \left[\nabla(\nabla \bullet \mathbf{G})\right]_{ij} - \left[\nabla^2 \mathbf{G}\right]_{ij} - k^2 G_{ij} = \delta_{ij}\delta(\mathbf{r}-\mathbf{r}')$$
$$\Rightarrow \partial_i (\nabla \bullet \mathbf{G})_j - \nabla^2 G_{ij} - k^2 G_{ij} = \delta_{ij}\delta(\mathbf{r}-\mathbf{r}') \quad , \qquad (3)$$
$$\Rightarrow \sum_k \partial_i \partial_k G_{kj} - \nabla^2 G_{ij} - k^2 G_{ij} = \delta_{ij}\delta(\mathbf{r}-\mathbf{r}')$$
$$\Rightarrow \sum_k \left[\partial_i \partial_k - k^2 \delta_{ki} - \delta_{ki}\nabla^2\right] G_{kj} = \delta_{ij}\delta(\mathbf{r}-\mathbf{r}')$$

where the vector identity $\nabla \times \nabla \times \mathbf{G} = \nabla(\nabla \bullet \mathbf{G}) - \nabla^2 \mathbf{G}$ is used. For obviously, we rewrite Eq. (3) in matrix form as follows:

$$\begin{pmatrix} \partial_x^2 - k^2 - \nabla^2 & \partial_x \partial_y & \partial_x \partial_z \\ \partial_y \partial_x & \partial_y^2 - k^2 - \nabla^2 & \partial_y \partial_z \\ \partial_z \partial_x & \partial_z \partial_y & \partial_z^2 - k^2 - \nabla^2 \end{pmatrix} \begin{pmatrix} G_{xx} & G_{xy} & G_{xz} \\ G_{yx} & G_{yy} & G_{yz} \\ G_{zx} & G_{zy} & G_{zz} \end{pmatrix} = \begin{pmatrix} 1 & 0 & 0 \\ 0 & 1 & 0 \\ 0 & 0 & 1 \end{pmatrix} \delta(\mathbf{r}-\mathbf{r}'),$$

(4)

Hence we obtain the explicit form of real space wave equation to the dyadic Green's function $\mathbf{G}(\mathbf{r},\mathbf{r}')$. However, in our goal, we need to obtain the explicit form of reciprocal space wave equation to the dyadic Green's function $\mathbf{g}_n(z,z')$. By the same method of the expansion of dyadic Green's function, we expand the delta function as:

$$\begin{cases} \delta(z-z') = \int d^2(\boldsymbol{\rho}-\boldsymbol{\rho}') e^{-i\mathbf{k}_n \bullet (\boldsymbol{\rho}-\boldsymbol{\rho}')} \delta(\mathbf{r}-\mathbf{r}') \\ \delta(\mathbf{r}-\mathbf{r}') = \dfrac{1}{(2\pi)^2} \int d^2\mathbf{k}_n e^{i\mathbf{k}_n \bullet (\boldsymbol{\rho}-\boldsymbol{\rho}')} \delta(z-z') \end{cases}. \qquad (5)$$

Hence we substitute Eq. (1), Eq. (5) into Eq. (4) and obtain the following explicit matrix form of reciprocal space wave equation to the dyadic Green's function $\mathbf{g}_n(z,z')$:

$$\begin{pmatrix} k_{ny}^2 - k^2 - \partial_z^2 & -k_{nx}k_{ny} & ik_{nx}\partial_z \\ -k_{nx}k_{ny} & k_{nx}^2 - k^2 - \partial_z^2 & ik_{ny}\partial_z \\ ik_{nx}\partial_z & ik_{ny}\partial_z & k_n^2 - k^2 \end{pmatrix} \begin{pmatrix} g_{nxx} & g_{nxy} & g_{nxz} \\ g_{nyx} & g_{nyy} & g_{nyz} \\ g_{nzx} & g_{nzy} & g_{nzz} \end{pmatrix} = \begin{pmatrix} 1 & 0 & 0 \\ 0 & 1 & 0 \\ 0 & 0 & 1 \end{pmatrix} \delta(z-z'),$$

(6)

where $k_n^2 = k_{nx}^2 + k_{ny}^2$.

In convenience, we rotate the coordinate to the new coordinate and then make the vector $(k_{nx}, k_{ny}, 0)$ to $(k_n, 0, 0)$, we have the following relation:

$$\begin{pmatrix} \cos\varphi_n & \sin\varphi_n & 0 \\ -\sin\varphi_n & \cos\varphi_n & 0 \\ 0 & 0 & 1 \end{pmatrix} \begin{pmatrix} k_{nx} \\ k_{ny} \\ 0 \end{pmatrix} \equiv \mathbf{S}_n \begin{pmatrix} k_{nx} \\ k_{ny} \\ 0 \end{pmatrix} = \begin{pmatrix} \cos\varphi_n k_{nx} + \sin\varphi_n k_{ny} \\ -\sin\varphi_n k_{nx} + \cos\varphi_n k_{ny} \\ 0 \end{pmatrix} = \begin{pmatrix} k_n \\ 0 \\ 0 \end{pmatrix}, \quad (7)$$

where $k_{nx} = k_n \cos\varphi_n$, $k_{ny} = k_n \sin\varphi_n$ and $\mathbf{S}_n$ is called a rotation matrix from an old coordinate to a new coordinate (Note that $\mathbf{S}_n^{-1} = \mathbf{S}_n^T$). We may construct the relation of Green tensor based on old and new coordinates and have the following relation:

$$\begin{cases} \mathbf{a}' = \mathbf{S}_n \mathbf{a} \Rightarrow \mathbf{a} = \mathbf{S}_n^{-1} \mathbf{a}' \\ \mathbf{b}' = \mathbf{S}_n \mathbf{b} \Rightarrow \mathbf{b} = \mathbf{S}_n^{-1} \mathbf{b}' \Rightarrow \mathbf{b}^T = \mathbf{b}'^T \mathbf{S}_n \end{cases}$$
$$\Rightarrow \mathbf{a}\mathbf{b}^T = \mathbf{S}_n^{-1} (\mathbf{a}'\mathbf{b}'^T) \mathbf{S}_n \qquad (8)$$
$$\Rightarrow \mathbf{B} = \mathbf{S}_n^{-1} \mathbf{B}' \mathbf{S}_n \Rightarrow \mathbf{B}' = \mathbf{S}_n \mathbf{B} \mathbf{S}_n^{-1}$$

Hence in the new frame, Eq. (6) deduces to the following simply form:

$$\begin{pmatrix} -\partial_z^2 - k^2 & 0 & ik_n \partial_z \\ 0 & -k^2 - \partial_z^2 + k_n^2 & 0 \\ ik_n \partial_z & 0 & -k^2 + k_n^2 \end{pmatrix} \begin{pmatrix} \tilde{g}_{nxx} & \tilde{g}_{nxy} & \tilde{g}_{nxz} \\ \tilde{g}_{nyx} & \tilde{g}_{nyy} & \tilde{g}_{nyz} \\ \tilde{g}_{nzx} & \tilde{g}_{nzy} & \tilde{g}_{nzz} \end{pmatrix} = \begin{pmatrix} 1 & 0 & 0 \\ 0 & 1 & 0 \\ 0 & 0 & 1 \end{pmatrix} \delta(z - z'), (9)$$

where $\tilde{\mathbf{g}}_n = \mathbf{S}_n \mathbf{g}_n \mathbf{S}_n^{-1}$.

## 3. Boundary conditions of components of the Green's function $\tilde{\mathbf{g}}_n$

In the beginning, according to Eq. (9), we have:

$$\begin{cases} \left(-k^2 - \partial_z^2 + k_n^2\right) \tilde{g}_{nyx} = 0 \\ \left(-k^2 - \partial_z^2 + k_n^2\right) \tilde{g}_{nyz} = 0 \end{cases}, \qquad (10)$$

for components $\tilde{g}_{nyx}$ and $\tilde{g}_{nyz}$,

$$\begin{cases} \left(-\partial_z^2 - k^2\right) \tilde{g}_{nxy} + ik_n \partial_z \tilde{g}_{nzy} = 0 \\ ik_n \partial_z \tilde{g}_{nxy} + \left(-k^2 + k_n^2\right) \tilde{g}_{nzy} = 0 \end{cases}, \qquad (11)$$

for components $\tilde{g}_{nxy}$ and $\tilde{g}_{nzy}$,

$$\left(-k^2 - \partial_z^2 + k_n^2\right) \tilde{g}_{nyy} = \delta(z - z'), \qquad (12)$$

for component $\tilde{g}_{nyy}$,

$$\begin{cases} \left(-\partial_z^2 - k^2\right)\tilde{g}_{nxx} + ik_n\partial_z\tilde{g}_{nzx} = \delta(z-z') \\ ik_n\partial_z\tilde{g}_{nxx} + \left(-k^2 + k_n^2\right)\tilde{g}_{nzx} = 0 \end{cases}, \quad (13)$$

for components $\tilde{g}_{nxx}$ and $\tilde{g}_{nzx}$ and

$$\begin{cases} \left(-\partial_z^2 - k^2\right)\tilde{g}_{nxz} + ik_n\partial_z\tilde{g}_{nzz} = 0 \\ ik_n\partial_z\tilde{g}_{nxz} + \left(-k^2 + k_n^2\right)\tilde{g}_{nzz} = \delta(z-z') \end{cases}, \quad (14)$$

for components $\tilde{g}_{nxz}$ and $\tilde{g}_{nzz}$.

Next we determine the boundary conditions of $\tilde{\mathbf{g}}_n$. Starting from Maxwell's equations, we have the following relation for electric field and dyadic Green's function [2,5,11]:

$$\begin{aligned}\mathbf{E}(\mathbf{r}) &= i\omega\mu_0 \int d^3\mathbf{r}'\, \mathbf{G}(\mathbf{r},\mathbf{r}') \bullet \mathbf{J}(\mathbf{r}') \\ \Rightarrow E_\alpha(\mathbf{r}) &= i\omega\mu_0 \sum_\beta \int d^3\mathbf{r}'\, G_{\alpha\beta}(\mathbf{r},\mathbf{r}') J_\beta(\mathbf{r}')\end{aligned}, \quad (15)$$

$$\begin{aligned}\mathbf{H}(\mathbf{r}) &= \frac{1}{i\omega\mu_0}\nabla\times\mathbf{E}(\mathbf{r}) = \frac{1}{i\omega\mu_0}\sum_{\beta\gamma}\varepsilon_{\alpha\beta\gamma}\partial_\beta E_\gamma(\mathbf{r}) \\ &= \sum_{\beta\gamma\delta}\varepsilon_{\alpha\beta\gamma}\partial_\beta \int d^3\mathbf{r}'\, G_{\gamma\delta}(\mathbf{r},\mathbf{r}') J_\delta(\mathbf{r}')\end{aligned}. \quad (16)$$

According to continuity of $\mathbf{E}_\parallel$, $\mathbf{H}_\parallel$, $\mathbf{D}_\perp$ and $\mathbf{B}_\perp$ at the boundary, we have:

(i) $E_x$ continuity $\Rightarrow$ $G_{x\delta}, \delta = x, y, z$ continuity $\Rightarrow$ $g_{nx\delta}, \delta = x, y, z$ continuity.

(ii) $E_y$ continuity $\Rightarrow$ $G_{y\delta}, \delta = x, y, z$ continuity $\Rightarrow$ $g_{ny\delta}, \delta = x, y, z$ continuity.

(iii) $D_z$ continuity $\Rightarrow$ $\varepsilon(\mathbf{r}) G_{z\delta}, \delta = x, y, z$ continuity $\Rightarrow$ $\varepsilon g_{nz\delta}, \delta = x, y, z$ continuity.

(iv) $H_x$ continuity $\Rightarrow$ $\partial_y G_{z\delta} - \partial_z G_{y\delta}, \delta = x, y, z$ continuity $\Rightarrow$ $ik_{ny} g_{nz\delta} - \partial_z g_{ny\delta}, \delta = x, y, z$ continuity.

(v) $H_y$ continuity $\Rightarrow$ $\partial_z G_{x\delta} - \partial_x G_{z\delta}, \delta = x, y, z$ continuity $\Rightarrow$ $\partial_z g_{nx\delta} - ik_{nx} g_{nz\delta}, \delta = x, y, z$ continuity.

(vi) $B_z$ continuity $\Rightarrow$ $\partial_x G_{y\delta} - \partial_y G_{x\delta}, \delta = x, y, z$ continuity $\Rightarrow$ $k_{nx} g_{ny\delta} - k_{ny} g_{nx\delta}, \delta = x, y, z$ continuity.

Use the relation $\tilde{\mathbf{g}}_n = \mathbf{S}_n \mathbf{g}_n \mathbf{S}_n^{-1}$, we have:

$$\tilde{g}_{nxx} = \frac{1}{k_n^2}\left(k_{nx}^2 g_{nxx} + k_{nx}k_{ny} g_{nxy} + k_{nx}k_{ny} g_{nyx} + k_{ny}^2 g_{nyy}\right). \tag{17}$$

Since $g_{nxx}$, $g_{nxy}$, $g_{nyx}$ and $g_{nyy}$ are continuous at the boundary, then we conclude that $\tilde{g}_{nxx}$ is continuous at the boundary. Next we calculate the derivation of $\tilde{g}_{nxx}$:

$$\partial_z \tilde{g}_{nxx} = \frac{k_{nx}}{k_n^2}\left(k_{nx}\partial_z g_{nxx} + k_{ny}\partial_z g_{nyx}\right) + \frac{k_{ny}}{k_n^2}\left(k_{nx}\partial_z g_{nxy} + k_{ny}\partial_z g_{nyy}\right). \tag{18}$$

According to Eq. (6), we have:

$$ik_{nx}\partial_z g_{nxx} + ik_{ny}\partial_z g_{nyx} + \left(-k^2 + k_n^2\right)g_{nzx} = 0, \tag{19}$$

$$ik_{nx}\partial_z g_{nxy} + ik_{ny}\partial_z g_{nyy} + \left(-k^2 + k_n^2\right)g_{nzy} = 0, \tag{20}$$

Eq. (18) becomes:

$$\begin{aligned}
\partial_z \tilde{g}_{nxx} &= \frac{k_{nx}}{k_n^2}\left(k_{nx}\partial_z g_{nxx} + k_{ny}\partial_z g_{nyx}\right) + \frac{k_{ny}}{k_n^2}\left(k_{nx}\partial_z g_{nxy} + k_{ny}\partial_z g_{nyy}\right) \\
&= \frac{ik_{nx}}{k_n^2}\left(-ik_{nx}\partial_z g_{nxx} - ik_{ny}\partial_z g_{nyx}\right) + \frac{ik_{ny}}{k_n^2}\left(-ik_{nx}\partial_z g_{nxy} - ik_{ny}\partial_z g_{nyy}\right) \\
&= \frac{ik_{nx}}{k_n^2}\left(-k^2 + k_n^2\right)g_{nzx} + \frac{ik_{ny}}{k_n^2}\left(-k^2 + k_n^2\right)g_{nzy} \\
&= \frac{i}{k_n^2}\left(k_{nx}g_{nzx} + k_{ny}g_{nzy}\right)\left(k_n^2 - k^2\right)
\end{aligned} \tag{21}$$

$$\Rightarrow \frac{\varepsilon}{q_n^2}\partial_z \tilde{g}_{nxx} = \frac{i}{k_n^2}\varepsilon\left(k_{nx}g_{nzx} + k_{ny}g_{nzy}\right)$$

where $k^2 = \varepsilon\frac{\omega^2}{c^2}$ and $q_n^2 = k_n^2 - \varepsilon\frac{\omega^2}{c^2} = k_n^2 - k^2$. Since $\varepsilon g_{nzx}$ and $\varepsilon g_{nzy}$ are continuous at the boundary, then we conclude that $\frac{\varepsilon}{q_n^2}\frac{d}{dz}\tilde{g}_{nxx}$ is continuous at the boundary. Next the component $\tilde{g}_{nyy}$ is:

$$\tilde{g}_{nyy} = \frac{1}{k_n^2}\left(k_{ny}^2 g_{nxx} - k_{ny}k_{nx} g_{nxy} - k_{nx}k_{ny} g_{nyx} + k_{nx}^2 g_{nyy}\right). \tag{22}$$

Since $g_{nxx}$, $g_{nxy}$, $g_{nyx}$ and $g_{nyy}$ are continuous at the boundary, then we conclude that $\tilde{g}_{nyy}$ is continuous at the boundary. Next we calculate the derivation of $\tilde{g}_{nyy}$:

$$\partial_z \tilde{g}_{n,yy} = \partial_z \left[ \frac{1}{k_n^2} \left( k_n^2 g_{nxx} + k_n^2 g_{nyy} \right) - \frac{1}{k_n^2} \left( k_{nx}^2 g_{nxx} + k_{nx} k_{ny} g_{nxy} + k_{nx} k_{ny} g_{nyx} + k_{ny}^2 g_{nyy} \right) \right]$$

$$= \partial_z \left[ g_{nxx} + g_{nyy} - \frac{1}{k_n^2} \left( k_{nx}^2 g_{nxx} + k_{nx} k_{ny} g_{nxy} + k_{nx} k_{ny} g_{nyx} + k_{ny}^2 g_{nyy} \right) \right]$$

$$= \partial_z g_{nxx} + \partial_z g_{nyy} - \frac{1}{k_n^2} k_{nx} \left( k_{nx} \partial_z g_{nxx} + k_{ny} \partial_z g_{nxy} \right) - \frac{1}{k_n^2} k_{ny} \left( k_{nx} \partial_z g_{nyx} + k_{ny} \partial_z g_{nyy} \right)$$

$$= \partial_z g_{nxx} + \partial_z g_{nyy} - \frac{i}{k_n^2} k_{nx} \left( k_n^2 - k^2 \right) g_{nzx} - \frac{i}{k_n^2} k_{ny} \left( k_n^2 - k^2 \right) g_{nzy}$$

$$= \partial_z g_{nxx} + \partial_z g_{nyy} - \frac{i}{k_n^2} \left( k_n^2 - k^2 \right) \left( k_{nx} g_{nzx} + k_{ny} g_{nzy} \right)$$

$$= \left( \partial_z g_{nxx} - i k_{nx} g_{nzx} \right) + \left( \partial_z g_{nyy} - i k_{ny} g_{nzy} \right) + \frac{i}{k_n^2} \varepsilon \frac{\omega^2}{c^2} \left( k_{nx} g_{nzx} + k_{ny} g_{nzy} \right)$$

(23)

where we use Eq. (19) and Eq. (20). Since $\partial_z g_{nxx} - i k_{nx} g_{nzx}$, $\partial_z g_{nyy} - i k_{ny} g_{nzy}$, $\varepsilon g_{nzx}$ and $\varepsilon g_{nzy}$ are continuous at the boundary, then we conclude that $\partial_z \tilde{g}_{nyy}$ is continuous at the boundary. Next the component $\tilde{g}_{nxz}$ is:

$$\tilde{g}_{nxz} = \frac{1}{k_n} \left( k_{nx} g_{nxz} + k_{ny} g_{nyz} \right). \tag{24}$$

Since $g_{nxz}$ and $g_{nyz}$ are continuous at the boundary, then we conclude that $\tilde{g}_{nxz}$ is continuous at the boundary. Next we calculate the derivation of $\tilde{g}_{nxz}$:

$$\partial_z \tilde{g}_{nxz} = \frac{1}{k_n} \left( k_{nx} \partial_z g_{nxz} + k_{ny} \partial_z g_{nyz} \right) = \frac{i}{k_n} \left( -i k_{nx} \partial_z g_{nxz} - i k_{ny} \partial_z g_{nyz} \right)$$

$$= \frac{i}{k_n} \left( k_n^2 - k^2 \right) g_{nzz} = \frac{i}{k_n} q_n^2 g_{nzz} \tag{25}$$

$$\Rightarrow \frac{\varepsilon}{q_n^2} \partial_z \tilde{g}_{nxz} = \frac{i}{k_n} \varepsilon g_{nzz}$$

Since $\varepsilon g_{nzz}$ is continuous at the boundary, then we conclude that $\frac{\varepsilon}{q_n^2} \partial_z \tilde{g}_{nxz}$ is continuous at the boundary.

In summary, we have the following boundary conditions for $\tilde{g}_{nxx}, \tilde{g}_{nyy}, \tilde{g}_{nxz}$ as follows [17]:

<1> For $\tilde{g}_{nxx}$: $\tilde{g}_{nxx}|_{b-} = \tilde{g}_{nxx}|_{b+}$ and $\left. \frac{\varepsilon}{q_n^2} \partial_z \tilde{g}_{nxx} \right|_{b-} = \left. \frac{\varepsilon}{q_n^2} \partial_z \tilde{g}_{nxx} \right|_{b+}$, (26)

<2> For $\tilde{g}_{nyy}$: $\tilde{g}_{nyy}\big|_{b-} = \tilde{g}_{nyy}\big|_{b+}$ and $\partial_z \tilde{g}_{nyy}\big|_{b-} = \partial_z \tilde{g}_{nyy}\big|_{b+}$,   (27)

<3> For $\tilde{g}_{nxz}$: $\tilde{g}_{nxz}\big|_{b-} = \tilde{g}_{nxz}\big|_{b+}$ and $\dfrac{\varepsilon}{q_n^2}\partial_z \tilde{g}_{nxz}\bigg|_{b-} = \dfrac{\varepsilon}{q_n^2}\partial_z \tilde{g}_{nxz}\bigg|_{b+}$.   (28)

When the boundary conditions are constructed, we can obtain explicit forms of $\tilde{g}_{nxx}$, $\tilde{g}_{nyy}$ and $\tilde{g}_{nxz}$. Then $\tilde{g}_{nzx}$ and $\tilde{g}_{nzz}$ are obtained by using Eq. (13) and Eq. (14).

In the following contexts, we will give all cases to construct the explicit form of the dyadic Green's function $\mathbf{g}_n(z,z')$:

## 4. The representation of how to obtain explicit forms of each component of the Green's function $\tilde{\mathbf{g}}_n$

### 4-1. Consider the substrate effect:

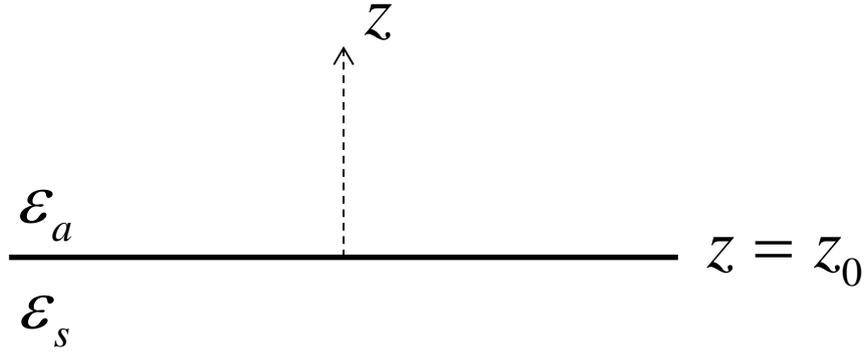

**Fig. 1** The geometry of considering the substrate effect

In the beginning, we consider the substrate effect as shown in Fig. 1. The boundary is only one which is at $z = z_0$. The dielectric function above and below $z = z_0$ are $\varepsilon_a$ and $\varepsilon_s$, respectively. According to the following well-known solutions of two simple differential equations [1,5,11]:

$$\left(\partial_z^2 + \alpha^2\right)g = \delta(z-z') \Rightarrow g = \frac{e^{i\alpha|z-z'|}}{2i\alpha},  \quad (29)$$

$$\left(\partial_z^2 + \alpha^2\right)g = \partial_z \delta(z-z') \Rightarrow g = \frac{1}{2}\operatorname{sgn}(z-z')e^{i\alpha|z-z'|}.  \quad (30)$$

Now we get ready enough to solve the explicit form of each component of $\tilde{\mathbf{g}}_n$. For the component $\tilde{g}_{nxz}$, starting from Eq. (14), we have:

$$\begin{cases} ik_n q_n^2 \partial_z \tilde{g}_{nzz} = \left(\partial_z^2 + k^2\right) q_n^2 \tilde{g}_{nxz} \\ -k_n^2 \partial_z^2 \tilde{g}_{nxz} + ik_n q_n^2 \partial_z \tilde{g}_{nzz} = ik_n \partial_z \delta(z-z') \end{cases}$$
$$\Rightarrow -k_n^2 \partial_z^2 \tilde{g}_{nxz} + \left(\partial_z^2 + k^2\right) q_n^2 \tilde{g}_{nxz} = ik_n \partial_z \delta(z-z')$$
$$\Rightarrow \left(-k_n^2 \partial_z^2 + \partial_z^2 q_n^2 + k^2 q_n^2\right) \tilde{g}_{nxz} = ik_n \partial_z \delta(z-z') \qquad (31)$$
$$\Rightarrow \left(\partial_z^2 - q_n^2\right) \tilde{g}_{nxz} = -\frac{ik_n}{k^2} \partial_z \delta(z-z')$$

Next we will consider the relative position between the field ($z$) and the source ($z'$). First we consider $z' < 0$ (without loss of generality, we may set $z_0 = 0$), the corresponding different equation in both regions $z > 0$ and $z < 0$ are:

$$\begin{cases} \left(\partial_z^2 - q_{na}^2\right) \tilde{g}_{nxz} = 0, z > 0 \\ \left(\partial_z^2 - q_{ns}^2\right) \tilde{g}_{nxz} = -\frac{ik_{ns}}{k_s^2} \partial_z \delta(z-z'), z < 0 \end{cases}, \qquad (32)$$

where $q_{na} = \sqrt{k_{na}^2 - \varepsilon_a \frac{\omega^2}{c^2}}$ and $q_{ns} = \sqrt{k_{ns}^2 - \varepsilon_s \frac{\omega^2}{c^2}}$. The solution of Eq. (32) is:

$$\tilde{g}_{nxz} = \begin{cases} A e^{-q_{na} z}, z > 0 \\ S_{nxz,s} \operatorname{sgn}(z-z') e^{-q_{ns}|z-z'|} + B e^{q_{ns} z}, z < 0 \end{cases}, \qquad (33)$$

where $S_{nxz,s} \equiv -\frac{ik_n}{2k_s^2}$. In order to match the boundary conditions of $\tilde{g}_{nxz}$ as discuss in Eq. (28):

$$\tilde{g}_{nxz}\big|_{0+} = \tilde{g}_{nxz}\big|_{0-} \quad \text{and} \quad \frac{\varepsilon_a}{q_{na}^2} \partial_z \tilde{g}_{nxz}\bigg|_{0+} = \frac{\varepsilon_s}{q_{ns}^2} \partial_z \tilde{g}_{nxz}\bigg|_{0-}, \qquad (34)$$

we have the following simulated equations:

$$A = S_{nxz,s} e^{q_{ns} z'} + B, \qquad (35)$$

and

$$A \frac{\varepsilon_a}{q_{na}^2}(-q_{na}) = S_{nxz,s} \frac{\varepsilon_s}{q_{ns}^2} \partial_z \left[\operatorname{sgn}(z-z') e^{-q_{ns}|z-z'|}\right]\bigg|_{0-} + B \frac{\varepsilon_s}{q_{ns}^2} q_{ns}$$
$$\Rightarrow -A \frac{\varepsilon_a}{q_{na}} = S_{nxz,s} \frac{\varepsilon_s}{q_{ns}^2}(-q_{ns}) e^{q_{ns} z'} + B \frac{\varepsilon_s}{q_{ns}} \qquad (36)$$
$$\Rightarrow -S_{nxz,s} \frac{\varepsilon_s}{q_{ns}} e^{q_{ns} z'} + B \frac{\varepsilon_s}{q_{ns}} = -A \frac{\varepsilon_a}{q_{na}}$$
$$\Rightarrow S_{nxz,s} \frac{\varepsilon_s q_{na}}{\varepsilon_a q_{ns}} e^{q_{ns} z'} - B \frac{\varepsilon_s q_{na}}{\varepsilon_a q_{ns}} = A$$

Combine Eq. (35) and Eq. (36), we obtain:

$$S_{nxz,s}e^{q_{ns}z'} + B = S_{nxz,s}\frac{\varepsilon_s q_{na}}{\varepsilon_a q_{ns}}e^{q_{ns}z'} - B\frac{\varepsilon_s q_{na}}{\varepsilon_a q_{ns}}$$

$$\Rightarrow \left(1 + \frac{\varepsilon_s q_{na}}{\varepsilon_a q_{ns}}\right)B = S_{nxz,s}e^{q_{ns}z'}\left(\frac{\varepsilon_s q_{na}}{\varepsilon_a q_{ns}} - 1\right)$$

$$\Rightarrow B = S_{nxz,s}e^{q_{ns}z'}\frac{\varepsilon_s q_{na} - \varepsilon_a q_{ns}}{\varepsilon_s q_{na} + \varepsilon_a q_{ns}} \equiv S_{nxz,s}e^{q_{ns}z'}R_{n//}^{sa} \qquad , \tag{37}$$

$$\Rightarrow A = S_{nxz,s}e^{q_{ns}z'} + B = S_{nxz,s}e^{q_{ns}z'} + S_{nxz,s}e^{q_{ns}z'}\frac{\varepsilon_s q_{na} - \varepsilon_a q_{ns}}{\varepsilon_s q_{na} + \varepsilon_a q_{ns}}$$

$$= S_{nxz,s}e^{q_{ns}z'}\frac{2\varepsilon_s q_{na}}{\varepsilon_s q_{na} + \varepsilon_a q_{ns}} \equiv S_{nxz,s}e^{q_{ns}z'}T_{n//}^{sa}$$

where $R_{n//}^{sa} \equiv \frac{\varepsilon_s q_{na} - \varepsilon_a q_{ns}}{\varepsilon_s q_{na} + \varepsilon_a q_{ns}}$ and $T_{n//}^{sa} \equiv \frac{2\varepsilon_s q_{na}}{\varepsilon_s q_{na} + \varepsilon_a q_{ns}}$ and we have $T_{n//}^{sa} - R_{n//}^{sa} = 1$.

Then,

$$\tilde{g}_{nxz} = \begin{cases} S_{nxz,s}T_{n//}^{sa}e^{q_{ns}z'}e^{-q_{na}z}, z>0 \\ S_{nxz,s}\,\text{sgn}(z-z')e^{-q_{ns}|z-z'|} + S_{nxz,s}R_{n//}^{sa}e^{q_{ns}z'}e^{q_{ns}z}, z<0 \end{cases}, \tag{38}$$

Second we consider $z' > 0$, the corresponding different equation in both regions $z > 0$ and $z < 0$ are:

$$\begin{cases} \left(\partial_z^2 - q_{na}^2\right)\tilde{g}_{nxz} = -\frac{ik_n}{k_a^2}\partial_z \delta(z-z'), z>0 \\ \left(\partial_z^2 - q_{ns}^2\right)\tilde{g}_{nxz} = 0, z<0 \end{cases}. \tag{39}$$

Furthermore, the solutions of Eq. (39) are:

$$\tilde{g}_{nxz} = \begin{cases} S_{nxz,a}\,\text{sgn}(z-z')e^{-q_{na}|z-z'|} + Ae^{-q_{na}z}, z>0 \\ Be^{q_{ns}z}, z<0 \end{cases}, \tag{40}$$

where $S_{nxz,a} \equiv -\frac{ik_n}{2k_a^2}$. In order to match the boundary conditions of $\tilde{g}_{nxz}$ as discuss in Eq. (34), we have the following simulated equations:

$$-S_{nxz,a}e^{-q_{na}z'} + A = B, \tag{41}$$

and

$$S_{nxz,a}\frac{\varepsilon_a}{q_{na}^2}\partial_z \text{sgn}(z-z')e^{-q_{na}|z-z'|}\bigg|_{0+} - \frac{\varepsilon_a}{q_{na}^2}Aq_{na} = \frac{\varepsilon_s}{q_{ns}^2}Bq_{ns}$$

$$\Rightarrow -S_{nxz,a}\frac{\varepsilon_a}{q_{na}^2}(q_{na})e^{-q_{na}z'} - \frac{\varepsilon_a}{q_{na}^2}Aq_{na} = \frac{\varepsilon_s}{q_{ns}^2}Bq_{ns} \qquad . \tag{42}$$

$$\Rightarrow -S_{nxz,a}\frac{\varepsilon_a q_{ns}}{\varepsilon_s q_{na}}e^{-q_{na}z'} - \frac{\varepsilon_a q_{ns}}{\varepsilon_s q_{na}}A = B$$

Combine Eq. (41) and Eq. (42), we obtain:

$$-S_{nxz,a}e^{-q_{na}z'} + A = -S_{nxz,a}\frac{\varepsilon_a q_{ns}}{\varepsilon_s q_{na}}e^{-q_{na}z'} - \frac{\varepsilon_a q_{ns}}{\varepsilon_s q_{na}}A$$

$$\Rightarrow \left(1 + \frac{\varepsilon_a q_{ns}}{\varepsilon_s q_{na}}\right)A = S_{nxz,a}e^{-q_{na}z'}\left(1 - \frac{\varepsilon_a q_{ns}}{\varepsilon_s q_{na}}\right)$$

$$\Rightarrow A = -S_{nxz,a}e^{-q_{na}z'}\frac{\varepsilon_a q_{ns} - \varepsilon_s q_{na}}{\varepsilon_a q_{ns} + \varepsilon_s q_{na}} \equiv -S_{nxz,a}e^{-q_{na}z'}R_{n//}^{as} \quad , \quad (43)$$

$$\Rightarrow B = -S_{nxz,a}e^{-q_{na}z'} + A = -S_{nxz,a}e^{-q_{na}z'} - S_{nxz,a}e^{-q_{na}z'}\frac{\varepsilon_a q_{ns} - \varepsilon_s q_{na}}{\varepsilon_a q_{ns} + \varepsilon_s q_{na}}$$

$$= -S_{nxz,a}e^{-q_{na}z'}\frac{2\varepsilon_a q_{ns}}{\varepsilon_s q_{na} + \varepsilon_a q_{ns}} \equiv -S_{nxz,a}e^{-q_{na}z'}T_{n//}^{as}$$

where $R_{n//}^{as} \equiv \frac{\varepsilon_a q_{ns} - \varepsilon_s q_{na}}{\varepsilon_a q_{ns} + \varepsilon_s q_{na}}$ and $T_{n//}^{as} \equiv \frac{2\varepsilon_a q_{ns}}{\varepsilon_s q_{na} + \varepsilon_a q_{ns}}$. Then

$$\tilde{g}_{nxz} = \begin{cases} S_{nxz,a}\,\text{sgn}(z-z')e^{-q_{na}|z-z'|} - S_{nxz,a}R_{n//}^{as}e^{-q_{na}z'}e^{-q_{na}z}, z > 0 \\ -S_{nxz,a}T_{n//}^{as}e^{-q_{na}z'}e^{q_{ns}z}, z < 0 \end{cases}. \quad (44)$$

Next we discuss the component $\tilde{g}_{nyy}$, starting from Eq. (12), we have:

$$\begin{aligned}(-k^2 - \partial_z^2 + k_n^2)\tilde{g}_{nyy} &= \delta(z-z') \\ \Rightarrow (\partial_z^2 - q_n^2)\tilde{g}_{nyy} &= -\delta(z-z')\end{aligned}. \quad (45)$$

Again, we will consider the relative position between the field ($z$) and the source ($z'$). First we consider $z' < 0$, the corresponding different equation in both regions $z > 0$ and $z < 0$ are:

$$\begin{cases} (\partial_z^2 - q_{na}^2)\tilde{g}_{nyy} = 0, z > 0 \\ (\partial_z^2 - q_{ns}^2)\tilde{g}_{nyy} = -\delta(z-z'), z < 0 \end{cases}, \quad (46)$$

Furthermore, the solutions of Eq. (46) are:

$$\tilde{g}_{nyy} = \begin{cases} Ae^{-q_{na}z}, z > 0 \\ S_{nyy,s}e^{-q_{ns}|z-z'|} + Be^{q_{ns}z}, z < 0 \end{cases}, \quad (47)$$

where $S_{nyy,s} \equiv \frac{1}{2q_{ns}}$. In order to match the boundary conditions of $\tilde{g}_{nyy}$ as discuss in Eq. (27):

$$\tilde{g}_{n,yy}\big|_{0+} = \tilde{g}_{n,yy}\big|_{0-} \text{ and } \partial_z\tilde{g}_{n,yy}\big|_{0+} = \partial_z\tilde{g}_{n,yy}\big|_{0-}, \quad (48)$$

we have the following simulated equations:

$$A = S_{nyy,s}e^{q_{ns}z'} + B, \quad (49)$$

and

$$-q_{na}A = S_{nyy,s}\partial_z e^{-q_{ns}|z-z'|} + q_{ns}B$$
$$\Rightarrow -q_{na}A = S_{nyy,s}e^{q_{ns}z'}(-q_{ns}) + q_{ns}B. \tag{50}$$
$$\Rightarrow A = S_{nyy,s}\frac{q_{ns}}{q_{na}}e^{q_{ns}z'} - \frac{q_{ns}}{q_{na}}B$$

Combine Eq. (49) and Eq. (50), we obtain:

$$S_{nyy,s}e^{q_{ns}z'} + B = S_{nyy,s}\frac{q_{ns}}{q_{na}}e^{q_{ns}z'} - \frac{q_{ns}}{q_{na}}B$$
$$\Rightarrow \left(1+\frac{q_{ns}}{q_{na}}\right)B = -S_{nyy,s}e^{q_{ns}z'}\left(1-\frac{q_{ns}}{q_{na}}\right)$$
$$\Rightarrow B = S_{nyy,s}e^{q_{ns}z'}\frac{q_{ns}-q_{na}}{q_{ns}+q_{na}} \equiv S_{nyy,s}e^{q_{ns}z'}R_{n\perp}^{sa}$$
$$\Rightarrow A = S_{nyy,s}e^{q_{ns}z'} + B = S_{nyy,s}e^{q_{ns}z'} + S_{nyy,s}e^{q_{ns}z'}\frac{q_{ns}-q_{na}}{q_{ns}+q_{na}} = S_{nyy,s}e^{q_{ns}z'}\frac{2q_{ns}}{q_{ns}+q_{na}} \equiv S_{nyy,s}e^{q_{ns}z'}T_{n\perp}^{sa}$$
$$\tag{51}$$

where $R_{n\perp}^{sa} \equiv \frac{q_{ns}-q_{na}}{q_{ns}+q_{na}}$ and $T_{n\perp}^{sa} \equiv \frac{2q_{ns}}{q_{ns}+q_{na}}$ and we have $T_{n\perp}^{sa} - R_{n\perp}^{sa} = 1$. Then,

$$\tilde{g}_{nyy} = \begin{cases} S_{nyy,s}T_{n\perp}^{sa}e^{q_{ns}z'}e^{-q_{na}z}, z > 0 \\ S_{nyy,s}e^{-q_{ns}|z-z'|} + S_{nyy,s}R_{n\perp}^{sa}e^{q_{ns}z'}e^{q_{ns}z}, z < 0 \end{cases}. \tag{52}$$

Second we consider $z' > 0$, the corresponding different equation in both regions $z > 0$ and $z < 0$ are:

$$\begin{cases} \left(\partial_z^2 - q_{na}^2\right)\tilde{g}_{nyy} = -\delta(z-z'), z > 0 \\ \left(\partial_z^2 - q_{ns}^2\right)\tilde{g}_{nyy} = 0, z < 0 \end{cases}, \tag{53}$$

Furthermore, the solutions of Eq. (53) are:

$$\tilde{g}_{nyy} = \begin{cases} S_{nyy,a}e^{-q_{na}|z-z'|} + Ae^{-q_{na}z}, z > 0 \\ Be^{q_{ns}z}, z < 0 \end{cases}, \tag{54}$$

where $S_{nyy,a} \equiv \frac{1}{2q_{na}}$. In order to match the boundary conditions of $\tilde{g}_{nyy}$ as discuss in Eq. (48), we have the following simulated equations:

$$S_{nyy,a}e^{-q_{na}z'} + A = B, \tag{55}$$

and

$$S_{nyy,a}\partial_z e^{-q_{na}|z-z'|} - q_{na}A = q_{ns}B$$
$$\Rightarrow S_{nyy,a}q_{na}e^{-q_{na}z'} - q_{na}A = q_{ns}B.$$ (56)
$$\Rightarrow B = S_{nyy,a}\frac{q_{na}}{q_{ns}}e^{-q_{na}z'} - \frac{q_{na}}{q_{ns}}A$$

Combine Eq. (55) and Eq. (56), we obtain:

$$S_{nyy,a}e^{-q_{na}z'} + A = S_{nyy,a}\frac{q_{na}}{q_{ns}}e^{-q_{na}z'} - \frac{q_{na}}{q_{ns}}A$$

$$\Rightarrow \left(1+\frac{q_{na}}{q_{ns}}\right)A = S_{nyy,a}\left(\frac{q_{na}}{q_{ns}}-1\right)e^{-q_{na}z'}$$

$$\Rightarrow A = S_{nyy,a}e^{-q_{na}z'}\frac{q_{na}-q_{ns}}{q_{na}+q_{ns}} \equiv S_{nyy,a}e^{-q_{na}z'}R_{n\perp}^{as} \quad , \quad (57)$$

$$\Rightarrow B = S_{nyy,a}e^{-q_{na}z'} + A = S_{nyy,a}e^{-q_{na}z'} + S_{nyy,a}e^{-q_{na}z'}\frac{q_{na}-q_{ns}}{q_{na}+q_{ns}}$$

$$= S_{nyy,a}e^{-q_{na}z'}\frac{2q_{na}}{q_{na}+q_{ns}} \equiv S_{nyy,a}e^{-q_{na}z'}T_{n\perp}^{as}$$

where $R_{n\perp}^{as} \equiv \frac{q_{na}-q_{ns}}{q_{na}+q_{ns}}$ and $T_{n\perp}^{as} \equiv \frac{2q_{na}}{q_{na}+q_{ns}}$. Then,

$$\tilde{g}_{nyy} = \begin{cases} S_{nyy,a}e^{-q_{na}|z-z'|} + S_{nyy,a}R_{n\perp}^{as}e^{-q_{na}z'}e^{-q_{na}z}, z>0 \\ S_{nyy,a}T_{n\perp}^{as}e^{-q_{na}z'}e^{q_{ns}z}, z<0 \end{cases}. \quad (58)$$

Next we discuss the component $\tilde{g}_{nxx}$, starting from Eq. (13), we have:

$$\begin{cases} \left(-\partial_z^2 - k^2\right)q_n^2\tilde{g}_{nxx} + ik_n q_n^2 \partial_z \tilde{g}_{nzx} = q^2\delta(z-z') \\ -k_n^2\partial_z^2 \tilde{g}_{nxx} + ik_n q_n^2 \partial_z \tilde{g}_{nzx} = 0 \end{cases}$$

$$\Rightarrow \left(-\partial_z^2 - k^2\right)q_n^2\tilde{g}_{nxx} + k_n^2\partial_z^2 \tilde{g}_{nxx} = q_n^2\delta(z-z') \quad . \quad (59)$$

$$\Rightarrow k^2\partial_z^2 \tilde{g}_{nxx} - k^2 q_n^2 \tilde{g}_{nxx} = q_n^2\delta(z-z')$$

$$\Rightarrow \left(\partial_z^2 - q_n^2\right)\tilde{g}_{nxx} = \frac{q_n^2}{k^2}\delta(z-z')$$

Again, we will consider the relative position between the field ($z$) and the source ($z'$). First we consider $z'<0$, the corresponding different equation in both regions $z>0$ and $z<0$ are:

$$\begin{cases} \left(\partial_z^2 - q_{na}^2\right)\tilde{g}_{nxx} = 0, z>0 \\ \left(\partial_z^2 - q_{ns}^2\right)\tilde{g}_{nxx} = \frac{q_{ns}^2}{k_s^2}\delta(z-z'), z<0 \end{cases}. \quad (60)$$

Furthermore, the solutions of Eq. (60) are:

$$\tilde{g}_{nxx} = \begin{cases} Ae^{-q_{na}z}, & z > 0 \\ S_{nxx,s}e^{-q_{ns}|z-z'|} + Be^{q_{ns}z}, & z < 0 \end{cases}, \tag{61}$$

where $S_{nxx,s} \equiv -\dfrac{q_{ns}}{2k_s^2}$. In order to match the boundary conditions of $\tilde{g}_{nxx}$ as discuss in Eq. (26):

$$\tilde{g}_{nxx}\Big|_{0+} = \tilde{g}_{nxx}\Big|_{0-} \quad \text{and} \quad \frac{\varepsilon_a}{q_{na}^2}\partial_z \tilde{g}_{nxx}\Big|_{0+} = \frac{\varepsilon_s}{q_{ns}^2}\partial_z \tilde{g}_{nxx}\Big|_{0-}, \tag{62}$$

we have the following simulated equations:

$$A = S_{nxx,s}e^{q_{ns}z'} + B, \tag{63}$$

and

$$-\frac{\varepsilon_a}{q_{na}^2}q_{na}A = S_{nxx,s}\frac{\varepsilon_s}{q_{ns}^2}(-q_{ns})e^{q_{ns}z'} + \frac{\varepsilon_s}{q_{ns}^2}q_{ns}B$$

$$\Rightarrow \frac{\varepsilon_a}{q_{na}}A = S_{nxx,s}\frac{\varepsilon_s}{q_{ns}}e^{q_{ns}z'} - \frac{\varepsilon_s}{q_{ns}}B \tag{64}$$

$$\Rightarrow A = S_{nxx,s}\frac{\varepsilon_s q_{na}}{\varepsilon_a q_{ns}}e^{q_{ns}z'} - \frac{\varepsilon_s q_{na}}{\varepsilon_a q_{ns}}B$$

Combine Eq. (63) and Eq. (64), we obtain:

$$S_{nxx,s}e^{q_{ns}z'} + B = S_{nxx,s}\frac{\varepsilon_s q_{na}}{\varepsilon_a q_{ns}}e^{q_{ns}z'} - \frac{\varepsilon_s q_{na}}{\varepsilon_a q_{ns}}B$$

$$\Rightarrow \left(1 + \frac{\varepsilon_s q_{na}}{\varepsilon_a q_{ns}}\right)B = S_{nxx,s}e^{q_{ns}z'}\left(\frac{\varepsilon_s q_{na}}{\varepsilon_a q_{ns}} - 1\right)$$

$$\Rightarrow B = S_{nxx,s}e^{q_{ns}z'}\frac{\varepsilon_s q_{na} - \varepsilon_a q_{ns}}{\varepsilon_s q_{na} + \varepsilon_a q_{ns}} \equiv S_{nxx,s}e^{q_{ns}z'}R_{n//}^{sa} \tag{65}$$

$$\Rightarrow A = S_{nxx,s}e^{q_{ns}z'} + B = S_{nxx,s}e^{q_{ns}z'} + S_{nxx,s}e^{q_{ns}z'}\frac{\varepsilon_s q_{na} - \varepsilon_a q_{ns}}{\varepsilon_s q_{na} + \varepsilon_a q_{ns}}$$

$$= S_{nxx,s}e^{q_{ns}z'}\frac{2\varepsilon_s q_{na}}{\varepsilon_a q_{ns} + \varepsilon_s q_{na}} \equiv S_{nxx,s}e^{q_{ns}z'}T_{n//}^{sa}$$

Then,

$$\tilde{g}_{nxx} = \begin{cases} S_{nxx,s}T_{n//}^{sa}e^{q_{ns}z'}e^{-q_{na}z}, & z > 0 \\ S_{nxx,s}e^{-q_{ns}|z-z'|} + S_{nxx,s}R_{n//}^{sa}e^{q_{ns}z'}e^{q_{ns}z}, & z < 0 \end{cases}. \tag{66}$$

Second we consider $z' > 0$, the corresponding different equation in both regions $z > 0$ and $z < 0$ are:

$$\begin{cases} \left(\partial_z^2 - q_{na}^2\right)\tilde{g}_{nxx} = \dfrac{q_{na}^2}{k_a^2}\delta(z-z'), z > 0 \\ \left(\partial_z^2 - q_{ns}^2\right)\tilde{g}_{nxx} = 0, z < 0 \end{cases} \tag{67}$$

Furthermore, the solutions of Eq. (67) are:

$$\tilde{g}_{nxx} = \begin{cases} S_{nxx,a} e^{-q_{na}|z-z'|} + A e^{-q_{na} z}, z > 0 \\ B e^{q_{ns} z}, z < 0 \end{cases}, \tag{68}$$

where $S_{nxx,a} \equiv -\dfrac{q_{na}}{2k_a^2}$. In order to match the boundary conditions of $\tilde{g}_{nxx}$ as discuss in Eq. (62), we have the following simulated equations:

$$S_{nxx,a} e^{-q_{na} z'} + A = B, \tag{69}$$

and

$$S_{nxx,a} \dfrac{\varepsilon_a}{q_{na}^2}\partial_z e^{-q_{na}|z-z'|} - q_{na}\dfrac{\varepsilon_a}{q_{na}^2} A = q_{ns}\dfrac{\varepsilon_s}{q_{ns}^2} B$$

$$\Rightarrow S_{nxx,a} \dfrac{\varepsilon_a}{q_{na}} e^{-q_{na} z'} - \dfrac{\varepsilon_a}{q_{na}} A = \dfrac{\varepsilon_s}{q_{ns}} B \tag{70}$$

$$\Rightarrow B = S_{nxx,a}\dfrac{\varepsilon_a q_{ns}}{\varepsilon_s q_{na}} e^{-q_{na} z'} - \dfrac{\varepsilon_a q_{ns}}{\varepsilon_s q_{na}} A$$

Combine Eq. (69) and Eq. (70), we obtain:

$$S_{nxx,a} e^{-q_{na} z'} + A = S_{nxx,a}\dfrac{\varepsilon_a q_{ns}}{\varepsilon_s q_{na}} e^{-q_{na} z'} - \dfrac{\varepsilon_a q_{ns}}{\varepsilon_s q_{na}} A$$

$$\Rightarrow \left(1 + \dfrac{\varepsilon_a q_{ns}}{\varepsilon_s q_{na}}\right) A = S_{nxx,a}\left(\dfrac{\varepsilon_a q_{ns}}{\varepsilon_s q_{na}} - 1\right) e^{-q_{na} z'}$$

$$\Rightarrow A = S_{nxx,a} e^{-q_{na} z'} \dfrac{\varepsilon_a q_{ns} - \varepsilon_s q_{na}}{\varepsilon_a q_{ns} + \varepsilon_s q_{na}} \equiv S_{nxx,a} e^{-q_{na} z'} R_{n//}^{as} \tag{71}$$

$$\Rightarrow B = S_{nxx,a} e^{-q_{na} z'} + A = S_{nxx,a} e^{-q_{na} z'} + S_{nxx,a} e^{-q_{na} z'}\dfrac{\varepsilon_a q_{ns} - \varepsilon_s q_{na}}{\varepsilon_a q_{ns} + \varepsilon_s q_{na}}$$

$$= S_{nxx,a} e^{-q_{na} z'}\dfrac{2\varepsilon_a q_{ns}}{\varepsilon_s q_{na} + \varepsilon_a q_{ns}} \equiv S_{nxx,a} e^{-q_{na} z'} T_{n//}^{as}$$

Then,

$$\tilde{g}_{nxx} = \begin{cases} S_{nxx,a} e^{-q_{na}|z-z'|} + S_{nxx,a} R_{n//}^{as} e^{-q_{na} z'} e^{-q_{na} z}, z > 0 \\ S_{nxx,a} T_{n//}^{as} e^{-q_{na} z'} e^{q_{ns} z}, z < 0 \end{cases}. \tag{72}$$

Next we discuss the component $\tilde{g}_{nzx}$, starting from Eq. (13), we have:

$$ik_n \partial_z \tilde{g}_{nxx} + \left(-k^2 + k_n^2\right)\tilde{g}_{nzx} = 0$$
$$\Rightarrow ik_n \partial_z \tilde{g}_{nxx} + q_n^2 \tilde{g}_{nzx} = 0 \quad , \tag{73}$$
$$\Rightarrow \tilde{g}_{nzx} = -\frac{ik_n}{q_n^2}\partial_z \tilde{g}_{nxx}$$

and then we obtain the explicit form of $\tilde{g}_{nzx}$ immediately. Again, we will consider the relative position between the field ($z$) and the source ($z'$). First we consider $z' < 0$, the corresponding different equation in both regions $z > 0$ and $z < 0$ are:

$$\tilde{g}_{nzx} = \begin{cases} -\dfrac{ik_n}{q_{na}^2} S_{nxx,s} T_{n//}^{sa} e^{q_{ns}z'} \partial_z e^{-q_{na}z}, & z > 0 \\[6pt] -\dfrac{ik_n}{q_{ns}^2} S_{nxx,s} \partial_z e^{-q_{ns}|z-z'|} - \dfrac{ik_n}{q_{ns}^2} S_{nxx,s} R_{n//}^{sa} e^{q_{ns}z'} \partial_z e^{q_{ns}z}, & z < 0 \end{cases}$$

$$= \begin{cases} \dfrac{ik_n}{q_{na}^2} S_{nxx,s} T_{n//}^{sa} e^{q_{ns}z'} q_{na} e^{-q_{na}z}, & z > 0 \\[6pt] -\dfrac{ik_n}{q_{ns}^2} S_{nxx,s}\left[-\operatorname{sgn}(z-z')\right] q_{ns} e^{-q_{ns}|z-z'|} - \dfrac{ik_n}{q_{ns}^2} S_{nxx,s} R_{n//}^{sa} e^{q_{ns}z'} q_{ns} e^{q_{ns}z}, & z < 0 \end{cases} \tag{74}$$

$$= \begin{cases} \dfrac{ik_n}{q_{na}} S_{nxx,s} T_{n//}^{sa} e^{q_{ns}z'} e^{-q_{na}z}, & z > 0 \\[6pt] \dfrac{ik_n}{q_{ns}} S_{nxx,s} \operatorname{sgn}(z-z') e^{-q_{ns}|z-z'|} - \dfrac{ik_n}{q_{ns}} S_{nxx,s} R_{n//}^{sa} e^{q_{ns}z'} e^{q_{ns}z}, & z < 0 \end{cases}$$

Second we consider $z' > 0$, the corresponding different equation in both regions $z > 0$ and $z < 0$ are:

$$\tilde{g}_{nzx} = \begin{cases} -\dfrac{ik_n}{q_{na}^2} S_{nxx,a} \partial_z e^{-q_{na}|z-z'|} - \dfrac{ik_n}{q_{na}^2} S_{nxx,a} R_{n//}^{as} e^{-q_{na}z'} \partial_z e^{-q_{na}z}, & z > 0 \\[6pt] -\dfrac{ik_n}{q_{ns}^2} S_{nxx,a} T_{n//}^{as} e^{-q_{na}z'} \partial_z e^{q_{ns}z}, & z < 0 \end{cases}$$

$$= \begin{cases} -\dfrac{ik_n}{q_{na}^2} S_{nxx,a}\left[-\operatorname{sgn}(z-z')\right] q_{na} e^{-q_{na}|z-z'|} - \dfrac{ik_n}{q_{na}^2} S_{nxx,a} R_{n//}^{as} e^{-q_{na}z'} (-q_{na}) e^{-q_{na}z}, & z > 0 \\[6pt] -\dfrac{ik_n}{q_{ns}^2} S_{nxx,a} T_{n//}^{as} e^{-q_{na}z'} q_{ns} e^{q_{ns}z}, & z < 0 \end{cases} \tag{75}$$

$$= \begin{cases} \dfrac{ik_n}{q_{na}} S_{nxx,a} \operatorname{sgn}(z-z') e^{-q_{na}|z-z'|} + \dfrac{ik_n}{q_{na}} S_{nxx,a} R_{n//}^{as} e^{-q_{na}z'} e^{-q_{na}z}, & z > 0 \\[6pt] -\dfrac{ik_n}{q_{ns}} S_{nxx,a} T_{n//}^{as} e^{-q_{na}z'} e^{q_{ns}z}, & z < 0 \end{cases}$$

Finally, we discuss the component $\tilde{g}_{nzz}$, starting from Eq. (14), we have:

$$ik_n \partial_z \tilde{g}_{nxz} + \left(-k^2 + k_n^2\right) \tilde{g}_{nzz} = \delta(z-z')$$
$$\Rightarrow ik_n \partial_z \tilde{g}_{nxz} + q_n^2 \tilde{g}_{nzz} = \delta(z-z') \qquad . \tag{76}$$
$$\Rightarrow \tilde{g}_{nzz} = \frac{1}{q_n^2} \delta(z-z') - \frac{1}{q_n^2} ik_n \partial_z \tilde{g}_{nxz}$$

Hence we obtain the explicit form of $\tilde{g}_{nzz}$ immediately. Again, we will consider the relative position between the field ($z$) and the source ($z'$). First we consider $z' < 0$, the corresponding different equation in both regions $z > 0$ and $z < 0$ are:

$$\tilde{g}_{nzz} = \begin{cases} -\dfrac{ik_n}{q_{na}^2} S_{nxz,s} T_{n//}^{sa} e^{q_{ns}z'} \partial_z e^{-q_{na}z}, & z > 0 \\ \dfrac{1}{q_{ns}^2} \delta(z-z') - \dfrac{ik_n}{q_{ns}^2} S_{nxz,s} \partial_z \left[ \mathrm{sgn}(z-z') e^{-q_{ns}|z-z'|} \right] - \dfrac{ik_n}{q_{ns}^2} S_{nxz,s} R_{n//}^{sa} e^{q_{ns}z'} \partial_z e^{q_{ns}z}, & z < 0 \end{cases}$$

$$= \begin{cases} \dfrac{ik_n}{q_{na}} S_{nxz,s} T_{n//}^{sa} e^{q_{ns}z'} e^{-q_{na}z}, & z > 0 \\ \dfrac{1}{q_{ns}^2} \delta(z-z') - \dfrac{ik_n}{q_{ns}^2} S_{nxz,s} \partial_z \left[ \mathrm{sgn}(z-z') e^{-q_{ns}|z-z'|} \right] - \dfrac{ik_n}{q_{ns}} S_{nxz,s} R_{n//}^{sa} e^{q_{ns}z'} e^{q_{ns}z}, & z < 0 \end{cases}$$
$$\tag{77}$$

Furthermore, we use the following identity:

$$\frac{d}{dz}\left[\mathrm{sgn}(z-z') e^{-q_{ns}|z-z'|}\right] = \frac{d}{dz}\left[\mathrm{sgn}(z-z')\right] e^{-q_{ns}|z-z'|} + \mathrm{sgn}(z-z') \frac{d}{dz} e^{-q_{ns}|z-z'|}$$
$$= 2\delta(z-z') e^{-q_{ns}|z-z'|} + \mathrm{sgn}(z-z') \left[-\mathrm{sgn}(z-z') q_{ns} e^{-q_{ns}|z-z'|}\right] \qquad . \tag{78}$$
$$= 2\delta(z-z') e^{-q_{ns}|z-z'|} - q_{ns} e^{-q_{ns}|z-z'|}$$

Eq. (77) becomes:

$$\tilde{g}_{nzz} = \begin{cases} \dfrac{ik_n}{q_{na}} S_{nxz,s} T_{n//}^{sa} e^{q_{ns}z'} e^{-q_{na}z}, & z > 0 \\ \dfrac{1}{q_{ns}^2} \delta(z-z') - \dfrac{ik_n}{q_{ns}^2} S_{nxz,s} \left[2\delta(z-z') e^{-q_{ns}|z-z'|} - q_{ns} e^{-q_{ns}|z-z'|}\right] - \dfrac{ik_n}{q_{ns}} S_{nxz,s} R_{n//}^{sa} e^{q_{ns}z'} e^{q_{ns}z}, & z < 0 \end{cases}$$

$$= \begin{cases} \dfrac{ik_n}{q_{na}} S_{nxz,s} T_{n//}^{sa} e^{q_{ns}z'} e^{-q_{na}z}, & z > 0 \\ \dfrac{1}{q_{ns}^2} \delta(z-z') - \dfrac{2ik_n}{q_{ns}^2} S_{nxz,s} \delta(z-z') + \dfrac{ik_n}{q_{ns}} S_{nxz,s} e^{-q_{ns}|z-z'|} - \dfrac{ik_n}{q_{ns}} S_{nxz,s} R_{n//}^{sa} e^{q_{ns}z'} e^{q_{ns}z}, & z < 0 \end{cases}$$

$$= \begin{cases} \dfrac{ik_n}{q_{na}} S_{nxz,s} T_{n//}^{sa} e^{q_{ns}z'} e^{-q_{na}z}, & z > 0 \\ -\dfrac{1}{k_s^2} \delta(z-z') + \dfrac{ik_n}{q_{ns}} S_{nxz,s} e^{-q_{ns}|z-z'|} - \dfrac{ik_n}{q_{ns}} S_{nxz,s} R_{n//}^{sa} e^{q_{ns}z'} e^{q_{ns}z}, & z < 0 \end{cases}$$
$$\tag{79}$$

Second we consider $z' > 0$, the corresponding different equation in both regions

$z > 0$ and $z < 0$ are:

$$\tilde{g}_{nzz} = \begin{cases} \dfrac{1}{q_{na}^2}\delta(z-z') - \dfrac{ik_n}{q_{na}^2}S_{nxz,a}\partial_z\left[\operatorname{sgn}(z-z')e^{-q_{na}|z-z'|}\right] + \dfrac{ik_n}{q_{na}^2}S_{nxz,a}R_{n//}^{as}e^{-q_{na}z'}\partial_z e^{-q_{na}z}, & z > 0 \\ \dfrac{ik_n}{q_{ns}^2}S_{nxz,a}T_{n//}^{as}e^{-q_{na}z'}\partial_z e^{q_{ns}z}, & z < 0 \end{cases}$$

$$= \begin{cases} \dfrac{1}{q_{na}^2}\delta(z-z') - \dfrac{ik_n}{q_{na}^2}S_{nxz,a}\left[2\delta(z-z') - q_{na}e^{-q_{na}|z-z'|}\right] - \dfrac{ik_n}{q_{na}}S_{nxz,a}R_{n//}^{as}e^{-q_{na}z'}e^{-q_{na}z}, & z > 0 \\ \dfrac{ik_n}{q_{ns}}S_{nxz,a}T_{n//}^{as}e^{-q_{na}z'}e^{q_{ns}z}, & z < 0 \end{cases}$$

$$= \begin{cases} -\dfrac{1}{k_a^2}\delta(z-z') + \dfrac{ik_n}{q_{na}}S_{nxz,a}e^{-q_{na}|z-z'|} - \dfrac{ik_n}{q_{na}}S_{nxz,a}R_{n//}^{as}e^{-q_{na}z'}e^{-q_{na}z}, & z > 0 \\ \dfrac{ik_n}{q_{ns}}S_{nxz,a}T_{n//}^{as}e^{-q_{na}z'}e^{q_{ns}z}, & z < 0 \end{cases}.$$

(80)

### 4-2. Consider the film over the substrate

Next we consider the film over the substrate and the geometry is shown in Fig. 2. The boundaries are at both $z = z_0$ and $z = z_1$. The dielectric function above $z = z_1$ is $\varepsilon_a$ and below $z = z_0$ is $\varepsilon_s$, respectively. The dielectric function of the film is $\varepsilon_f$.

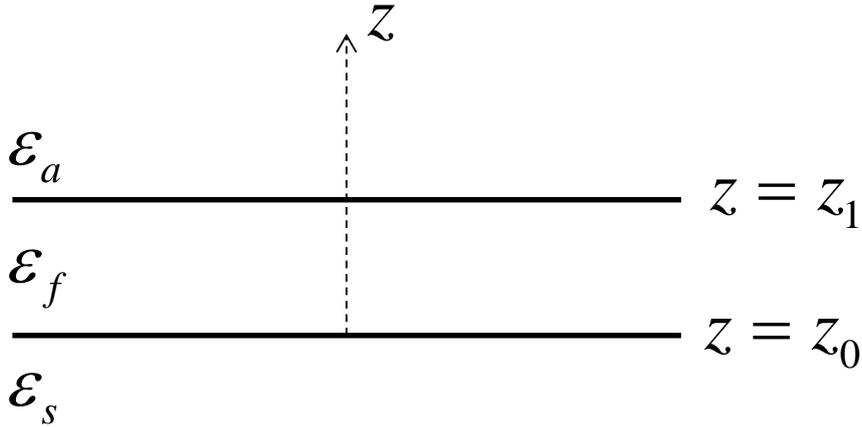

**Fig. 2** The geometry of considering the film over the substrate

For the component $\tilde{g}_{nxz}$, starting from Eq. (14) or Eq. (31), we have:

$$\left(\partial_z^2 - q_n^2\right)\tilde{g}_{nxz} = -\dfrac{ik_n}{k^2}\partial_z\delta(z-z'). \tag{81}$$

Next we will consider the relative position between the field ($z$) and the source ($z'$). First we consider $z' < 0$ (without loss of generality, we may set $z_0 = 0$), the corresponding different equation in regions $z > z_1$, $0 < z < z_1$ and $z < 0$ are:

$$\begin{cases} \left(\partial_z^2 - q_{na}^2\right)\tilde{g}_{nxz} = 0, z > z_1 \\ \left(\partial_z^2 - q_{nf}^2\right)\tilde{g}_{nxz} = 0, 0 < z < z_1 \\ \left(\partial_z^2 - q_{ns}^2\right)\tilde{g}_{nxz} = -\frac{ik_n}{k_s^2}\partial_z\delta(z-z'), z < 0 \end{cases} \quad (82)$$

Furthermore, the solutions of Eq. (82) are:

$$\tilde{g}_{nxz} = \begin{cases} Ae^{-q_{na}z}, z > z_1 \\ Be^{q_{nf}z} + Ce^{-q_{nf}z}, 0 < z < z_1 \\ S_{nxz,s}\operatorname{sgn}(z-z')e^{-q_{ns}|z-z'|} + De^{q_{ns}z}, z < 0 \end{cases}, \quad (83)$$

where $S_{nxz,s} \equiv -\frac{ik_n}{2k_s^2}$. In order to match the boundary conditions of $\tilde{g}_{nxz}$ as discuss in Eq. (28):

$$\begin{cases} \tilde{g}_{nxz}\big|_{z_{1+}} = \tilde{g}_{nxz}\big|_{z_{1-}}, \frac{\varepsilon_a}{q_{na}^2}\partial_z\tilde{g}_{nxz}\bigg|_{z_{1+}} = \frac{\varepsilon_f}{q_{nf}^2}\partial_z\tilde{g}_{nxz}\bigg|_{z_{1-}} \\ \tilde{g}_{nxz}\big|_{0+} = \tilde{g}_{nxz}\big|_{0-}, \frac{\varepsilon_f}{q_{nf}^2}\partial_z\tilde{g}_{nxz}\bigg|_{0+} = \frac{\varepsilon_s}{q_{ns}^2}\partial_z\tilde{g}_{nxz}\bigg|_{0-} \end{cases}, \quad (84)$$

we have the following simulated equations:

$$\begin{cases} Ae^{-q_{na}z_1} = Be^{q_{nf}z_1} + Ce^{-q_{nf}z_1} \\ B + C = S_{nxz,s}e^{q_{ns}z'} + D \end{cases}, \quad (85)$$

and

$$\begin{cases} -\frac{\varepsilon_a}{q_{na}}Ae^{-q_{na}z_1} = \frac{\varepsilon_f}{q_{nf}}\left(Be^{q_{nf}z_1} - Ce^{-q_{nf}z_1}\right) \\ \frac{\varepsilon_f}{q_{nf}}(B-C) = -\frac{\varepsilon_s}{q_{ns}}S_{nxz,s}e^{q_{ns}z'} + \frac{\varepsilon_s}{q_{ns}}D \end{cases}. \quad (86)$$

Furthermore, we obtain:

$$-\frac{\varepsilon_a}{q_{na}}\left(Be^{q_{nf}z_1} + Ce^{-q_{nf}z_1}\right) = \frac{\varepsilon_f}{q_{nf}}\left(Be^{q_{nf}z_1} - Ce^{-q_{nf}z_1}\right)$$

$$\Rightarrow -\left(\frac{\varepsilon_a}{q_{na}} + \frac{\varepsilon_f}{q_{nf}}\right)B = \left(\frac{\varepsilon_a}{q_{na}} - \frac{\varepsilon_f}{q_{nf}}\right)Ce^{-2q_{nf}z_1} \quad , \quad (87)$$

$$\Rightarrow B = \frac{\varepsilon_f q_{na} - \varepsilon_a q_{nf}}{\varepsilon_f q_{na} + \varepsilon_a q_{nf}}Ce^{-2q_{nf}z_1} = R_{n//}^{fa}e^{-2q_{nf}z_1}C$$

$$\Rightarrow A = R_{n//}^{fa}e^{q_{nf}z_1}e^{q_{na}z_1}C + e^{-q_{nf}z_1}e^{q_{na}z_1}C$$

and

$$R_{n//}^{fa}e^{-2q_{nf}z_1}C + C = S_{nxz,s}e^{q_{ns}z'} + D$$
$$\Rightarrow D = \left(R_{n//}^{fa}e^{-2q_{nf}z_1} + 1\right)C - S_{nxz,s}e^{q_{ns}z'}.$$
(88)

Furthermore, we get the following form to solve the coefficient $C$:

$$\frac{\varepsilon_f}{q_{nf}}\left(R_{n//}^{fa}e^{-2q_{nf}z_1} - 1\right)C = -\frac{\varepsilon_s}{q_{ns}}S_{nxz,s}e^{q_{ns}z'} + \frac{\varepsilon_s}{q_{ns}}\left[\left(R_{n//}^{fa}e^{-2q_{nf}z_1} + 1\right)C - S_{nxz,s}e^{q_{ns}z'}\right]$$

$$\Rightarrow \left[\left(\frac{\varepsilon_f}{q_{nf}} - \frac{\varepsilon_s}{q_{ns}}\right)R_{n//}^{fa}e^{-2q_{nf}z_1} - \left(\frac{\varepsilon_f}{q_{nf}} + \frac{\varepsilon_s}{q_{ns}}\right)\right]C = -\frac{2\varepsilon_s}{q_{ns}}S_{nxz,s}e^{q_{ns}z'}$$

$$\Rightarrow C = \frac{-\dfrac{2\varepsilon_s}{q_{ns}}S_{nxz,s}e^{q_{ns}z'}}{\dfrac{\varepsilon_f q_{ns} - \varepsilon_s q_{nf}}{q_{nf}q_{ns}}R_{n//}^{fa}e^{-2q_{nf}z_1} - \dfrac{\varepsilon_f q_{ns} + \varepsilon_s q_{nf}}{q_{nf}q_{ns}}} = \frac{-\dfrac{2\varepsilon_s q_{nf}}{\varepsilon_f q_{ns} + \varepsilon_s q_{nf}}S_{nxz,s}e^{q_{ns}z'}}{\dfrac{\varepsilon_f q_{ns} - \varepsilon_s q_{nf}}{\varepsilon_f q_{ns} + \varepsilon_s q_{nf}}R_{n//}^{fa}e^{-2q_{nf}z_1} - 1}$$
(89)

$$= \frac{T_{n//}^{sf}S_{nxz,s}e^{q_{ns}z'}}{1 - R_{n//}^{fs}R_{n//}^{fa}e^{-2q_{nf}z_1}}$$

Hence we can get the other three coefficients as the following forms:

$$B = R_{n//}^{fa}e^{-2q_{nf}z_1}C = \frac{R_{n//}^{fa}T_{n//}^{sf}S_{nxz,s}e^{q_{ns}z'}e^{-2q_{nf}z_1}}{1 - R_{n//}^{fs}R_{n//}^{fa}e^{-2q_{nf}z_1}},$$
(90)

$$A = R_{n//}^{fa}e^{q_{nf}z_1}e^{q_{na}z_1}C + e^{-q_{nf}z_1}e^{q_{na}z_1}C = S_{nxz,s}T_{n//}^{sf}\frac{R_{n//}^{fa}e^{q_{nf}z_1} + e^{-q_{nf}z_1}}{1 - R_{n//}^{fs}R_{n//}^{fa}e^{-2q_{nf}z_1}}e^{q_{ns}z'}e^{q_{na}z_1},$$
(91)

$$D = \left(R_{n//}^{fa}e^{-2q_{nf}z_1} + 1\right)C - S_{nxz,s}e^{q_{ns}z'} = \frac{\left(R_{n//}^{fa}e^{-2q_{nf}z_1} + 1\right)T_{n//}^{sf}S_{nxz,s}e^{q_{ns}z'}}{1 - R_{n//}^{fs}R_{n//}^{fa}e^{-2q_{nf}z_1}} - S_{nxz,s}e^{q_{ns}z'}$$
$$= S_{nxz,s}e^{q_{ns}z'}\frac{R_{n//}^{fa}e^{-2q_{nf}z_1} + T_{n//}^{sf} - 1}{1 - R_{n//}^{fs}R_{n//}^{fa}e^{-2q_{nf}z_1}}$$
(92)

Hence we solve the coefficients in Eq. (83) completely. Next we consider $0 < z' < z_1$, the corresponding different equation in regions $z > z_1$, $0 < z < z_1$ and $z < 0$ are:

$$\begin{cases}\left(\partial_z^2 - q_{na}^2\right)\tilde{g}_{nxz} = 0, z > z_1 \\ \left(\partial_z^2 - q_{nf}^2\right)\tilde{g}_{nxz} = -\dfrac{ik_n}{k_f^2}\partial_z\delta(z - z'), 0 < z < z_1 \\ \left(\partial_z^2 - q_{ns}^2\right)\tilde{g}_{nxz} = 0, z < 0\end{cases}$$
(93)

Furthermore, the solutions of Eq. (93) are:

$$\tilde{g}_{nxz} = \begin{cases} Ae^{-q_{na}z}, z > z_1 \\ S_{nxz,f}\,\text{sgn}(z-z')e^{-q_{nf}|z-z'|} + Be^{q_{nf}z} + Ce^{-q_{nf}z}, 0 < z < z_1, \\ De^{q_{ns}z}, z < 0 \end{cases} \qquad (94)$$

where $S_{nxz,f} \equiv -\dfrac{ik_n}{2k_f^2}$. In order to match the boundary conditions of $\tilde{g}_{nxz}$ as discuss in Eq. (84), we have the following simulated equations:

$$\begin{cases} Ae^{-q_{na}z_1} = S_{nxz,f}e^{-q_{nf}z_1}e^{q_{nf}z'} + Be^{q_{nf}z_1} + Ce^{-q_{nf}z_1} \\ -S_{nxz,f}e^{-q_{nf}z'} + B + C = D \end{cases}, \qquad (95)$$

and

$$\begin{cases} -\dfrac{\varepsilon_a}{q_{na}}Ae^{-q_{na}z_1} = -\dfrac{\varepsilon_f}{q_{nf}}S_{nxz,f}e^{q_{nf}z'}e^{-q_{nf}z_1} + \dfrac{\varepsilon_f}{q_{nf}}\left(Be^{q_{nf}z_1} - Ce^{-q_{nf}z_1}\right) \\ -\dfrac{\varepsilon_f}{q_{nf}}S_{nxz,f}e^{-q_{nf}z'} + \dfrac{\varepsilon_f}{q_{nf}}(B-C) = \dfrac{\varepsilon_s}{q_{ns}}D \end{cases}. \qquad (96)$$

Furthermore, we obtain:

$$-\dfrac{\varepsilon_a}{q_{na}}\left(S_{nxz,f}e^{-q_{nf}z_1}e^{q_{nf}z'} + Be^{q_{nf}z_1} + Ce^{-q_{nf}z_1}\right) = -\dfrac{\varepsilon_f}{q_{nf}}S_{nxz,f}e^{q_{nf}z'}e^{-q_{nf}z_1} + \dfrac{\varepsilon_f}{q_{nf}}\left(Be^{q_{nf}z_1} - Ce^{-q_{nf}z_1}\right)$$

$$\Rightarrow -\left(\dfrac{\varepsilon_a}{q_{na}} + \dfrac{\varepsilon_f}{q_{nf}}\right)B = \left(\dfrac{\varepsilon_a}{q_{na}} - \dfrac{\varepsilon_f}{q_{nf}}\right)e^{-2q_{nf}z_1}C + \left(\dfrac{\varepsilon_a}{q_{na}} - \dfrac{\varepsilon_f}{q_{nf}}\right)S_{nxz,f}e^{q_{nf}z'}e^{-2q_{nf}z_1}$$

$$\Rightarrow B = -\dfrac{\varepsilon_a q_{nf} - \varepsilon_f q_{na}}{\varepsilon_a q_{nf} + \varepsilon_f q_{na}}e^{-2q_{nf}z_1}C - \dfrac{\varepsilon_a q_{nf} - \varepsilon_f q_{na}}{\varepsilon_a q_{nf} + \varepsilon_f q_{na}}S_{nxz,f}e^{q_{nf}z'}e^{-2q_{nf}z_1}$$

$$= R_{n//}^{fa}e^{-2q_{nf}z_1}C + R_{n//}^{fa}S_{nxz,f}e^{q_{nf}z'}e^{-2q_{nf}z_1}$$

$$\Rightarrow A = S_{nxz,f}e^{-q_{nf}z_1}e^{q_{na}z_1}e^{q_{nf}z'} + Be^{q_{nf}z_1}e^{q_{na}z_1} + Ce^{-q_{nf}z_1}e^{q_{na}z_1}$$

$$= S_{nxz,f}e^{-q_{nf}z_1}e^{q_{na}z_1}e^{q_{nf}z'} + R_{n//}^{fa}S_{nxz,f}e^{-q_{nf}z_1}e^{q_{na}z_1}e^{q_{nf}z'} + R_{n//}^{fa}e^{-q_{nf}z_1}e^{q_{na}z_1}C + e^{-q_{nf}z_1}e^{q_{na}z_1}C$$

$$= \left(1 + R_{n//}^{fa}\right)\left(S_{nxz,f}e^{-q_{nf}z_1}e^{q_{na}z_1}e^{q_{nf}z'} + e^{-q_{nf}z_1}e^{q_{na}z_1}C\right)$$

$$= T_{n//}^{fa}S_{nxz,f}e^{-q_{nf}z_1}e^{q_{na}z_1}e^{q_{nf}z'} + T_{n//}^{fa}e^{-q_{nf}z_1}e^{q_{na}z_1}C$$

, (97)

and

$$D = -S_{nxz,f}e^{-q_{nf}z'} + B + C = -S_{nxz,f}e^{-q_{nf}z'} + R_{n//}^{fa}S_{nxz,f}e^{q_{nf}z'}e^{-2q_{nf}z_1} + \left(1 + R_{n//}^{fa}e^{-2q_{nf}z_1}\right)C. \quad (98)$$

Furthermore, we get the following form to solve the coefficient $C$:

$$-\frac{\varepsilon_f}{q_{nf}}S_{nxz,f}e^{-q_{nf}z'} + \frac{\varepsilon_f}{q_{nf}}\left(R_{n//}^{fa}e^{-2q_{nf}z_1}C + R_{n//}^{fa}S_{nxz,f}e^{q_{nf}z'}e^{-2q_{nf}z_1} - C\right)$$

$$= \frac{\varepsilon_s}{q_{ns}}\left[-S_{nxz,f}e^{-q_{nf}z'} + R_{n//}^{fa}S_{nxz,f}e^{q_{nf}z'}e^{-2q_{nf}z_1} + \left(1 + R_{n//}^{fa}e^{-2q_{nf}z_1}\right)C\right]$$

$$\Rightarrow \left[\left(\frac{\varepsilon_f}{q_{nf}} - \frac{\varepsilon_s}{q_{ns}}\right)R_{n//}^{fa}e^{-2q_{nf}z_1} - \left(\frac{\varepsilon_f}{q_{nf}} + \frac{\varepsilon_s}{q_{ns}}\right)\right]C = \left(\frac{\varepsilon_f}{q_{nf}} - \frac{\varepsilon_s}{q_{ns}}\right)\left(e^{-q_{nf}z'} - R_{n//}^{fa}e^{q_{nf}z'}e^{-2q_{nf}z_1}\right)S_{nxz,f}.$$

$$C = \frac{\left(\dfrac{\varepsilon_f q_{ns} - \varepsilon_s q_{nf}}{\varepsilon_f q_{ns} + \varepsilon_s q_{nf}}\right)\left(e^{-q_{nf}z'} - R_{n//}^{fa}e^{q_{nf}z'}e^{-2q_{nf}z_1}\right)S_{nxz,f}}{\left(\dfrac{\varepsilon_f q_{ns} - \varepsilon_s q_{nf}}{\varepsilon_f q_{ns} + \varepsilon_s q_{nf}}\right)R_{n//}^{fa}e^{-2q_{nf}z_1} - 1} = \frac{R_{n//}^{fa}e^{q_{nf}z'}e^{-2q_{nf}z_1} - e^{-q_{nf}z'}}{1 - R_{n//}^{fs}R_{n//}^{fa}e^{-2q_{nf}z_1}}R_{n//}^{fs}S_{nxz,f}$$

(99)

Hence we can get the other three coefficients as the following forms:

$$B = R_{n//}^{fa}e^{-2q_{nf}z_1}C + R_{n//}^{fa}S_{nxz,f}e^{q_{nf}z'}e^{-2q_{nf}z_1}$$

$$= R_{n//}^{fa}e^{-2q_{nf}z_1}\frac{R_{n//}^{fa}e^{q_{nf}z'}e^{-2q_{nf}z_1} - e^{-q_{nf}z'}}{1 - R_{n//}^{fs}R_{n//}^{fa}e^{-2q_{nf}z_1}}R_{n//}^{fs}S_{nxz,f} + R_{n//}^{fa}S_{nxz,f}e^{q_{nf}z'}e^{-2q_{nf}z_1}, \quad (100)$$

$$= S_{nxz,f}R_{n//}^{fa}\frac{e^{q_{nf}z'} - R_{n//}^{fs}e^{-q_{nf}z'}}{1 - R_{n//}^{fs}R_{n//}^{fa}e^{-2q_{nf}z_1}}e^{-2q_{nf}z_1}$$

$$A = T_{n//}^{fa}S_{nxz,f}e^{-q_{nf}z_1}e^{q_{na}z_1}e^{q_{nf}z'} + T_{n//}^{fa}e^{-q_{nf}z_1}e^{q_{na}z_1}C$$

$$= T_{n//}^{fa}S_{nxz,f}e^{-q_{nf}z_1}e^{q_{na}z_1}e^{q_{nf}z'} + T_{n//}^{fa}e^{-q_{nf}z_1}e^{q_{na}z_1}\frac{R_{n//}^{fa}e^{q_{nf}z'}e^{-2q_{nf}z_1} - e^{-q_{nf}z'}}{1 - R_{n//}^{fs}R_{n//}^{fa}e^{-2q_{nf}z_1}}R_{n//}^{fs}S_{nxz,f}, \quad (101)$$

$$= S_{nxz,f}T_{n//}^{fa}\frac{e^{q_{nf}z'} - e^{-q_{nf}z'}R_{n//}^{fs}}{1 - R_{n//}^{fs}R_{n//}^{fa}e^{-2q_{nf}z_1}}e^{-q_{nf}z_1}e^{q_{na}z_1}$$

$$D = -S_{nxz,f}e^{-q_{nf}z'} + R_{n//}^{fa}S_{nxz,f}e^{q_{nf}z'}e^{-2q_{nf}z_1} + \left(1 + R_{n//}^{fa}e^{-2q_{nf}z_1}\right)C$$

$$= -S_{nxz,f}e^{-q_{nf}z'} + R_{n//}^{fa}S_{nxz,f}e^{q_{nf}z'}e^{-2q_{nf}z_1} + \frac{R_{n//}^{fa}e^{q_{nf}z'}e^{-2q_{nf}z_1} - e^{-q_{nf}z'}}{1 - R_{n//}^{fs}R_{n//}^{fa}e^{-2q_{nf}z_1}}\left(1 + R_{n//}^{fa}e^{-2q_{nf}z_1}\right)R_{n//}^{fs}S_{nxz,f}$$

$$= S_{nxz,f}\frac{-e^{-q_{nf}z'}\left(1 + R_{n//}^{fs}\right) + R_{n//}^{fa}e^{q_{nf}z'}e^{-2q_{nf}z_1}\left(1 + R_{n//}^{fs}\right)}{1 - R_{n//}^{fs}R_{n//}^{fa}e^{-2q_{nf}z_1}}.$$

$$= S_{nxz,f}T_{n//}^{fs}\frac{R_{n//}^{fa}e^{q_{nf}z'}e^{-2q_{nf}z_1} - e^{-q_{nf}z'}}{1 - R_{n//}^{fs}R_{n//}^{fa}e^{-2q_{nf}z_1}}$$

(102)

Hence we solve the coefficients in Eq. (95) completely. Next we consider $z' > z_1$, the corresponding different equation in regions $z > z_1$, $0 < z < z_1$ and $z < 0$ are:

$$\begin{cases} \left(\partial_z^2 - q_{na}^2\right)\tilde{g}_{nxz} = -\dfrac{ik_n}{k_a^2}\partial_z \delta(z-z'), z > z_1 \\ \left(\partial_z^2 - q_{nf}^2\right)\tilde{g}_{nxz} = 0, 0 < z < z_1 \\ \phantom{xxx}\left(\partial_z^2 - q_{ns}^2\right)\tilde{g}_{nxz} = 0, z < 0 \end{cases} \quad (103)$$

Furthermore, the solutions of Eq. (103) are:

$$\tilde{g}_{nxz} = \begin{cases} S_{nxz,a}\,\mathrm{sgn}(z-z')e^{-q_{na}|z-z'|} + Ae^{-q_{na}z}, z > z_1 \\ Be^{q_{nf}z} + Ce^{-q_{nf}z}, 0 < z < z_1 \\ \phantom{xxx}De^{q_{ns}z}, z < 0 \end{cases}, \quad (104)$$

where $S_{nxz,a} \equiv -\dfrac{ik_n}{2k_a^2}$. In order to match the boundary conditions of $\tilde{g}_{nxz}$ as discuss in Eq. (84), we have the following simulated equations:

$$\begin{cases} -S_{nxz,a}e^{-q_{na}z'}e^{q_{na}z_1} + Ae^{-q_{na}z_1} = Be^{q_{nf}z_1} + Ce^{-q_{nf}z_1} \\ \phantom{xxxxxxxxxx} B + C = D \end{cases}, \quad (105)$$

and

$$\begin{cases} -S_{nxz,a}\dfrac{\varepsilon_a}{q_{na}}e^{-q_{na}z'}e^{q_{na}z_1} - \dfrac{\varepsilon_a}{q_{na}}Ae^{-q_{na}z_1} = \dfrac{\varepsilon_f}{q_{nf}}\left(Be^{q_{na}z_1} - Ce^{-q_{na}z_1}\right) \\ \phantom{xxxxxxxxx}\dfrac{\varepsilon_f}{q_{nf}}(B - C) = \dfrac{\varepsilon_s}{q_{ns}}D \end{cases}. \quad (106)$$

Furthermore, we obtain:

$$-S_{nxz,a}\dfrac{\varepsilon_a}{q_{na}}e^{-q_{na}z'}e^{q_{na}z_1} - \dfrac{\varepsilon_a}{q_{na}}\left(S_{nxz,a}e^{-q_{na}z'}e^{q_{na}z_1} + Be^{q_{nf}z_1} + Ce^{-q_{nf}z_1}\right) = \dfrac{\varepsilon_f}{q_{nf}}\left(Be^{q_{na}z_1} - Ce^{-q_{na}z_1}\right)$$

$$\Rightarrow -\left(\dfrac{\varepsilon_a}{q_{na}} + \dfrac{\varepsilon_f}{q_{nf}}\right)B = \left(\dfrac{\varepsilon_a}{q_{na}} - \dfrac{\varepsilon_f}{q_{nf}}\right)Ce^{-q_{na}z_1}e^{-q_{nf}z_1} + S_{nxz,a}\dfrac{2\varepsilon_a}{q_{na}}e^{-q_{na}z'}e^{q_{na}z_1}e^{-q_{nf}z_1}$$

$$\Rightarrow B = \dfrac{\varepsilon_f q_{na} - \varepsilon_a q_{nf}}{\varepsilon_a q_{nf} + \varepsilon_f q_{na}}Ce^{-q_{na}z_1}e^{-q_{nf}z_1} - \dfrac{2\varepsilon_a q_{nf}}{\varepsilon_a q_{nf} + \varepsilon_f q_{na}}S_{nxz,a}e^{-q_{na}z'}e^{q_{na}z_1}e^{-q_{nf}z_1}$$

$$= R_{n//}^{fa}e^{-q_{na}z_1}e^{-q_{nf}z_1}C - T_{n//}^{af}S_{nxz,a}e^{-q_{na}z'}e^{q_{na}z_1}e^{-q_{nf}z_1}$$

$$\Rightarrow A = Be^{q_{nf}z_1}e^{q_{na}z_1} + Ce^{-q_{nf}z_1}e^{q_{na}z_1} + S_{nxz,a}e^{-q_{na}z'}e^{2q_{na}z_1}$$

$$= R_{n//}^{fa}C - T_{n//}^{af}S_{nxz,a}e^{-q_{na}z'}e^{2q_{na}z_1} + Ce^{-q_{nf}z_1}e^{q_{na}z_1} + S_{nxz,a}e^{-q_{na}z'}e^{2q_{na}z_1}$$

$$= \left(R_{n//}^{fa} + e^{-q_{nf}z_1}e^{q_{na}z_1}\right)C + S_{nxz,a}e^{-q_{na}z'}e^{2q_{na}z_1}R_{n//}^{fa}$$

(107)

and

$$D = B + C = \left( R_{n//}^{fa} e^{-q_{na}z_1} e^{-q_{nf}z_1} + 1 \right) C - T_{n//}^{af} S_{nxz,a} e^{-q_{na}z'} e^{q_{na}z_1} e^{-q_{nf}z_1} . \tag{108}$$

Furthermore, we get the following form to solve the coefficient $C$:

$$\frac{\varepsilon_f}{q_{nf}} \left( R_{n//}^{fa} e^{-q_{na}z_1} e^{-q_{nf}z_1} C - C - T_{n//}^{af} S_{nxz,a} e^{-q_{na}z'} e^{q_{na}z_1} e^{-q_{nf}z_1} \right)$$

$$= \frac{\varepsilon_s}{q_{ns}} \left[ R_{n//}^{fa} e^{-q_{na}z_1} e^{-q_{nf}z_1} C + C - T_{n//}^{af} S_{nxz,a} e^{-q_{na}z'} e^{q_{na}z_1} e^{-q_{nf}z_1} \right]$$

$$\Rightarrow \left[ \left( \frac{\varepsilon_f}{q_{nf}} - \frac{\varepsilon_s}{q_{ns}} \right) R_{n//}^{fa} e^{-q_{na}z_1} e^{-q_{nf}z_1} - \left( \frac{\varepsilon_f}{q_{nf}} + \frac{\varepsilon_s}{q_{ns}} \right) \right] C = \left( \frac{\varepsilon_f}{q_{nf}} - \frac{\varepsilon_s}{q_{ns}} \right) T_{n//}^{af} S_{nxz,a} e^{-q_{na}z'} e^{q_{na}z_1} e^{-q_{nf}z_1} .$$

$$\Rightarrow C = \frac{\left( \frac{\varepsilon_f}{q_{nf}} - \frac{\varepsilon_s}{q_{ns}} \right) T_{n//}^{af} S_{nxz,a} e^{-q_{na}z'} e^{q_{na}z_1} e^{-q_{nf}z_1}}{\left( \frac{\varepsilon_f}{q_{nf}} - \frac{\varepsilon_s}{q_{ns}} \right) R_{n//}^{fa} e^{-q_{na}z_1} e^{-q_{nf}z_1} - \left( \frac{\varepsilon_f}{q_{nf}} + \frac{\varepsilon_s}{q_{ns}} \right)} = \frac{-R_{n//}^{fs} T_{n//}^{af} S_{nxz,a} e^{-q_{na}z'} e^{q_{na}z_1} e^{-q_{nf}z_1}}{1 - R_{n//}^{fs} R_{n//}^{fa} e^{-q_{na}z_1} e^{-q_{nf}z_1}}$$

$$\tag{109}$$

Hence we can get the other three coefficients as the following forms:

$$B = R_{n//}^{fa} e^{-q_{na}z_1} e^{-q_{nf}z_1} C - T_{n//}^{af} S_{nxz,a} e^{-q_{na}z'} e^{q_{na}z_1} e^{-q_{nf}z_1}$$

$$= \frac{-R_{n//}^{fa} R_{n//}^{fs} T_{n//}^{af} S_{nxz,a} e^{-q_{na}z'} e^{-2q_{nf}z_1}}{1 - R_{n//}^{fs} R_{n//}^{fa} e^{-q_{na}z_1} e^{-q_{nf}z_1}} - T_{n//}^{af} S_{nxz,a} e^{-q_{na}z'} e^{q_{na}z_1} e^{-q_{nf}z_1} = \frac{-T_{n//}^{af} S_{nxz,a} e^{-q_{na}z'} e^{q_{na}z_1} e^{-q_{nf}z_1}}{1 - R_{n//}^{fs} R_{n//}^{fa} e^{-q_{na}z_1} e^{-q_{nf}z_1}} ,$$

$$\tag{110}$$

$$A = \left( R_{n//}^{fa} + e^{-q_{nf}z_1} e^{q_{na}z_1} \right) C + S_{nxz,a} e^{-q_{na}z'} e^{2q_{na}z_1} R_{n//}^{fa}$$

$$= \frac{-R_{n//}^{fs} T_{n//}^{af} S_{nxz,a} e^{-q_{na}z'} e^{q_{na}z_1} e^{-q_{nf}z_1}}{1 - R_{n//}^{fs} R_{n//}^{fa} e^{-q_{na}z_1} e^{-q_{nf}z_1}} \left( R_{n//}^{fa} + e^{-q_{nf}z_1} e^{q_{na}z_1} \right) + S_{nxz,a} e^{-q_{na}z'} e^{2q_{na}z_1} R_{n//}^{fa} , \tag{111}$$

$$= S_{nxz,a} \frac{-R_{n//}^{fs} e^{-q_{na}z'} e^{q_{na}z_1} e^{-q_{nf}z_1} R_{n//}^{fa} - R_{n//}^{fs} T_{n//}^{af} e^{-q_{na}z'} e^{2q_{na}z_1} e^{-2q_{nf}z_1} + e^{-q_{na}z'} e^{2q_{na}z_1} R_{n//}^{fa}}{1 - R_{n//}^{fs} R_{n//}^{fa} e^{-q_{na}z_1} e^{-q_{nf}z_1}}$$

$$D = \left( R_{n//}^{fa} e^{-q_{na}z_1} e^{-q_{nf}z_1} + 1 \right) C - T_{n//}^{af} S_{nxz,a} e^{-q_{na}z'} e^{q_{na}z_1} e^{-q_{nf}z_1}$$

$$= \frac{-R_{n//}^{fs} T_{n//}^{af} S_{nxz,a} e^{-q_{na}z'} e^{q_{na}z_1} e^{-q_{nf}z_1}}{1 - R_{n//}^{fs} R_{n//}^{fa} e^{-q_{na}z_1} e^{-q_{nf}z_1}} \left( R_{n//}^{fa} e^{-q_{na}z_1} e^{-q_{nf}z_1} + 1 \right) - T_{n//}^{af} S_{nxz,a} e^{-q_{na}z'} e^{q_{na}z_1} e^{-q_{nf}z_1} . \tag{112}$$

$$= S_{nxz,a} \frac{-T_{n//}^{fs} T_{n//}^{af} e^{-q_{na}z'} e^{q_{na}z_1} e^{-q_{nf}z_1}}{1 - R_{n//}^{fs} R_{n//}^{fa} e^{-q_{na}z_1} e^{-q_{nf}z_1}}$$

Hence we solve the coefficients in Eq. (104) completely. Next we discuss the component $\tilde{g}_{nyy}$, starting from Eq. (12) or Eq. (45), we have:

$$\left( \partial_z^2 - q_n^2 \right) \tilde{g}_{nyy} = -\delta(z - z'). \tag{113}$$

Next we will consider the relative position between the field ($z$) and the source ($z'$). First we consider $z' < 0$, the corresponding different equation in regions $z > z_1$, $0 < z < z_1$ and $z < 0$ are:

$$\begin{cases} \left(\partial_z^2 - q_{na}^2\right)\tilde{g}_{nyy} = 0, z > z_1 \\ \left(\partial_z^2 - q_{nf}^2\right)\tilde{g}_{nyy} = 0, 0 < z < z_1 \\ \left(\partial_z^2 - q_{ns}^2\right)\tilde{g}_{nyy} = -\delta(z - z'), z < 0 \end{cases} \tag{114}$$

Furthermore, the solutions of Eq. (114) are:

$$\tilde{g}_{nyy} = \begin{cases} Ae^{-q_{na}z}, z > z_1 \\ Be^{q_{nf}z} + Ce^{-q_{nf}z}, 0 < z < z_1 \\ S_{nyy,s}e^{-q_{ns}|z-z'|} + De^{q_{ns}z}, z < 0 \end{cases} \tag{115}$$

where $S_{nyy,s} \equiv \dfrac{1}{2q_{ns}}$. In order to match the boundary conditions of $\tilde{g}_{nyy}$ as discuss in Eq. (27):

$$\begin{cases} \tilde{g}_{nyy}\big|_{z_{1+}} = \tilde{g}_{nyy}\big|_{z_{1-}}, \partial_z \tilde{g}_{nyy}\big|_{z_{1+}} = \partial_z \tilde{g}_{nyy}\big|_{z_{1-}} \\ \tilde{g}_{nyy}\big|_{0+} = \tilde{g}_{nyy}\big|_{0-}, \partial_z \tilde{g}_{nyy}\big|_{0+} = \partial_z \tilde{g}_{nyy}\big|_{0-} \end{cases}, \tag{116}$$

we have the following simulated equations:

$$\begin{cases} Ae^{-q_{na}z_1} = Be^{q_{nf}z_1} + Ce^{-q_{nf}z_1} \\ B + C = S_{nyy,s}e^{q_{ns}z'} + D \end{cases}, \tag{117}$$

and

$$\begin{cases} -q_{na}Ae^{-q_{na}z_1} = q_{nf}Be^{q_{nf}z_1} - q_{nf}Ce^{-q_{nf}z_1} \\ q_{nf}B - q_{nf}C = -S_{nyy,s}e^{q_{ns}z'}q_{ns} + q_{ns}D \end{cases}. \tag{118}$$

Furthermore, we obtain:

$$-q_{na}\left(Be^{q_{nf}z_1} + Ce^{-q_{nf}z_1}\right) = q_{nf}Be^{q_{nf}z_1} - q_{nf}Ce^{-q_{nf}z_1}$$
$$\Rightarrow (q_{na} + q_{nf})B = (q_{nf} - q_{na})Ce^{-2q_{nf}z_1}$$
$$\Rightarrow B = \frac{q_{nf} - q_{na}}{q_{na} + q_{nf}}Ce^{-2q_{nf}z_1} = R_{n\perp}^{fa}e^{-2q_{nf}z_1}C \tag{119}$$
$$\Rightarrow A = Be^{q_{nf}z_1}e^{q_{na}z_1} + Ce^{-q_{nf}z_1}e^{q_{na}z_1} = R_{n\perp}^{fa}Ce^{-q_{nf}z_1}e^{q_{na}z_1} + Ce^{-q_{nf}z_1}e^{q_{na}z_1}$$
$$= \left(R_{n\perp}^{fa} + 1\right)e^{-q_{nf}z_1}e^{q_{na}z_1}C = T_{n\perp}^{fa}e^{-q_{nf}z_1}e^{q_{na}z_1}C$$

and

$$R_{n\perp}^{fa} e^{-2q_{nf} z_1} C + C = S_{nyy,s} e^{q_{ns} z'} + D$$
$$\Rightarrow D = \left( R_{n\perp}^{fa} e^{-2q_{nf} z_1} + 1 \right) C - S_{nyy,s} e^{q_{ns} z'} \quad . \tag{120}$$

Furthermore, we get the following form to solve the coefficient $C$:

$$q_{nf} R_{n\perp}^{fa} e^{-2q_{nf} z_1} C - q_{nf} C = -S_{nyy,s} e^{q_{ns} z'} q_{ns} + q_{ns} \left[ \left( R_{n\perp}^{fa} e^{-2q_{nf} z_1} + 1 \right) C - S_{nyy,s} e^{q_{ns} z'} \right]$$

$$\Rightarrow \left[ (q_{nf} - q_{ns}) R_{n\perp}^{fa} e^{-2q_{nf} z_1} - (q_{nf} + q_{ns}) \right] C = -2 S_{nyy,s} e^{q_{ns} z'} q_{ns}$$

$$\Rightarrow C = \frac{-2 q_{ns} S_{nyy,s} e^{q_{ns} z'}}{(q_{nf} - q_{ns}) R_{n\perp}^{fa} e^{-2q_{nf} z_1} - (q_{nf} + q_{ns})} = \frac{-\dfrac{2 q_{ns}}{q_{nf} + q_{ns}} S_{nyy,s} e^{q_{ns} z'}}{\dfrac{q_{nf} - q_{ns}}{q_{nf} + q_{ns}} R_{n\perp}^{fa} e^{-2q_{nf} z_1} - 1} \tag{121}$$

$$= \frac{T_{n\perp}^{sf} S_{nyy,s} e^{q_{ns} z'}}{1 - R_{n\perp}^{fs} R_{n\perp}^{fa} e^{-2q_{nf} z_1}}$$

Hence we can get the other three coefficients as the following forms:

$$B = R_{n\perp}^{fa} e^{-2q_{nf} z_1} C = \frac{R_{n\perp}^{fa} T_{n\perp}^{sf} S_{nyy,s} e^{q_{ns} z'} e^{-2q_{nf} z_1}}{1 - R_{n\perp}^{fs} R_{n\perp}^{fa} e^{-2q_{nf} z_1}}, \tag{122}$$

$$A = T_{n\perp}^{fa} e^{-q_{nf} z_1} e^{q_{na} z_1} C = \frac{T_{n\perp}^{fa} T_{n\perp}^{sf} S_{nyy,s} e^{q_{ns} z'} e^{-q_{nf} z_1} e^{q_{na} z_1}}{1 - R_{n\perp}^{fs} R_{n\perp}^{fa} e^{-2q_{nf} z_1}}, \tag{123}$$

$$D = \left( R_{n\perp}^{fa} e^{-2q_{nf} z_1} + 1 \right) C - S_{nyy,s} e^{q_{ns} z'} = \frac{T_{n\perp}^{sf} S_{nyy,s} e^{q_{ns} z'} \left( R_{n\perp}^{fa} e^{-2q_{nf} z_1} + 1 \right)}{1 - R_{n\perp}^{fs} R_{n\perp}^{fa} e^{-2q_{nf} z_1}} - S_{nyy,s} e^{q_{ns} z'}$$

$$= \frac{T_{n\perp}^{sf} R_{n\perp}^{fa} e^{-2q_{nf} z_1} + T_{n\perp}^{sf} - 1 + R_{n\perp}^{fs} R_{n\perp}^{fa} e^{-2q_{nf} z_1}}{1 - R_{n\perp}^{fs} R_{n\perp}^{fa} e^{-2q_{nf} z_1}} S_{nyy,s} e^{q_{ns} z'} = \frac{R_{n\perp}^{fa} e^{-2q_{nf} z_1} + T_{n\perp}^{sf} - 1}{1 - R_{n\perp}^{fs} R_{n\perp}^{fa} e^{-2q_{nf} z_1}} S_{nyy,s} e^{q_{ns} z'} \tag{124}$$

Hence we solve the coefficients in Eq. (115) completely. Next we consider $0 < z' < z_1$, the corresponding different equation in regions $z > z_1$, $0 < z < z_1$ and $z < 0$ are:

$$\begin{cases} \left( \partial_z^2 - q_{na}^2 \right) \tilde{g}_{nyy} = 0, z > z_1 \\ \left( \partial_z^2 - q_{nf}^2 \right) \tilde{g}_{nyy} = -\delta(z - z'), 0 < z < z_1 \\ \left( \partial_z^2 - q_{ns}^2 \right) \tilde{g}_{nyy} = 0, z < 0 \end{cases} \tag{125}$$

Furthermore, the solutions of Eq. (125) are:

$$\tilde{g}_{nyy} = \begin{cases} A e^{-q_{na} z}, z > z_1 \\ S_{nyy,f} e^{-q_{nf} |z - z'|} + B e^{q_{nf} z} + C e^{-q_{nf} z}, 0 < z < z_1 \\ D e^{q_{ns} z}, z < 0 \end{cases} \tag{126}$$

where $S_{nyy,f} \equiv \dfrac{1}{2 q_{nf}}$. In order to match the boundary conditions of $\tilde{g}_{nyy}$ as discuss in

Eq. (116), we have the following simulated equations:

$$\begin{cases} Ae^{-q_{na}z_1} = S_{nyy,f}e^{q_{nf}z'}e^{-q_{nf}z_1} + Be^{q_{nf}z_1} + Ce^{-q_{nf}z_1} \\ S_{nyy,f}e^{-q_{nf}z'} + B + C = D \end{cases}, \quad (127)$$

and

$$\begin{cases} -q_{na}Ae^{-q_{na}z_1} = -S_{nyy,f}e^{q_{nf}z'}e^{-q_{nf}z_1}q_{nf} + q_{nf}Be^{q_{nf}z_1} - q_{nf}Ce^{-q_{nf}z_1} \\ S_{nyy,f}e^{-q_{nf}z'}q_{nf} + q_{nf}B - q_{nf}C = q_{ns}D \end{cases}. \quad (128)$$

Furthermore, we obtain:

$$-q_{na}\left(S_{nyy,f}e^{q_{nf}z'}e^{-q_{nf}z_1} + Be^{q_{nf}z_1} + Ce^{-q_{nf}z_1}\right) = -S_{nyy,f}e^{q_{nf}z'}e^{-q_{nf}z_1}q_{nf} + q_{nf}Be^{q_{nf}z_1} - q_{nf}Ce^{-q_{nf}z_1}$$

$$\Rightarrow (q_{na} + q_{nf})B = (q_{nf} - q_{na})Ce^{-2q_{nf}z_1} + (q_{nf} - q_{na})S_{nyy,f}e^{q_{nf}z'}e^{-2q_{nf}z_1}$$

$$\Rightarrow B = \frac{q_{nf} - q_{na}}{q_{na} + q_{nf}}Ce^{-2q_{nf}z_1} + \frac{q_{nf} - q_{na}}{q_{na} + q_{nf}}S_{nyy,f}e^{q_{nf}z'}e^{-2q_{nf}z_1}$$

$$= R_{n\perp}^{fa}Ce^{-2q_{nf}z_1} + R_{n\perp}^{fa}S_{nyy,f}e^{q_{nf}z'}e^{-2q_{nf}z_1}$$

$$\Rightarrow A = S_{nyy,f}e^{q_{nf}z'}e^{-q_{nf}z_1}e^{q_{na}z_1} + Be^{q_{nf}z_1}e^{q_{na}z_1} + Ce^{-q_{nf}z_1}e^{q_{na}z_1}$$

$$= S_{nyy,f}e^{q_{nf}z'}e^{-q_{nf}z_1}e^{q_{na}z_1} + \left(R_{n\perp}^{fa}Ce^{-2q_{nf}z_1} + R_{n\perp}^{fa}S_{nyy,f}e^{q_{nf}z'}e^{-2q_{nf}z_1}\right)e^{q_{nf}z_1}e^{q_{na}z_1} + Ce^{-q_{nf}z_1}e^{q_{na}z_1}$$

$$= T_{n\perp}^{fa}Ce^{-q_{nf}z_1}e^{q_{na}z_1} + T_{n\perp}^{fa}S_{nyy,f}e^{q_{nf}z'}e^{-q_{nf}z_1}e^{q_{na}z_1}$$

, (129)

and

$$D = S_{nyy,f}e^{-q_{nf}z'} + R_{n\perp}^{fa}Ce^{-2q_{nf}z_1} + R_{n\perp}^{fa}S_{nyy,f}e^{q_{nf}z'}e^{-2q_{nf}z_1} + C$$

$$= \left(R_{n\perp}^{fa}e^{-2q_{nf}z_1} + 1\right)C + S_{nyy,f}e^{-q_{nf}z'} + R_{n\perp}^{fa}S_{nyy,f}e^{q_{nf}z'}e^{-2q_{nf}z_1}$$

. (130)

Furthermore, we get the following form to solve the coefficient $C$:

$$S_{nyy,f}e^{-q_{nf}z'}q_{nf} + q_{nf}\left(R_{n\perp}^{fa}Ce^{-2q_{nf}z_1} + R_{n\perp}^{fa}S_{nyy,f}e^{q_{nf}z'}e^{-2q_{nf}z_1}\right) - q_{nf}C$$

$$= q_{ns}\left[\left(R_{n\perp}^{fa}e^{-2q_{nf}z_1} + 1\right)C + S_{nyy,f}e^{-q_{nf}z'} + R_{n\perp}^{fa}S_{nyy,f}e^{q_{nf}z'}e^{-2q_{nf}z_1}\right]$$

$$\Rightarrow \left[(q_{nf} - q_{ns})R_{n\perp}^{fa}e^{-2q_{nf}z_1} - (q_{nf} + q_{ns})\right]C = -S_{nyy,f}\left[(q_{nf} - q_{ns})e^{-q_{nf}z'} + (q_{nf} - q_{ns})R_{n\perp}^{fa}e^{q_{nf}z'}e^{-2q_{nf}z_1}\right]$$

$$\Rightarrow C = \frac{-(q_{nf} - q_{ns})e^{-q_{nf}z'} - (q_{nf} - q_{ns})R_{n\perp}^{fa}e^{q_{nf}z'}e^{-2q_{nf}z_1}}{(q_{nf} - q_{ns})R_{n\perp}^{fa}e^{-2q_{nf}z_1} - (q_{nf} + q_{ns})}S_{nyy,f}$$

$$= R_{n\perp}^{fs}\frac{e^{-q_{nf}z'} + R_{n\perp}^{fa}e^{q_{nf}z'}e^{-2q_{nf}z_1}}{1 - R_{n\perp}^{fs}R_{n\perp}^{fa}e^{-2q_{nf}z_1}}S_{nyy,f}$$

. (131)

Hence we can get the other three coefficients as the following forms:

$$B = R_{n\perp}^{fa}R_{n\perp}^{fs}\frac{e^{-q_{nf}z'}+R_{n\perp}^{fa}e^{q_{nf}z'}e^{-2q_{nf}z_1}}{1-R_{n\perp}^{fs}R_{n\perp}^{fa}e^{-2q_{nf}z_1}}S_{nyy,f}e^{-2q_{nf}z_1}+R_{n\perp}^{fa}e^{q_{nf}z'}S_{nyy,f}e^{-2q_{nf}z_1}$$
$$= S_{nyy,f}e^{-2q_{nf}z_1}R_{n\perp}^{fa}\frac{R_{n\perp}^{fs}e^{-q_{nf}z'}+e^{q_{nf}z'}}{1-R_{n\perp}^{fs}R_{n\perp}^{fa}e^{-2q_{nf}z_1}}$$
(132)

$$A = T_{n\perp}^{fa}R_{n\perp}^{fs}\frac{e^{-q_{nf}z'}+R_{n\perp}^{fa}e^{q_{nf}z'}e^{-2q_{nf}z_1}}{1-R_{n\perp}^{fs}R_{n\perp}^{fa}e^{-2q_{nf}z_1}}S_{nyy,f}e^{-q_{nf}z_1}e^{q_{na}z_1}+T_{n\perp}^{fa}e^{q_{nf}z'}S_{nyy,f}e^{-q_{nf}z_1}e^{q_{na}z_1}$$
$$= S_{nyy,f}e^{-q_{nf}z_1}e^{q_{na}z_1}T_{n\perp}^{fa}\frac{R_{n\perp}^{fs}e^{-q_{nf}z'}+e^{q_{nf}z'}}{1-R_{n\perp}^{fs}R_{n\perp}^{fa}e^{-2q_{nf}z_1}}$$
(133)

$$D = \left(R_{n\perp}^{fa}e^{-2q_{nf}z_1}+1\right)R_{n\perp}^{fs}\frac{e^{-q_{nf}z'}+R_{n\perp}^{fa}e^{q_{nf}z'}e^{-2q_{nf}z_1}}{1-R_{n\perp}^{fs}R_{n\perp}^{fa}e^{-2q_{nf}z_1}}S_{nyy,f}+S_{nyy,f}e^{-q_{nf}z'}+S_{nyy,f}R_{n\perp}^{fa}e^{q_{nf}z'}e^{-2q_{nf}z_1}$$
$$= \frac{R_{n\perp}^{fa}R_{n\perp}^{fs}e^{-2q_{nf}z_1}+T_{n\perp}^{fs}e^{-q_{nf}z'}+T_{n\perp}^{fs}R_{n\perp}^{fa}e^{q_{nf}z'}e^{-2q_{nf}z_1}-R_{n\perp}^{fs}R_{n\perp}^{fa}e^{-q_{nf}z'}e^{-2q_{nf}z_1}}{1-R_{n\perp}^{fs}R_{n\perp}^{fa}e^{-2q_{nf}z_1}}$$
(134)

Hence we solve the coefficients in Eq. (126) completely. Next we consider $z' > z_1$, the corresponding different equation in regions $z > z_1$, $0 < z < z_1$ and $z < 0$ are:

$$\begin{cases} \left(\partial_z^2 - q_{na}^2\right)\tilde{g}_{nyy} = -\delta(z-z'), z > z_1 \\ \left(\partial_z^2 - q_{nf}^2\right)\tilde{g}_{nyy} = 0, 0 < z < z_1 \\ \left(\partial_z^2 - q_{ns}^2\right)\tilde{g}_{nyy} = 0, z < 0 \end{cases}$$
(135)

Furthermore, the solutions of Eq. (135) are:

$$\tilde{g}_{nyy} = \begin{cases} S_{nyy,a}e^{-q_{na}|z-z'|}+Ae^{-q_{na}z}, z > z_1 \\ Be^{q_{nf}z}+Ce^{-q_{nf}z}, 0 < z < z_1 \\ De^{q_{ns}z}, z < 0 \end{cases}$$
(136)

where $S_{nyy,a} \equiv \frac{1}{2q_{na}}$. In order to match the boundary conditions of $\tilde{g}_{nyy}$ as discuss in Eq. (116), we have the following simulated equations:

$$\begin{cases} S_{nyy,a}e^{-q_{na}z'}e^{q_{na}z_1}+Ae^{-q_{na}z_1} = Be^{q_{nf}z_1}+Ce^{-q_{nf}z_1} \\ B+C = D \end{cases}$$
(137)

and

$$\begin{cases} S_{nyy,a}e^{-q_{na}z'}q_{na}e^{q_{na}z_1}-q_{na}Ae^{-q_{na}z_1} = q_{nf}Be^{q_{nf}z_1}-q_{nf}Ce^{-q_{nf}z_1} \\ q_{nf}B-q_{nf}C = q_{ns}D \end{cases}$$
(138)

Furthermore, we obtain:

$$S_{nyy,a}e^{-q_{na}z'}q_{na}e^{q_{na}z_1} - q_{na}\left(Be^{q_{nf}z_1} + Ce^{-q_{nf}z_1} - S_{nyy,a}e^{-q_{na}z'}e^{q_{na}z_1}\right) = q_{nf}Be^{q_{nf}z_1} - q_{nf}Ce^{-q_{nf}z_1}$$

$$\Rightarrow (q_{nf} + q_{na})B = (q_{nf} - q_{na})e^{-2q_{nf}z_1}C + 2q_{na}S_{nyy,a}e^{-q_{na}z'}e^{q_{na}z_1}e^{-q_{nf}z_1}$$

$$\Rightarrow B = \frac{q_{nf} - q_{na}}{q_{nf} + q_{na}}e^{-2q_{nf}z_1}C + \frac{2q_{na}}{q_{nf} + q_{na}}S_{nyy,a}e^{-q_{na}z'}e^{q_{na}z_1}e^{-q_{nf}z_1}$$

$$= R_{n\perp}^{fa}e^{-2q_{nf}z_1}C + T_{n\perp}^{af}S_{nyy,a}e^{-q_{na}z'}e^{q_{na}z_1}e^{-q_{nf}z_1}$$

$$\Rightarrow A = \left(R_{n\perp}^{fa}e^{-2q_{nf}z_1}C + T_{n\perp}^{af}S_{nyy,a}e^{-q_{na}z'}e^{q_{na}z_1}e^{-q_{nf}z_1}\right)e^{q_{nf}z_1}e^{q_{na}z_1} + Ce^{-q_{nf}z_1}e^{q_{na}z_1} - S_{nyy,a}e^{-q_{na}z'}e^{2q_{na}z_1}$$

$$= T_{n\perp}^{fa}e^{-q_{nf}z_1}e^{q_{na}z_1}C - R_{n\perp}^{fa}S_{nyy,a}e^{-q_{na}z'}e^{2q_{na}z_1}$$

(139)

and

$$D = B + C = \left(R_{n\perp}^{fa}e^{-2q_{nf}z_1} + 1\right)C + T_{n\perp}^{af}S_{nyy,a}e^{-q_{na}z'}e^{q_{na}z_1}e^{-q_{nf}z_1}.\tag{140}$$

Furthermore, we get the following form to solve the coefficient $C$:

$$q_{nf}\left(R_{n\perp}^{fa}e^{-2q_{nf}z_1}C + T_{n\perp}^{af}S_{nyy,a}e^{-q_{na}z'}e^{q_{na}z_1}e^{-q_{nf}z_1}\right) - q_{nf}C$$

$$= q_{ns}\left[\left(R_{n\perp}^{fa}e^{-2q_{nf}z_1} + 1\right)C + T_{n\perp}^{af}S_{nyy,a}e^{-q_{na}z'}e^{q_{na}z_1}e^{-q_{nf}z_1}\right]$$

$$\Rightarrow \left[(q_{nf} - q_{ns})R_{n\perp}^{fa}e^{-2q_{nf}z_1} - (q_{nf} + q_{ns})\right]C = -(q_{nf} - q_{ns})T_{n\perp}^{af}S_{nyy,a}e^{-q_{na}z'}e^{q_{na}z_1}e^{-q_{nf}z_1}.\tag{141}$$

$$\Rightarrow C = \frac{\dfrac{q_{nf} - q_{ns}}{q_{nf} + q_{ns}}T_{n\perp}^{af}S_{nyy,a}e^{-q_{na}z'}e^{q_{na}z_1}e^{-q_{nf}z_1}}{1 - \dfrac{q_{nf} - q_{ns}}{q_{nf} + q_{ns}}R_{n\perp}^{fa}e^{-2q_{nf}z_1}} = \frac{R_{n\perp}^{fs}T_{n\perp}^{af}S_{nyy,a}e^{-q_{na}z'}e^{q_{na}z_1}e^{-q_{nf}z_1}}{1 - R_{n\perp}^{fs}R_{n\perp}^{fa}e^{-2q_{nf}z_1}}$$

Hence we can get the other three coefficients as the following forms:

$$B = R_{n\perp}^{fa}e^{-2q_{nf}z_1}\frac{R_{n\perp}^{fs}T_{n\perp}^{af}S_{nyy,a}e^{-q_{na}z'}e^{q_{na}z_1}e^{-q_{nf}z_1}}{1 - R_{n\perp}^{fs}R_{n\perp}^{fa}e^{-2q_{nf}z_1}} + T_{n\perp}^{af}S_{nyy,a}e^{-q_{na}z'}e^{q_{na}z_1}e^{-q_{nf}z_1}$$

$$= \frac{T_{n\perp}^{af}S_{nyy,a}e^{-q_{na}z'}e^{q_{na}z_1}e^{-q_{nf}z_1}}{1 - R_{n\perp}^{fs}R_{n\perp}^{fa}e^{-2q_{nf}z_1}}$$

(142)

$$A = T_{n\perp}^{fa}e^{-q_{nf}z_1}e^{q_{na}z_1}\frac{R_{n\perp}^{fs}T_{n\perp}^{af}S_{nyy,a}e^{-q_{na}z'}e^{q_{na}z_1}e^{-q_{nf}z_1}}{1 - R_{n\perp}^{fs}R_{n\perp}^{fa}e^{-2q_{nf}z_1}} - R_{n\perp}^{fa}S_{nyy,a}e^{-q_{na}z'}e^{2q_{na}z_1}$$

$$= \frac{R_{n\perp}^{fs}\left(T_{n\perp}^{fa}T_{n\perp}^{af} + R_{n\perp}^{fa}R_{n\perp}^{fa}\right)S_{nyy,a}e^{-q_{na}z'}e^{2q_{na}z_1}e^{-2q_{nf}z_1} - R_{n\perp}^{fa}S_{nyy,a}e^{-q_{na}z'}e^{2q_{na}z_1}}{1 - R_{n\perp}^{fs}R_{n\perp}^{fa}e^{-2q_{nf}z_1}},$$

(143)

$$= \frac{R_{n\perp}^{fs}\left(T_{n\perp}^{fa} + R_{n\perp}^{af}\right)S_{nyy,a}e^{-q_{na}z'}e^{2q_{na}z_1}e^{-2q_{nf}z_1} - R_{n\perp}^{fa}S_{nyy,a}e^{-q_{na}z'}e^{2q_{na}z_1}}{1 - R_{n\perp}^{fs}R_{n\perp}^{fa}e^{-2q_{nf}z_1}}$$

$$D = \left(R_{n\perp}^{fa}e^{-2q_{nf}z_1} + 1\right)\frac{R_{n\perp}^{fs}T_{n\perp}^{af}S_{nyy,a}e^{-q_{na}z'}e^{q_{na}z_1}e^{-q_{nf}z_1}}{1 - R_{n\perp}^{fs}R_{n\perp}^{fa}e^{-2q_{nf}z_1}} + T_{n\perp}^{af}S_{nyy,a}e^{-q_{na}z'}e^{q_{na}z_1}e^{-q_{nf}z_1}$$

$$= \frac{T_{n\perp}^{af}T_{n\perp}^{fs}S_{nyy,a}e^{-q_{na}z'}e^{q_{na}z_1}e^{-q_{nf}z_1}}{1 - R_{n\perp}^{fs}R_{n\perp}^{fa}e^{-2q_{nf}z_1}}$$

(144)

Hence we solve the coefficients in Eq. (136) completely. Next we discuss the component $\tilde{g}_{nxx}$, starting from Eq. (13) or Eq. (59), we have:

$$\left(\partial_z^2 - q_n^2\right)\tilde{g}_{nxx} = \frac{q_n^2}{k^2}\delta(z - z'). \tag{145}$$

Next we will consider the relative position between the field ($z$) and the source ($z'$). First we consider $z' < 0$, the corresponding different equation in regions $z > z_1$, $0 < z < z_1$ and $z < 0$ are:

$$\begin{cases} \left(\partial_z^2 - q_{na}^2\right)\tilde{g}_{nxx} = 0, z > z_1 \\ \left(\partial_z^2 - q_{nf}^2\right)\tilde{g}_{nxx} = 0, 0 < z < z_1 \\ \left(\partial_z^2 - q_{ns}^2\right)\tilde{g}_{nxx} = \frac{q_{ns}^2}{k_s^2}\delta(z - z'), z < 0 \end{cases} \tag{146}$$

Furthermore, the solutions of Eq. (146) are:

$$\tilde{g}_{nxx} = \begin{cases} Ae^{-q_{na}z}, z > z_1 \\ Be^{q_{nf}z} + Ce^{-q_{nf}z}, 0 < z < z_1 \\ S_{nxx,s}e^{-q_{ns}|z-z'|} + De^{q_{ns}z}, z < 0 \end{cases} \tag{147}$$

where $S_{nxx,s} \equiv -\frac{q_{ns}}{2k_s^2}$. In order to match the boundary conditions of $\tilde{g}_{nxx}$ as discuss in Eq. (26):

$$\begin{cases} \tilde{g}_{nxx}\big|_{z_{1+}} = \tilde{g}_{nxx}\big|_{z_{1-}}, \frac{\varepsilon_a}{q_{na}^2}\partial_z\tilde{g}_{nxx}\big|_{z_{1+}} = \frac{\varepsilon_f}{q_{nf}^2}\partial_z\tilde{g}_{nxx}\big|_{z_{1-}} \\ \tilde{g}_{nxx}\big|_{0+} = \tilde{g}_{nxx}\big|_{0-}, \frac{\varepsilon_f}{q_{nf}^2}\partial_z\tilde{g}_{nxx}\big|_{0+} = \frac{\varepsilon_s}{q_{ns}^2}\partial_z\tilde{g}_{nxx}\big|_{0-} \end{cases} \tag{148}$$

we have the following simulated equations:

$$\begin{cases} Ae^{-q_{na}z_1} = Be^{q_{nf}z_1} + Ce^{-q_{nf}z_1} \\ B + C = S_{nxx,s}e^{q_{ns}z'} + D \end{cases} \tag{149}$$

and

$$\begin{cases} -\dfrac{\varepsilon_a}{q_{na}} A e^{-q_{na} z_1} = \dfrac{\varepsilon_f}{q_{nf}} B e^{q_{nf} z_1} - \dfrac{\varepsilon_f}{q_{nf}} C e^{-q_{nf} z_1} \\ \dfrac{\varepsilon_f}{q_{nf}} B - \dfrac{\varepsilon_f}{q_{nf}} C = -\dfrac{\varepsilon_s}{q_{ns}} S_{nxx,s} e^{q_{ns} z'} + \dfrac{\varepsilon_s}{q_{ns}} D \end{cases}. \tag{150}$$

Furthermore, we obtain:

$$-\dfrac{\varepsilon_a}{q_{na}} \left( B e^{q_{nf} z_1} + C e^{-q_{nf} z_1} \right) = \dfrac{\varepsilon_f}{q_{nf}} B e^{q_{nf} z_1} - \dfrac{\varepsilon_f}{q_{nf}} C e^{-q_{nf} z_1}$$

$$\Rightarrow \left( \dfrac{\varepsilon_f}{q_{nf}} + \dfrac{\varepsilon_a}{q_{na}} \right) B = \left( \dfrac{\varepsilon_f}{q_{nf}} - \dfrac{\varepsilon_a}{q_{na}} \right) C e^{-2 q_{nf} z_1}$$

$$\Rightarrow B = \dfrac{\varepsilon_f q_{na} - \varepsilon_a q_{nf}}{\varepsilon_f q_{na} + \varepsilon_a q_{nf}} C e^{-2 q_{nf} z_1} = R_{n//}^{fa} C e^{-2 q_{nf} z_1}$$

$$\Rightarrow A = R_{n//}^{fa} C e^{-q_{nf} z_1} e^{q_{na} z_1} + C e^{-q_{nf} z_1} e^{q_{na} z_1} = \left( R_{n//}^{fa} + 1 \right) C e^{-q_{nf} z_1} e^{q_{na} z_1} = T_{n//}^{fa} C e^{-q_{nf} z_1} e^{q_{na} z_1}$$

$$\tag{151}$$

and

$$D = \left( R_{n//}^{fa} e^{-2 q_{nf} z_1} + 1 \right) C - S_{nxx,s} e^{q_{ns} z'}. \tag{152}$$

Furthermore, we get the following form to solve the coefficient $C$:

$$\dfrac{\varepsilon_f}{q_{nf}} R_{n//}^{fa} C e^{-2 q_{nf} z_1} - \dfrac{\varepsilon_f}{q_{nf}} C = -\dfrac{\varepsilon_s}{q_{ns}} S_{nxx,s} e^{q_{ns} z'} + \dfrac{\varepsilon_s}{q_{ns}} \left[ \left( R_{n//}^{fa} e^{-2 q_{nf} z_1} + 1 \right) C - S_{nxx,s} e^{q_{ns} z'} \right]$$

$$\Rightarrow \left[ \left( \dfrac{\varepsilon_f}{q_{nf}} + \dfrac{\varepsilon_s}{q_{ns}} \right) - \left( \dfrac{\varepsilon_f}{q_{nf}} - \dfrac{\varepsilon_s}{q_{ns}} \right) R_{n//}^{fa} e^{-2 q_{nf} z_1} \right] C = \dfrac{2 \varepsilon_s}{q_{ns}} S_{nxx,s} e^{q_{ns} z'}$$

$$\tag{153}$$

$$\Rightarrow C = \dfrac{\dfrac{2 \varepsilon_s q_{nf}}{\varepsilon_f q_{ns} + \varepsilon_s q_{nf}} S_{nxx,s} e^{q_{ns} z'}}{1 - \left( \dfrac{\varepsilon_f q_{ns} - \varepsilon_s q_{nf}}{\varepsilon_f q_{ns} + \varepsilon_s q_{nf}} \right) R_{n//}^{fa} e^{-2 q_{nf} z_1}} = \dfrac{T_{n//}^{sf} S_{nxx,s} e^{q_{ns} z'}}{1 - R_{n//}^{fs} R_{n//}^{fa} e^{-2 q_{nf} z_1}}$$

Hence we can get the other three coefficients as the following forms:

$$B = \dfrac{R_{n//}^{fa} T_{n//}^{sf} S_{nxx,s} e^{q_{ns} z'} e^{-2 q_{nf} z_1}}{1 - R_{n//}^{fs} R_{n//}^{fa} e^{-2 q_{nf} z_1}}, \tag{154}$$

$$A = \dfrac{T_{n//}^{fa} T_{n//}^{sf} S_{nxx,s} e^{q_{ns} z'} e^{-q_{nf} z_1} e^{q_{na} z_1}}{1 - R_{n//}^{fs} R_{n//}^{fa} e^{-2 q_{nf} z_1}}, \tag{155}$$

$$D = \dfrac{\left( R_{n//}^{fa} e^{-2 q_{nf} z_1} + 1 \right) T_{n//}^{sf} S_{nxx,s} e^{q_{ns} z'}}{1 - R_{n//}^{fs} R_{n//}^{fa} e^{-2 q_{nf} z_1}} - S_{nxx,s} e^{q_{ns} z'} = \dfrac{R_{n//}^{fa} S_{nxx,s} e^{q_{ns} z'} e^{-2 q_{nf} z_1} - R_{n//}^{fs} S_{nxx,s} e^{q_{ns} z'}}{1 - R_{n//}^{fs} R_{n//}^{fa} e^{-2 q_{nf} z_1}}. \tag{156}$$

Hence we solve the coefficients in Eq. (147) completely. Next we consider $0 < z' < z_1$,

the corresponding different equation in regions $z > z_1$, $0 < z < z_1$ and $z < 0$ are:

$$\begin{cases} \left(\partial_z^2 - q_{na}^2\right)\tilde{g}_{nxx} = 0, z > z_1 \\ \left(\partial_z^2 - q_{nf}^2\right)\tilde{g}_{nxx} = \dfrac{q_{nf}^2}{k_f^2}\delta(z - z'), 0 < z < z_1 \\ \left(\partial_z^2 - q_{ns}^2\right)\tilde{g}_{nxx} = 0, z < 0 \end{cases} \quad (157)$$

Furthermore, the solutions of Eq. (157) are:

$$\tilde{g}_{nxx} = \begin{cases} Ae^{-q_{na}z}, z > z_1 \\ S_{nxx,f}e^{-q_{nf}|z-z'|} + Be^{q_{nf}z} + Ce^{-q_{nf}z}, 0 < z < z_1 \\ De^{q_{ns}z}, z < 0 \end{cases} \quad (158)$$

where $S_{nxx,f} \equiv -\dfrac{q_{nf}}{2k_f^2}$. In order to match the boundary conditions of $\tilde{g}_{nxx}$ as discuss in Eq. (148), we have the following simulated equations:

$$\begin{cases} Ae^{-q_{na}z_1} = S_{nxx,f}e^{q_{nf}z'}e^{-q_{nf}z_1} + Be^{q_{nf}z_1} + Ce^{-q_{nf}z_1} \\ S_{nxx,f}e^{-q_{nf}z'} + B + C = D \end{cases}, \quad (159)$$

and

$$\begin{cases} -\dfrac{\varepsilon_a}{q_{na}}Ae^{-q_{na}z_1} = -\dfrac{\varepsilon_f}{q_{nf}}S_{nxx,f}e^{q_{nf}z'}e^{-q_{nf}z_1} + \dfrac{\varepsilon_f}{q_{nf}}Be^{q_{nf}z_1} - \dfrac{\varepsilon_f}{q_{nf}}Ce^{-q_{nf}z_1} \\ \dfrac{\varepsilon_f}{q_{nf}}S_{nxx,f}e^{-q_{nf}z'} + \dfrac{\varepsilon_f}{q_{nf}}B - \dfrac{\varepsilon_f}{q_{nf}}C = \dfrac{\varepsilon_s}{q_{ns}}D \end{cases}. \quad (160)$$

Furthermore, we obtain:

$$-\dfrac{\varepsilon_a}{q_{na}}\left(S_{nxx,f}e^{q_{nf}z'}e^{-q_{nf}z_1} + Be^{q_{nf}z_1} + Ce^{-q_{nf}z_1}\right) = -\dfrac{\varepsilon_f}{q_{nf}}S_{nxx,f}e^{q_{nf}z'}e^{-q_{nf}z_1} + \dfrac{\varepsilon_f}{q_{nf}}Be^{q_{nf}z_1} - \dfrac{\varepsilon_f}{q_{nf}}Ce^{-q_{nf}z_1}$$

$$\Rightarrow \left(\dfrac{\varepsilon_f}{q_{nf}} + \dfrac{\varepsilon_a}{q_{na}}\right)B = \left(\dfrac{\varepsilon_f}{q_{nf}} - \dfrac{\varepsilon_a}{q_{na}}\right)Ce^{-2q_{nf}z_1} + \left(\dfrac{\varepsilon_f}{q_{nf}} - \dfrac{\varepsilon_a}{q_{na}}\right)S_{nxx,f}e^{q_{nf}z'}e^{-2q_{nf}z_1}$$

$$\Rightarrow B = \dfrac{\varepsilon_f q_{na} - \varepsilon_a q_{nf}}{\varepsilon_f q_{na} + \varepsilon_a q_{nf}}Ce^{-2q_{nf}z_1} + \dfrac{\varepsilon_f q_{na} - \varepsilon_a q_{nf}}{\varepsilon_f q_{na} + \varepsilon_a q_{nf}}S_{nxx,f}e^{q_{nf}z'}e^{-2q_{nf}z_1}$$

$$= R_{n//}^{fa}Ce^{-2q_{nf}z_1} + R_{n//}^{fa}S_{nxx,f}e^{q_{nf}z'}e^{-2q_{nf}z_1}$$

$$\Rightarrow A = S_{nxx,f}e^{q_{nf}z'}e^{-q_{nf}z_1}e^{q_{na}z_1} + \left(R_{n//}^{fa}Ce^{-2q_{nf}z_1} + R_{n//}^{fa}S_{nxx,f}e^{q_{nf}z'}e^{-2q_{nf}z_1}\right)e^{q_{nf}z_1}e^{q_{na}z_1} + Ce^{-q_{nf}z_1}e^{q_{na}z_1}$$

$$= T_{n//}^{fa}Ce^{-q_{nf}z_1}e^{q_{na}z_1} + T_{n//}^{fa}S_{nxx,f}e^{q_{nf}z'}e^{-q_{nf}z_1}e^{q_{na}z_1}$$

, (161)

and

$$D = \left(R_{n//}^{fa}e^{-2q_{nf}z_1} + 1\right)C + S_{nxx,f}e^{-q_{nf}z'} + R_{n//}^{fa}S_{nxx,f}e^{q_{nf}z'}e^{-2q_{nf}z_1}. \qquad (162)$$

Furthermore, we get the following form to solve the coefficient $C$:

$$\frac{\varepsilon_f}{q_{nf}}S_{nxx,f}e^{-q_{nf}z'} + \frac{\varepsilon_f}{q_{nf}}\left[R_{n//}^{fa}Ce^{-2q_{nf}z_1} + R_{n//}^{fa}S_{nxx,f}e^{q_{nf}z'}e^{-2q_{nf}z_1}\right] - \frac{\varepsilon_f}{q_{nf}}C$$

$$= \frac{\varepsilon_s}{q_{ns}}\left[\left(R_{n//}^{fa}e^{-2q_{nf}z_1} + 1\right)C + S_{nxx,f}e^{-q_{nf}z'} + R_{n//}^{fa}S_{nxx,f}e^{q_{nf}z'}e^{-2q_{nf}z_1}\right]$$

$$\Rightarrow \left[\left(\frac{\varepsilon_f}{q_{nf}} + \frac{\varepsilon_s}{q_{ns}}\right) - \left(\frac{\varepsilon_f}{q_{nf}} - \frac{\varepsilon_s}{q_{ns}}\right)R_{n//}^{fa}e^{-2q_{nf}z_1}\right]C \qquad (163)$$

$$= \left(\frac{\varepsilon_f}{q_{nf}} - \frac{\varepsilon_s}{q_{ns}}\right)S_{nxx,f}e^{-q_{nf}z'} + \left(\frac{\varepsilon_f}{q_{nf}} - \frac{\varepsilon_s}{q_{ns}}\right)R_{n//}^{fa}S_{nxx,f}e^{q_{nf}z'}e^{-2q_{nf}z_1}$$

$$\Rightarrow C = \frac{R_{n//}^{fs}S_{nxx,f}e^{-q_{nf}z'} + R_{n//}^{fs}R_{n//}^{fa}S_{nxx,f}e^{q_{nf}z'}e^{-2q_{nf}z_1}}{1 - R_{n//}^{fs}R_{n//}^{fa}e^{-2q_{nf}z_1}}$$

Hence we can get the other three coefficients as the following forms:

$$B = R_{n//}^{fa}\frac{R_{n//}^{fs}S_{nxx,f}e^{-q_{nf}z'} + R_{n//}^{fs}R_{n//}^{fa}S_{nxx,f}e^{q_{nf}z'}e^{-2q_{nf}z_1}}{1 - R_{n//}^{fs}R_{n//}^{fa}e^{-2q_{nf}z_1}}e^{-2q_{nf}z_1} + R_{n//}^{fa}S_{nxx,f}e^{q_{nf}z'}e^{-2q_{nf}z_1}$$

$$= \frac{R_{n//}^{fa}R_{n//}^{fs}S_{nxx,f}e^{-q_{nf}z'}e^{-2q_{nf}z_1} + R_{n//}^{fa}S_{nxx,f}e^{q_{nf}z'}e^{-2q_{nf}z_1}}{1 - R_{n//}^{fs}R_{n//}^{fa}e^{-2q_{nf}z_1}}, \qquad (164)$$

$$A = T_{n//}^{fa}\frac{R_{n//}^{fs}S_{nxx,f}e^{-q_{nf}z'} + R_{n//}^{fs}R_{n//}^{fa}S_{nxx,f}e^{q_{nf}z'}e^{-2q_{nf}z_1}}{1 - R_{n//}^{fs}R_{n//}^{fa}e^{-2q_{nf}z_1}}e^{-q_{nf}z_1}e^{q_{na}z_1} + T_{n//}^{fa}S_{nxx,f}e^{q_{nf}z'}e^{-q_{nf}z_1}e^{q_{na}z_1}$$

$$= \frac{T_{n//}^{fa}R_{n//}^{fs}S_{nxx,f}e^{-q_{nf}z'}e^{-q_{nf}z_1}e^{q_{na}z_1} + T_{n//}^{fa}S_{nxx,f}e^{q_{nf}z'}e^{-q_{nf}z_1}e^{q_{na}z_1}}{1 - R_{n//}^{fs}R_{n//}^{fa}e^{-2q_{nf}z_1}},$$

$$(165)$$

$$D = \left(R_{n//}^{fa}e^{-2q_{nf}z_1} + 1\right)\frac{R_{n//}^{fs}S_{nxx,f}e^{-q_{nf}z'} + R_{n//}^{fs}R_{n//}^{fa}S_{nxx,f}e^{q_{nf}z'}e^{-2q_{nf}z_1}}{1 - R_{n//}^{fs}R_{n//}^{fa}e^{-2q_{nf}z_1}} + S_{nxx,f}e^{-q_{nf}z'} + R_{n//}^{fa}S_{nxx,f}e^{q_{nf}z'}e^{-2q_{nf}z_1}$$

$$= S_{nxx,f}\frac{T_{n//}^{fs}e^{-q_{nf}z'} + T_{n//}^{fs}R_{n//}^{fa}e^{q_{nf}z'}e^{-2q_{nf}z_1}}{1 - R_{n//}^{fs}R_{n//}^{fa}e^{-2q_{nf}z_1}}$$

$$. \qquad (166)$$

Hence we solve the coefficients in Eq. (158) completely. Next we consider $z' > z_1$, the corresponding different equation in regions $z > z_1$, $0 < z < z_1$ and $z < 0$ are:

$$\begin{cases} \left(\partial_z^2 - q_{na}^2\right)\tilde{g}_{nxx} = \dfrac{q_{na}^2}{k_a^2}\delta(z-z'), z > z_1 \\ \left(\partial_z^2 - q_{nf}^2\right)\tilde{g}_{nxx} = 0, 0 < z < z_1 \\ \left(\partial_z^2 - q_{ns}^2\right)\tilde{g}_{nxx} = 0, z < 0 \end{cases} \qquad (167)$$

Then the solutions of Eq. (167) are:

$$\tilde{g}_{nxx} = \begin{cases} S_{nxx,a}e^{-q_{na}|z-z'|} + Ae^{-q_{na}z}, z > z_1 \\ Be^{q_{nf}z} + Ce^{-q_{nf}z}, 0 < z < z_1 \\ De^{q_{ns}z}, z < 0 \end{cases} \qquad (168)$$

where $S_{nxx,a} \equiv -\dfrac{q_{na}}{2k_a^2}$. In order to match the boundary conditions of $\tilde{g}_{nxx}$ as discuss in Eq. (148), we have the following simulated equations:

$$\begin{cases} S_{nxx,a}e^{-q_{na}z'}e^{q_{na}z_1} + Ae^{-q_{na}z_1} = Be^{q_{nf}z_1} + Ce^{-q_{nf}z_1} \\ B + C = D \end{cases} \qquad (169)$$

and

$$\begin{cases} \dfrac{\varepsilon_a}{q_{na}}S_{nxx,a}e^{-q_{na}z'}e^{q_{na}z_1} - \dfrac{\varepsilon_a}{q_{na}}Ae^{-q_{na}z_1} = \dfrac{\varepsilon_f}{q_{nf}}Be^{q_{nf}z_1} - \dfrac{\varepsilon_f}{q_{nf}}Ce^{-q_{nf}z_1} \\ \dfrac{\varepsilon_f}{q_{nf}}B - \dfrac{\varepsilon_f}{q_{nf}}C = \dfrac{\varepsilon_s}{q_{ns}}D \end{cases} \qquad (170)$$

Furthermore, we obtain:

$$\dfrac{\varepsilon_a}{q_{na}}S_{nxx,a}e^{-q_{na}z'}e^{q_{na}z_1} - \dfrac{\varepsilon_a}{q_{na}}\left(Be^{q_{nf}z_1} + Ce^{-q_{nf}z_1} - S_{nxx,a}e^{-q_{na}z'}e^{q_{na}z_1}\right) = \dfrac{\varepsilon_f}{q_{nf}}Be^{q_{nf}z_1} - \dfrac{\varepsilon_f}{q_{nf}}Ce^{-q_{nf}z_1}$$

$$\Rightarrow \left(\dfrac{\varepsilon_f}{q_{nf}} + \dfrac{\varepsilon_a}{q_{na}}\right)B = \left(\dfrac{\varepsilon_f}{q_{nf}} - \dfrac{\varepsilon_a}{q_{na}}\right)Ce^{-2q_{nf}z_1} + \dfrac{2\varepsilon_a}{q_{na}}S_{nxx,a}e^{-q_{na}z'}e^{q_{na}z_1}e^{-q_{nf}z_1}$$

$$\Rightarrow B = \dfrac{\varepsilon_f q_{na} - \varepsilon_a q_{nf}}{\varepsilon_f q_{na} + \varepsilon_a q_{nf}}Ce^{-2q_{nf}z_1} + \dfrac{2\varepsilon_a q_{nf}}{\varepsilon_f q_{na} + \varepsilon_a q_{nf}}S_{nxx,a}e^{-q_{na}z'}e^{q_{na}z_1}e^{-q_{nf}z_1},$$

$$= R_{n//}^{fa}e^{-2q_{nf}z_1}C + T_{n//}^{af}S_{nxx,a}e^{-q_{na}z'}e^{q_{na}z_1}e^{-q_{nf}z_1}$$

$$\Rightarrow A = \left(R_{n//}^{fa}e^{-2q_{nf}z_1}C + T_{n//}^{af}S_{nxx,a}e^{-q_{na}z'}e^{q_{na}z_1}e^{-q_{nf}z_1}\right)e^{q_{nf}z_1}e^{q_{na}z_1} + Ce^{-q_{nf}z_1}e^{q_{na}z_1} - S_{nxx,a}e^{-q_{na}z'}e^{2q_{na}z_1}$$

$$= T_{n//}^{fa}e^{-q_{nf}z_1}e^{q_{na}z_1}C - R_{n//}^{fa}S_{nxx,a}e^{-q_{na}z'}e^{2q_{na}z_1}$$

$$(171)$$

and

$$D = \left(R_{n//}^{fa}e^{-2q_{nf}z_1} + 1\right)C + T_{n//}^{af}S_{nxx,a}e^{-q_{na}z'}e^{q_{na}z_1}e^{-q_{nf}z_1}. \qquad (172)$$

Furthermore, we get the following form to solve the coefficient $C$:

$$\frac{\varepsilon_f}{q_{nf}}\left[R_{n//}^{fa}e^{-2q_{nf}z_1}C + T_{n//}^{af}S_{nxx,a}e^{-q_{na}z'}e^{q_{na}z_1}e^{-q_{nf}z_1}\right] - \frac{\varepsilon_f}{q_{nf}}C$$

$$= \frac{\varepsilon_s}{q_{ns}}\left[\left(R_{n//}^{fa}e^{-2q_{nf}z_1} + 1\right)C + T_{n//}^{af}S_{nxx,a}e^{-q_{na}z'}e^{q_{na}z_1}e^{-q_{nf}z_1}\right]$$

$$\Rightarrow \left[\left(\frac{\varepsilon_f}{q_{nf}} - \frac{\varepsilon_s}{q_{ns}}\right)R_{n//}^{fa}e^{-2q_{nf}z_1} - \left(\frac{\varepsilon_f}{q_{nf}} + \frac{\varepsilon_s}{q_{ns}}\right)\right]C = -\left(\frac{\varepsilon_f}{q_{nf}} - \frac{\varepsilon_s}{q_{ns}}\right)T_{n//}^{af}S_{nxx,a}e^{-q_{na}z'}e^{q_{na}z_1}e^{-q_{nf}z_1}.$$

$$\Rightarrow C = \frac{\frac{\varepsilon_f q_{ns} - \varepsilon_s q_{nf}}{\varepsilon_f q_{ns} + \varepsilon_s q_{nf}}T_{n//}^{af}S_{nxx,a}e^{-q_{na}z'}e^{q_{na}z_1}e^{-q_{nf}z_1}}{1 - \frac{\varepsilon_f q_{ns} - \varepsilon_s q_{nf}}{\varepsilon_f q_{ns} + \varepsilon_s q_{nf}}R_{n//}^{fa}e^{-2q_{nf}z_1}} = \frac{R_{n//}^{fs}T_{n//}^{af}S_{nxx,a}e^{-q_{na}z'}e^{q_{na}z_1}e^{-q_{nf}z_1}}{1 - R_{n//}^{fs}R_{n//}^{fa}e^{-2q_{nf}z_1}}$$

(173)

Hence we can get the other three coefficients as the following forms:

$$B = R_{n//}^{fa}e^{-2q_{nf}z_1}\frac{R_{n//}^{fs}T_{n//}^{af}S_{nxx,a}e^{-q_{na}z'}e^{q_{na}z_1}e^{-q_{nf}z_1}}{1 - R_{n//}^{fs}R_{n//}^{fa}e^{-2q_{nf}z_1}} + T_{n//}^{af}S_{nxx,a}e^{-q_{na}z'}e^{q_{na}z_1}e^{-q_{nf}z_1}$$

$$= \frac{T_{n//}^{af}S_{nxx,a}e^{-q_{na}z'}e^{q_{na}z_1}e^{-q_{nf}z_1}}{1 - R_{n//}^{fs}R_{n//}^{fa}e^{-2q_{nf}z_1}}$$

(174)

$$A = T_{n//}^{fa}e^{-q_{nf}z_1}e^{q_{na}z_1}C - R_{n//}^{fa}S_{nxx,a}e^{-q_{na}z'}e^{2q_{na}z_1}$$

$$= T_{n//}^{fa}e^{-q_{nf}z_1}e^{q_{na}z_1}\frac{R_{n//}^{fs}T_{n//}^{af}S_{nxx,a}e^{-q_{na}z'}e^{q_{na}z_1}e^{-q_{nf}z_1}}{1 - R_{n//}^{fs}R_{n//}^{fa}e^{-2q_{nf}z_1}} - R_{n//}^{fa}S_{nxx,a}e^{-q_{na}z'}e^{2q_{na}z_1},$$

(175)

$$= \frac{R_{n//}^{fs}S_{nxx,a}e^{-2q_{nf}z_1}e^{2q_{na}z_1}e^{-q_{na}z'} - R_{n//}^{fa}S_{nxx,a}e^{-q_{na}z'}e^{2q_{na}z_1}}{1 - R_{n//}^{fs}R_{n//}^{fa}e^{-2q_{nf}z_1}}$$

$$D = \left(R_{n//}^{fa}e^{-2q_{nf}z_1} + 1\right)\frac{R_{n//}^{fs}T_{n//}^{af}S_{nxx,a}e^{-q_{na}z'}e^{q_{na}z_1}e^{-q_{nf}z_1}}{1 - R_{n//}^{fs}R_{n//}^{fa}e^{-2q_{nf}z_1}} + T_{n//}^{af}S_{nxx,a}e^{-q_{na}z'}e^{q_{na}z_1}e^{-q_{nf}z_1}$$

$$= \frac{T_{n//}^{fs}T_{n//}^{af}S_{nxx,a}e^{-q_{na}z'}e^{q_{na}z_1}e^{-q_{nf}z_1}}{1 - R_{n//}^{fs}R_{n//}^{fa}e^{-2q_{nf}z_1}}$$

(176)

Hence we solve the coefficients in Eq. (168) completely. Next we discuss the component $\tilde{g}_{nzx}$, writing down the second equation of Eq. (13):

$$\tilde{g}_{nzx} = -\frac{ik_n}{q_n^2}\partial_z \tilde{g}_{nxx}.$$

(177)

Next we will consider the relative position between the field ($z$) and the source ($z'$). First we consider $z' < 0$, the corresponding different equation in regions $z < z_1$, $0 < z < z_1$ and $z < 0$ are:

$$\tilde{g}_{nzx} = \begin{cases} -\dfrac{ik_n}{q_{na}^2} A \partial_z e^{-q_{na} z}, z > z_1 \\ -\dfrac{ik_n}{q_{nf}^2} B \partial_z e^{q_{nf} z} - \dfrac{ik_n}{q_{nf}^2} C \partial_z e^{-q_{nf} z}, 0 < z < z_1 \\ -\dfrac{ik_n}{q_{ns}^2} S_{nxx,s} \partial_z e^{-q_{ns}|z-z'|} - \dfrac{ik_n}{q_{ns}^2} D \partial_z e^{q_{ns} z}, z < 0 \end{cases},$$

(178)

$$= \begin{cases} A \dfrac{ik_n}{q_{na}} e^{-q_{na} z}, z > z_1 \\ -B \dfrac{ik_n}{q_{nf}} e^{q_{nf} z} + C \dfrac{ik_n}{q_{nf}} e^{-q_{nf} z}, 0 < z < z_1 \\ S_{nxx,s} \dfrac{ik_n}{q_{ns}} \operatorname{sgn}(z-z') e^{-q_{ns}|z-z'|} - D \dfrac{ik_n}{q_{ns}} e^{q_{ns} z}, z < 0 \end{cases}$$

where the coefficients $A$, $B$, $C$ and $D$ have already defined in Eq. (155), (154), (153), (156) as the following explicit forms:

$$A = \frac{T_{n//}^{fa} T_{n//}^{sf} S_{nxx,s} e^{q_{ns} z'} e^{-q_{nf} z_1} e^{q_{na} z_1}}{1 - R_{n//}^{fs} R_{n//}^{fa} e^{-2 q_{nf} z_1}},$$

(179)

$$B = \frac{R_{n//}^{fa} T_{n//}^{sf} S_{nxx,s} e^{q_{ns} z'} e^{-2 q_{nf} z_1}}{1 - R_{n//}^{fs} R_{n//}^{fa} e^{-2 q_{nf} z_1}},$$

(180)

$$C = \frac{T_{n//}^{sf} S_{nxx,s} e^{q_{ns} z'}}{1 - R_{n//}^{fs} R_{n//}^{fa} e^{-2 q_{nf} z_1}},$$

(181)

$$D = \frac{R_{n//}^{fa} S_{nxx,s} e^{q_{ns} z'} e^{-2 q_{nf} z_1} - R_{n//}^{fs} S_{nxx,s} e^{q_{ns} z'}}{1 - R_{n//}^{fs} R_{n//}^{fa} e^{-2 q_{nf} z_1}}.$$

(182)

Next we consider $0 < z' < z_1$, the corresponding different equation in regions $z > z_1$, $0 < z < z_1$ and $z < 0$ are:

$$\tilde{g}_{nzx} = \begin{cases} -\dfrac{ik_n}{q_{na}^2} A\partial_z e^{-q_{na}z}, z > z_1 \\ -\dfrac{ik_n}{q_{nf}^2} S_{nxx,f}\partial_z e^{-q_{nf}|z-z'|} - \dfrac{ik_n}{q_{nf}^2} B\partial_z e^{q_{nf}z} - \dfrac{ik_n}{q_{nf}^2} C\partial_z e^{-q_{nf}z}, 0 < z < z_1 \\ -\dfrac{ik_n}{q_{ns}^2} D\partial_z e^{q_{ns}z}, z < 0 \end{cases}, \quad (183)$$

$$= \begin{cases} A\dfrac{ik_n}{q_{na}} e^{-q_{na}z}, z > z_1 \\ S_{nxx,f}\dfrac{ik_n}{q_{nf}} \operatorname{sgn}(z-z') e^{-q_{nf}|z-z'|} - B\dfrac{ik_n}{q_{nf}} e^{q_{nf}z} + C\dfrac{ik_n}{q_{nf}} e^{-q_{nf}z}, 0 < z < z_1 \\ -D\dfrac{ik_n}{q_{ns}} e^{q_{ns}z}, z < 0 \end{cases}$$

where the coefficients $A$, $B$, $C$ and $D$ have already defined in Eq. (165), (164), (163), (166) as the following explicit forms:

$$A = \frac{T_{n//}^{fa} R_{n//}^{fs} S_{nxx,f} e^{-q_{nf}z'} e^{-q_{nf}z_1} e^{q_{na}z_1} + T_{n//}^{fa} S_{nxx,f} e^{q_{nf}z'} e^{-q_{nf}z_1} e^{q_{na}z_1}}{1 - R_{n//}^{fs} R_{n//}^{fa} e^{-2q_{nf}z_1}}, \quad (184)$$

$$B = \frac{R_{n//}^{fa} R_{n//}^{fs} S_{nxx,f} e^{-q_{nf}z'} e^{-2q_{nf}z_1} + R_{n//}^{fa} S_{nxx,f} e^{q_{nf}z'} e^{-2q_{nf}z_1}}{1 - R_{n//}^{fs} R_{n//}^{fa} e^{-2q_{nf}z_1}}, \quad (185)$$

$$C = \frac{R_{n//}^{fs} S_{nxx,f} e^{-q_{nf}z'} + R_{n//}^{fs} R_{n//}^{fa} S_{nxx,f} e^{q_{nf}z'} e^{-2q_{nf}z_1}}{1 - R_{n//}^{fs} R_{n//}^{fa} e^{-2q_{nf}z_1}}, \quad (186)$$

$$D = S_{nxx,f} \frac{T_{n//}^{fs} e^{-q_{nf}z'} + T_{n//}^{fs} R_{n//}^{fa} e^{q_{nf}z'} e^{-2q_{nf}z_1}}{1 - R_{n//}^{fs} R_{n//}^{fa} e^{-2q_{nf}z_1}}. \quad (187)$$

Next we consider $z' > z_1$, the corresponding different equation in regions $z > z_1$, $0 < z < z_1$ and $z < 0$ are:

$$\tilde{g}_{nzx} = \begin{cases} -\dfrac{ik_n}{q_{na}^2} S_{nxx,a} \partial_z e^{-q_{na}|z-z'|} - \dfrac{ik_n}{q_{na}^2} A \partial_z e^{-q_{na}z}, & z > z_1 \\ -\dfrac{ik_n}{q_{nf}^2} B \partial_z e^{q_{nf}z} - \dfrac{ik_n}{q_{nf}^2} C \partial_z e^{-q_{nf}z}, & 0 < z < z_1 \\ -\dfrac{ik_n}{q_{ns}^2} D \partial_z e^{q_{ns}z}, & z < 0 \end{cases}$$

$$= \begin{cases} S_{nxx,a} \dfrac{ik_n}{q_{na}} \mathrm{sgn}(z-z') e^{-q_{na}|z-z'|} + A \dfrac{ik_n}{q_{na}} e^{-q_{na}z}, & z > z_1 \\ -B \dfrac{ik_n}{q_{nf}} e^{q_{nf}z} + C \dfrac{ik_n}{q_{nf}} e^{-q_{nf}z}, & 0 < z < z_1 \\ -D \dfrac{ik_n}{q_{ns}} e^{q_{ns}z}, & z < 0 \end{cases} \quad (188)$$

where the coefficients $A$, $B$, $C$ and $D$ have already defined in Eq. (175), (174), (173), (176) as the following explicit forms:

$$A = \frac{R_{n//}^{fs} S_{nxx,a} e^{-2q_{nf}z_1} e^{2q_{na}z_1} e^{-q_{na}z'} - R_{n//}^{fa} S_{nxx,a} e^{-q_{na}z'} e^{2q_{na}z_1}}{1 - R_{n//}^{fs} R_{n//}^{fa} e^{-2q_{nf}z_1}}, \tag{189}$$

$$B = \frac{T_{n//}^{af} S_{nxx,a} e^{-q_{na}z'} e^{q_{na}z_1} e^{-q_{nf}z_1}}{1 - R_{n//}^{fs} R_{n//}^{fa} e^{-2q_{nf}z_1}}, \tag{190}$$

$$C = \frac{R_{n//}^{fs} T_{n//}^{af} S_{nxx,a} e^{-q_{na}z'} e^{q_{na}z_1} e^{-q_{nf}z_1}}{1 - R_{n//}^{fs} R_{n//}^{fa} e^{-2q_{nf}z_1}}, \tag{191}$$

$$D = \frac{T_{n//}^{fs} T_{n//}^{af} S_{nxx,a} e^{-q_{na}z'} e^{q_{na}z_1} e^{-q_{nf}z_1}}{1 - R_{n//}^{fs} R_{n//}^{fa} e^{-2q_{nf}z_1}}. \tag{192}$$

Hence we obtain the explicit form of $\tilde{g}_{nzx}$ directly. Next we discuss the component $\tilde{g}_{nzz}$, writing down the second equation of Eq. (14):

$$\tilde{g}_{nzz} = \frac{1}{q_n^2} \delta(z-z') - \frac{ik_n}{q_n^2} \partial_z \tilde{g}_{nxz}. \tag{193}$$

Next we will consider the relative position between the field ($z$) and the source ($z'$). First we consider $z' < 0$, the corresponding different equation in regions $z > z_1$, $0 < z < z_1$ and $z < 0$ are:

$$\tilde{g}_{nzz} = \begin{cases} -\dfrac{ik_n}{q_{na}^2} A \partial_z e^{-q_{na}z}, & z > z_1 \\ -\dfrac{ik_n}{q_{nf}^2} B \partial_z e^{q_{nf}z} - \dfrac{ik_n}{q_{nf}^2} C \partial_z e^{-q_{nf}z}, & 0 < z < z_1 \\ \dfrac{1}{q_{ns}^2}\delta(z-z') - \dfrac{ik_n}{q_{ns}^2} S_{nxz,s} \partial_z\left[\operatorname{sgn}(z-z')e^{-q_{ns}|z-z'|}\right] - \dfrac{ik_n}{q_{ns}^2} D \partial_z e^{q_{ns}z}, & z < 0 \end{cases}$$

$$= \begin{cases} \dfrac{ik_n}{q_{na}} A e^{-q_{na}z}, & z > z_1 \\ -\dfrac{ik_n}{q_{nf}} B e^{q_{nf}z} + \dfrac{ik_n}{q_{nf}} C e^{-q_{nf}z}, & 0 < z < z_1 \\ \dfrac{1}{q_{ns}^2}\delta(z-z') - \dfrac{ik_n}{q_{ns}^2} S_{nxz,s}\left[2\delta(z-z')e^{-q_{ns}|z-z'|} - q_{ns}e^{-q_{ns}|z-z'|}\right] - \dfrac{ik_n}{q_{ns}} D e^{q_{ns}z}, & z < 0 \end{cases}$$

$$= \begin{cases} \dfrac{ik_n}{q_{na}} A e^{-q_{na}z}, & z > z_1 \\ -\dfrac{ik_n}{q_{nf}} B e^{q_{nf}z} + \dfrac{ik_n}{q_{nf}} C e^{-q_{nf}z}, & 0 < z < z_1 \\ -\dfrac{1}{k_s^2}\delta(z-z') + \dfrac{ik_n}{q_{ns}} S_{nxz,s} e^{-q_{ns}|z-z'|} - \dfrac{ik_n}{q_{ns}} D e^{q_{ns}z}, & z < 0 \end{cases}, \quad (194)$$

where we use the following equation:

$$\dfrac{d}{dz}\left[\operatorname{sgn}(z-z')e^{-q_{ns}|z-z'|}\right] = 2\delta(z-z')e^{-q_{ns}|z-z'|} - q_{ns}e^{-q_{ns}|z-z'|} = 2\delta(z-z') - q_{ns}e^{-q_{ns}|z-z'|}, \tag{195}$$

where the coefficients $A$, $B$, $C$ and $D$ have already defined in Eq. (91), (90), (89), (92) as the following explicit forms:

$$A = S_{nxz,s} T_{n//}^{sf} \dfrac{R_{n//}^{fa} e^{q_{nf}z_1} + e^{-q_{nf}z_1}}{1 - R_{n//}^{fs} R_{n//}^{fa} e^{-2q_{nf}z_1}} e^{q_{ns}z'} e^{q_{na}z_1}, \tag{196}$$

$$B = \dfrac{R_{n//}^{fa} T_{n//}^{sf} S_{nxz,s} e^{q_{ns}z'} e^{-2q_{nf}z_1}}{1 - R_{n//}^{fs} R_{n//}^{fa} e^{-2q_{nf}z_1}}, \tag{197}$$

$$C = \dfrac{T_{n//}^{sf} S_{nxz,s} e^{q_{ns}z'}}{1 - R_{n//}^{fs} R_{n//}^{fa} e^{-2q_{nf}z_1}}, \tag{198}$$

$$D = S_{nxz,s} e^{q_{ns}z'} \dfrac{R_{n//}^{fa} e^{-2q_{nf}z_1} + T_{n//}^{sf} - 1}{1 - R_{n//}^{fs} R_{n//}^{fa} e^{-2q_{nf}z_1}}. \tag{199}$$

Next we consider $0 < z' < z_1$, the corresponding different equation in regions $z > z_1$, $0 < z < z_1$ and $z < 0$ are:

$$\tilde{g}_{nzz} = \begin{cases} -\dfrac{ik_n}{q_{na}^2} A\partial_z e^{-q_{na}z}, z > z_1 \\ \dfrac{1}{q_{nf}^2}\delta(z-z') - \dfrac{ik_n}{q_{nf}^2} S_{nxz,f}\partial_z\left[\text{sgn}(z-z')e^{-q_{nf}|z-z'|}\right] - \dfrac{ik_n}{q_{nf}^2} B\partial_z e^{q_{nf}z} - \dfrac{ik_n}{q_{nf}^2} C\partial_z e^{-q_{nf}z}, 0 < z < z_1 \\ -\dfrac{ik_n}{q_{ns}^2} D\partial_z e^{q_{ns}z}, z < 0 \end{cases}$$

$$= \begin{cases} \dfrac{ik_n}{q_{na}} Ae^{-q_{na}z}, z > z_1 \\ \dfrac{1}{q_{nf}^2}\delta(z-z') - \dfrac{ik_n}{q_{nf}^2} S_{nxz,f}\left[2\delta(z-z')e^{-q_{nf}|z-z'|} - q_{nf}e^{-q_{nf}|z-z'|}\right] - \dfrac{ik_n}{q_{nf}} Be^{q_{nf}z} + \dfrac{ik_n}{q_{nf}} Ce^{-q_{nf}z}, 0 < z < z_1 \\ -\dfrac{ik_n}{q_{ns}} De^{q_{ns}z}, z < 0 \end{cases}$$

$$= \begin{cases} \dfrac{ik_n}{q_{na}} Ae^{-q_{na}z}, z > z_1 \\ -\dfrac{1}{k_f^2}\delta(z-z') + \dfrac{ik_n}{q_{nf}} S_{nxz,f} e^{-q_{nf}|z-z'|} - \dfrac{ik_n}{q_{nf}} Be^{q_{nf}z} + \dfrac{ik_n}{q_{nf}} Ce^{-q_{nf}z}, 0 < z < z_1 \\ -\dfrac{ik_n}{q_{ns}} De^{q_{ns}z}, z < 0 \end{cases}$$

, (200)

where the coefficients $A$, $B$, $C$ and $D$ have already defined in Eq. (101), (100), (99), (102) as the following explicit forms:

$$A = S_{nxz,f} T_{n//}^{fa} \frac{e^{q_{nf}z'} - e^{-q_{nf}z'} R_{n//}^{fs}}{1 - R_{n//}^{fs} R_{n//}^{fa} e^{-2q_{nf}z_1}} e^{-q_{nf}z_1} e^{q_{na}z_1}, \tag{201}$$

$$B = S_{nxz,f} R_{n//}^{fa} \frac{e^{q_{nf}z'} - R_{n//}^{fs} e^{-q_{nf}z'}}{1 - R_{n//}^{fs} R_{n//}^{fa} e^{-2q_{nf}z_1}} e^{-2q_{nf}z_1}, \tag{202}$$

$$C = \frac{R_{n//}^{fa} e^{q_{nf}z'} e^{-2q_{nf}z_1} - e^{-q_{nf}z'}}{1 - R_{n//}^{fs} R_{n//}^{fa} e^{-2q_{nf}z_1}} R_{n//}^{fs} S_{nxz,f}, \tag{203}$$

$$D = S_{nxz,f} T_{n//}^{fs} \frac{R_{n//}^{fa} e^{q_{nf}z'} e^{-2q_{nf}z_1} - e^{-q_{nf}z'}}{1 - R_{n//}^{fs} R_{n//}^{fa} e^{-2q_{nf}z_1}}. \tag{204}$$

Next we consider $z' > z_1$, the corresponding different equation in regions $z > z_1$, $0 < z < z_1$ and $z < 0$ are:

$$\tilde{g}_{nzz} = \begin{cases} \dfrac{1}{q_{na}^2}\delta(z-z') - \dfrac{ik_n}{q_{na}^2} S_{nxz,a}\partial_z\left[\text{sgn}(z-z')e^{-q_{na}|z-z'|}\right] - \dfrac{ik_n}{q_{na}^2} A\partial_z e^{-q_{na}z}, & z > z_1 \\ -\dfrac{ik_n}{q_{nf}^2} B\partial_z e^{q_{nf}z} - \dfrac{ik_n}{q_{nf}^2} C\partial_z e^{-q_{nf}z}, & 0 < z < z_1 \\ -\dfrac{ik_n}{q_{ns}^2} D\partial_z e^{q_{ns}z}, & z < 0 \end{cases}$$

$$= \begin{cases} \dfrac{1}{q_{na}^2}\delta(z-z') - \dfrac{ik_n}{q_{na}^2} S_{nxz,a}\left[2\delta(z-z')e^{-q_{na}|z-z'|} - q_{na}e^{-q_{na}|z-z'|}\right] + \dfrac{ik_n}{q_{na}} A e^{-q_{na}z}, & z > z_1 \\ -\dfrac{ik_n}{q_{nf}} Be^{q_{nf}z} + \dfrac{ik_n}{q_{nf}} Ce^{-q_{nf}z}, & 0 < z < z_1 \\ -\dfrac{ik_n}{q_{ns}} De^{q_{ns}z}, & z < 0 \end{cases}$$

$$= \begin{cases} -\dfrac{1}{k_a^2}\delta(z-z') + \dfrac{ik_n}{q_{na}} S_{nxz,a} e^{-q_{na}|z-z'|} + \dfrac{ik_n}{q_{na}} A e^{-q_{na}z}, & z > z_1 \\ -\dfrac{ik_n}{q_{nf}} Be^{q_{nf}z} + \dfrac{ik_n}{q_{nf}} Ce^{-q_{nf}z}, & 0 < z < z_1 \\ -\dfrac{ik_n}{q_{ns}} De^{q_{ns}z}, & z < 0 \end{cases}, \quad (205)$$

where the coefficients $A$, $B$, $C$ and $D$ have already defined in Eq. (111), (110), (109), (112) as the following explicit forms:

$$A = S_{nxz,a} \frac{-R_{n//}^{fs} e^{-q_{na}z'} e^{q_{na}z_1} e^{-q_{nf}z_1} R_{n//}^{fa} - R_{n//}^{fs} T_{n//}^{af} e^{-q_{na}z'} e^{2q_{na}z_1} e^{-2q_{nf}z_1} + e^{-q_{na}z'} e^{2q_{na}z_1} R_{n//}^{fa}}{1 - R_{n//}^{fs} R_{n//}^{fa} e^{-q_{na}z_1} e^{-q_{nf}z_1}}, \quad (206)$$

$$B = \frac{-T_{n//}^{af} S_{nxz,a} e^{-q_{na}z'} e^{q_{na}z_1} e^{-q_{nf}z_1}}{1 - R_{n//}^{fs} R_{n//}^{fa} e^{-q_{na}z_1} e^{-q_{nf}z_1}}, \quad (207)$$

$$C = \frac{-R_{n//}^{fs} T_{n//}^{af} S_{nxz,a} e^{-q_{na}z'} e^{q_{na}z_1} e^{-q_{nf}z_1}}{1 - R_{n//}^{fs} R_{n//}^{fa} e^{-q_{na}z_1} e^{-q_{nf}z_1}}, \quad (208)$$

$$D = S_{nxz,a} \frac{-T_{n//}^{fs} T_{n//}^{af} e^{-q_{na}z'} e^{q_{na}z_1} e^{-q_{nf}z_1}}{1 - R_{n//}^{fs} R_{n//}^{fa} e^{-q_{na}z_1} e^{-q_{nf}z_1}}. \quad (209)$$

Hence we obtain the explicit form of $\tilde{g}_{nzz}$ directly.

### 4-3. Consider the multi-planer films over the substrate

To discuss the general case, we consider the multi-planer films over the substrate and the geometry is shown in Fig. 3. The boundaries are at $z = z_0$ and $z = z_j$, $j = 0, 1, 2, ..., N$. The dielectric function above $z = z_N$ is $\varepsilon_a$ and below $z = z_0$ is $\varepsilon_s$, respectively. The dielectric function of each film is $\varepsilon_{f_j}, j = 1, 2, ..., N$.

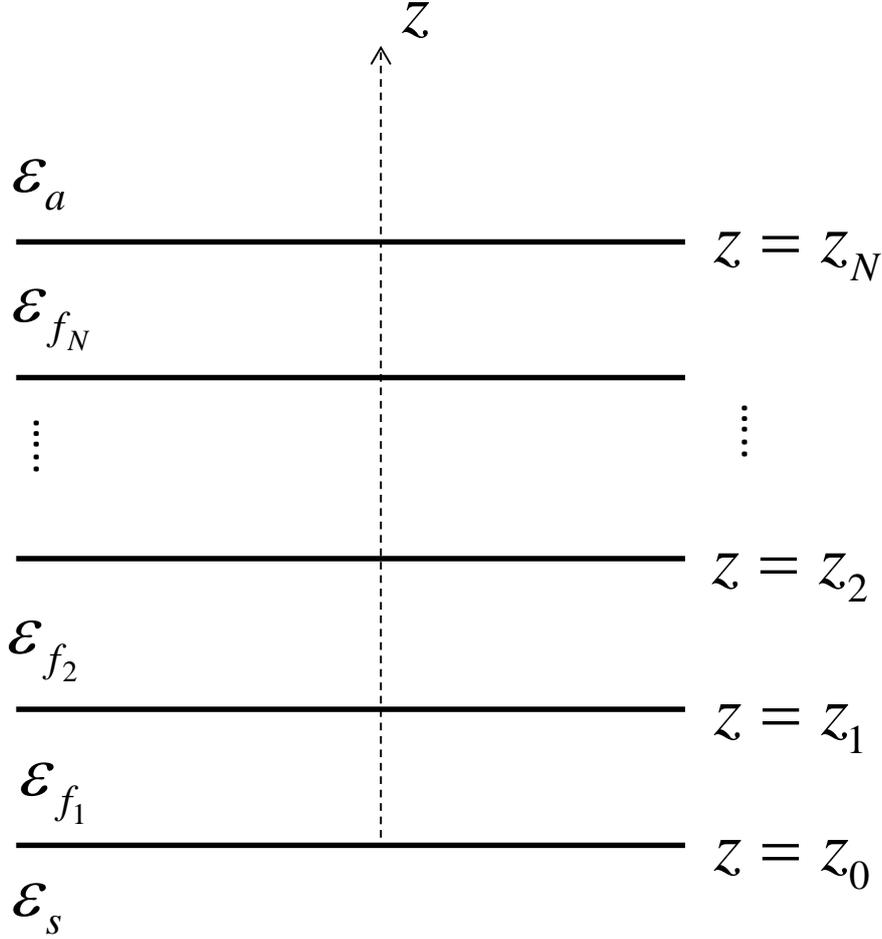

**Fig. 3** The geometry of considering the multi-planer films over the substrate

For the component $\tilde{g}_{nxz}$, starting from Eq. (81), we have:

$$\left(\partial_z^2 - q_n^2\right)\tilde{g}_{nxz} = -\frac{ik_n}{k^2}\partial_z\delta(z-z'). \tag{210}$$

Next we will consider the relative position between the field ($z$) and the source ($z'$). First we consider $z' < 0$ (without loss of generality, we may set $z_0 = 0$), the corresponding different equation in regions $z > z_N$, $z_{j-1} < z < z_j, j = N, N-1, ..., 2, 1$ and $z < 0$ are:

$$\begin{cases} \left(\partial_z^2 - q_{na}^2\right)\tilde{g}_{nxz} = 0, z > z_N \\ \left(\partial_z^2 - q_{nf_j}^2\right)\tilde{g}_{nxz} = 0, z_{j-1} < z < z_j, j = 1, 2, ..., N \\ \left(\partial_z^2 - q_{ns}^2\right)\tilde{g}_{nxz} = -\frac{ik_n}{k_s^2}\partial_z\delta(z-z'), z < 0 \end{cases}. \tag{211}$$

Furthermore, the solutions of Eq. (211) are:

$$\tilde{g}_{nxz} = \begin{cases} Ae^{-q_{na}z}, z > z_N \\ B_j e^{q_{nf_j}z} + C_j e^{-q_{nf_j}z}, z_{j-1} < z < z_j, j=1,2,...,N, \\ S_{nxz,s}\,\text{sgn}(z-z')e^{-q_{ns}|z-z'|} + De^{q_{ns}z}, z < 0 \end{cases} \quad (212)$$

where $S_{nxz,s} \equiv -\dfrac{ik_n}{2k_s^2}$. In order to match the boundary conditions of $\tilde{g}_{nxz}$ as discuss in Eq. (84):

$$\begin{cases} \tilde{g}_{nxz}\big|_{z_{N+}} = \tilde{g}_{nxz}\big|_{z_{N-}}, \dfrac{\varepsilon_a}{q_{na}^2}\partial_z \tilde{g}_{nxz}\bigg|_{z_{N+}} = \dfrac{\varepsilon_{f_1}}{q_{nf_1}^2}\partial_z \tilde{g}_{nxz}\bigg|_{z_{N-}} \\ \tilde{g}_{nxz}\big|_{z_{j+}} = \tilde{g}_{nxz}\big|_{z_{j-}}, \dfrac{\varepsilon_{f_j}}{q_{nf_j}^2}\partial_z \tilde{g}_{nxz}\bigg|_{z_{j+}} = \dfrac{\varepsilon_{f_{j+1}}}{q_{nf_{j+1}}^2}\partial_z \tilde{g}_{nxz}\bigg|_{z_{j-}}, j=1,2,...,N-1, \\ \tilde{g}_{nxz}\big|_{0+} = \tilde{g}_{nxz}\big|_{0-}, \dfrac{\varepsilon_{f_N}}{q_{nf_N}^2}\partial_z \tilde{g}_{nxz}\bigg|_{0+} = \dfrac{\varepsilon_s}{q_{ns}^2}\partial_z \tilde{g}_{nxz}\bigg|_{0-} \end{cases} \quad (213)$$

we have the following simulated equations:

$$\begin{cases} Ae^{-q_{na}z_N} = B_N e^{q_{nf_N}z_N} + C_N e^{-q_{nf_N}z_N}, z=z_N \\ B_j e^{q_{nf_j}z_j} + C_j e^{-q_{nf_j}z_j} = B_{j+1}e^{q_{nf_{j+1}}z_j} + C_{j+1}e^{-q_{nf_{j+1}}z_j}, z=z_j, j=1,2,...,N-1, \\ B_1 + C_1 = S_{nxz,s}e^{q_{ns}z'} + D, z=0 \end{cases} \quad (214)$$

and

$$\begin{cases} -\dfrac{\varepsilon_a}{q_{na}}Ae^{-q_{na}z_N} = \dfrac{\varepsilon_{f_N}}{q_{nf_N}}\left(B_N e^{q_{nf_N}z_N} - C_N e^{-q_{nf_N}z_N}\right), z=z_N \\ \dfrac{\varepsilon_{f_j}}{q_{nf_j}}\left[B_j e^{q_{nf_j}z_j} - C_j e^{-q_{nf_j}z_j}\right] = \dfrac{\varepsilon_{f_{j+1}}}{q_{nf_{j+1}}}\left[B_{j+1}e^{q_{nf_{j+1}}z_j} - C_{j+1}e^{-q_{nf_{j+1}}z_j}\right], z=z_j, j=1,2,...,N-1. \\ \dfrac{\varepsilon_{f_1}}{q_{nf_1}}(B_1 - C_1) = -\dfrac{\varepsilon_s}{q_{ns}}S_{nxz,s}e^{q_{ns}z'} + \dfrac{\varepsilon_s}{q_{ns}}D, z=0 \end{cases}$$

$$(215)$$

Furthermore, we obtain:

$$-\dfrac{\varepsilon_a}{q_{na}}\left(B_N e^{q_{nf_N}z_N} + C_N e^{-q_{nf_N}z_N}\right) = \dfrac{\varepsilon_{f_N}}{q_{nf_N}}\left(B_N e^{q_{nf_N}z_N} - C_N e^{-q_{nf_N}z_N}\right)$$

$$\Rightarrow \left(\dfrac{\varepsilon_{f_N}}{q_{nf_N}} + \dfrac{\varepsilon_a}{q_{na}}\right)B_N = \left(\dfrac{\varepsilon_{f_N}}{q_{nf_N}} - \dfrac{\varepsilon_a}{q_{na}}\right)C_N e^{-2q_{nf_N}z_N} \quad , \quad (216)$$

$$\Rightarrow B_N = \dfrac{\varepsilon_{f_N}q_{na} - \varepsilon_a q_{nf_N}}{\varepsilon_{f_N}q_{na} + \varepsilon_a q_{nf_N}}C_N e^{-2q_{nf_N}z_N} = R_{n//}^{f_N a}C_N e^{-2q_{nf_N}z_N}$$

$$\Rightarrow A = R_{n//}^{f_N a}C_N e^{-q_{nf_N}z_N}e^{q_{na}z_N} + C_N e^{-q_{nf_N}z_N}e^{q_{na}z_N} = T_{n//}^{f_N a}C_N e^{-q_{nf_N}z_N}e^{q_{na}z_N}$$

and

$$\begin{cases} B_j e^{q_{nf_j} z_j} + C_j e^{-q_{nf_j} z_j} = B_{j+1} e^{q_{nf_{j+1}} z_j} + C_{j+1} e^{-q_{nf_{j+1}} z_j}, j=1,2,...,N-1 \\ \dfrac{\varepsilon_{f_j}}{q_{nf_j}} \left[ B_j e^{q_{nf_j} z_j} - C_j e^{-q_{nf_j} z_j} \right] = \dfrac{\varepsilon_{f_{j+1}}}{q_{nf_{j+1}}} \left[ B_{j+1} e^{q_{nf_{j+1}} z_j} - C_{j+1} e^{-q_{nf_{j+1}} z_j} \right], j=1,2,...,N-1 \end{cases}$$

$$\Rightarrow \begin{pmatrix} e^{q_{nf_j} z_j} & e^{-q_{nf_j} z_j} \\ \dfrac{\varepsilon_{f_j}}{q_{nf_j}} e^{q_{nf_j} z_j} & -\dfrac{\varepsilon_{f_j}}{q_{nf_j}} e^{-q_{nf_j} z_j} \end{pmatrix} \begin{pmatrix} B_j \\ C_j \end{pmatrix} = \begin{pmatrix} e^{q_{nf_{j+1}} z_j} & e^{-q_{nf_{j+1}} z_j} \\ \dfrac{\varepsilon_{f_{j+1}}}{q_{nf_{j+1}}} e^{q_{nf_{j+1}} z_j} & -\dfrac{\varepsilon_{f_{j+1}}}{q_{nf_{j+1}}} e^{-q_{nf_{j+1}} z_j} \end{pmatrix} \begin{pmatrix} B_{j+1} \\ C_{j+1} \end{pmatrix}, j=1,...,N-1$$

$$\Rightarrow \begin{pmatrix} B_j \\ C_j \end{pmatrix} = \begin{pmatrix} e^{q_{nf_j} z_j} & e^{-q_{nf_j} z_j} \\ \dfrac{\varepsilon_{f_j}}{q_{nf_j}} e^{q_{nf_j} z_j} & -\dfrac{\varepsilon_{f_j}}{q_{nf_j}} e^{-q_{nf_j} z_j} \end{pmatrix}^{-1} \begin{pmatrix} e^{q_{nf_{j+1}} z_j} & e^{-q_{nf_{j+1}} z_j} \\ \dfrac{\varepsilon_{f_{j+1}}}{q_{nf_{j+1}}} e^{q_{nf_{j+1}} z_j} & -\dfrac{\varepsilon_{f_{j+1}}}{q_{nf_{j+1}}} e^{-q_{nf_{j+1}} z_j} \end{pmatrix} \begin{pmatrix} B_{j+1} \\ C_{j+1} \end{pmatrix}$$

$$\equiv \mathrm{T}_j^{//} \begin{pmatrix} B_{j+1} \\ C_{j+1} \end{pmatrix}, j=1,...,N-1$$

$$\Rightarrow \begin{pmatrix} B_1 \\ C_1 \end{pmatrix} = \mathrm{T}_1^{//} \begin{pmatrix} B_2 \\ C_2 \end{pmatrix} = \mathrm{T}_1^{//} \mathrm{T}_2^{//} \begin{pmatrix} B_3 \\ C_3 \end{pmatrix} = ... = \mathrm{T}_1^{//} \mathrm{T}_2^{//} \bullet\bullet\bullet \mathrm{T}_{N-1}^{//} \begin{pmatrix} B_N \\ C_N \end{pmatrix} \equiv \prod_{j=1}^{N-1} \mathrm{T}_j^{//} \begin{pmatrix} B_N \\ C_N \end{pmatrix}$$

$$\equiv \begin{pmatrix} M_{11}^{//} & M_{12}^{//} \\ M_{21}^{//} & M_{22}^{//} \end{pmatrix} \begin{pmatrix} B_N \\ C_N \end{pmatrix} = \begin{pmatrix} M_{11}^{//} B_N + M_{12}^{//} C_N \\ M_{21}^{//} B_N + M_{22}^{//} C_N \end{pmatrix}$$

$$\Rightarrow \begin{pmatrix} B_j \\ C_j \end{pmatrix} = \mathrm{T}_j^{//} \begin{pmatrix} B_{j+1} \\ C_{j+1} \end{pmatrix} = \mathrm{T}_j^{//} \mathrm{T}_{j+1}^{//} \begin{pmatrix} B_{j+2} \\ C_{j+2} \end{pmatrix} = ... = \mathrm{T}_j^{//} \mathrm{T}_{j+1}^{//} \bullet\bullet\bullet \mathrm{T}_{N-1}^{//} \begin{pmatrix} B_N \\ C_N \end{pmatrix}$$

$$\equiv \begin{pmatrix} M_{j,11}^{//} & M_{j,12}^{//} \\ M_{j,21}^{//} & M_{j,22}^{//} \end{pmatrix} \begin{pmatrix} B_N \\ C_N \end{pmatrix} = \begin{pmatrix} M_{j,11}^{//} B_N + M_{j,12}^{//} C_N \\ M_{j,21}^{//} B_N + M_{j,22}^{//} C_N \end{pmatrix}, j=2,3,...,N-1$$

, (217)

where we define a transfer matrix $\mathrm{T}_j^{//}, j=1,...,N-1$ as the following form:

$$\mathrm{T}_{j}^{//} = \begin{pmatrix} e^{q_{nf_j}z_j} & e^{-q_{nf_j}z_j} \\ \dfrac{\varepsilon_{f_j}}{q_{nf_j}}e^{q_{nf_j}z_j} & -\dfrac{\varepsilon_{f_j}}{q_{nf_j}}e^{-q_{nf_j}z_j} \end{pmatrix}^{-1} \begin{pmatrix} e^{q_{nf_{j+1}}z_j} & e^{-q_{nf_{j+1}}z_j} \\ \dfrac{\varepsilon_{f_{j+1}}}{q_{nf_{j+1}}}e^{q_{nf_{j+1}}z_j} & -\dfrac{\varepsilon_{f_{j+1}}}{q_{nf_{j+1}}}e^{-q_{nf_{j+1}}z_j} \end{pmatrix}$$

$$= -\dfrac{q_{nf_j}}{2\varepsilon_{f_j}} \begin{pmatrix} -\dfrac{\varepsilon_{f_j}}{q_{nf_j}}e^{-q_{nf_j}z_j} & -e^{-q_{nf_j}z_j} \\ -\dfrac{\varepsilon_{f_j}}{q_{nf_j}}e^{q_{nf_j}z_j} & e^{q_{nf_j}z_j} \end{pmatrix} \begin{pmatrix} e^{q_{nf_{j+1}}z_j} & e^{-q_{nf_{j+1}}z_j} \\ \dfrac{\varepsilon_{f_{j+1}}}{q_{nf_{j+1}}}e^{q_{nf_{j+1}}z_j} & -\dfrac{\varepsilon_{f_{j+1}}}{q_{nf_{j+1}}}e^{-q_{nf_{j+1}}z_j} \end{pmatrix}$$

$$= \dfrac{\varepsilon_{f_j}q_{nf_{j+1}}+\varepsilon_{f_{j+1}}q_{nf_j}}{2\varepsilon_{f_j}q_{nf_{j+1}}} \begin{pmatrix} e^{-q_{nf_j}z_j}e^{q_{nf_{j+1}}z_j} & \dfrac{\varepsilon_{f_j}q_{nf_{j+1}}-\varepsilon_{f_{j+1}}q_{nf_j}}{\varepsilon_{f_j}q_{nf_{j+1}}+\varepsilon_{f_{j+1}}q_{nf_j}}e^{-q_{nf_j}z_j}e^{-q_{nf_{j+1}}z_j} \\ \dfrac{\varepsilon_{f_j}q_{nf_{j+1}}-\varepsilon_{f_{j+1}}q_{nf_j}}{\varepsilon_{f_j}q_{nf_{j+1}}+\varepsilon_{f_{j+1}}q_{nf_j}}e^{q_{nf_j}z_j}e^{q_{nf_{j+1}}z_j} & e^{q_{nf_j}z_j}e^{-q_{nf_{j+1}}z_j} \end{pmatrix}$$

$$= \dfrac{1}{T_{n//}^{f_j f_{j+1}}} \begin{pmatrix} e^{-q_{nf_j}z_j}e^{q_{nf_{j+1}}z_j} & R_{n//}^{f_j f_{j+1}}e^{-q_{nf_j}z_j}e^{-q_{nf_{j+1}}z_j} \\ R_{n//}^{f_j f_{j+1}}e^{q_{nf_j}z_j}e^{q_{nf_{j+1}}z_j} & e^{q_{nf_j}z_j}e^{-q_{nf_{j+1}}z_j} \end{pmatrix},$$

(218)

$$\prod_{j=1}^{N-1} \mathrm{T}_j^{//} = \prod_{j=1}^{N-1} \dfrac{1}{T_{n//}^{f_j f_{j+1}}} \begin{pmatrix} e^{-q_{nf_j}z_j}e^{q_{nf_{j+1}}z_j} & R_{n//}^{f_j f_{j+1}}e^{-q_{nf_j}z_j}e^{-q_{nf_{j+1}}z_j} \\ R_{n//}^{f_j f_{j+1}}e^{q_{nf_j}z_j}e^{q_{nf_{j+1}}z_j} & e^{q_{nf_j}z_j}e^{-q_{nf_{j+1}}z_j} \end{pmatrix} \equiv \begin{pmatrix} M_{11}^{//} & M_{12}^{//} \\ M_{21}^{//} & M_{22}^{//} \end{pmatrix}, \quad (219)$$

and

$$\begin{aligned} D &= B_1 + C_1 - S_{nxz,s}e^{q_{ns}z'} = M_{11}^{//}B_N + M_{12}^{//}C_N + M_{21}^{//}B_N + M_{22}^{//}C_N - S_{nxz,s}e^{q_{ns}z'} \\ &= \left(M_{11}^{//}R_{n//}^{f_N a}e^{-2q_{nf_N}z_N} + M_{21}^{//}R_{n//}^{f_N a}e^{-2q_{nf_N}z_N} + M_{12}^{//} + M_{22}^{//}\right)C_N - S_{nxz,s}e^{q_{ns}z'} \end{aligned}.$$

(220)

Furthermore, we get the following form to solve the coefficient $C_N$:

$$\frac{\varepsilon_{f_1}}{q_{nf_1}}\left(M_{11}^{//}R_{n//}^{f_Na}e^{-2q_{nfN}z_N} + M_{12}^{//} - M_{21}^{//}R_{n//}^{f_Na}e^{-2q_{nfN}z_N} - M_{22}^{//}\right)C_N$$

$$= -\frac{\varepsilon_s}{q_{ns}}S_{nxz,s}e^{q_{ns}z'} + \frac{\varepsilon_s}{q_{ns}}\left(M_{11}^{//}R_{n//}^{f_Na}e^{-2q_{nfN}z_N} + M_{21}^{//}R_{n//}^{f_Na}e^{-2q_{nfN}z_N} + M_{12}^{//} + M_{22}^{//}\right)C_N - \frac{\varepsilon_s}{q_{ns}}S_{nxz,s}e^{q_{ns}z'}$$

$$\Rightarrow \left[\left(\frac{\varepsilon_s}{q_{ns}} - \frac{\varepsilon_{f_1}}{q_{nf_1}}\right)\left(M_{11}^{//}R_{n//}^{f_Na}e^{-2q_{nfN}z_N} + M_{12}^{//}\right) + \left(\frac{\varepsilon_s}{q_{ns}} + \frac{\varepsilon_{f_1}}{q_{nf_1}}\right)\left(M_{21}^{//}R_{n//}^{f_Na}e^{-2q_{nfN}z_N} + M_{22}^{//}\right)\right]C_N$$

$$= -\frac{2\varepsilon_s}{q_{ns}}S_{nxz,s}e^{q_{ns}z'}$$

$$\Rightarrow C_N = \frac{-\dfrac{2\varepsilon_s q_{nf_1}}{\varepsilon_s q_{nf_1} + \varepsilon_{f_1}q_{ns}}S_{nxz,s}e^{q_{ns}z'}}{\dfrac{\varepsilon_s q_{nf_1} - \varepsilon_{f_1}q_{ns}}{\varepsilon_s q_{nf_1} + \varepsilon_{f_1}q_{ns}}\left(M_{11}^{//}R_{n//}^{f_Na}e^{-2q_{nfN}z_N} + M_{12}^{//}\right) + \left(M_{21}^{//}R_{n//}^{f_Na}e^{-2q_{nfN}z_N} + M_{22}^{//}\right)}$$

$$= \frac{-T_{n//}^{sf_1}S_{nxz,s}e^{q_{ns}z'}}{R_{n//}^{sf_1}\left(M_{11}^{//}R_{n//}^{f_Na}e^{-2q_{nfN}z_N} + M_{12}^{//}\right) + \left(M_{21}^{//}R_{n//}^{f_Na}e^{-2q_{nfN}z_N} + M_{22}^{//}\right)}$$

. (221)

Hence we can get the other three coefficients as the following forms:

$$B_N = R_{n//}^{f_Na}C_N e^{-2q_{nfN}z_N} = \frac{-R_{n//}^{f_Na}T_{n//}^{sf_1}S_{nxz,s}e^{q_{ns}z'}e^{-2q_{nfN}z_N}}{R_{n//}^{sf_1}\left(M_{11}^{//}R_{n//}^{f_Na}e^{-2q_{nfN}z_N} + M_{12}^{//}\right) + \left(M_{21}^{//}R_{n//}^{f_Na}e^{-2q_{nfN}z_N} + M_{22}^{//}\right)}, \quad (222)$$

$$A = T_{n//}^{f_Na}C_N e^{-q_{nfN}z_N}e^{q_{na}z_N} = \frac{-T_{n//}^{f_Na}T_{n//}^{sf_1}S_{nxz,s}e^{q_{ns}z'}e^{-q_{nfN}z_N}e^{q_{na}z_N}}{R_{n//}^{sf_1}\left(M_{11}^{//}R_{n//}^{f_Na}e^{-2q_{nfN}z_N} + M_{12}^{//}\right) + \left(M_{21}^{//}R_{n//}^{f_Na}e^{-2q_{nfN}z_N} + M_{22}^{//}\right)},$$

(223)

$$D = \left(M_{11}^{//}R_{n//}^{f_Na}e^{-2q_{nfN}z_N} + M_{21}^{//}R_{n//}^{f_Na}e^{-2q_{nfN}z_N} + M_{12}^{//} + M_{22}^{//}\right)C_N - S_{nxz,s}e^{q_{ns}z'}$$

$$= \frac{-T_{n//}^{sf_1}S_{nxz,s}e^{q_{ns}z'}\left(M_{11}^{//}R_{n//}^{f_Na}e^{-2q_{nfN}z_N} + M_{21}^{//}R_{n//}^{f_Na}e^{-2q_{nfN}z_N} + M_{12}^{//} + M_{22}^{//}\right)}{R_{n//}^{sf_1}\left(M_{11}^{//}R_{n//}^{f_Na}e^{-2q_{nfN}z_N} + M_{12}^{//}\right) + \left(M_{21}^{//}R_{n//}^{f_Na}e^{-2q_{nfN}z_N} + M_{22}^{//}\right)} - S_{nxz,s}e^{q_{ns}z'}$$

$$= -\frac{\left(T_{n//}^{sf_1} + R_{n//}^{sf_1}\right)M_{11}^{//}R_{n//}^{f_Na}e^{-2q_{nfN}z_N} + \left(T_{n//}^{sf_1} + R_{n//}^{sf_1}\right)M_{12}^{//} + \left(T_{n//}^{sf_1} + 1\right)M_{21}^{//}R_{n//}^{f_Na}e^{-2q_{nfN}z_N} + \left(T_{n//}^{sf_1} + 1\right)M_{22}^{//}}{R_{n//}^{sf_1}\left(M_{11}^{//}R_{n//}^{f_Na}e^{-2q_{nfN}z_N} + M_{12}^{//}\right) + \left(M_{21}^{//}R_{n//}^{f_Na}e^{-2q_{nfN}z_N} + M_{22}^{//}\right)}S_{nxz,s}e^{q_{ns}z'}$$

. (224)

According to Eq. (217), the coefficients $B_j$ and $C_j, j = 1,2,...,N-1$ are obtained and then we solve the coefficients in Eq. (212) completely. Next we consider $z' > z_N$, the corresponding different equation in regions $z > z_N$, $z_{j-1} < z < z_j, j = N, N-1,...,2,1$ and $z < 0$ are:

$$\begin{cases} \left(\partial_z^2 - q_{na}^2\right)\tilde{g}_{nxz} = -\frac{ik_n}{k_a^2}\partial_z\delta(z-z'), z > z_N \\ \left(\partial_z^2 - q_{nf_j}^2\right)\tilde{g}_{nxz} = 0, z_{j-1} < z < z_j, j = 1,2,...,N \\ \left(\partial_z^2 - q_{ns}^2\right)\tilde{g}_{nxz} = 0, z < 0 \end{cases} \quad (225)$$

Furthermore, the solutions of Eq. (225) are:

$$\tilde{g}_{nxz} = \begin{cases} S_{nxz,a}\,\text{sgn}(z-z')e^{-q_{na}|z-z'|} + Ae^{-q_{na}z}, z > z_N \\ B_j e^{q_{nf_j}z} + C_j e^{-q_{nf_j}z}, z_{j-1} < z < z_j, j = 1,2,...,N \\ De^{q_{ns}z}, z < 0 \end{cases} \quad (226)$$

where $S_{nxz,a} \equiv -\frac{ik_n}{2k_a^2}$. In order to match the boundary conditions of $\tilde{g}_{nxz}$ as discuss in Eq. (213), we have the following simulated equations:

$$\begin{cases} -S_{nxz,a}e^{-q_{na}z'}e^{q_{na}z_N} + Ae^{-q_{na}z_N} = B_N e^{q_{nf_N}z_N} + C_N e^{-q_{nf_N}z_N}, z = z_N \\ B_j e^{q_{nf_j}z_j} + C_j e^{-q_{nf_j}z_j} = B_{j+1}e^{q_{nf_{j+1}}z_j} + C_{j+1}e^{-q_{nf_{j+1}}z_j}, z = z_j, j = 1,2,...,N-1 \\ B_1 + C_1 = D, z = 0 \end{cases} \quad (227)$$

and

$$\begin{cases} -\dfrac{\varepsilon_a}{q_{na}}S_{nxz,a}e^{-q_{na}z'}e^{q_{na}z_N} - \dfrac{\varepsilon_a}{q_{na}}Ae^{-q_{na}z_N} = \dfrac{\varepsilon_{f_N}}{q_{nf_N}}\left(B_N e^{q_{nf_N}z_N} - C_N e^{-q_{nf_N}z_N}\right), z = z_N \\ \dfrac{\varepsilon_{f_j}}{q_{nf_j}}\left[B_j e^{q_{nf_j}z_j} - C_j e^{-q_{nf_j}z_j}\right] = \dfrac{\varepsilon_{f_{j+1}}}{q_{nf_{j+1}}}\left[B_{j+1}e^{q_{nf_{j+1}}z_j} - C_{j+1}e^{-q_{nf_{j+1}}z_j}\right], z = z_j, j = 1,2,...,N-1 \\ \dfrac{\varepsilon_{f_1}}{q_{nf_1}}(B_1 - C_1) = \dfrac{\varepsilon_s}{q_{ns}}D, z = 0 \end{cases} \quad (228)$$

Furthermore, we obtain:

$$-\frac{\varepsilon_a}{q_{na}}S_{nxz,a}e^{-q_{na}z'}e^{q_{na}z_N} - \frac{\varepsilon_a}{q_{na}}\left(B_N e^{q_{nf_N}z_N} + C_N e^{-q_{nf_N}z_N} + S_{nxz,a}e^{-q_{na}z'}e^{q_{na}z_N}\right) = \frac{\varepsilon_{f_N}}{q_{nf_N}}\left(B_N e^{q_{nf_N}z_N} - C_N e^{-q_{nf_N}z_N}\right)$$

$$\Rightarrow \left(\frac{\varepsilon_{f_N}}{q_{nf_N}} + \frac{\varepsilon_a}{q_{na}}\right)B_N = \left(\frac{\varepsilon_{f_N}}{q_{nf_N}} - \frac{\varepsilon_a}{q_{na}}\right)C_N e^{-2q_{nf_N}z_N} - \frac{2\varepsilon_a}{q_{na}}S_{nxz,a}e^{-q_{na}z'}e^{q_{na}z_N}e^{-q_{nf_N}z_N}$$

$$\Rightarrow B_N = \frac{\varepsilon_{f_N}q_{na} - \varepsilon_a q_{nf_N}}{\varepsilon_{f_N}q_{na} + \varepsilon_a q_{nf_N}}C_N e^{-2q_{nf_N}z_N} - \frac{2\varepsilon_a q_{nf_N}}{\varepsilon_{f_N}q_{na} + \varepsilon_a q_{nf_N}}S_{nxz,a}e^{-q_{na}z'}e^{q_{na}z_N}e^{-q_{nf_N}z_N}$$

$$= R_{n//}^{f_N a}C_N e^{-2q_{nf_N}z_N} - S_{nxz,a}T_{n//}^{af_N}e^{-q_{na}z'}e^{q_{na}z_N}e^{-q_{nf_N}z_N}$$

$$\Rightarrow A = B_N e^{q_{nf_N}z_N}e^{q_{na}z_N} + C_N e^{-q_{nf_N}z_N}e^{q_{na}z_N} + S_{nxz,a}e^{-q_{na}z'}e^{2q_{na}z_N}$$

$$= \left(R_{n//}^{f_N a}C_N e^{-2q_{nf_N}z_N} - S_{nxz,a}T_{n//}^{af_N}e^{-q_{na}z'}e^{q_{na}z_N}e^{-q_{nf_N}z_N}\right)e^{q_{nf_N}z_N}e^{q_{na}z_N} + C_N e^{-q_{nf_N}z_N}e^{q_{na}z_N} + S_{nxz,a}e^{-q_{na}z'}e^{2q_{na}z_N}$$

$$= T_{n//}^{f_N a}C_N e^{-q_{nf_N}z_N}e^{q_{na}z_N} + R_{n//}^{f_N a}S_{nxz,a}e^{-q_{na}z'}e^{2q_{na}z_N}$$

,
$$(229)$$

and

$$D = B_1 + C_1 = M_{11}^{//}B_N + M_{12}^{//}C_N + M_{21}^{//}B_N + M_{22}^{//}C_N$$

$$= M_{11}^{//}\left(R_{n//}^{f_N a}C_N e^{-2q_{nf_N} z_N} - S_{nxz,a}T_{n//}^{af_N} e^{-q_{na}z'} e^{q_{na}z_N} e^{-q_{nf_N} z_N}\right) + M_{12}^{//}C_N$$

$$+ M_{21}^{//}\left(R_{n//}^{f_N a}C_N e^{-2q_{nf_N} z_N} - S_{nxz,a}T_{n//}^{af_N} e^{-q_{na}z'} e^{q_{na}z_N} e^{-q_{nf_N} z_N}\right) + M_{22}^{//}C_N \quad . \tag{230}$$

$$= \left(M_{11}^{//}R_{n//}^{f_N a}e^{-2q_{nf_N} z_N} + M_{12}^{//} + M_{21}^{//}R_{n//}^{f_N a}e^{-2q_{nf_N} z_N} + M_{22}^{//}\right)C_N$$

$$- M_{11}^{//}S_{nxz,a}T_{n//}^{af_N} e^{-q_{na}z'} e^{q_{na}z_N} e^{-q_{nf_N} z_N} - M_{21}^{//}S_{nxz,a}T_{n//}^{af_N} e^{-q_{na}z'} e^{q_{na}z_N} e^{-q_{nf_N} z_N}$$

Furthermore, we get the following form to solve the coefficient $C_N$:

$$\frac{\varepsilon_{f_1}}{q_{nf_1}}\left(M_{11}^{//}R_{n//}^{f_N a}e^{-2q_{nf_N} z_N} + M_{12}^{//} - M_{21}^{//}R_{n//}^{f_N a}e^{-2q_{nf_N} z_N} - M_{22}^{//}\right)C_N$$

$$- \frac{\varepsilon_{f_1}}{q_{nf_1}}S_{nxz,a}T_{n//}^{af_N}M_{11}^{//}e^{-q_{na}z'}e^{q_{na}z_N}e^{-q_{nf_N} z_N} + \frac{\varepsilon_{f_1}}{q_{nf_1}}S_{nxz,a}T_{n//}^{af_N}M_{21}^{//}e^{-q_{na}z'}e^{q_{na}z_N}e^{-q_{nf_N} z_N}$$

$$= \frac{\varepsilon_s}{q_{ns}}\left(M_{11}^{//}R_{n//}^{f_N a}e^{-2q_{nf_N} z_N} + M_{12}^{//} + M_{21}^{//}R_{n//}^{f_N a}e^{-2q_{nf_N} z_N} + M_{22}^{//}\right)C_N$$

$$- \frac{\varepsilon_s}{q_{ns}}M_{11}^{//}S_{nxz,a}T_{n//}^{af_N}e^{-q_{na}z'}e^{q_{na}z_N}e^{-q_{nf_N} z_N} - \frac{\varepsilon_s}{q_{ns}}M_{21}^{//}S_{nxz,a}T_{n//}^{af_N}e^{-q_{na}z'}e^{q_{na}z_N}e^{-q_{nf_N} z_N}$$

$$\Rightarrow \left[\left(\frac{\varepsilon_s}{q_{ns}} - \frac{\varepsilon_{f_1}}{q_{nf_1}}\right)\left(M_{11}^{//}R_{n//}^{f_N a}e^{-2q_{nf_N} z_N} + M_{12}^{//}\right) + \left(\frac{\varepsilon_s}{q_{ns}} + \frac{\varepsilon_{f_1}}{q_{nf_1}}\right)\left(M_{21}^{//}R_{n//}^{f_N a}e^{-2q_{nf_N} z_N} + M_{22}^{//}\right)\right]C_N$$

$$= \left(\frac{\varepsilon_s}{q_{ns}} - \frac{\varepsilon_{f_1}}{q_{nf_1}}\right)S_{nxz,a}T_{n//}^{af_N}M_{11}^{//}e^{-q_{na}z'}e^{q_{na}z_N}e^{-q_{nf_N} z_N} + \left(\frac{\varepsilon_s}{q_{ns}} + \frac{\varepsilon_{f_1}}{q_{nf_1}}\right)S_{nxz,a}T_{n//}^{af_N}M_{21}^{//}e^{-q_{na}z'}e^{q_{na}z_N}e^{-q_{nf_N} z_N}$$

$$\Rightarrow C_N = \frac{S_{nxz,a}T_{n//}^{af_N}\left(R_{n//}^{sf_1}M_{11}^{//} + M_{21}^{//}\right)e^{-q_{na}z'}e^{q_{na}z_N}e^{-q_{nf_N} z_N}}{R_{n//}^{sf_1}\left(M_{11}^{//}R_{n//}^{f_N a}e^{-2q_{nf_N} z_N} + M_{12}^{//}\right) + \left(M_{21}^{//}R_{n//}^{f_N a}e^{-2q_{nf_N} z_N} + M_{22}^{//}\right)}$$

. $\tag{231}$

Hence we can get the other three coefficients as the following forms:

$$B_N = R_{n//}^{f_N a}C_N e^{-2q_{nf_N} z_N} - S_{nxz,a}T_{n//}^{af_N}e^{-q_{na}z'}e^{q_{na}z_N}e^{-q_{nf_N} z_N}$$

$$= \frac{R_{n//}^{f_N a}S_{nxz,a}T_{n//}^{af_N}\left(R_{n//}^{sf_1}M_{11}^{//} + M_{21}^{//}\right)e^{-q_{na}z'}e^{q_{na}z_N}e^{-q_{nf_N} z_N}}{R_{n//}^{sf_1}\left(M_{11}^{//}R_{n//}^{f_N a}e^{-2q_{nf_N} z_N} + M_{12}^{//}\right) + \left(M_{21}^{//}R_{n//}^{f_N a}e^{-2q_{nf_N} z_N} + M_{22}^{//}\right)}e^{-2q_{nf_N} z_N} - S_{nxz,a}T_{n//}^{af_N}e^{-q_{na}z'}e^{q_{na}z_N}e^{-q_{nf_N} z_N}$$

$$= \frac{-S_{nxz,a}e^{-q_{na}z'}e^{q_{na}z_N}e^{-q_{nf_N} z_N}T_{n//}^{af_N}\left(R_{n//}^{sf_1}M_{12}^{//} + M_{22}^{//}\right)}{R_{n//}^{sf_1}\left(M_{11}^{//}R_{n//}^{f_N a}e^{-2q_{nf_N} z_N} + M_{12}^{//}\right) + \left(M_{21}^{//}R_{n//}^{f_N a}e^{-2q_{nf_N} z_N} + M_{22}^{//}\right)}$$

, $\tag{232}$

$$A = T_{n//}^{f_N a} C_N e^{-q_{nf_N} z_N} e^{q_{na} z_N} + R_{n//}^{f_N a} S_{nxz,a} e^{-q_{na} z'} e^{2 q_{na} z_N}$$

$$= \frac{T_{n//}^{f_N a} T_{n//}^{af_N} S_{nxz,a} \left( R_{n//}^{sf_1} M_{11}^{//} + M_{21}^{//} \right) e^{-q_{na} z'} e^{2 q_{na} z_N} e^{-2 q_{nf_N} z_N}}{R_{n//}^{sf_1} \left( M_{11}^{//} R_{n//}^{f_N a} e^{-2 q_{nf_N} z_N} + M_{12}^{//} \right) + \left( M_{21}^{//} R_{n//}^{f_N a} e^{-2 q_{nf_N} z_N} + M_{22}^{//} \right)} + R_{n//}^{f_N a} S_{nxz,a} e^{-q_{na} z'} e^{2 q_{na} z_N}, \quad (233)$$

$$= S_{nxz,a} e^{-q_{na} z'} e^{2 q_{na} z_N} \frac{R_{n//}^{sf_1} \left( M_{11}^{//} + M_{21}^{//} \right) e^{-2 q_{nf_N} z_N} + R_{n//}^{f_N a} \left( R_{n//}^{sf_1} M_{12}^{//} + M_{22}^{//} \right)}{R_{n//}^{sf_1} \left( M_{11}^{//} R_{n//}^{f_N a} e^{-2 q_{nf_N} z_N} + M_{12}^{//} \right) + \left( M_{21}^{//} R_{n//}^{f_N a} e^{-2 q_{nf_N} z_N} + M_{22}^{//} \right)}$$

$$D = \left( M_{11}^{//} R_{n//}^{f_N a} e^{-2 q_{nf_N} z_N} + M_{12}^{//} + M_{21}^{//} R_{n//}^{f_N a} e^{-2 q_{nf_N} z_N} + M_{22}^{//} \right) C_N$$

$$- M_{11}^{//} S_{nxz,a} T_{n//}^{af_N} e^{-q_{na} z'} e^{q_{na} z_N} e^{-q_{nf_N} z_N} - M_{21}^{//} S_{nxz,a} T_{n//}^{af_N} e^{-q_{na} z'} e^{q_{na} z_N} e^{-q_{nf_N} z_N}$$

$$= \frac{S_{nxz,a} T_{n//}^{af_N} \left( M_{11}^{//} R_{n//}^{f_N a} e^{-2 q_{nf_N} z_N} + M_{12}^{//} + M_{21}^{//} R_{n//}^{f_N a} e^{-2 q_{nf_N} z_N} + M_{22}^{//} \right) \left( R_{n//}^{sf_1} M_{11}^{//} + M_{21}^{//} \right) e^{-q_{na} z'} e^{q_{na} z_N} e^{-q_{nf_N} z_N}}{R_{n//}^{sf_1} \left( M_{11}^{//} R_{n//}^{f_N a} e^{-2 q_{nf_N} z_N} + M_{12}^{//} \right) + \left( M_{21}^{//} R_{n//}^{f_N a} e^{-2 q_{nf_N} z_N} + M_{22}^{//} \right)}$$

$$- M_{11}^{//} S_{nxz,a} T_{n//}^{af_N} e^{-q_{na} z'} e^{q_{na} z_N} e^{-q_{nf_N} z_N} - M_{21}^{//} S_{nxz,a} T_{n//}^{af_N} e^{-q_{na} z'} e^{q_{na} z_N} e^{-q_{nf_N} z_N}$$

$$= \frac{-S_{nxz,a} T_{n//}^{af_N} T_{n//}^{f_1 s} \left( M_{22}^{//} M_{11}^{//} - M_{21}^{//} M_{12}^{//} \right) e^{-q_{na} z'} e^{q_{na} z_N} e^{-q_{nf_N} z_N}}{R_{n//}^{sf_1} \left( M_{11}^{//} R_{n//}^{f_N a} e^{-2 q_{nf_N} z_N} + M_{12}^{//} \right) + \left( M_{21}^{//} R_{n//}^{f_N a} e^{-2 q_{nf_N} z_N} + M_{22}^{//} \right)}$$

. (234)

According to Eq. (217), the coefficients $B_j$ and $C_j, j = 1, 2, ..., N-1$ are obtained and then we solve the coefficients in Eq. (226) completely. Next we consider the location of a point source is inside the film layer, that is $z_{k-1} < z' < z_k, k = 1, 2, ..., N$. Fixed the index $k$, the corresponding different equation in regions $z > z_N$, $z_{j-1} < z < z_j, j = N, N-1, ..., 2, 1$ and $z < 0$ are:

$$\begin{cases} \left( \partial_z^2 - q_{na}^2 \right) \tilde{g}_{nxz} = 0, z > z_N \\ \left( \partial_z^2 - q_{nf_j}^2 \right) \tilde{g}_{nxz} = -\frac{i k_n}{k_{f_k}^2} \delta_{jk} \partial_z \delta(z - z'), z_{j-1} < z < z_j, j = 1, 2, ..., N \\ \left( \partial_z^2 - q_{ns}^2 \right) \tilde{g}_{nxz} = 0, z < 0 \end{cases} \quad (235)$$

Furthermore, the solutions of Eq. (235) are:

$$\tilde{g}_{nxz} = \begin{cases} A e^{-q_{na} z}, z > z_N \\ S_{nxz, f_k} \operatorname{sgn}(z - z') e^{-q_{nf_k} |z - z'|} \delta_{jk} + B_j e^{q_{nf_j} z} + C_j e^{-q_{nf_j} z}, z_{j-1} < z < z_j, j = 1, 2, ..., N, \\ D e^{q_{ns} z}, z < 0 \end{cases}$$

(236)

where $S_{nxz, f_k} = -\frac{i k_n}{2 k_{f_k}^2}$. Using the boundary conditions of $\tilde{g}_{nxz}$ as shown in Eq. (213),

we have the following simulated equations:

<1> If $k = N$:

$$\begin{cases} Ae^{-q_{na}z_N} = S_{nxz,f_N} e^{q_{nf_N} z'} e^{-q_{nf_N} z_N} + B_N e^{q_{nf_N} z_N} + C_N e^{-q_{nf_N} z_N} \\ -S_{nxz,f_N} e^{-q_{nf_N} z'} e^{q_{nf_N} z_{N-1}} + B_N e^{q_{nf_N} z_{N-1}} + C_N e^{-q_{nf_N} z_{N-1}} = B_{N-1} e^{q_{nf_{N-1}} z_{N-1}} + C_{N-1} e^{-q_{nf_{N-1}} z_{N-1}} \\ B_j e^{q_{nf_j} z_j} + C_j e^{-q_{nf_j} z_j} = B_{j+1} e^{q_{nf_{j+1}} z_j} + C_{j+1} e^{-q_{nf_{j+1}} z_j}, \quad j = 1,2,...,N-2 \\ B_1 + C_1 = D \end{cases}$$ (237)

and

$$\begin{cases} -\dfrac{\varepsilon_a}{q_{na}} Ae^{-q_{na}z_N} = -\dfrac{\varepsilon_{f_N}}{q_{nf_N}} S_{nxz,f_N} e^{q_{nf_N} z'} e^{-q_{nf_N} z_N} + \dfrac{\varepsilon_{f_N}}{q_{nf_N}} \left( B_N e^{q_{nf_N} z_N} - C_N e^{-q_{nf_N} z_N} \right) \\ -\dfrac{\varepsilon_{f_N}}{q_{nf_N}} S_{nxz,f_N} e^{-q_{nf_N} z'} e^{q_{nf_N} z_{N-1}} + \dfrac{\varepsilon_{f_N}}{q_{nf_N}} \left[ B_N e^{q_{nf_N} z_{N-1}} - C_N e^{-q_{nf_N} z_{N-1}} \right] = \dfrac{\varepsilon_{f_{N-1}}}{q_{nf_{N-1}}} \left[ B_{N-1} e^{q_{nf_{N-1}} z_{N-1}} - C_{N-1} e^{-q_{nf_{N-1}} z_{N-1}} \right] \\ \dfrac{\varepsilon_{f_j}}{q_{nf_j}} \left[ B_j e^{q_{nf_j} z_j} - C_j e^{-q_{nf_j} z_j} \right] = \dfrac{\varepsilon_{f_{j+1}}}{q_{nf_{j+1}}} \left[ B_{j+1} e^{q_{nf_{j+1}} z_j} - C_{j+1} e^{-q_{nf_{j+1}} z_j} \right], \quad j = 1,2,...,N-2 \\ \dfrac{\varepsilon_{f_1}}{q_{nf_1}} \left( B_1 - C_1 \right) = \dfrac{\varepsilon_{ns}}{q_{ns}} D \end{cases}$$

. (238)

<2> If $k = N-1, N-2, ..., 2$:

$$\begin{cases} Ae^{-q_{na}z_N} = B_N e^{q_{nf_N} z_N} + C_N e^{-q_{nf_N} z_N} \\ B_{k+1} e^{q_{nf_{k+1}} z_k} + C_{k+1} e^{-q_{nf_{k+1}} z_k} = S_{nxz,f_k} e^{q_{nf_k} z'} e^{-q_{nf_k} z_k} + B_k e^{q_{nf_k} z_k} + C_k e^{-q_{nf_k} z_k} \\ -S_{nxz,f_k} e^{-q_{nf_k} z'} e^{q_{nf_k} z_{k-1}} + B_k e^{q_{nf_k} z_{k-1}} + C_k e^{-q_{nf_k} z_{k-1}} = B_{k-1} e^{q_{nf_{k-1}} z_{k-1}} + C_{k-1} e^{-q_{nf_{k-1}} z_{k-1}} \\ B_j e^{q_{nf_j} z_j} + C_j e^{-q_{nf_j} z_j} = B_{j+1} e^{q_{nf_{j+1}} z_j} + C_{j+1} e^{-q_{nf_{j+1}} z_j}, \quad j = 1,...,N-1, j \neq k-1, k \\ B_1 + C_1 = D \end{cases}$$ (239)

and

$$\begin{cases} -\dfrac{\varepsilon_a}{q_{na}} Ae^{-q_{na}z_N} = \dfrac{\varepsilon_{f_N}}{q_{nf_N}} \left( B_N e^{q_{nf_N} z_N} - C_N e^{-q_{nf_N} z_N} \right) \\ \dfrac{\varepsilon_{f_{k+1}}}{q_{nf_{k+1}}} \left[ B_{k+1} e^{q_{nf_{k+1}} z_k} - C_{k+1} e^{-q_{nf_{k+1}} z_k} \right] = -\dfrac{\varepsilon_{f_k}}{q_{nf_k}} S_{nxz,f_k} e^{q_{nf_k} z'} e^{-q_{nf_k} z_k} + \dfrac{\varepsilon_{f_k}}{q_{nf_k}} \left[ B_k e^{q_{nf_k} z_k} - C_k e^{-q_{nf_k} z_k} \right] \\ -\dfrac{\varepsilon_{f_k}}{q_{nf_k}} S_{nxz,f_k} e^{-q_{nf_k} z'} e^{q_{nf_k} z_{k-1}} + \dfrac{\varepsilon_{f_k}}{q_{nf_k}} \left[ B_k e^{q_{nf_k} z_{k-1}} - C_k e^{-q_{nf_k} z_{k-1}} \right] = \dfrac{\varepsilon_{f_{k-1}}}{q_{nf_{k-1}}} \left[ B_{k-1} e^{q_{nf_{k-1}} z_{k-1}} - C_{k-1} e^{-q_{nf_{k-1}} z_{k-1}} \right] \\ \dfrac{\varepsilon_{f_j}}{q_{nf_j}} \left[ B_j e^{q_{nf_j} z_j} - C_j e^{-q_{nf_j} z_j} \right] = \dfrac{\varepsilon_{f_{j+1}}}{q_{nf_{j+1}}} \left[ B_{j+1} e^{q_{nf_{j+1}} z_j} - C_{j+1} e^{-q_{nf_{j+1}} z_j} \right], \quad j = 1,...,N-1, j \neq k-1, k \\ \dfrac{\varepsilon_{f_1}}{q_{nf_1}} \left( B_1 - C_1 \right) = \dfrac{\varepsilon_s}{q_{ns}} D \end{cases}$$

. (240)

<3> If $k = 1$:

$$\begin{cases} Ae^{-q_{na}z_N} = B_N e^{q_{nf_N}z_N} + C_N e^{-q_{nf_N}z_N} \\ B_j e^{q_{nf_j}z_j} + C_j e^{-q_{nf_j}z_j} = B_{j+1} e^{q_{nf_{j+1}}z_j} + C_{j+1} e^{-q_{nf_{j+1}}z_j}, j = 2,...,N-1 \\ B_2 e^{q_{nf_2}z_1} + C_2 e^{-q_{nf_2}z_1} = S_{nxz,f_1} e^{q_{nf_1}z'} e^{-q_{nf_1}z_1} + B_1 e^{q_{nf_1}z_1} + C_1 e^{-q_{nf_1}z_1} \\ \qquad\qquad -S_{nxz,f_1} e^{-q_{nf_1}z'} + B_1 + C_1 = D \end{cases}, \quad (241)$$

and

$$\begin{cases} -\dfrac{\varepsilon_a}{q_{na}} A e^{-q_{na}z_N} = \dfrac{\varepsilon_{f_N}}{q_{nf_N}} \left( B_N e^{q_{nf_N}z_N} - C_N e^{-q_{nf_N}z_N} \right) \\ \dfrac{\varepsilon_{f_j}}{q_{nf_j}} \left[ B_j e^{q_{nf_j}z_j} - C_j e^{-q_{nf_j}z_j} \right] = \dfrac{\varepsilon_{f_{j+1}}}{q_{nf_{j+1}}} \left[ B_{j+1} e^{q_{nf_{j+1}}z_j} - C_{j+1} e^{-q_{nf_{j+1}}z_j} \right], j = 2,...,N-1 \\ \dfrac{\varepsilon_{f_2}}{q_{nf_2}} \left[ B_2 e^{q_{nf_2}z_1} - C_2 e^{-q_{nf_2}z_1} \right] = -\dfrac{\varepsilon_{f_1}}{q_{nf_1}} S_{nxz,f_1} e^{q_{nf_1}z'} e^{-q_{nf_1}z_1} + \dfrac{\varepsilon_{f_1}}{q_{nf_1}} \left[ B_1 e^{q_{nf_1}z_1} - C_1 e^{-q_{nf_1}z_1} \right] \\ \qquad -\dfrac{\varepsilon_{f_1}}{q_{nf_1}} S_{nxz,f_1} e^{-q_{nf_1}z'} + \dfrac{\varepsilon_{f_1}}{q_{nf_1}} (B_1 - C_1) = \dfrac{\varepsilon_s}{q_{ns}} D \end{cases}. \quad (242)$$

Furthermore, we simplify these above equations and show as the following forms:

<1> If $k = N$:

$$-\dfrac{\varepsilon_a}{q_{na}} \left( S_{nxz,f_N} e^{q_{nf_N}z'} e^{-q_{nf_N}z_N} + B_N e^{q_{nf_N}z_N} + C_N e^{-q_{nf_N}z_N} \right)$$

$$= -\dfrac{\varepsilon_{f_N}}{q_{nf_N}} S_{nxz,f_N} e^{q_{nf_N}z'} e^{-q_{nf_N}z_N} + \dfrac{\varepsilon_{f_N}}{q_{nf_N}} \left( B_N e^{q_{nf_N}z_N} - C_N e^{-q_{nf_N}z_N} \right)$$

$$\Rightarrow \left( \dfrac{\varepsilon_{f_N}}{q_{nf_N}} + \dfrac{\varepsilon_a}{q_{na}} \right) B_N = \left( \dfrac{\varepsilon_{f_N}}{q_{nf_N}} - \dfrac{\varepsilon_a}{q_{na}} \right) C_N e^{-2q_{nf_N}z_N} + \left( \dfrac{\varepsilon_{f_N}}{q_{nf_N}} - \dfrac{\varepsilon_a}{q_{na}} \right) S_{nxz,f_N} e^{q_{nf_N}z'} e^{-2q_{nf_N}z_N}$$

$$\Rightarrow B_N = \dfrac{\varepsilon_{f_N} q_{na} - \varepsilon_a q_{nf_N}}{\varepsilon_{f_N} q_{na} + \varepsilon_a q_{nf_N}} C_N e^{-2q_{nf_N}z_N} + \dfrac{\varepsilon_{f_N} q_{na} - \varepsilon_a q_{nf_N}}{\varepsilon_{f_N} q_{na} + \varepsilon_a q_{nf_N}} S_{nxz,f_N} e^{q_{nf_N}z'} e^{-2q_{nf_N}z_N} \quad , \quad (243)$$

$$= R_{n//}^{f_N a} C_N e^{-2q_{nf_N}z_N} + R_{n//}^{f_N a} S_{nxz,f_N} e^{q_{nf_N}z'} e^{-2q_{nf_N}z_N}$$

$$\Rightarrow A = S_{nxz,f_N} e^{q_{nf_N}z'} e^{-q_{nf_N}z_N} e^{q_{na}z_N} + B_N e^{q_{nf_N}z_N} e^{q_{na}z_N} + C_N e^{-q_{nf_N}z_N} e^{q_{na}z_N}$$

$$= S_{nxz,f_N} \left(1 + R_{n//}^{f_N a}\right) e^{q_{nf_N}z'} e^{-q_{nf_N}z_N} e^{q_{na}z_N} + \left(1 + R_{n//}^{f_N a}\right) C_N e^{-q_{nf_N}z_N} e^{q_{na}z_N}$$

$$= S_{nxz,f_N} T_{n//}^{f_N a} e^{q_{nf_N}z'} e^{-q_{nf_N}z_N} e^{q_{na}z_N} + T_{n//}^{f_N a} C_N e^{-q_{nf_N}z_N} e^{q_{na}z_N}$$

and

$$\begin{cases} -S_{nxz,f_N} e^{-q_{nf_N} z'} e^{q_{nf_N} z_{N-1}} + B_N e^{q_{nf_N} z_{N-1}} + C_N e^{-q_{nf_N} z_{N-1}} = B_{N-1} e^{q_{nf_{N-1}} z_{N-1}} + C_{N-1} e^{-q_{nf_{N-1}} z_{N-1}} \\ -\dfrac{\varepsilon_{f_N}}{q_{nf_N}} S_{nxz,f_N} e^{-q_{nf_N} z'} e^{q_{nf_N} z_{N-1}} + \dfrac{\varepsilon_{f_N}}{q_{nf_N}} \left[ B_N e^{q_{nf_N} z_{N-1}} - C_N e^{-q_{nf_N} z_{N-1}} \right] = \dfrac{\varepsilon_{f_{N-1}}}{q_{nf_{N-1}}} \left[ B_{N-1} e^{q_{nf_{N-1}} z_{N-1}} - C_{N-1} e^{-q_{nf_{N-1}} z_{N-1}} \right] \end{cases}$$

$$\Rightarrow \begin{pmatrix} B_N \\ C_N \end{pmatrix} = \left[ T''_{N-1} \right]^{-1} \begin{pmatrix} B_{N-1} \\ C_{N-1} \end{pmatrix} + \begin{pmatrix} e^{q_{nf_N} z_{N-1}} & e^{-q_{nf_N} z_{N-1}} \\ \dfrac{\varepsilon_{f_N}}{q_{nf_N}} e^{q_{nf_N} z_{N-1}} & -\dfrac{\varepsilon_{f_N}}{q_{nf_N}} e^{-q_{nf_N} z_{N-1}} \end{pmatrix}^{-1} \begin{pmatrix} 1 \\ \dfrac{\varepsilon_{f_N}}{q_{nf_N}} \end{pmatrix} S_{nxz,f_N} e^{-q_{nf_N} z'} e^{q_{nf_N} z_{N-1}}$$

$$= \left[ T''_{N-1} \right]^{-1} \begin{pmatrix} B_{N-1} \\ C_{N-1} \end{pmatrix} + \dfrac{q_{nf_N}}{2\varepsilon_{f_N}} \begin{pmatrix} \dfrac{\varepsilon_{f_N}}{q_{nf_N}} e^{-q_{nf_N} z_{N-1}} & e^{-q_{nf_N} z_{N-1}} \\ \dfrac{\varepsilon_{f_N}}{q_{nf_N}} e^{q_{nf_N} z_{N-1}} & -e^{q_{nf_N} z_{N-1}} \end{pmatrix} \begin{pmatrix} 1 \\ \dfrac{\varepsilon_{f_N}}{q_{nf_N}} \end{pmatrix} S_{nxz,f_N} e^{-q_{nf_N} z'} e^{q_{nf_N} z_{N-1}}$$

$$= \left[ T''_{N-1} \right]^{-1} \begin{pmatrix} B_{N-1} \\ C_{N-1} \end{pmatrix} + \begin{pmatrix} 1 \\ 0 \end{pmatrix} S_{nxz,f_N} e^{-q_{nf_N} z'}$$

$$\Rightarrow \begin{pmatrix} B_{N-1} \\ C_{N-1} \end{pmatrix} = T''_{N-1} \begin{pmatrix} B_N \\ C_N \end{pmatrix} - T''_{N-1} \begin{pmatrix} 1 \\ 0 \end{pmatrix} S_{nxz,f_N} e^{-q_{nf_N} z'}$$

. (244)

Combining Eq. (244) with the transfer matrix method which is described in Eqs. (217)-(219), we obtain:

$$\begin{pmatrix} B_1 \\ C_1 \end{pmatrix} = T''_1 T''_2 \bullet \bullet \bullet T''_{N-2} \begin{pmatrix} B_{N-1} \\ C_{N-1} \end{pmatrix} = T''_1 T''_2 \bullet \bullet \bullet T''_{N-2} T''_{N-1} \begin{pmatrix} B_N \\ C_N \end{pmatrix} - T''_1 T''_2 \bullet \bullet \bullet T''_{N-2} T''_{N-1} \begin{pmatrix} 1 \\ 0 \end{pmatrix} S_{nxz,f_N} e^{-q_{nf_N} z'}$$

$$\equiv \begin{pmatrix} M''_{11} & M''_{12} \\ M''_{21} & M''_{22} \end{pmatrix} \begin{pmatrix} B_N \\ C_N \end{pmatrix} - \begin{pmatrix} M''_{11} & M''_{12} \\ M''_{21} & M''_{22} \end{pmatrix} \begin{pmatrix} 1 \\ 0 \end{pmatrix} S_{nxz,f_N} e^{-q_{nf_N} z'}$$

$$= \begin{pmatrix} M''_{11} B_N + M''_{12} C_N \\ M''_{21} B_N + M''_{22} C_N \end{pmatrix} - \begin{pmatrix} M''_{11} \\ M''_{21} \end{pmatrix} S_{nxz,f_N} e^{-q_{nf_N} z'}$$

, (245)

$$\begin{pmatrix} B_j \\ C_j \end{pmatrix} = T''_j \bullet \bullet \bullet T''_{N-2} \begin{pmatrix} B_{N-1} \\ C_{N-1} \end{pmatrix} = T''_j \bullet \bullet \bullet T''_{N-2} T''_{N-1} \begin{pmatrix} B_N \\ C_N \end{pmatrix} - T''_j \bullet \bullet \bullet T''_{N-2} T''_{N-1} \begin{pmatrix} 1 \\ 0 \end{pmatrix} S_{nxz,f_N} e^{-q_{nf_N} z'}$$

$$\equiv \begin{pmatrix} M''_{j,11} & M''_{j,12} \\ M''_{j,21} & M''_{j,22} \end{pmatrix} \begin{pmatrix} B_N \\ C_N \end{pmatrix} - \begin{pmatrix} M''_{j,11} & M''_{j,12} \\ M''_{j,21} & M''_{j,22} \end{pmatrix} \begin{pmatrix} 1 \\ 0 \end{pmatrix} S_{nxz,f_N} e^{-q_{nf_N} z'}$$

$$= \begin{pmatrix} M''_{j,11} B_N + M''_{j,12} C_N \\ M''_{j,21} B_N + M''_{j,22} C_N \end{pmatrix} - \begin{pmatrix} M''_{j,11} \\ M''_{j,21} \end{pmatrix} S_{nxz,f_N} e^{-q_{nf_N} z'}, \; j = 2, 3, ..., N-2$$

, (246)

and we obtain:

$$D = B_1 + C_1 = M_{11}'' R_{n//}^{f_N a} C_N e^{-2q_{nf_N} z_N} + M_{11}'' R_{n//}^{f_N a} S_{nxz,f_N} e^{q_{nf_N} z'} e^{-2q_{nf_N} z_N} + M_{12}'' C_N - M_{11}'' S_{nxz,f_N} e^{-q_{nf_N} z'}$$

$$+ M_{21}'' R_{n//}^{f_N a} C_N e^{-2q_{nf_N} z_N} + M_{21}'' R_{n//}^{f_N a} S_{nxz,f_N} e^{q_{nf_N} z'} e^{-2q_{nf_N} z_N} + M_{22}'' C_N - M_{21}'' S_{nxz,f_N} e^{-q_{nf_N} z'}$$

$$= \left( M_{11}'' R_{n//}^{f_N a} e^{-2q_{nf_N} z_N} + M_{12}'' + M_{21}'' R_{n//}^{f_N a} e^{-2q_{nf_N} z_N} + M_{22}'' \right) C_N + M_{11}'' R_{n//}^{f_N a} S_{nxz,f_N} e^{q_{nf_N} z'} e^{-2q_{nf_N} z_N}$$

$$- M_{11}'' S_{nxz,f_N} e^{-q_{nf_N} z'} + M_{21}'' R_{n//}^{f_N a} S_{nxz,f_N} e^{q_{nf_N} z'} e^{-2q_{nf_N} z_N} - M_{21}'' S_{nxz,f_N} e^{-q_{nf_N} z'}$$

. (247)

Furthermore, we get the following form to solve the coefficient $C_N$:

$$\frac{\varepsilon_{f_1}}{q_{nf_1}} (B_1 - C_1) = \frac{\varepsilon_{ns}}{q_{ns}} D = \frac{\varepsilon_{ns}}{q_{ns}} (B_1 + C_1)$$

$$\Rightarrow \left( \frac{\varepsilon_{ns}}{q_{ns}} - \frac{\varepsilon_{f_1}}{q_{nf_1}} \right) B_1 = -\left( \frac{\varepsilon_{ns}}{q_{ns}} + \frac{\varepsilon_{f_1}}{q_{nf_1}} \right) C_1 \Rightarrow C_1 = -\frac{\varepsilon_{ns} q_{nf_1} - \varepsilon_{f_1} q_{ns}}{\varepsilon_{ns} q_{nf_1} + \varepsilon_{f_1} q_{ns}} B_1 = R_{n//}^{sf_1} B_1$$

$$\Rightarrow M_{21}'' R_{n//}^{f_N a} C_N e^{-2q_{nf_N} z_N} + M_{21}'' R_{n//}^{f_N a} S_{nxz,f_N} e^{q_{nf_N} z'} e^{-2q_{nf_N} z_N} + M_{22}'' C_N - M_{21}'' S_{nxz,f_N} e^{-q_{nf_N} z'}$$

$$= R_{n//}^{sf_1} \left( M_{11}'' R_{n//}^{f_N a} C_N e^{-2q_{nf_N} z_N} + M_{11}'' R_{n//}^{f_N a} S_{nxz,f_N} e^{q_{nf_N} z'} e^{-2q_{nf_N} z_N} + M_{12}'' C_N - M_{11}'' S_{nxz,f_N} e^{-q_{nf_N} z'} \right)$$

$$\Rightarrow \left( M_{21}'' R_{n//}^{f_N a} e^{-2q_{nf_N} z_N} + M_{22}'' - R_{n//}^{sf_1} M_{11}'' R_{n//}^{f_N a} e^{-2q_{nf_N} z_N} - R_{n//}^{sf_1} M_{12}'' \right) C_N$$

$$= -M_{21}'' R_{n//}^{f_N a} S_{nxz,f_N} e^{q_{nf_N} z'} e^{-2q_{nf_N} z_N} + M_{21}'' S_{nxz,f_N} e^{-q_{nf_N} z'} + R_{n//}^{sf_1} M_{11}'' R_{n//}^{f_N a} S_{nxz,f_N} e^{q_{nf_N} z'} e^{-2q_{nf_N} z_N}$$

$$- R_{n//}^{sf_1} M_{11}'' S_{nxz,f_N} e^{-q_{nf_N} z'}$$

$$\Rightarrow C_N = S_{nxz,f_N} \frac{R_{n//}^{sf_1} R_{n//}^{f_N a} M_{11}'' e^{q_{nf_N} z'} e^{-2q_{nf_N} z_N} - R_{n//}^{f_N a} M_{21}'' e^{q_{nf_N} z'} e^{-2q_{nf_N} z_N} + M_{21}'' e^{-q_{nf_N} z'} - R_{n//}^{sf_1} M_{11}'' e^{-q_{nf_N} z'}}{R_{n//}^{f_N a} M_{21}'' e^{-2q_{nf_N} z_N} + M_{22}'' - R_{n//}^{sf_1} R_{n//}^{f_N a} M_{11}'' e^{-2q_{nf_N} z_N} - R_{n//}^{sf_1} M_{12}''}$$

$$= S_{nxz,f_N} \frac{\left( R_{n//}^{sf_1} M_{11}'' - M_{21}'' \right) \left( R_{n//}^{f_N a} e^{q_{nf_N} z'} e^{-2q_{nf_N} z_N} - e^{-q_{nf_N} z'} \right)}{R_{n//}^{f_N a} M_{21}'' e^{-2q_{nf_N} z_N} + M_{22}'' - R_{n//}^{sf_1} R_{n//}^{f_N a} M_{11}'' e^{-2q_{nf_N} z_N} - R_{n//}^{sf_1} M_{12}''}$$

. (248)

Hence we can get the other three coefficients as the following forms:

$$B_N = \frac{R_{n//}^{f_N a} S_{nxz,f_N} \left( R_{n//}^{sf_1} M_{11}'' - M_{21}'' \right) \left( R_{n//}^{f_N a} e^{q_{nf_N} z'} e^{-2q_{nf_N} z_N} - e^{-q_{nf_N} z'} \right) e^{-2q_{nf_N} z_N}}{R_{n//}^{f_N a} M_{21}'' e^{-2q_{nf_N} z_N} + M_{22}'' - R_{n//}^{sf_1} R_{n//}^{f_N a} M_{11}'' e^{-2q_{nf_N} z_N} - R_{n//}^{sf_1} M_{12}''} + R_{n//}^{f_N a} S_{nxz,f_N} e^{q_{nf_N} z'} e^{-2q_{nf_N} z_N}$$

$$= R_{n//}^{f_N a} S_{nxz,f_N} e^{-2q_{nf_N} z_N} \frac{\left( M_{21}'' - R_{n//}^{sf_1} M_{11}'' \right) e^{-q_{nf_N} z'} + \left( M_{22}'' - R_{n//}^{sf_1} M_{12}'' \right) e^{q_{nf_N} z'}}{R_{n//}^{f_N a} \left( M_{21}'' - R_{n//}^{sf_1} M_{11}'' \right) e^{-2q_{nf_N} z_N} + M_{22}'' - R_{n//}^{sf_1} M_{12}''}$$

, (249)

$$A = \frac{T_{n//}^{f_N a} S_{nxz,f_N} \left( R_{n//}^{sf_1} M_{11}'' - M_{21}'' \right) \left( R_{n//}^{f_N a} e^{q_{nf_N} z'} e^{-2q_{nf_N} z_N} - e^{-q_{nf_N} z'} \right) e^{-q_{nf_N} z_N} e^{q_{na} z_N}}{R_{n//}^{f_N a} M_{21}'' e^{-2q_{nf_N} z_N} + M_{22}'' - R_{n//}^{sf_1} R_{n//}^{f_N a} M_{11}'' e^{-2q_{nf_N} z_N} - R_{n//}^{sf_1} M_{12}''} + S_{nxz,f_N} T_{n//}^{f_N a} e^{q_{nf_N} z'} e^{-q_{nf_N} z_N} e^{q_{na} z_N}$$

$$= T_{n//}^{f_N a} S_{nxz,f_N} e^{-q_{nf_N} z_N} e^{q_{na} z_N} \frac{\left( M_{21}'' - R_{n//}^{sf_1} M_{11}'' \right) e^{-q_{nf_N} z'} + \left( M_{22}'' - R_{n//}^{sf_1} M_{12}'' \right) e^{q_{nf_N} z'}}{R_{n//}^{f_N a} \left( M_{21}'' - R_{n//}^{sf_1} M_{11}'' \right) e^{-2q_{nf_N} z_N} + M_{22}'' - R_{n//}^{sf_1} M_{12}''}$$

, (250)

$$D = \left(M_{11}^{//}R_{n//}^{f_N a}e^{-2q_{nf_N}z_N} + M_{12}^{//} + M_{21}^{//}R_{n//}^{f_N a}e^{-2q_{nf_N}z_N} + M_{22}^{//}\right)C_N + M_{11}^{//}R_{n//}^{f_N a}S_{nxz,f_N}e^{q_{nf_N}z'}e^{-2q_{nf_N}z_N}$$

$$-M_{11}^{//}S_{nxz,f_N}e^{-q_{nf_N}z'} + M_{21}^{//}R_{n//}^{f_N a}S_{nxz,f_N}e^{q_{nf_N}z'}e^{-2q_{nf_N}z_N} - M_{21}^{//}S_{nxz,f_N}e^{-q_{nf_N}z'}$$

$$= S_{nxz,f_N}\frac{\left(M_{11}^{//}R_{n//}^{f_N a}e^{-2q_{nf_N}z_N} + M_{12}^{//} + M_{21}^{//}R_{n//}^{f_N a}e^{-2q_{nf_N}z_N} + M_{22}^{//}\right)\left(R_{n//}^{sf_1}M_{11}^{//} - M_{21}^{//}\right)\left(R_{n//}^{f_N a}e^{q_{nf_N}z'}e^{-2q_{nf_N}z_N} - e^{-q_{nf_N}z'}\right)}{R_{n//}^{f_N a}M_{21}^{//}e^{-2q_{nf_N}z_N} + M_{22}^{//} - R_{n//}^{sf_1}R_{n//}^{f_N a}M_{11}^{//}e^{-2q_{nf_N}z_N} - R_{n//}^{sf_1}M_{12}^{//}}$$

$$+M_{11}^{//}R_{n//}^{f_N a}S_{nxz,f_N}e^{q_{nf_N}z'}e^{-2q_{nf_N}z_N} - M_{11}^{//}S_{nxz,f_N}e^{-q_{nf_N}z'} + M_{21}^{//}R_{n//}^{f_N a}S_{nxz,f_N}e^{q_{nf_N}z'}e^{-2q_{nf_N}z_N} - M_{21}^{//}S_{nxz,f_N}e^{-q_{nf_N}z'}$$

$$= S_{nxz,f_N}T_{n//}^{sf_1}\frac{\left(M_{22}^{//}M_{11}^{//} - M_{12}^{//}M_{21}^{//}\right)R_{n//}^{f_N a}e^{q_{nf_N}z'}e^{-2q_{nf_N}z_N} - \left(M_{22}^{//}M_{11}^{//} - M_{12}^{//}M_{21}^{//}\right)e^{-q_{nf_N}z'}}{R_{n//}^{f_N a}M_{21}^{//}e^{-2q_{nf_N}z_N} + M_{22}^{//} - R_{n//}^{sf_1}R_{n//}^{f_N a}M_{11}^{//}e^{-2q_{nf_N}z_N} - R_{n//}^{sf_1}M_{12}^{//}}$$

. (251)

Then we solve the coefficients in Eq. (236) completely.

<2> If $k = 2, 3, ..., N-1$

$$-\frac{\varepsilon_a}{q_{na}}\left(B_N e^{q_{nf_N}z_N} + C_N e^{-q_{nf_N}z_N}\right) = \frac{\varepsilon_{f_N}}{q_{nf_N}}\left(B_N e^{q_{nf_N}z_N} - C_N e^{-q_{nf_N}z_N}\right)$$

$$\Rightarrow \left(\frac{\varepsilon_{f_N}}{q_{nf_N}} + \frac{\varepsilon_a}{q_{na}}\right)B_N = \left(\frac{\varepsilon_{f_N}}{q_{nf_N}} - \frac{\varepsilon_a}{q_{na}}\right)C_N e^{-2q_{nf_N}z_N} \quad , \quad (252)$$

$$\Rightarrow B_N = \frac{\varepsilon_{f_N}q_{na} - \varepsilon_a q_{nf_N}}{\varepsilon_{f_N}q_{na} + \varepsilon_a q_{nf_N}}C_N e^{-2q_{nf_N}z_N} = R_{n//}^{f_N a}C_N e^{-2q_{nf_N}z_N}$$

$$\Rightarrow A = R_{n//}^{f_N a}C_N e^{-q_{nf_N}z_N}e^{q_{na}z_N} + C_N e^{-q_{nf_N}z_N}e^{q_{na}z_N} = T_{n//}^{f_N a}C_N e^{-q_{nf_N}z_N}e^{q_{na}z_N}$$

which is the same as Eq. (216) and

$$\begin{cases} B_{k+1}e^{q_{nf_{k+1}}z_k} + C_{k+1}e^{-q_{nf_{k+1}}z_k} = S_{nxz,f_k}e^{q_{nf_k}z'}e^{-q_{nf_k}z_k} + B_k e^{q_{nf_k}z_k} + C_k e^{-q_{nf_k}z_k} \\ \frac{\varepsilon_{f_{k+1}}}{q_{nf_{k+1}}}\left[B_{k+1}e^{q_{nf_{k+1}}z_k} - C_{k+1}e^{-q_{nf_{k+1}}z_k}\right] = -\frac{\varepsilon_{f_k}}{q_{nf_k}}S_{nxz,f_k}e^{q_{nf_k}z'}e^{-q_{nf_k}z_k} + \frac{\varepsilon_{f_k}}{q_{nf_k}}\left[B_k e^{q_{nf_k}z_k} - C_k e^{-q_{nf_k}z_k}\right] \end{cases}$$

$$\Rightarrow \begin{pmatrix} B_k \\ C_k \end{pmatrix} = T_k^{//}\begin{pmatrix} B_{k+1} \\ C_{k+1} \end{pmatrix} + \begin{pmatrix} e^{q_{nf_k}z_k} & e^{-q_{nf_k}z_k} \\ \frac{\varepsilon_{f_k}}{q_{nf_k}}e^{q_{nf_k}z_k} & -\frac{\varepsilon_{f_k}}{q_{nf_k}}e^{-q_{nf_k}z_k} \end{pmatrix}^{-1}\begin{pmatrix} -1 \\ \frac{\varepsilon_{f_k}}{q_{nf_k}} \end{pmatrix}S_{nxz,f_k}e^{q_{nf_k}z'}e^{-q_{nf_k}z_k}$$

$$= T_k^{//}\begin{pmatrix} B_{k+1} \\ C_{k+1} \end{pmatrix} + \frac{q_{nf_k}}{2\varepsilon_{f_k}}\begin{pmatrix} \frac{\varepsilon_{f_k}}{q_{nf_k}}e^{-q_{nf_k}z_k} & e^{-q_{nf_k}z_k} \\ \frac{\varepsilon_{f_k}}{q_{nf_k}}e^{q_{nf_k}z_k} & -e^{q_{nf_k}z_k} \end{pmatrix}\begin{pmatrix} -1 \\ \frac{\varepsilon_{f_k}}{q_{nf_k}} \end{pmatrix}S_{nxz,f_k}e^{q_{nf_k}z'}e^{-q_{nf_k}z_k}$$

$$= T_k^{//}\begin{pmatrix} B_{k+1} \\ C_{k+1} \end{pmatrix} - \begin{pmatrix} 0 \\ 1 \end{pmatrix}S_{nxz,f_k}e^{q_{nf_k}z'}$$

, (253)

and

$$\begin{cases} -S_{nxz,f_k} e^{-q_{nf_k} z'} e^{q_{nf_k} z_{k-1}} + B_k e^{q_{nf_k} z_{k-1}} + C_k e^{-q_{nf_k} z_{k-1}} = B_{k-1} e^{q_{nf_{k-1}} z_{k-1}} + C_{k-1} e^{-q_{nf_{k-1}} z_{k-1}} \\ -\dfrac{\varepsilon_{f_k}}{q_{nf_k}} S_{nxz,f_k} e^{-q_{nf_k} z'} e^{q_{nf_k} z_{k-1}} + \dfrac{\varepsilon_{f_k}}{q_{nf_k}} \left[ B_k e^{q_{nf_k} z_{k-1}} - C_k e^{-q_{nf_k} z_{k-1}} \right] = \dfrac{\varepsilon_{f_{k-1}}}{q_{nf_{k-1}}} \left[ B_{k-1} e^{q_{nf_{k-1}} z_{k-1}} - C_{k-1} e^{-q_{nf_{k-1}} z_{k-1}} \right] \end{cases}$$

$$\begin{pmatrix} B_k \\ C_k \end{pmatrix} = \left[ T''_{k-1} \right]^{-1} \begin{pmatrix} B_{k-1} \\ C_{k-1} \end{pmatrix} + \begin{pmatrix} e^{q_{nf_k} z_{k-1}} & e^{-q_{nf} z_{k-1}} \\ \dfrac{\varepsilon_{f_k}}{q_{nf_k}} e^{q_{nf_k} z_{k-1}} & -\dfrac{\varepsilon_{f_k}}{q_{nf_k}} e^{-q_{nf_k} z_{k-1}} \end{pmatrix}^{-1} \begin{pmatrix} 1 \\ \dfrac{\varepsilon_{f_k}}{q_{nf_k}} \end{pmatrix} S_{nxz,f_k} e^{-q_{nf_k} z'} e^{q_{nf_k} z_{k-1}}$$

$$= \left[ T''_{k-1} \right]^{-1} \begin{pmatrix} B_{k-1} \\ C_{k-1} \end{pmatrix} + \dfrac{q_{nf_k}}{2\varepsilon_{f_k}} \begin{pmatrix} \dfrac{\varepsilon_{f_k}}{q_{nf_k}} e^{-q_{nf_k} z_{k-1}} & e^{-q_{nf} z_{k-1}} \\ \dfrac{\varepsilon_{f_k}}{q_{nf_k}} e^{q_{nf_k} z_{k-1}} & -e^{q_{nf_k} z_{k-1}} \end{pmatrix} \begin{pmatrix} 1 \\ \dfrac{\varepsilon_{f_k}}{q_{nf_k}} \end{pmatrix} S_{nxz,f_k} e^{-q_{nf_k} z'} e^{q_{nf_k} z_{k-1}}$$

$$= \left[ T''_{k-1} \right]^{-1} \begin{pmatrix} B_{k-1} \\ C_{k-1} \end{pmatrix} + \begin{pmatrix} 1 \\ 0 \end{pmatrix} S_{nxz,f_k} e^{-q_{nf_k} z'}$$

$$\Rightarrow \begin{pmatrix} B_{k-1} \\ C_{k-1} \end{pmatrix} = T''_{k-1} \begin{pmatrix} B_k \\ C_k \end{pmatrix} - T''_{k-1} \begin{pmatrix} 1 \\ 0 \end{pmatrix} S_{nxz,f_k} e^{-q_{nf_k} z'}$$

.
(254)

Combining Eqs. (253)-(254) with the transfer matrix method which is described in Eqs. (217)-(219), we obtain:

$$\begin{pmatrix} B_1 \\ C_1 \end{pmatrix} = T''_1 T''_2 \bullet \bullet T''_{k-2} \begin{pmatrix} B_{k-1} \\ C_{k-1} \end{pmatrix} = T''_1 T''_2 \bullet \bullet T''_{k-2} T''_{k-1} \begin{pmatrix} B_k \\ C_k \end{pmatrix} - T''_1 T''_2 \bullet \bullet T''_{k-2} T''_{k-1} \begin{pmatrix} 1 \\ 0 \end{pmatrix} S_{nxz,f_k} e^{-q_{nf_k} z'}$$

$$= T''_1 T''_2 \bullet \bullet T''_{k-2} T''_{k-1} T''_k \begin{pmatrix} B_{k+1} \\ C_{k+1} \end{pmatrix} - T''_1 T''_2 \bullet \bullet T''_{k-2} T''_{k-1} \begin{pmatrix} 0 \\ 1 \end{pmatrix} S_{nxz,f_k} e^{q_{nf_k} z'} - T''_1 T''_2 \bullet \bullet T''_{k-2} T''_{k-1} \begin{pmatrix} 1 \\ 0 \end{pmatrix} S_{nxz,f_k} e^{-q_{nf_k} z'}$$

$$\equiv \begin{pmatrix} M''_{11} & M''_{12} \\ M''_{21} & M''_{22} \end{pmatrix} \begin{pmatrix} B_N \\ C_N \end{pmatrix} - \begin{pmatrix} M''_{11,k-1} & M''_{12,k-1} \\ M''_{21,k-1} & M''_{22,k-1} \end{pmatrix} \begin{pmatrix} 0 \\ 1 \end{pmatrix} S_{nxz,f_k} e^{q_{nf_k} z'} - \begin{pmatrix} M''_{11,k-1} & M''_{12,k-1} \\ M''_{21,k-1} & M''_{22,k-1} \end{pmatrix} \begin{pmatrix} 1 \\ 0 \end{pmatrix} S_{nxz,f_k} e^{-q_{nf_k} z'}$$

$$= \begin{pmatrix} M''_{11} B_N + M''_{12} C_N \\ M''_{21} B_N + M''_{22} C_N \end{pmatrix} - \begin{pmatrix} M''_{12,k-1} \\ M''_{22,k-1} \end{pmatrix} S_{nxz,f_k} e^{q_{nf_k} z'} - \begin{pmatrix} M''_{11,k-1} \\ M''_{21,k-1} \end{pmatrix} S_{nxz,f_k} e^{-q_{nf_k} z'}$$

, (255)

$$\begin{pmatrix} B_j \\ C_j \end{pmatrix} = T''_j \bullet \bullet T''_{k-2} \begin{pmatrix} B_{k-1} \\ C_{k-1} \end{pmatrix} = T''_j \bullet \bullet T''_{k-2} T''_{k-1} \begin{pmatrix} B_k \\ C_k \end{pmatrix} - T''_j \bullet \bullet T''_{k-2} T''_{k-1} \begin{pmatrix} 1 \\ 0 \end{pmatrix} S_{nxz,f_k} e^{-q_{nf_k} z'}$$

$$= T''_j \bullet \bullet T''_{k-2} T''_{k-1} T''_k \begin{pmatrix} B_{k+1} \\ C_{k+1} \end{pmatrix} - T''_j \bullet \bullet T''_{k-2} T''_{k-1} \begin{pmatrix} 0 \\ 1 \end{pmatrix} S_{nxz,f_k} e^{q_{nf_k} z'} - T''_j \bullet \bullet T''_{k-2} T''_{k-1} \begin{pmatrix} 1 \\ 0 \end{pmatrix} S_{nxz,f_k} e^{-q_{nf_k} z'}$$

$$\equiv \begin{pmatrix} M''_{j,11} & M''_{j,12} \\ M''_{j,21} & M''_{j,22} \end{pmatrix} \begin{pmatrix} B_N \\ C_N \end{pmatrix} - \begin{pmatrix} M''_{j,11,k-1} & M''_{j,12,k-1} \\ M''_{j,21,k-1} & M''_{j,22,k-1} \end{pmatrix} \begin{pmatrix} 0 \\ 1 \end{pmatrix} S_{nxz,f_k} e^{q_{nf_k} z'} - \begin{pmatrix} M''_{j,11,k-1} & M''_{j,12,k-1} \\ M''_{j,21,k-1} & M''_{j,22,k-1} \end{pmatrix} \begin{pmatrix} 1 \\ 0 \end{pmatrix} S_{nxz,f_k} e^{-q_{nf_k} z'}$$

$$= \begin{pmatrix} M''_{j,11} B_N + M''_{j,12} C_N \\ M''_{j,21} B_N + M''_{j,22} C_N \end{pmatrix} - \begin{pmatrix} M''_{j,12,k-1} \\ M''_{j,22,k-1} \end{pmatrix} S_{nxz,f_k} e^{q_{nf_k} z'} - \begin{pmatrix} M''_{j,11,k-1} \\ M''_{j,21,k-1} \end{pmatrix} S_{nxz,f_k} e^{-q_{nf_k} z'}, \; j = 2,3,\ldots,N-1, \; j \leq k-1$$

$$\begin{pmatrix} B_k \\ C_k \end{pmatrix} = T_k^{//} \begin{pmatrix} B_{k+1} \\ C_{k+1} \end{pmatrix} - \begin{pmatrix} 0 \\ 1 \end{pmatrix} S_{nxz,f_k} e^{q_{nf_k} z'} = T_k^{//} \bullet \bullet T_{N-1}^{//} \begin{pmatrix} B_N \\ C_N \end{pmatrix} - \begin{pmatrix} 0 \\ 1 \end{pmatrix} S_{nxz,f_k} e^{q_{nf_k} z'} \quad (256)$$

$$\equiv \begin{pmatrix} M_{k,11}^{//} & M_{k,12}^{//} \\ M_{k,21}^{//} & M_{k,22}^{//} \end{pmatrix} \begin{pmatrix} B_N \\ C_N \end{pmatrix} - \begin{pmatrix} 0 \\ 1 \end{pmatrix} S_{nxz,f_k} e^{q_{nf_k} z'} \quad (257)$$

$$= \begin{pmatrix} M_{k,11}^{//} B_N + M_{k,12}^{//} C_N \\ M_{k,21}^{//} B_N + M_{k,22}^{//} C_N \end{pmatrix} - \begin{pmatrix} 0 \\ 1 \end{pmatrix} S_{nxz,f_k} e^{q_{nf_k} z'}, \ j = 2,3,...,N-1, j = k$$

$$\begin{pmatrix} B_j \\ C_j \end{pmatrix} = T_j^{//} \begin{pmatrix} B_{j+1} \\ C_{j+1} \end{pmatrix} = T_j^{//} T_{j+1}^{//} \bullet \bullet \bullet T_{N-1}^{//} \begin{pmatrix} B_N \\ C_N \end{pmatrix} \equiv \begin{pmatrix} M_{j,11}^{//} & M_{j,12}^{//} \\ M_{j,21}^{//} & M_{j,22}^{//} \end{pmatrix} \begin{pmatrix} B_N \\ C_N \end{pmatrix},$$

$$= \begin{pmatrix} M_{j,11}^{//} B_N + M_{j,12}^{//} C_N \\ M_{j,21}^{//} B_N + M_{j,22}^{//} C_N \end{pmatrix}, \ j = 2,3,...,N-1, j \geq k+1 \quad (258)$$

and we obtain:

$$D = B_1 + C_1 = M_{11}^{//} R_{n//}^{f_N a} C_N e^{-2q_{nf_N} z_N} + M_{12}^{//} C_N + M_{21}^{//} R_{n//}^{f_N a} C_N e^{-2q_{nf_N} z_N} + M_{22}^{//} C_N$$
$$- M_{12,k-1}^{//} S_{nxz,f_k} e^{q_{nf_k} z'} - M_{11,k-1}^{//} S_{nxz,f_k} e^{-q_{nf_k} z'} - M_{22,k-1}^{//} S_{nxz,f_k} e^{q_{nf_k} z'} - M_{21,k-1}^{//} S_{nxz,f_k} e^{-q_{nf_k} z'}. \quad (259)$$

Furthermore, we get the following form to solve the coefficient $C_N$:

$$\frac{\varepsilon_{f_1}}{q_{nf_1}} \left( M_{11}^{//} R_{n//}^{f_N a} C_N e^{-2q_{nf_N} z_N} + M_{12}^{//} C_N - M_{12,k-1}^{//} S_{nxz,f_k} e^{q_{nf_k} z'} - M_{11,k-1}^{//} S_{nxz,f_k} e^{-q_{nf_k} z'} \right)$$

$$- \frac{\varepsilon_{f_1}}{q_{nf_1}} \left( M_{21}^{//} R_{n//}^{f_N a} C_N e^{-2q_{nf_N} z_N} + M_{22}^{//} C_N - M_{22,k-1}^{//} S_{nxz,f_k} e^{q_{nf_k} z'} - M_{21,k-1}^{//} S_{nxz,f_k} e^{-q_{nf_k} z'} \right)$$

$$= \frac{\varepsilon_s}{q_{ns}} \left( M_{11}^{//} R_{n//}^{f_N a} C_N e^{-2q_{nf_N} z_N} + M_{12}^{//} C_N + M_{21}^{//} R_{n//}^{f_N a} C_N e^{-2q_{nf_N} z_N} + M_{22}^{//} C_N \right)$$

$$- \frac{\varepsilon_s}{q_{ns}} \left( M_{12,k-1}^{//} S_{nxz,f_k} e^{q_{nf_k} z'} + M_{11,k-1}^{//} S_{nxz,f_k} e^{-q_{nf_k} z'} + M_{22,k-1}^{//} S_{nxz,f_k} e^{q_{nf_k} z'} + M_{21,k-1}^{//} S_{nxz,f_k} e^{-q_{nf_k} z'} \right)$$

$$\Rightarrow \left[ \left( \frac{\varepsilon_s}{q_{ns}} - \frac{\varepsilon_{f_1}}{q_{nf_1}} \right) \left( M_{11}^{//} R_{n//}^{f_N a} e^{-2q_{nf_N} z_N} + M_{12}^{//} \right) + \left( \frac{\varepsilon_s}{q_{ns}} + \frac{\varepsilon_{f_1}}{q_{nf_1}} \right) \left( M_{21}^{//} R_{n//}^{f_N a} e^{-2q_{nf_N} z_N} + M_{22}^{//} \right) \right] C_N$$

$$= \left[ \left( \frac{\varepsilon_s}{q_{ns}} - \frac{\varepsilon_{f_1}}{q_{nf_1}} \right) \left( M_{12,k-1}^{//} e^{q_{nf_k} z'} + M_{11,k-1}^{//} e^{-q_{nf_k} z'} \right) + \left( \frac{\varepsilon_s}{q_{ns}} + \frac{\varepsilon_{f_1}}{q_{nf_1}} \right) \left( M_{22,k-1}^{//} e^{q_{nf_k} z'} + M_{21,k-1}^{//} e^{-q_{nf_k} z'} \right) \right] S_{nxz,f_k}$$

$$\Rightarrow C_N = S_{nxz,f_k} \frac{\left( \frac{\varepsilon_s}{q_{ns}} - \frac{\varepsilon_{f_1}}{q_{nf_1}} \right) \left( M_{12,k-1}^{//} e^{q_{nf_k} z'} + M_{11,k-1}^{//} e^{-q_{nf_k} z'} \right) + \left( \frac{\varepsilon_s}{q_{ns}} + \frac{\varepsilon_{f_1}}{q_{nf_1}} \right) \left( M_{22,k-1}^{//} e^{q_{nf_k} z'} + M_{21,k-1}^{//} e^{-q_{nf_k} z'} \right)}{\left( \frac{\varepsilon_s}{q_{ns}} - \frac{\varepsilon_{f_1}}{q_{nf_1}} \right) \left( M_{11}^{//} R_{n//}^{f_N a} e^{-2q_{nf_N} z_N} + M_{12}^{//} \right) + \left( \frac{\varepsilon_s}{q_{ns}} + \frac{\varepsilon_{f_1}}{q_{nf_1}} \right) \left( M_{21}^{//} R_{n//}^{f_N a} e^{-2q_{nf_N} z_N} + M_{22}^{//} \right)}$$

$$= S_{nxz,f_k} \frac{R_{n//}^{sf_1} \left( M_{12,k-1}^{//} e^{q_{nf_k} z'} + M_{11,k-1}^{//} e^{-q_{nf_k} z'} \right) + \left( M_{22,k-1}^{//} e^{q_{nf_k} z'} + M_{21,k-1}^{//} e^{-q_{nf_k} z'} \right)}{R_{n//}^{sf_1} \left( M_{11}^{//} R_{n//}^{f_N a} e^{-2q_{nf_N} z_N} + M_{12}^{//} \right) + \left( M_{21}^{//} R_{n//}^{f_N a} e^{-2q_{nf_N} z_N} + M_{22}^{//} \right)}$$

.
(260)

Hence we can get the other three coefficients as the following forms:

$$B_N = R_{n//}^{f_N a} S_{nxz, f_k} e^{-2q_{nf_N} z_N} \frac{R_{n//}^{sf_1} \left( M_{12,k-1}^{//} e^{q_{nf_k} z'} + M_{11,k-1}^{//} e^{-q_{nf_k} z'} \right) + \left( M_{22,k-1}^{//} e^{q_{nf_k} z'} + M_{21,k-1}^{//} e^{-q_{nf_k} z'} \right)}{R_{n//}^{sf_1} \left( M_{11}^{//} R_{n//}^{f_N a} e^{-2q_{nf_N} z_N} + M_{12}^{//} \right) + \left( M_{21}^{//} R_{n//}^{f_N a} e^{-2q_{nf_N} z_N} + M_{22}^{//} \right)},$$

(261)

$$A = T_{n//}^{f_N a} S_{nxz, f_k} e^{-q_{nf_N} z_N} e^{q_{na} z_N} \frac{R_{n//}^{sf_1} \left( M_{12,k-1}^{//} e^{q_{nf_k} z'} + M_{11,k-1}^{//} e^{-q_{nf_k} z'} \right) + \left( M_{22,k-1}^{//} e^{q_{nf_k} z'} + M_{21,k-1}^{//} e^{-q_{nf_k} z'} \right)}{R_{n//}^{sf_1} \left( M_{11}^{//} R_{n//}^{f_N a} e^{-2q_{nf_N} z_N} + M_{12}^{//} \right) + \left( M_{21}^{//} R_{n//}^{f_N a} e^{-2q_{nf_N} z_N} + M_{22}^{//} \right)}$$

,
(262)

$$\begin{aligned}
D &= \left( M_{11}^{//} R_{n//}^{f_N a} e^{-2q_{nf_N} z_N} + M_{12}^{//} + M_{21}^{//} R_{n//}^{f_N a} e^{-2q_{nf_N} z_N} + M_{22}^{//} \right) C_N \\
&\quad - M_{12,k-1}^{//} S_{nxz, f_k} e^{q_{nf_k} z'} - M_{11,k-1}^{//} S_{nxz, f_k} e^{-q_{nf_k} z'} - M_{22,k-1}^{//} S_{nxz, f_k} e^{q_{nf_k} z'} - M_{21,k-1}^{//} S_{nxz, f_k} e^{-q_{nf_k} z'} \\
&= \frac{R_{n//}^{sf_1} \left( M_{12,k-1}^{//} e^{q_{nf_k} z'} + M_{11,k-1}^{//} e^{-q_{nf_k} z'} \right) + \left( M_{22,k-1}^{//} e^{q_{nf_k} z'} + M_{21,k-1}^{//} e^{-q_{nf_k} z'} \right)}{R_{n//}^{sf_1} \left( M_{11}^{//} R_{n//}^{f_N a} e^{-2q_{nf_N} z_N} + M_{12}^{//} \right) + \left( M_{21}^{//} R_{n//}^{f_N a} e^{-2q_{nf_N} z_N} + M_{22}^{//} \right)} M_{11}^{//} R_{n//}^{f_N a} e^{-2q_{nf_N} z_N} S_{nxz, f_k} \\
&\quad + \frac{R_{n//}^{sf_1} \left( M_{12,k-1}^{//} e^{q_{nf_k} z'} + M_{11,k-1}^{//} e^{-q_{nf_k} z'} \right) + \left( M_{22,k-1}^{//} e^{q_{nf_k} z'} + M_{21,k-1}^{//} e^{-q_{nf_k} z'} \right)}{R_{n//}^{sf_1} \left( M_{11}^{//} R_{n//}^{f_N a} e^{-2q_{nf_N} z_N} + M_{12}^{//} \right) + \left( M_{21}^{//} R_{n//}^{f_N a} e^{-2q_{nf_N} z_N} + M_{22}^{//} \right)} M_{12}^{//} S_{nxz, f_k} \\
&\quad + \frac{R_{n//}^{sf_1} \left( M_{12,k-1}^{//} e^{q_{nf_k} z'} + M_{11,k-1}^{//} e^{-q_{nf_k} z'} \right) + \left( M_{22,k-1}^{//} e^{q_{nf_k} z'} + M_{21,k-1}^{//} e^{-q_{nf_k} z'} \right)}{R_{n//}^{sf_1} \left( M_{11}^{//} R_{n//}^{f_N a} e^{-2q_{nf_N} z_N} + M_{12}^{//} \right) + \left( M_{21}^{//} R_{n//}^{f_N a} e^{-2q_{nf_N} z_N} + M_{22}^{//} \right)} M_{21}^{//} R_{n//}^{f_N a} e^{-2q_{nf_N} z_N} S_{nxz, f_k} \\
&\quad + \frac{R_{n//}^{sf_1} \left( M_{12,k-1}^{//} e^{q_{nf_k} z'} + M_{11,k-1}^{//} e^{-q_{nf_k} z'} \right) + \left( M_{22,k-1}^{//} e^{q_{nf_k} z'} + M_{21,k-1}^{//} e^{-q_{nf_k} z'} \right)}{R_{n//}^{sf_1} \left( M_{11}^{//} R_{n//}^{f_N a} e^{-2q_{nf_N} z_N} + M_{12}^{//} \right) + \left( M_{21}^{//} R_{n//}^{f_N a} e^{-2q_{nf_N} z_N} + M_{22}^{//} \right)} M_{22}^{//} S_{nxz, f_k} \\
&\quad - M_{12,k-1}^{//} S_{nxz, f_k} e^{q_{nf_k} z'} - M_{11,k-1}^{//} S_{nxz, f_k} e^{-q_{nf_k} z'} - M_{22,k-1}^{//} S_{nxz, f_k} e^{q_{nf_k} z'} - M_{21,k-1}^{//} S_{nxz, f_k} e^{-q_{nf_k} z'}
\end{aligned}$$

.
(263)

Then we solve the coefficients in Eq. (236) completely.

<3> If $k = 1$

$$-\frac{\varepsilon_a}{q_{na}} \left( B_N e^{q_{nf_N} z_N} + C_N e^{-q_{nf_N} z_N} \right) = \frac{\varepsilon_{f_N}}{q_{nf_N}} \left( B_N e^{q_{nf_N} z_N} - C_N e^{-q_{nf_N} z_N} \right)$$

$$\Rightarrow \left( \frac{\varepsilon_{f_N}}{q_{nf_N}} + \frac{\varepsilon_a}{q_{na}} \right) B_N = \left( \frac{\varepsilon_{f_N}}{q_{nf_N}} - \frac{\varepsilon_a}{q_{na}} \right) C_N e^{-2q_{nf_N} z_N}$$

, (264)

$$\Rightarrow B_N = \frac{\varepsilon_{f_N} q_{na} - \varepsilon_a q_{nf_N}}{\varepsilon_{f_N} q_{na} + \varepsilon_a q_{nf_N}} C_N e^{-2q_{nf_N} z_N} = R_{n//}^{f_N a} C_N e^{-2q_{nf_N} z_N}$$

$$\Rightarrow A = R_{n//}^{f_N a} C_N e^{-q_{nf_N} z_N} e^{q_{na} z_N} + C_N e^{-q_{nf_N} z_N} e^{q_{na} z_N} = T_{n//}^{f_N a} C_N e^{-q_{nf_N} z_N} e^{q_{na} z_N}$$

and

$$\begin{cases} B_2 e^{q_{nf_2}z_1} + C_2 e^{-q_{nf_2}z_1} = S_{nxz,f_1} e^{q_{nf_1}z'} e^{-q_{nf_1}z_1} + B_1 e^{q_{nf_1}z_1} + C_1 e^{-q_{nf_1}z_1} \\ \dfrac{\varepsilon_{f_2}}{q_{nf_2}}\left[ B_2 e^{q_{nf_2}z_1} - C_2 e^{-q_{nf_2}z_1} \right] = -\dfrac{\varepsilon_{f_1}}{q_{nf_1}} S_{nxz,f_1} e^{q_{nf_1}z'} e^{-q_{nf_1}z_1} + \dfrac{\varepsilon_{f_1}}{q_{nf_1}}\left[ B_1 e^{q_{nf_1}z_1} - C_1 e^{-q_{nf_1}z_1} \right] \end{cases}$$

$$\Rightarrow \begin{pmatrix} B_1 \\ C_1 \end{pmatrix} = T_1'' \begin{pmatrix} B_2 \\ C_2 \end{pmatrix} + \begin{pmatrix} e^{q_{nf_1}z_1} & e^{-q_{nf_1}z_1} \\ \dfrac{\varepsilon_{f_1}}{q_{nf_1}} e^{q_{nf_1}z_1} & -\dfrac{\varepsilon_{f_1}}{q_{nf_1}} e^{-q_{nf_1}z_1} \end{pmatrix}^{-1} \begin{pmatrix} -1 \\ \dfrac{\varepsilon_{f_1}}{q_{nf_1}} \end{pmatrix} S_{nxz,f_1} e^{q_{nf_1}z'} e^{-q_{nf_1}z_1}$$ 
(265)

$$= T_1'' \begin{pmatrix} B_2 \\ C_2 \end{pmatrix} + \dfrac{q_{nf_1}}{2\varepsilon_{f_1}} \begin{pmatrix} \dfrac{\varepsilon_{f_1}}{q_{nf_1}} e^{-q_{nf_1}z_1} & e^{-q_{nf_1}z_1} \\ \dfrac{\varepsilon_{f_1}}{q_{nf_1}} e^{q_{nf_1}z_1} & -e^{q_{nf_1}z_1} \end{pmatrix} \begin{pmatrix} -1 \\ \dfrac{\varepsilon_{f_1}}{q_{nf_1}} \end{pmatrix} S_{nxz,f_1} e^{q_{nf_1}z'} e^{-q_{nf_1}z_1}$$

$$= T_1'' \begin{pmatrix} B_2 \\ C_2 \end{pmatrix} - \begin{pmatrix} 0 \\ 1 \end{pmatrix} S_{nxz,f_1} e^{q_{nf_1}z'}$$

Combining Eq. (265) with the transfer matrix method which is described in Eqs. (217)-(219), we obtain:

$$\begin{pmatrix} B_1 \\ C_1 \end{pmatrix} = T_1'' \begin{pmatrix} B_2 \\ C_2 \end{pmatrix} - \begin{pmatrix} 0 \\ 1 \end{pmatrix} S_{nxz,f_1} e^{q_{nf_1}z'} = \begin{pmatrix} M_{11}'' & M_{12}'' \\ M_{21}'' & M_{22}'' \end{pmatrix} \begin{pmatrix} B_N \\ C_N \end{pmatrix} - \begin{pmatrix} 0 \\ 1 \end{pmatrix} S_{nxz,f_1} e^{q_{nf_1}z'}$$

$$= \begin{pmatrix} M_{11}'' B_N + M_{12}'' C_N \\ M_{21}'' B_N + M_{22}'' C_N \end{pmatrix} - \begin{pmatrix} 0 \\ 1 \end{pmatrix} S_{nxz,f_1} e^{q_{nf_1}z'}$$
(266)

$$\begin{pmatrix} B_j \\ C_j \end{pmatrix} = T_j'' \bullet \bullet \bullet T_{N-1}'' \begin{pmatrix} B_N \\ C_N \end{pmatrix} = \begin{pmatrix} M_{j,11}'' & M_{j,12}'' \\ M_{j,21}'' & M_{j,22}'' \end{pmatrix} \begin{pmatrix} B_N \\ C_N \end{pmatrix} = \begin{pmatrix} M_{j,11}'' B_N + M_{j,11}'' C_N \\ M_{j,21}'' B_N + M_{j,22}'' C_N \end{pmatrix}, j = 2,...,N-1$$
(267)

,
and we obtain:

$$D = -S_{nxz,f_1} e^{-q_{nf_1}z'} + B_1 + C_1 = -S_{nxz,f_1} e^{-q_{nf_1}z'} + M_{11}'' R_{n//}^{f_N a} C_N e^{-2q_{nf_N}z_N} + M_{12}'' C_N$$

$$+ M_{21}'' R_{n//}^{f_N a} C_N e^{-2q_{nf_N}z_N} + M_{22}'' C_N - S_{nxz,f_1} e^{q_{nf_1}z'}$$
(268)

$$= \left( M_{11}'' R_{n//}^{f_N a} e^{-2q_{nf_N}z_N} + M_{12}'' + M_{21}'' R_{n//}^{f_N a} e^{-2q_{nf_N}z_N} + M_{22}'' \right) C_N - S_{nxz,f_1} e^{-q_{nf_1}z'} - S_{nxz,f_1} e^{q_{nf_1}z'}$$

Furthermore, we get the following form to solve the coefficient $C_N$:

$$-\frac{\varepsilon_{f_1}}{q_{nf_1}} S_{nxz,f_1} e^{-q_{nf_1} z'} + \frac{\varepsilon_{f_1}}{q_{nf_1}} (B_1 - C_1) = \frac{\varepsilon_s}{q_{ns}} \left( -S_{nxz,f_1} e^{-q_{nf_1} z'} + B_1 + C_1 \right)$$

$$\Rightarrow \left( \frac{\varepsilon_s}{q_{ns}} - \frac{\varepsilon_{f_1}}{q_{nf_1}} \right) B_1 = -\left( \frac{\varepsilon_s}{q_{ns}} + \frac{\varepsilon_{f_1}}{q_{nf_1}} \right) C_1 + \left( \frac{\varepsilon_s}{q_{ns}} - \frac{\varepsilon_{f_1}}{q_{nf_1}} \right) S_{nxz,f_1} e^{-q_{nf_1} z'}$$

$$\Rightarrow M_{11}^{//} R_{n//}^{f_N a} C_N e^{-2 q_{nf_N} z_N} + M_{12}^{//} C_N = -\frac{1}{R_{n//}^{sf_1}} \left( M_{21}^{//} R_{n//}^{f_N a} C_N e^{-2 q_{nf_N} z_N} + M_{22}^{//} C_N - S_{nxz,f_1} e^{q_{nf_1} z'} \right) + S_{nxz,f_1} e^{-q_{nf_1} z'}$$

$$\Rightarrow \left( M_{11}^{//} R_{n//}^{sf_1} R_{n//}^{f_N a} e^{-2 q_{nf_N} z_N} + M_{12}^{//} R_{n//}^{sf_1} + M_{21}^{//} R_{n//}^{f_N a} e^{-2 q_{nf_N} z_N} + M_{22}^{//} \right) C_N = S_{nxz,f_1} e^{q_{nf_1} z'} + S_{nxz,f_1} R_{n//}^{sf_1} e^{-q_{nf_1} z'}$$

$$\Rightarrow C_N = S_{nxz,f_1} \frac{e^{q_{nf_1} z'} + R_{n//}^{sf_1} e^{-q_{nf_1} z'}}{M_{11}^{//} R_{n//}^{sf_1} R_{n//}^{f_N a} e^{-2 q_{nf_N} z_N} + M_{12}^{//} R_{n//}^{sf_1} + M_{21}^{//} R_{n//}^{f_N a} e^{-2 q_{nf_N} z_N} + M_{22}^{//}}$$

(269)

Hence we can get the other three coefficients as the following forms:

$$B_N = R_{n//}^{f_N a} S_{nxz,f_1} \frac{e^{q_{nf_1} z'} + R_{n//}^{sf_1} e^{-q_{nf_1} z'}}{M_{11}^{//} R_{n//}^{sf_1} R_{n//}^{f_N a} e^{-2 q_{nf_N} z_N} + M_{12}^{//} R_{n//}^{sf_1} + M_{21}^{//} R_{n//}^{f_N a} e^{-2 q_{nf_N} z_N} + M_{22}^{//}} e^{-2 q_{nf_N} z_N}, \quad (270)$$

$$A = T_{n//}^{f_N a} S_{nxz,f_1} \frac{e^{q_{nf_1} z'} + R_{n//}^{sf_1} e^{-q_{nf_1} z'}}{M_{11}^{//} R_{n//}^{sf_1} R_{n//}^{f_N a} e^{-2 q_{nf_N} z_N} + M_{12}^{//} R_{n//}^{sf_1} + M_{21}^{//} R_{n//}^{f_N a} e^{-2 q_{nf_N} z_N} + M_{22}^{//}} e^{-q_{nf_N} z_N} e^{q_{na} z_N},$$

(271)

$$D = \left( M_{11}^{//} R_{n//}^{f_N a} e^{-2 q_{nf_N} z_N} + M_{12}^{//} + M_{21}^{//} R_{n//}^{f_N a} e^{-2 q_{nf_N} z_N} + M_{22}^{//} \right) C_N - S_{nxz,f_1} e^{-q_{nf_1} z'} - S_{nxz,f_1} e^{q_{nf_1} z'}$$

$$= S_{nxz,f_1} \frac{\left( M_{11}^{//} R_{n//}^{f_N a} e^{-2 q_{nf_N} z_N} + M_{12}^{//} + M_{21}^{//} R_{n//}^{f_N a} e^{-2 q_{nf_N} z_N} + M_{22}^{//} \right) \left( e^{q_{nf_1} z'} + R_{n//}^{sf_1} e^{-q_{nf_1} z'} \right)}{M_{11}^{//} R_{n//}^{sf_1} R_{n//}^{f_N a} e^{-2 q_{nf_N} z_N} + M_{12}^{//} R_{n//}^{sf_1} + M_{21}^{//} R_{n//}^{f_N a} e^{-2 q_{nf_N} z_N} + M_{22}^{//}} - S_{nxz,f_1} e^{-q_{nf_1} z'} - S_{nxz,f_1} e^{q_{nf_1} z'}$$

$$= S_{nxz,f_1} T_{n//}^{f_1 s} \frac{e^{q_{nf_1} z'} \left( M_{11}^{//} R_{n//}^{f_N a} e^{-2 q_{nf_N} z_N} + M_{12}^{//} \right) - e^{-q_{nf_1} z'} \left( M_{21}^{//} R_{n//}^{f_N a} e^{-2 q_{nf_N} z_N} + M_{22}^{//} \right)}{M_{11}^{//} R_{n//}^{sf_1} R_{n//}^{f_N a} e^{-2 q_{nf_N} z_N} + M_{12}^{//} R_{n//}^{sf_1} + M_{21}^{//} R_{n//}^{f_N a} e^{-2 q_{nf_N} z_N} + M_{22}^{//}}$$

(272)

Then we solve the coefficients in Eq. (236) completely. Next we discuss the component $\tilde{g}_{nyy}$, starting from Eq. (113), we have:

$$\left( \partial_z^2 - q_n^2 \right) \tilde{g}_{nyy} = -\delta(z - z'). \tag{273}$$

Next we will consider the relative position between the field ($z$) and the source ($z'$). First we consider $z' < 0$ (without loss of generality, we may set $z_0 = 0$), the corresponding different equation in regions $z > z_N$, $z_{j-1} < z < z_j, j = N, N-1, ..., 2, 1$ and $z < 0$ are:

$$\begin{cases} \left(\partial_z^2 - q_{na}^2\right)\tilde{g}_{nyy} = 0, z > z_N \\ \left(\partial_z^2 - q_{nf_j}^2\right)\tilde{g}_{nyy} = 0, z_{j-1} < z < z_j, j = 1,2,...,N \\ \left(\partial_z^2 - q_{ns}^2\right)\tilde{g}_{nyy} = -\delta(z-z'), z < 0 \end{cases}. \tag{274}$$

Furthermore, the solutions of Eq. (274) are:

$$\tilde{g}_{nyy} = \begin{cases} Ae^{-q_{na}z}, z > z_N \\ B_j e^{q_{nf_j}z} + C_j e^{-q_{nf_j}z}, z_{j-1} < z < z_j, j = 1,2,...,N \\ S_{nyy,s}e^{-q_{ns}|z-z'|} + De^{q_{ns}z}, z < 0 \end{cases}, \tag{275}$$

where $S_{nyy,s} \equiv \dfrac{1}{2q_{ns}}$. In order to match the boundary conditions of $\tilde{g}_{nyy}$ as discuss in Eq. (116):

$$\tilde{g}_{nyy}\big|_{z_{j+}} = \tilde{g}_{nyy}\big|_{z_{j-}}, \partial_z \tilde{g}_{nyy}\big|_{z_{j+}} = \partial_z \tilde{g}_{nyy}\big|_{z_{j-}}, j = 0,1,...,N, \tag{276}$$

we have the following simulated equations:

$$\begin{cases} Ae^{-q_{na}z_N} = B_N e^{q_{nf_N}z_N} + C_N e^{-q_{nf_N}z_N}, z = z_N \\ B_j e^{q_{nf_j}z_j} + C_j e^{-q_{nf_j}z_j} = B_{j+1} e^{q_{nf_{j+1}}z_j} + C_{j+1} e^{-q_{nf_{j+1}}z_j}, z = z_j, j = 1,...,N-1 \\ B_1 + C_1 = S_{nyy,s} e^{q_{ns}z'} + D, z = 0 \end{cases}, \tag{277}$$

and

$$\begin{cases} -q_{na}Ae^{-q_{na}z_N} = B_N q_{nf_N} e^{q_{nf_N}z_N} - C_N q_{nf_N} e^{-q_{nf_N}z_N}, z = z_N \\ B_j q_{nf_j} e^{q_{nf_j}z_j} - C_j q_{nf_j} e^{-q_{nf_j}z_j} = B_{j+1} q_{nf_{j+1}} e^{q_{nf_{j+1}}z_j} - C_{j+1} q_{nf_{j+1}} e^{-q_{nf_{j+1}}z_j}, z = z_j, j = 1,...,N-1 \\ B_1 q_{nf_1} - C_1 q_{nf_1} = -S_{nyy,s} q_{ns} e^{q_{ns}z'} + Dq_{ns}, z = 0 \end{cases}. \tag{278}$$

Furthermore, we obtain:

$$-q_{na}\left(B_N e^{q_{nf_N}z_N} + C_N e^{-q_{nf_N}z_N}\right) = B_N q_{nf_N} e^{q_{nf_N}z_N} - C_N q_{nf_N} e^{-q_{nf_N}z_N}$$

$$\Rightarrow \left(q_{nf_N} + q_{na}\right)B_N = \left(q_{nf_N} - q_{na}\right)C_N e^{-2q_{nf_N}z_N}$$

$$\Rightarrow B_N = \frac{q_{nf_N} - q_{na}}{q_{nf_N} + q_{na}} C_N e^{-2q_{nf_N}z_N} = R_{n\perp}^{f_N a} C_N e^{-2q_{nf_N}z_N} \tag{279}$$

$$\Rightarrow A = R_{n\perp}^{f_N a} C_N e^{-q_{nf_N}z_N} e^{q_{na}z_N} + C_N e^{-q_{nf_N}z_N} e^{q_{na}z_N} = T_{n\perp}^{f_N a} e^{-q_{nf_N}z_N} e^{q_{na}z_N} C_N$$

and

$$\begin{cases} B_j e^{q_{nf_j} z_j} + C_j e^{-q_{nf_j} z_j} = B_{j+1} e^{q_{nf_{j+1}} z_j} + C_{j+1} e^{-q_{nf_{j+1}} z_j}, j = 1,...,N-1 \\ B_j q_{nf_j} e^{q_{nf_j} z_j} - C_j q_{nf_j} e^{-q_{nf_j} z_j} = B_{j+1} q_{nf_{j+1}} e^{q_{nf_{j+1}} z_j} - C_{j+1} q_{nf_{j+1}} e^{-q_{nf_{j+1}} z_j}, j = 1,...,N-1 \end{cases}$$

$$\Rightarrow \begin{pmatrix} e^{q_{nf_j} z_j} & e^{-q_{nf_j} z_j} \\ q_{nf_j} e^{q_{nf_j} z_j} & -q_{nf_j} e^{-q_{nf_j} z_j} \end{pmatrix} \begin{pmatrix} B_j \\ C_j \end{pmatrix} = \begin{pmatrix} e^{q_{nf_{j+1}} z_j} & e^{-q_{nf_{j+1}} z_j} \\ q_{nf_{j+1}} e^{q_{nf_{j+1}} z_j} & -q_{nf_{j+1}} e^{-q_{nf_{j+1}} z_j} \end{pmatrix} \begin{pmatrix} B_{j+1} \\ C_{j+1} \end{pmatrix}, j = 1,...,N-1$$

$$\Rightarrow \begin{pmatrix} B_j \\ C_j \end{pmatrix} = \begin{pmatrix} e^{q_{nf_j} z_j} & e^{-q_{nf_j} z_j} \\ q_{nf_j} e^{q_{nf_j} z_j} & -q_{nf_j} e^{-q_{nf_j} z_j} \end{pmatrix}^{-1} \begin{pmatrix} e^{q_{nf_{j+1}} z_j} & e^{-q_{nf_{j+1}} z_j} \\ q_{nf_{j+1}} e^{q_{nf_{j+1}} z_j} & -q_{nf_{j+1}} e^{-q_{nf_{j+1}} z_j} \end{pmatrix} \begin{pmatrix} B_{j+1} \\ C_{j+1} \end{pmatrix}$$

$$\equiv T_j^\perp \begin{pmatrix} B_{j+1} \\ C_{j+1} \end{pmatrix}, j = 1,...,N-1$$

$$\Rightarrow \begin{pmatrix} B_1 \\ C_1 \end{pmatrix} = T_1^\perp \begin{pmatrix} B_2 \\ C_2 \end{pmatrix} = T_1^\perp T_2^\perp \begin{pmatrix} B_3 \\ C_3 \end{pmatrix} = ... = T_1^\perp T_2^\perp \bullet \bullet \bullet T_{N-1}^\perp \begin{pmatrix} B_N \\ C_N \end{pmatrix} \equiv \prod_{j=1}^{N-1} T_j^\perp \begin{pmatrix} B_N \\ C_N \end{pmatrix}$$

$$\equiv \begin{pmatrix} M_{11}^\perp & M_{12}^\perp \\ M_{21}^\perp & M_{22}^\perp \end{pmatrix} \begin{pmatrix} B_N \\ C_N \end{pmatrix} = \begin{pmatrix} M_{11}^\perp B_N + M_{12}^\perp C_N \\ M_{21}^\perp B_N + M_{22}^\perp C_N \end{pmatrix}$$

$$\Rightarrow \begin{pmatrix} B_j \\ C_j \end{pmatrix} = T_j^\perp \begin{pmatrix} B_{j+1} \\ C_{j+1} \end{pmatrix} = T_j^\perp T_{j+1}^\perp \begin{pmatrix} B_{j+2} \\ C_{j+2} \end{pmatrix} = ... = T_j^\perp T_{j+1}^\perp \bullet \bullet \bullet T_{N-1}^\perp \begin{pmatrix} B_N \\ C_N \end{pmatrix}$$

$$\equiv \begin{pmatrix} M_{j,11}^\perp & M_{j,12}^\perp \\ M_{j,21}^\perp & M_{j,22}^\perp \end{pmatrix} \begin{pmatrix} B_N \\ C_N \end{pmatrix} = \begin{pmatrix} M_{j,11}^\perp B_N + M_{j,12}^\perp C_N \\ M_{j,21}^\perp B_N + M_{j,22}^\perp C_N \end{pmatrix}, j = 2,3,...,N-1$$

(280)

where we define a transfer matrix $T_j^\perp, j = 1,...,N-1$ as the following form:

$$\begin{aligned} T_j^\perp &= \begin{pmatrix} e^{q_{nf_j} z_j} & e^{-q_{nf_j} z_j} \\ q_{nf_j} e^{q_{nf_j} z_j} & -q_{nf_j} e^{-q_{nf_j} z_j} \end{pmatrix}^{-1} \begin{pmatrix} e^{q_{nf_{j+1}} z_j} & e^{-q_{nf_{j+1}} z_j} \\ q_{nf_{j+1}} e^{q_{nf_{j+1}} z_j} & -q_{nf_{j+1}} e^{-q_{nf_{j+1}} z_j} \end{pmatrix} \\ &= \frac{\begin{pmatrix} -q_{nf_j} e^{-q_{nf_j} z_j} & -e^{-q_{nf_j} z_j} \\ -q_{nf_j} e^{q_{nf_j} z_j} & e^{q_{nf_j} z_j} \end{pmatrix} \begin{pmatrix} e^{q_{nf_{j+1}} z_j} & e^{-q_{nf_{j+1}} z_j} \\ q_{nf_{j+1}} e^{q_{nf_{j+1}} z_j} & -q_{nf_{j+1}} e^{-q_{nf_{j+1}} z_j} \end{pmatrix}}{-2q_{nf_j}} \\ &= \begin{pmatrix} \frac{q_{nf_j} + q_{nf_{j+1}}}{2q_{nf_j}} e^{-q_{nf_j} z_j} e^{q_{nf_{j+1}} z_j} & \frac{q_{nf_j} - q_{nf_{j+1}}}{2q_{nf_j}} e^{-q_{nf_j} z_j} e^{-q_{nf_{j+1}} z_j} \\ \frac{q_{nf_j} - q_{nf_{j+1}}}{2q_{nf_j}} e^{q_{nf_j} z_j} e^{q_{nf_{j+1}} z_j} & \frac{q_{nf_j} + q_{nf_{j+1}}}{2q_{nf_j}} e^{q_{nf_j} z_j} e^{-q_{nf_{j+1}} z_j} \end{pmatrix} \\ &= \frac{1}{T_{n\perp}^{f_j f_{j+1}}} \begin{pmatrix} e^{-q_{nf_j} z_j} e^{q_{nf_{j+1}} z_j} & R_{n\perp}^{f_j f_{j+1}} e^{-q_{nf_j} z_j} e^{-q_{nf_{j+1}} z_j} \\ R_{n\perp}^{f_j f_{j+1}} e^{q_{nf_j} z_j} e^{q_{nf_{j+1}} z_j} & e^{q_{nf_j} z_j} e^{-q_{nf_{j+1}} z_j} \end{pmatrix} \end{aligned}$$

(281)

$$\prod_{j=1}^{N-1} T_j^\perp = \prod_{j=1}^{N-1} \frac{1}{T_{n\perp}^{f_j f_{j+1}}} \begin{pmatrix} e^{-q_{nf_j} z_j} e^{q_{nf_{j+1}} z_j} & R_{n\perp}^{f_j f_{j+1}} e^{-q_{nf_j} z_j} e^{-q_{nf_{j+1}} z_j} \\ R_{n\perp}^{f_j f_{j+1}} e^{q_{nf_j} z_j} e^{q_{nf_{j+1}} z_j} & e^{q_{nf_j} z_j} e^{-q_{nf_{j+1}} z_j} \end{pmatrix} \equiv \begin{pmatrix} M_{11}^\perp & M_{12}^\perp \\ M_{21}^\perp & M_{22}^\perp \end{pmatrix}, \quad (282)$$

and

$$D = B_1 + C_1 - S_{nyy,s} e^{q_{ns} z'}$$
$$\Rightarrow D = \left( M_{11}^\perp R_{n\perp}^{f_N a} e^{-2q_{nf_N} z_N} + M_{12}^\perp + M_{21}^\perp R_{n\perp}^{f_N a} e^{-2q_{nf_N} z_N} + M_{22}^\perp \right) C_N - S_{nyy,s} e^{q_{ns} z'}. \quad (283)$$

Furthermore, we get the following form to solve the coefficient $C_N$:

$$B_1 q_{nf_1} - C_1 q_{nf_1} = -S_{nyy,s} q_{ns} e^{q_{ns} z'} + D q_{ns}$$
$$\Rightarrow \left( M_{11}^\perp R_{n\perp}^{f_N a} e^{-2q_{nf_N} z_N} + M_{12}^\perp - M_{21}^\perp R_{n\perp}^{f_N a} C_N e^{-2q_{nf_N} z_N} - M_{22}^\perp \right) q_{nf_1} C_N = -S_{nyy,s} q_{ns} e^{q_{ns} z'}$$
$$+ \left( M_{11}^\perp R_{n\perp}^{f_N a} e^{-2q_{nf_N} z_N} + M_{12}^\perp + M_{21}^\perp R_{n\perp}^{f_N a} e^{-2q_{nf_N} z_N} + M_{22}^\perp \right) q_{ns} C_N - S_{nyy,s} q_{ns} e^{q_{ns} z'}$$
$$\Rightarrow \left[ (q_{ns} - q_{nf_1}) \left( M_{11}^\perp R_{n\perp}^{f_N a} e^{-2q_{nf_N} z_N} + M_{12}^\perp \right) + (q_{ns} + q_{nf_1}) \left( M_{21}^\perp R_{n\perp}^{f_N a} e^{-2q_{nf_N} z_N} + M_{22}^\perp \right) \right] C_N = 2 q_{ns} S_{nyy,s} e^{q_{ns} z'}$$
$$\Rightarrow C_N = \frac{2 q_{ns} S_{nyy,s} e^{q_{ns} z'}}{(q_{ns} - q_{nf_1}) \left( M_{11}^\perp R_{n\perp}^{f_N a} e^{-2q_{nf_N} z_N} + M_{12}^\perp \right) + (q_{ns} + q_{nf_1}) \left( M_{21}^\perp R_{n\perp}^{f_N a} e^{-2q_{nf_N} z_N} + M_{22}^\perp \right)}$$
$$= \frac{T_{n\perp}^{sf_1} S_{nyy,s} e^{q_{ns} z'}}{R_{n\perp}^{sf_1} \left( M_{11}^\perp R_{n\perp}^{f_N a} e^{-2q_{nf_N} z_N} + M_{12}^\perp \right) + \left( M_{21}^\perp R_{n\perp}^{f_N a} e^{-2q_{nf_N} z_N} + M_{22}^\perp \right)}. \quad (284)$$

Hence we can get the other three coefficients as the following forms:

$$B_N = R_{n\perp}^{f_N a} C_N e^{-2q_{nf_N} z_N} = \frac{R_{n\perp}^{f_N a} T_{n\perp}^{sf_1} S_{nyy,s} e^{q_{ns} z'} e^{-2q_{nf_N} z_N}}{R_{n\perp}^{sf_1} \left( M_{11}^\perp R_{n\perp}^{f_N a} e^{-2q_{nf_N} z_N} + M_{12}^\perp \right) + \left( M_{21}^\perp R_{n\perp}^{f_N a} e^{-2q_{nf_N} z_N} + M_{22}^\perp \right)}, \quad (285)$$

$$A = T_{n\perp}^{f_N a} e^{-q_{nf_N} z_N} e^{q_{na} z_N} C_N = \frac{T_{n\perp}^{f_N a} T_{n\perp}^{sf_1} S_{nyy,s} e^{q_{ns} z'} e^{-q_{nf_N} z_N} e^{q_{na} z_N}}{R_{n\perp}^{sf_1} \left( M_{11}^\perp R_{n\perp}^{f_N a} e^{-2q_{nf_N} z_N} + M_{12}^\perp \right) + \left( M_{21}^\perp R_{n\perp}^{f_N a} e^{-2q_{nf_N} z_N} + M_{22}^\perp \right)}, \quad (286)$$

$$D = \left( M_{11}^\perp R_{n\perp}^{f_N a} e^{-2q_{nf_N} z_N} + M_{12}^\perp + M_{21}^\perp R_{n\perp}^{f_N a} e^{-2q_{nf_N} z_N} + M_{22}^\perp \right) C_N - S_{nyy,s} e^{q_{ns} z'}$$
$$= \frac{\left( M_{11}^\perp R_{n\perp}^{f_N a} e^{-2q_{nf_N} z_N} + M_{12}^\perp + M_{21}^\perp R_{n\perp}^{f_N a} e^{-2q_{nf_N} z_N} + M_{22}^\perp \right) T_{n\perp}^{sf_1} S_{nyy,s} e^{q_{ns} z'}}{R_{n\perp}^{sf_1} \left( M_{11}^\perp R_{n\perp}^{f_N a} e^{-2q_{nf_N} z_N} + M_{12}^\perp \right) + \left( M_{21}^\perp R_{n\perp}^{f_N a} e^{-2q_{nf_N} z_N} + M_{22}^\perp \right)} - S_{nyy,s} e^{q_{ns} z'}. \quad (287)$$
$$= \frac{R_{n\perp}^{f_N a} \left( M_{11}^\perp + M_{21}^\perp R_{n\perp}^{sf_1} \right) e^{-2q_{nf_N} z_N} + M_{12}^\perp + M_{22}^\perp R_{n\perp}^{sf_1}}{R_{n\perp}^{sf_1} \left( M_{11}^\perp R_{n\perp}^{f_N a} e^{-2q_{nf_N} z_N} + M_{12}^\perp \right) + \left( M_{21}^\perp R_{n\perp}^{f_N a} e^{-2q_{nf_N} z_N} + M_{22}^\perp \right)} S_{nyy,s} e^{q_{ns} z'}$$

Hence we solve the coefficients in Eq. (275) completely. Next we consider $z' > z_N$, the corresponding different equation in regions $z > z_N$, $z_{j-1} < z < z_j$, $j = 1, 2, ..., N$

and $z < 0$ are:

$$\begin{cases} \left(\partial_z^2 - q_{na}^2\right)\tilde{g}_{nyy} = -\delta(z-z'), z > z_N \\ \left(\partial_z^2 - q_{nf_j}^2\right)\tilde{g}_{nyy} = 0, z_{j-1} < z < z_j, j = 1,2,...,N \\ \left(\partial_z^2 - q_{ns}^2\right)\tilde{g}_{nyy} = 0, z < 0 \end{cases} \quad (288)$$

Furthermore, the solutions of Eq. (288) are:

$$\tilde{g}_{nyy} = \begin{cases} S_{nyy,a} e^{-q_{na}|z-z'|} + A e^{-q_{na}z}, z > z_N \\ B_j e^{q_{nf_j}z} + C_j e^{-q_{nf_j}z}, z_{j-1} < z < z_j, j = 1,2,...,N \\ D e^{q_{ns}z}, z < 0 \end{cases} \quad (289)$$

where $S_{nyy,a} = \dfrac{1}{2q_{na}}$. In order to match the boundary conditions of $\tilde{g}_{nyy}$ as discuss in Eq. (276), we have the following simulated equations:

$$\begin{cases} S_{nyy,a} e^{-q_{na}z'} e^{q_{na}z_N} + A e^{-q_{na}z_N} = B_N e^{q_{nf_N}z_N} + C_N e^{-q_{nf_N}z_N}, z = z_N \\ B_j e^{q_{nf_j}z_j} + C_j e^{-q_{nf_j}z_j} = B_{j+1} e^{q_{nf_{j+1}}z_j} + C_{j+1} e^{-q_{nf_{j+1}}z_j}, z = z_j, j = 1,...,N-1 \\ B_1 + C_1 = D, z = 0 \end{cases} \quad (290)$$

and

$$\begin{cases} S_{nyy,a} q_{na} e^{-q_{na}z'} e^{q_{na}z_N} - q_{na} A e^{-q_{na}z} = B_N q_{nf_N} e^{q_{nf_N}z_N} - C_N q_{nf_N} e^{-q_{nf_N}z_N}, z = z_N \\ B_j q_{nf_j} e^{q_{nf_j}z_j} - C_j q_{nf_j} e^{-q_{nf_j}z_j} = B_{j+1} q_{nf_{j+1}} e^{q_{nf_{j+1}}z_j} - C_{j+1} q_{nf_{j+1}} e^{-q_{nf_{j+1}}z_j}, z = z_j, j = 1,...,N-1 \\ B_1 q_{nf_1} - C_1 q_{nf_1} = D q_{ns}, z = 0 \end{cases}$$

$$(291)$$

Furthermore, we obtain:

$$S_{nyy,a} q_{na} e^{-q_{na}z'} e^{q_{na}z_N} - q_{na}\left(B_N e^{q_{nf_N}z_N} + C_N e^{-q_{nf_N}z_N} - S_{nyy,a} e^{-q_{na}z'} e^{q_{na}z_N}\right) = B_N q_{nf_N} e^{q_{nf_N}z_N} - C_N q_{nf_N} e^{-q_{nf_N}z_N}$$

$$\Rightarrow \left(q_{nf_N} + q_{na}\right) B_N = \left(q_{nf_N} - q_{na}\right) C_N e^{-2q_{nf_N}z_N} + 2 S_{nyy,a} q_{na} e^{-q_{na}z'} e^{q_{na}z_N} e^{-q_{nf_N}z_N}$$

$$\Rightarrow B_N = \frac{q_{nf_N} - q_{na}}{q_{nf_N} + q_{na}} C_N e^{-2q_{nf_N}z_N} + \frac{2q_{na}}{q_{nf_N} + q_{na}} S_{nyy,a} e^{-q_{na}z'} e^{q_{na}z_N} e^{-q_{nf_N}z_N}$$

$$= R_{n\perp}^{f_N a} C_N e^{-2q_{nf_N}z_N} + T_{n\perp}^{af_N} S_{nyy,a} e^{-q_{na}z'} e^{q_{na}z_N} e^{-q_{nf_N}z_N}$$

$$\Rightarrow A = B_N e^{q_{nf_N}z_N} e^{q_{na}z_N} + C_N e^{-q_{nf_N}z_N} e^{q_{na}z_N} - S_{nyy,a} e^{-q_{na}z'} e^{2q_{na}z_N}$$

$$= R_{n\perp}^{f_N a} C_N e^{-q_{nf_N}z_N} e^{q_{na}z_N} + T_{n\perp}^{af_N} S_{nyy,a} e^{-q_{na}z'} e^{2q_{na}z_N} + C_N e^{-q_{nf_N}z_N} e^{q_{na}z_N} - S_{nyy,a} e^{-q_{na}z'} e^{2q_{na}z_N}$$

$$= T_{n\perp}^{f_N a} C_N e^{-q_{nf_N}z_N} e^{q_{na}z_N} - R_{n\perp}^{f_N a} S_{nyy,a} e^{-q_{na}z'} e^{2q_{na}z_N}$$

, (292)

and

$$D = B_1 + C_1 = \left(M_{11}^\perp R_{n\perp}^{f_N a} e^{-2q_{nf_N} z_N} + M_{12}^\perp + M_{21}^\perp R_{n\perp}^{f_N a} e^{-2q_{nf_N} z_N} + M_{22}^\perp\right) C_N$$
$$+ M_{11}^\perp T_{n\perp}^{af_N} S_{nyy,a} e^{-q_{na} z'} e^{q_{na} z_N} e^{-q_{nf_N} z_N} + M_{21}^\perp T_{n\perp}^{af_N} S_{nyy,a} e^{-q_{na} z'} e^{q_{na} z_N} e^{-q_{nf_N} z_N}$$
(293)

Furthermore, we get the following form to solve the coefficient $C_N$:

$$B_1 q_{nf_1} - C_1 q_{nf_1} = (B_1 + C_1) q_{ns}$$
$$\Rightarrow (q_{ns} - q_{nf_1}) B_1 = -(q_{ns} + q_{nf_1}) C_1$$
$$\Rightarrow C_1 = -R_{n\perp}^{sf_1} B_1$$
$$\Rightarrow \left(M_{21}^\perp R_{n\perp}^{f_N a} e^{-2q_{nf_N} z_N} + M_{22}^\perp\right) C_N + M_{21}^\perp T_{n\perp}^{af_N} S_{nyy,a} e^{-q_{na} z'} e^{q_{na} z_N} e^{-q_{nf_N} z_N}$$
$$= -R_{n\perp}^{sf_1} \left(M_{11}^\perp R_{n\perp}^{f_N a} e^{-2q_{nf_N} z_N} + M_{12}^\perp\right) C_N - R_{n\perp}^{sf_1} M_{11}^\perp T_{n\perp}^{af_N} S_{nyy,a} e^{-q_{na} z'} e^{q_{na} z_N} e^{-q_{nf_N} z_N}$$
(294)
$$\Rightarrow \left[M_{21}^\perp R_{n\perp}^{f_N a} e^{-2q_{nf_N} z_N} + M_{22}^\perp + R_{n\perp}^{sf_1}\left(M_{11}^\perp R_{n\perp}^{f_N a} e^{-2q_{nf_N} z_N} + M_{12}^\perp\right)\right] C_N$$
$$= -M_{21}^\perp T_{n\perp}^{af_N} S_{nyy,a} e^{-q_{na} z'} e^{q_{na} z_N} e^{-q_{nf_N} z_N} - R_{n\perp}^{sf_1} M_{11}^\perp T_{n\perp}^{af_N} S_{nyy,a} e^{-q_{na} z'} e^{q_{na} z_N} e^{-q_{nf_N} z_N}$$
$$\Rightarrow C_N = \frac{-T_{n\perp}^{af_N} S_{nyy,a} \left(M_{21}^\perp + R_{n\perp}^{sf_1} M_{11}^\perp\right) e^{-q_{na} z'} e^{q_{na} z_N} e^{-q_{nf_N} z_N}}{M_{21}^\perp R_{n\perp}^{f_N a} e^{-2q_{nf_N} z_N} + M_{22}^\perp + R_{n\perp}^{sf_1}\left(M_{11}^\perp R_{n\perp}^{f_N a} e^{-2q_{nf_N} z_N} + M_{12}^\perp\right)}$$

Hence we can get the other three coefficients as the following forms:

$$B_N = \frac{-R_{n\perp}^{f_N a} T_{n\perp}^{af_N} S_{nyy,a} \left(M_{21}^\perp + R_{n\perp}^{sf_1} M_{11}^\perp\right) e^{-q_{na} z'} e^{q_{na} z_N} e^{-q_{nf_N} z_N}}{M_{21}^\perp R_{n\perp}^{f_N a} e^{-2q_{nf_N} z_N} + M_{22}^\perp + R_{n\perp}^{sf_1}\left(M_{11}^\perp R_{n\perp}^{f_N a} e^{-2q_{nf_N} z_N} + M_{12}^\perp\right)} e^{-2q_{nf_N} z_N} + T_{n\perp}^{af_N} S_{nyy,a} e^{-q_{na} z'} e^{q_{na} z_N} e^{-q_{nf_N} z_N}$$
$$= \frac{S_{nyy,a} T_{n\perp}^{af_N} \left(M_{22}^\perp + R_{n\perp}^{sf_1} M_{12}^\perp\right) e^{-q_{na} z'} e^{q_{na} z_N} e^{-q_{nf_N} z_N}}{M_{21}^\perp R_{n\perp}^{f_N a} e^{-2q_{nf_N} z_N} + M_{22}^\perp + R_{n\perp}^{sf_1}\left(M_{11}^\perp R_{n\perp}^{f_N a} e^{-2q_{nf_N} z_N} + M_{12}^\perp\right)}$$

, (295)

$$A = T_{n\perp}^{f_N a} C_N e^{-q_{nf_N} z_N} e^{q_{na} z_N} - R_{n\perp}^{f_N a} S_{nyy,a} e^{-q_{na} z'} e^{2q_{na} z_N}$$
$$= \frac{-T_{n\perp}^{f_N a} T_{n\perp}^{af_N} S_{nyy,a} \left(M_{21}^\perp + R_{n\perp}^{sf_1} M_{11}^\perp\right) e^{-q_{na} z'} e^{q_{na} z_N} e^{-q_{nf_N} z_N}}{M_{21}^\perp R_{n\perp}^{f_N a} e^{-2q_{nf_N} z_N} + M_{22}^\perp + R_{n\perp}^{sf_1}\left(M_{11}^\perp R_{n\perp}^{f_N a} e^{-2q_{nf_N} z_N} + M_{12}^\perp\right)} e^{-q_{nf_N} z_N} e^{q_{na} z_N} - R_{n\perp}^{f_N a} S_{nyy,a} e^{-q_{na} z'} e^{2q_{na} z_N}$$
$$= -S_{nyy,a} \frac{M_{21}^\perp e^{-2q_{nf_N} z_N} + R_{n\perp}^{sf_1} M_{11}^\perp e^{-2q_{nf_N} z_N} + R_{n\perp}^{f_N a} M_{22}^\perp + R_{n\perp}^{f_N a} R_{n\perp}^{sf_1} M_{12}^\perp}{M_{21}^\perp R_{n\perp}^{f_N a} e^{-2q_{nf_N} z_N} + M_{22}^\perp + R_{n\perp}^{sf_1}\left(M_{11}^\perp R_{n\perp}^{f_N a} e^{-2q_{nf_N} z_N} + M_{12}^\perp\right)} e^{-q_{na} z'} e^{2q_{na} z_N}$$

, (296)

$$
\begin{aligned}
D &= \left( M_{11}^{\perp} R_{n\perp}^{f_N a} e^{-2q_{nf_N} z_N} + M_{12}^{\perp} + M_{21}^{\perp} R_{n\perp}^{f_N a} e^{-2q_{nf_N} z_N} + M_{22}^{\perp} \right) C_N \\
&\quad + M_{11}^{\perp} T_{n\perp}^{af_N} S_{nyy,a} e^{-q_{na} z'} e^{q_{na} z_N} e^{-q_{nf_N} z_N} + M_{21}^{\perp} T_{n\perp}^{af_N} S_{nyy,a} e^{-q_{na} z'} e^{q_{na} z_N} e^{-q_{nf_N} z_N} \\
&= \frac{ -T_{n\perp}^{af_N} S_{nyy,a} \left( M_{11}^{\perp} R_{n\perp}^{f_N a} e^{-2q_{nf_N} z_N} + M_{12}^{\perp} + M_{21}^{\perp} R_{n\perp}^{f_N a} e^{-2q_{nf_N} z_N} + M_{22}^{\perp} \right) \left( M_{21}^{\perp} + R_{n\perp}^{sf_1} M_{11}^{\perp} \right) e^{-q_{na} z'} e^{q_{na} z_N} e^{-q_{nf_N} z_N} }{ M_{21}^{\perp} R_{n\perp}^{f_N a} e^{-2q_{nf_N} z_N} + M_{22}^{\perp} + R_{n\perp}^{sf_1} \left( M_{11}^{\perp} R_{n\perp}^{f_N a} e^{-2q_{nf_N} z_N} + M_{12}^{\perp} \right) } \\
&\quad + M_{11}^{\perp} T_{n\perp}^{af_N} S_{nyy,a} e^{-q_{na} z'} e^{q_{na} z_N} e^{-q_{nf_N} z_N} + M_{21}^{\perp} T_{n\perp}^{af_N} S_{nyy,a} e^{-q_{na} z'} e^{q_{na} z_N} e^{-q_{nf_N} z_N} \\
&= \frac{ T_{n\perp}^{af_N} S_{nyy,a} \left( 1 - R_{n\perp}^{sf_1} \right) \left( M_{11}^{\perp} M_{22}^{\perp} - M_{12}^{\perp} M_{21}^{\perp} \right) e^{-q_{na} z'} e^{q_{na} z_N} e^{-q_{nf_N} z_N} }{ M_{21}^{\perp} R_{n\perp}^{f_N a} e^{-2q_{nf_N} z_N} + M_{22}^{\perp} + R_{n\perp}^{sf_1} \left( M_{11}^{\perp} R_{n\perp}^{f_N a} e^{-2q_{nf_N} z_N} + M_{12}^{\perp} \right) }
\end{aligned}
$$

. (297)

According to Eq. (280), the coefficients $B_j$ and $C_j, j = 1, 2, ..., N-1$ are obtained and then we solve the coefficients in Eq. (289) completely. Next we consider the location of a point source is inside the film layer, that is $z_{k-1} < z' < z_k, k = 1, 2, ..., N$. Fixed the index $k$, the corresponding different equation in regions $z > z_N$, $z_{j-1} < z < z_j, j = N, N-1, ..., 2, 1$ and $z < 0$ are:

$$
\begin{cases}
\left( \partial_z^2 - q_{na}^2 \right) \tilde{g}_{nyy} = 0, z > z_N \\
\left( \partial_z^2 - q_{nf_j}^2 \right) \tilde{g}_{nyy} = -\delta(z - z') \delta_{jk}, z_{j-1} < z < z_j, j = 1, 2, ..., N \\
\left( \partial_z^2 - q_{ns}^2 \right) \tilde{g}_{nyy} = 0, z < 0
\end{cases}
\quad (298)
$$

Furthermore, the solutions of Eq. (298) are:

$$
\tilde{g}_{nyy} = \begin{cases}
A e^{-q_{na} z}, z > z_N \\
S_{nyy, f_k} e^{-q_{nf_k} |z - z'|} \delta_{jk} + B_j e^{q_{nf_j} z} + C_j e^{-q_{nf_j} z}, z_{j-1} < z < z_j, j = 1, 2, ..., N, \\
D e^{q_{ns} z}, z < 0
\end{cases}
\quad (299)
$$

where $S_{nyy, f_k} = \dfrac{1}{2 q_{nf_k}}$. Using the boundary conditions of $\tilde{g}_{nyy}$ as shown in Eq. (276), we have the following simulated equations:

<1> If $k = N$:

$$
\begin{cases}
A e^{-q_{na} z_N} = S_{nyy, f_N} e^{q_{nf_N} z'} e^{-q_{nf_N} z_N} + B_N e^{q_{nf_N} z_N} + C_N e^{-q_{nf_N} z_N} \\
S_{nyy, f_N} e^{-q_{nf_N} z'} e^{q_{nf_N} z_{N-1}} + B_N e^{q_{nf_N} z_{N-1}} + C_N e^{-q_{nf_N} z_{N-1}} = B_{N-1} e^{q_{nf_{N-1}} z_{N-1}} + C_{N-1} e^{-q_{nf_{N-1}} z_{N-1}} \\
B_j e^{q_{nf_j} z_j} + C_j e^{-q_{nf_j} z_j} = B_{j+1} e^{q_{nf_{j+1}} z_j} + C_{j+1} e^{-q_{nf_{j+1}} z_j}, j = 1, ..., N-2 \\
B_1 + C_1 = D
\end{cases}
\quad (300)
$$

and

$$\begin{cases} -q_{na}Ae^{-q_{na}z_N} = -S_{nyy,f_N}q_{nf_N}e^{q_{nf_N}z'}e^{-q_{nf_N}z_N} + q_{nf_N}\left(B_N e^{q_{nf_N}z_N} - C_N e^{-q_{nf_N}z_N}\right) \\ S_{nyy,f_N}q_{nf_N}e^{-q_{nf_N}z'}e^{q_{nf_N}z_{N-1}} + q_{nf_N}\left[B_N e^{q_{nf_N}z_{N-1}} - C_N e^{-q_{nf_N}z_{N-1}}\right] = q_{nf_{N-1}}\left[B_{N-1}e^{q_{nf_{N-1}}z_{N-1}} - C_{N-1}e^{-q_{nf_{N-1}}z_{N-1}}\right] \\ q_{nf_j}\left[B_j e^{q_{nf_j}z_j} - C_j e^{-q_{nf_j}z_j}\right] = q_{nf_{j+1}}\left[B_{j+1}e^{q_{nf_{j+1}}z_j} - C_{j+1}e^{-q_{nf_{j+1}}z_j}\right], j=1,...,N-2 \\ q_{nf_1}(B_1 - C_1) = Dq_{ns} \end{cases}$$

(301)

<2> If $k = 2, 3, ..., N-1$

$$\begin{cases} Ae^{-q_{na}z_N} = B_N e^{q_{nf_N}z_N} + C_N e^{-q_{nf_N}z_N} \\ B_{k+1}e^{q_{nf_{k+1}}z_k} + C_{k+1}e^{-q_{nf_{k+1}}z_k} = S_{nyy,f_k}e^{q_{nf_k}z'}e^{-q_{nf_k}z_k} + B_k e^{q_{nf_k}z_k} + C_k e^{-q_{nf_k}z_k} \\ S_{nyy,f_k}e^{-q_{nf_k}z'}e^{q_{nf_k}z_{k-1}} + B_k e^{q_{nf_k}z_{k-1}} + C_k e^{-q_{nf_k}z_{k-1}} = B_{k-1}e^{q_{nf_{k-1}}z_{k-1}} + C_{k-1}e^{-q_{nf_{k-1}}z_{k-1}} \\ B_j e^{q_{nf_j}z_j} + C_j e^{-q_{nf_j}z_j} = B_{j+1}e^{q_{nf_{j+1}}z_j} + C_{j+1}e^{-q_{nf_{j+1}}z_j}, j=1,...,N-1, j \neq k-1, k \\ B_1 + C_1 = D \end{cases}$$

(302)

and

$$\begin{cases} -q_{na}Ae^{-q_{na}z_N} = q_{nf_N}\left(B_N e^{q_{nf_N}z_N} - C_N e^{-q_{nf_N}z_N}\right) \\ q_{nf_{k+1}}\left[B_{k+1}e^{q_{nf_{k+1}}z_k} - C_{k+1}e^{-q_{nf_{k+1}}z_k}\right] = -q_{nf_k}S_{nyy,f_k}e^{q_{nf_k}z'}e^{-q_{nf_k}z_k} + q_{nf_k}\left[B_k e^{q_{nf_k}z_k} - C_k e^{-q_{nf_k}z_k}\right] \\ q_{nf_k}S_{nyy,f_k}e^{-q_{nf_k}z'}e^{q_{nf_k}z_{k-1}} + q_{nf_k}\left[B_k e^{q_{nf_k}z_{k-1}} - C_k e^{-q_{nf_k}z_{k-1}}\right] = q_{nf_{k-1}}\left[B_{k-1}e^{q_{nf_{k-1}}z_{k-1}} - C_{k-1}e^{-q_{nf_{k-1}}z_{k-1}}\right] \\ q_{nf_j}\left[B_j e^{q_{nf_j}z_j} - C_j e^{-q_{nf_j}z_j}\right] = q_{nf_{j+1}}\left[B_{j+1}e^{q_{nf_{j+1}}z_j} - C_{j+1}e^{-q_{nf_{j+1}}z_j}\right], j=1,...,N-1, j \neq k-1, k \\ q_{nf_1}(B_1 - C_1) = q_{ns}D \end{cases}$$

(303)

<3> If $k = 1$

$$\begin{cases} Ae^{-q_{na}z_N} = B_N e^{q_{nf_N}z_N} + C_N e^{-q_{nf_N}z_N} \\ B_j e^{q_{nf_j}z_j} + C_j e^{-q_{nf_j}z_j} = B_{j+1}e^{q_{nf_{j+1}}z_j} + C_{j+1}e^{-q_{nf_{j+1}}z_j}, j=2,...,N-1 \\ B_2 e^{q_{nf_2}z_1} + C_2 e^{-q_{nf_2}z_1} = S_{nyy,f_1}e^{q_{nf_1}z'}e^{-q_{nf_1}z_1} + B_1 e^{q_{nf_1}z_1} + C_1 e^{-q_{nf_1}z_1} \\ S_{nyy,f_1}e^{-q_{nf_1}z'} + B_1 + C_1 = D \end{cases}$$

(304)

and

$$\begin{cases} -q_{na}Ae^{-q_{na}z_N} = q_{nf_N}\left(B_N e^{q_{nf_N}z_N} - C_N e^{-q_{nf_N}z_N}\right) \\ q_{nf_j}\left[B_j e^{q_{nf_j}z_j} - C_j e^{-q_{nf_j}z_j}\right] = q_{nf_{j+1}}\left[B_{j+1}e^{q_{nf_{j+1}}z_j} - C_{j+1}e^{-q_{nf_{j+1}}z_j}\right], j=2,...,N-1 \\ q_{nf_2}\left[B_2 e^{q_{nf_2}z_1} - C_2 e^{-q_{nf_2}z_1}\right] = -q_{nf_1}S_{nyy,f_1}e^{q_{nf_1}z'}e^{-q_{nf_1}z_1} + q_{nf_1}\left[B_1 e^{q_{nf_1}z_1} - C_1 e^{-q_{nf_1}z_1}\right] \\ q_{nf_1}S_{nyy,f_1}e^{-q_{nf_1}z'} + q_{nf_1}(B_1 - C_1) = q_{ns}D \end{cases} \quad (305)$$

Furthermore, we simplify these above equations and show as the following forms:

<1> $k = N$

$$-q_{na}\left(S_{nyy,f_N}e^{q_{nf_N}z'}e^{-q_{nf_N}z_N} + B_N e^{q_{nf_N}z_N} + C_N e^{-q_{nf_N}z_N}\right)$$

$$= -S_{nyy,f_N}q_{nf_N}e^{q_{nf_N}z'}e^{-q_{nf_N}z_N} + q_{nf_N}\left(B_N e^{q_{nf_N}z_N} - C_N e^{-q_{nf_N}z_N}\right)$$

$$\Rightarrow \left(q_{nf_N} + q_{na}\right)B_N = \left(q_{nf_N} - q_{na}\right)C_N e^{-2q_{nf_N}z_N} + \left(q_{nf_N} - q_{na}\right)S_{nyy,f_N}e^{q_{nf_N}z'}e^{-2q_{nf_N}z_N}$$

$$\Rightarrow B_N = \frac{q_{nf_N} - q_{na}}{q_{nf_N} + q_{na}}C_N e^{-2q_{nf_N}z_N} + \frac{q_{nf_N} - q_{na}}{q_{nf_N} + q_{na}}S_{nyy,f_N}e^{q_{nf_N}z'}e^{-2q_{nf_N}z_N} \quad (306)$$

$$= R_{n\perp}^{f_N a}C_N e^{-2q_{nf_N}z_N} + R_{n\perp}^{f_N a}S_{nyy,f_N}e^{q_{nf_N}z'}e^{-2q_{nf_N}z_N}$$

$$\Rightarrow A = S_{nyy,f_N}e^{q_{nf_N}z'}e^{-q_{nf_N}z_N}e^{q_{na}z_N} + B_N e^{q_{nf_N}z_N}e^{q_{na}z_N} + C_N e^{-q_{nf_N}z_N}e^{q_{na}z_N}$$

$$= T_{n\perp}^{f_N a}C_N e^{-q_{nf_N}z_N}e^{q_{na}z_N} + T_{n\perp}^{f_N a}S_{nyy,f_N}e^{q_{nf_N}z'}e^{-q_{nf_N}z_N}e^{q_{na}z_N}$$

and

$$\begin{cases} S_{nyy,f_N}e^{-q_{nf_N}z'}e^{q_{nf_N}z_{N-1}} + B_N e^{q_{nf_N}z_{N-1}} + C_N e^{-q_{nf_N}z_{N-1}} = B_{N-1}e^{q_{nf_{N-1}}z_{N-1}} + C_{N-1}e^{-q_{nf_{N-1}}z_{N-1}} \\ S_{nyy,f_N}q_{nf_N}e^{-q_{nf_N}z'}e^{q_{nf_N}z_{N-1}} + q_{nf_N}\left[B_N e^{q_{nf_N}z_{N-1}} - C_N e^{-q_{nf_N}z_{N-1}}\right] = q_{nf_{N-1}}\left[B_{N-1}e^{q_{nf_{N-1}}z_{N-1}} - C_{N-1}e^{-q_{nf_{N-1}}z_{N-1}}\right] \end{cases}$$

$$\Rightarrow \begin{pmatrix} B_N \\ C_N \end{pmatrix} = \left[T_{N-1}^\perp\right]^{-1}\begin{pmatrix} B_{N-1} \\ C_{N-1} \end{pmatrix} - \begin{pmatrix} e^{q_{nf_N}z_{N-1}} & e^{-q_{nf_N}z_{N-1}} \\ q_{nf_N}e^{q_{nf_N}z_{N-1}} & -q_{nf_N}e^{-q_{nf_N}z_{N-1}} \end{pmatrix}^{-1}\begin{pmatrix} 1 \\ q_{nf_N} \end{pmatrix}S_{nyy,f_N}e^{-q_{nf_N}z'}e^{q_{nf_N}z_{N-1}}$$

$$= \left[T_{N-1}^\perp\right]^{-1}\begin{pmatrix} B_{N-1} \\ C_{N-1} \end{pmatrix} - \frac{1}{2q_{nf_N}}\begin{pmatrix} q_{nf_N}e^{-q_{nf_N}z_{N-1}} & e^{-q_{nf_N}z_{N-1}} \\ q_{nf_N}e^{q_{nf_N}z_{N-1}} & -e^{q_{nf_N}z_{N-1}} \end{pmatrix}\begin{pmatrix} 1 \\ q_{nf_N} \end{pmatrix}S_{nyy,f_N}e^{-q_{nf_N}z'}e^{q_{nf_N}z_{N-1}}$$

$$= \left[T_{N-1}^\perp\right]^{-1}\begin{pmatrix} B_{N-1} \\ C_{N-1} \end{pmatrix} - \begin{pmatrix} 1 \\ 0 \end{pmatrix}S_{nyy,f_N}e^{-q_{nf_N}z'}$$

$$\Rightarrow \begin{pmatrix} B_{N-1} \\ C_{N-1} \end{pmatrix} = T_{N-1}^\perp \begin{pmatrix} B_N \\ C_N \end{pmatrix} + T_{N-1}^\perp \begin{pmatrix} 1 \\ 0 \end{pmatrix}S_{nyy,f_N}e^{-q_{nf_N}z'}$$

. (307)

Combining Eq. (307) with the transfer matrix method which is described in Eqs. (280)-(282), we obtain:

$$\begin{pmatrix} B_1 \\ C_1 \end{pmatrix} = T_1^\perp T_2^\perp \bullet \bullet \bullet T_{N-2}^\perp \begin{pmatrix} B_{N-1} \\ C_{N-1} \end{pmatrix} = T_1^\perp T_2^\perp \bullet \bullet \bullet T_{N-2}^\perp T_{N-1}^\perp \begin{pmatrix} B_N \\ C_N \end{pmatrix} + T_1^\perp T_2^\perp \bullet \bullet \bullet T_{N-2}^\perp T_{N-1}^\perp \begin{pmatrix} 1 \\ 0 \end{pmatrix} S_{nyy,f_N} e^{-q_{nf_N} z'}$$

$$\equiv \begin{pmatrix} M_{11}^\perp & M_{12}^\perp \\ M_{21}^\perp & M_{22}^\perp \end{pmatrix} \begin{pmatrix} B_N \\ C_N \end{pmatrix} + \begin{pmatrix} M_{11}^\perp & M_{12}^\perp \\ M_{21}^\perp & M_{22}^\perp \end{pmatrix} \begin{pmatrix} 1 \\ 0 \end{pmatrix} S_{nyy,f_N} e^{-q_{nf_N} z'}$$

$$= \begin{pmatrix} M_{11}^\perp B_N + M_{12}^\perp C_N \\ M_{21}^\perp B_N + M_{22}^\perp C_N \end{pmatrix} + \begin{pmatrix} M_{11}^\perp \\ M_{21}^\perp \end{pmatrix} S_{nyy,f_N} e^{-q_{nf_N} z'}$$

, (308)

$$\begin{pmatrix} B_j \\ C_j \end{pmatrix} = T_j^\perp \bullet \bullet \bullet T_{N-2}^\perp \begin{pmatrix} B_{N-1} \\ C_{N-1} \end{pmatrix} = T_j^\perp \bullet \bullet \bullet T_{N-2}^\perp T_{N-1}^\perp \begin{pmatrix} B_N \\ C_N \end{pmatrix} + T_j^\perp \bullet \bullet \bullet T_{N-2}^\perp T_{N-1}^\perp \begin{pmatrix} 1 \\ 0 \end{pmatrix} S_{nyy,f_N} e^{-q_{nf_N} z'}$$

$$\equiv \begin{pmatrix} M_{j,11}^\perp & M_{j,12}^\perp \\ M_{j,21}^\perp & M_{j,22}^\perp \end{pmatrix} \begin{pmatrix} B_N \\ C_N \end{pmatrix} + \begin{pmatrix} M_{j,11}^\perp & M_{j,12}^\perp \\ M_{j,21}^\perp & M_{j,22}^\perp \end{pmatrix} \begin{pmatrix} 1 \\ 0 \end{pmatrix} S_{nyy,f_N} e^{-q_{nf_N} z'}$$

$$= \begin{pmatrix} M_{j,11}^\perp B_N + M_{j,12}^\perp C_N \\ M_{j,21}^\perp B_N + M_{j,22}^\perp C_N \end{pmatrix} + \begin{pmatrix} M_{j,11}^\perp \\ M_{j,21}^\perp \end{pmatrix} S_{nyy,f_N} e^{-q_{nf_N} z'}, \quad j = 2, 3, \dots, N-2$$

, (309)

and we obtain:

$$D = B_1 + C_1 = \left( M_{11}^\perp R_{n\perp}^{f_N a} e^{-2q_{nf_N} z_N} + M_{12}^\perp + M_{21}^\perp R_{n\perp}^{f_N a} e^{-2q_{nf_N} z_N} + M_{22}^\perp \right) C_N$$

$$+ M_{11}^\perp R_{n\perp}^{f_N a} S_{nyy,f_N} e^{q_{nf_N} z'} e^{-2q_{nf_N} z_N} + M_{11}^\perp S_{nyy,f_N} e^{-q_{nf_N} z'} + M_{21}^\perp R_{n\perp}^{f_N a} S_{nyy,f_N} e^{q_{nf_N} z'} e^{-2q_{nf_N} z_N} + M_{21}^\perp S_{nyy,f_N} e^{-q_{nf_N} z'}$$

. (310)

Furthermore, we get the following form to solve the coefficient $C_N$:

$$q_{nf_1} (B_1 - C_1) = (B_1 + C_1) q_{ns}$$

$$\Rightarrow (q_{ns} - q_{nf_1}) B_1 = -(q_{ns} + q_{nf_1}) C_1$$

$$\Rightarrow C_1 = -R_{n\perp}^{sf_1} B_1$$

$$\Rightarrow \left( M_{21}^\perp R_{n\perp}^{f_N a} e^{-2q_{nf_N} z_N} + M_{22}^\perp \right) C_N + M_{21}^\perp R_{n\perp}^{f_N a} S_{nyy,f_N} e^{q_{nf_N} z'} e^{-2q_{nf_N} z_N} + M_{21}^\perp S_{nyy,f_N} e^{-q_{nf_N} z'}$$

$$= -R_{n\perp}^{sf_1} \left( M_{11}^\perp R_{n\perp}^{f_N a} e^{-2q_{nf_N} z_N} + M_{12}^\perp \right) C_N - R_{n\perp}^{sf_1} \left( M_{11}^\perp R_{n\perp}^{f_N a} S_{nyy,f_N} e^{q_{nf_N} z'} e^{-2q_{nf_N} z_N} + M_{11}^\perp S_{nyy,f_N} e^{-q_{nf_N} z'} \right)$$

$$\Rightarrow \left[ M_{21}^\perp R_{n\perp}^{f_N a} e^{-2q_{nf_N} z_N} + M_{22}^\perp + R_{n\perp}^{sf_1} \left( M_{11}^\perp R_{n\perp}^{f_N a} e^{-2q_{nf_N} z_N} + M_{12}^\perp \right) \right] C_N$$

$$= -S_{nyy,f_N} \left[ M_{21}^\perp R_{n\perp}^{f_N a} e^{q_{nf_N} z'} e^{-2q_{nf_N} z_N} + M_{21}^\perp e^{-q_{nf_N} z'} + R_{n\perp}^{sf_1} \left( M_{11}^\perp R_{n\perp}^{f_N a} e^{q_{nf_N} z'} e^{-2q_{nf_N} z_N} + M_{11}^\perp e^{-q_{nf_N} z'} \right) \right]$$

$$\Rightarrow C_N = \frac{-S_{nyy,f_N} \left[ M_{21}^\perp R_{n\perp}^{f_N a} e^{q_{nf_N} z'} e^{-2q_{nf_N} z_N} + M_{21}^\perp e^{-q_{nf_N} z'} + R_{n\perp}^{sf_1} \left( M_{11}^\perp R_{n\perp}^{f_N a} e^{q_{nf_N} z'} e^{-2q_{nf_N} z_N} + M_{11}^\perp e^{-q_{nf_N} z'} \right) \right]}{M_{21}^\perp R_{n\perp}^{f_N a} e^{-2q_{nf_N} z_N} + M_{22}^\perp + R_{n\perp}^{sf_1} \left( M_{11}^\perp R_{n\perp}^{f_N a} e^{-2q_{nf_N} z_N} + M_{12}^\perp \right)}$$

. (311)

Hence we can get the other three coefficients as the following forms:

$$B_N = R_{n\perp}^{f_N a} e^{-2q_{nf_N} z_N} C_N + R_{n\perp}^{f_N a} S_{nyy,f_N} e^{q_{nf_N} z'} e^{-2q_{nf_N} z_N}$$

$$= -\frac{R_{n\perp}^{f_N a} e^{-2q_{nf_N} z_N} S_{nyy,f_N} \left[ M_{21}^{\perp} R_{n\perp}^{f_N a} e^{q_{nf_N} z'} e^{-2q_{nf_N} z_N} + M_{21}^{\perp} e^{-q_{nf_N} z'} + R_{n\perp}^{sf_1} \left( M_{11}^{\perp} R_{n\perp}^{f_N a} e^{q_{nf_N} z'} e^{-2q_{nf_N} z_N} + M_{11}^{\perp} e^{-q_{nf_N} z'} \right) \right]}{M_{21}^{\perp} R_{n\perp}^{f_N a} e^{-2q_{nf_N} z_N} + M_{22}^{\perp} + R_{n\perp}^{sf_1} \left( M_{11}^{\perp} R_{n\perp}^{f_N a} e^{-2q_{nf_N} z_N} + M_{12}^{\perp} \right)}$$

$$+ R_{n\perp}^{f_N a} S_{nyy,f_N} e^{q_{nf_N} z'} e^{-2q_{nf_N} z_N}$$

$$= R_{n\perp}^{f_N a} e^{-2q_{nf_N} z_N} S_{nyy,f_N} \frac{-M_{21}^{\perp} e^{-q_{nf_N} z'} - R_{n\perp}^{sf_1} M_{11}^{\perp} e^{-q_{nf_N} z'} + M_{22}^{\perp} e^{q_{nf_N} z'} + M_{12}^{\perp} e^{q_{nf_N} z'}}{M_{21}^{\perp} R_{n\perp}^{f_N a} e^{-2q_{nf_N} z_N} + M_{22}^{\perp} + R_{n\perp}^{sf_1} \left( M_{11}^{\perp} R_{n\perp}^{f_N a} e^{-2q_{nf_N} z_N} + M_{12}^{\perp} \right)}$$

, (312)

$$A = T_{n\perp}^{f_N a} e^{-q_{nf_N} z_N} e^{q_{na} z_N} C_N + T_{n\perp}^{f_N a} S_{nyy,f_N} e^{q_{nf_N} z'} e^{-q_{nf_N} z_N} e^{q_{na} z_N}$$

$$= T_{n\perp}^{f_N a} e^{-q_{nf_N} z_N} e^{q_{na} z_N} \frac{-S_{nyy,f_N} \left[ M_{21}^{\perp} R_{n\perp}^{f_N a} e^{q_{nf_N} z'} e^{-2q_{nf_N} z_N} + M_{21}^{\perp} e^{-q_{nf_N} z'} + R_{n\perp}^{sf_1} \left( M_{11}^{\perp} R_{n\perp}^{f_N a} e^{q_{nf_N} z'} e^{-2q_{nf_N} z_N} + M_{11}^{\perp} e^{-q_{nf_N} z'} \right) \right]}{M_{21}^{\perp} R_{n\perp}^{f_N a} e^{-2q_{nf_N} z_N} + M_{22}^{\perp} + R_{n\perp}^{sf_1} \left( M_{11}^{\perp} R_{n\perp}^{f_N a} e^{-2q_{nf_N} z_N} + M_{12}^{\perp} \right)}$$

$$+ T_{n\perp}^{f_N a} S_{nyy,f_N} e^{q_{nf_N} z'} e^{-q_{nf_N} z_N} e^{q_{na} z_N}$$

$$= T_{n\perp}^{f_N a} S_{nyy,f_N} e^{-q_{nf_N} z_N} e^{q_{na} z_N} \frac{-M_{21}^{\perp} e^{-q_{nf_N} z'} - R_{n\perp}^{sf_1} M_{11}^{\perp} e^{-q_{nf_N} z'} + e^{q_{nf_N} z'} M_{22}^{\perp} + R_{n\perp}^{sf_1} M_{12}^{\perp} e^{q_{nf_N} z'}}{M_{21}^{\perp} R_{n\perp}^{f_N a} e^{-2q_{nf_N} z_N} + M_{22}^{\perp} + R_{n\perp}^{sf_1} \left( M_{11}^{\perp} R_{n\perp}^{f_N a} e^{-2q_{nf_N} z_N} + M_{12}^{\perp} \right)}$$

, (313)

$$D = \left( M_{11}^{\perp} R_{n\perp}^{f_N a} e^{-2q_{nf_N} z_N} + M_{12}^{\perp} + M_{21}^{\perp} R_{n\perp}^{f_N a} e^{-2q_{nf_N} z_N} + M_{22}^{\perp} \right) C_N$$

$$+ M_{11}^{\perp} R_{n\perp}^{f_N a} S_{nyy,f_N} e^{q_{nf_N} z'} e^{-2q_{nf_N} z_N} + M_{11}^{\perp} S_{nyy,f_N} e^{-q_{nf_N} z'} + M_{21}^{\perp} R_{n\perp}^{f_N a} S_{nyy,f_N} e^{q_{nf_N} z'} e^{-2q_{nf_N} z_N} + M_{21}^{\perp} S_{nyy,f_N} e^{-q_{nf_N} z'}$$

$$= -S_{nyy,f_N} T_{n\perp}^{f_1 s} \frac{\left( M_{12}^{\perp} M_{21}^{\perp} + M_{11}^{\perp} M_{22}^{\perp} \right) \left( R_{n\perp}^{f_N a} e^{q_{nf_N} z'} e^{-2q_{nf_N} z_N} + e^{-q_{nf_N} z'} \right)}{M_{21}^{\perp} R_{n\perp}^{f_N a} e^{-2q_{nf_N} z_N} + M_{22}^{\perp} + R_{n\perp}^{sf_1} \left( M_{11}^{\perp} R_{n\perp}^{f_N a} e^{-2q_{nf_N} z_N} + M_{12}^{\perp} \right)}$$

. (314)

Hence we solve the coefficients in Eq. (299) completely.

<2> If $k = 2, 3, ..., N-1$

$$-q_{na} \left( B_N e^{q_{nf_N} z_N} + C_N e^{-q_{nf_N} z_N} \right) = q_{nf_N} \left( B_N e^{q_{nf_N} z_N} - C_N e^{-q_{nf_N} z_N} \right)$$

$$\Rightarrow \left( q_{nf_N} + q_{na} \right) B_N = \left( q_{nf_N} - q_{na} \right) C_N e^{-2q_{nf_N} z_N}$$

$$\Rightarrow B_N = \frac{q_{nf_N} - q_{na}}{q_{nf_N} + q_{na}} C_N e^{-2q_{nf_N} z_N} = R_{n\perp}^{f_N a} C_N e^{-2q_{nf_N} z_N}$$

, (315)

$$\Rightarrow A = R_{n\perp}^{f_N a} C_N e^{-q_{nf_N} z_N} e^{q_{na} z_N} + C_N e^{-q_{nf_N} z_N} e^{q_{na} z_N} = T_{n\perp}^{f_N a} C_N e^{-q_{nf_N} z_N} e^{q_{na} z_N}$$

and

$$\begin{cases} B_{k+1}e^{q_{nf_{k+1}}z_k} + C_{k+1}e^{-q_{nf_{k+1}}z_k} = S_{nyy,f_k}e^{q_{nf_k}z'}e^{-q_{nf_k}z_k} + B_k e^{q_{nf_k}z_k} + C_k e^{-q_{nf_k}z_k} \\ q_{nf_{k+1}}\left[B_{k+1}e^{q_{nf_{k+1}}z_k} - C_{k+1}e^{-q_{nf_{k+1}}z_k}\right] = -q_{nf_k}S_{nyy,f_k}e^{q_{nf_k}z'}e^{-q_{nf_k}z_k} + q_{nf_k}\left[B_k e^{q_{nf_k}z_k} - C_k e^{-q_{nf_k}z_k}\right] \end{cases}$$

$$\Rightarrow \begin{pmatrix} B_k \\ C_k \end{pmatrix} = T_k^\perp \begin{pmatrix} B_{k+1} \\ C_{k+1} \end{pmatrix} + \begin{pmatrix} e^{q_{nf_k}z_k} & e^{-q_{nf_k}z_k} \\ q_{nf_k}e^{q_{nf_k}z_k} & -q_{nf_k}e^{-q_{nf_k}z_k} \end{pmatrix}^{-1} \begin{pmatrix} -1 \\ q_{nf_k} \end{pmatrix} S_{nyy,f_k} e^{q_{nf_k}z'} e^{-q_{nf_k}z_k}$$

$$= T_k^\perp \begin{pmatrix} B_{k+1} \\ C_{k+1} \end{pmatrix} + \frac{1}{2q_{nf_k}} \begin{pmatrix} q_{nf_k}e^{-q_{nf_k}z_k} & e^{-q_{nf_k}z_k} \\ q_{nf_k}e^{q_{nf_k}z_k} & -e^{q_{nf_k}z_k} \end{pmatrix} \begin{pmatrix} -1 \\ q_{nf_k} \end{pmatrix} S_{nyy,f_k} e^{q_{nf_k}z'} e^{-q_{nf_k}z_k}$$

$$= T_k^\perp \begin{pmatrix} B_{k+1} \\ C_{k+1} \end{pmatrix} - \begin{pmatrix} 0 \\ 1 \end{pmatrix} S_{nyy,f_k} e^{q_{nf_k}z'}$$

(316)

and

$$\begin{cases} S_{nyy,f_k}e^{-q_{nf_k}z'}e^{q_{nf_k}z_{k-1}} + B_k e^{q_{nf_k}z_{k-1}} + C_k e^{-q_{nf_k}z_{k-1}} = B_{k-1}e^{q_{nf_{k-1}}z_{k-1}} + C_{k-1}e^{-q_{nf_{k-1}}z_{k-1}} \\ q_{nf_k}S_{nyy,f_k}e^{-q_{nf_k}z'}e^{q_{nf_k}z_{k-1}} + q_{nf_k}\left[B_k e^{q_{nf_k}z_{k-1}} - C_k e^{-q_{nf_k}z_{k-1}}\right] = q_{nf_{k-1}}\left[B_{k-1}e^{q_{nf_{k-1}}z_{k-1}} - C_{k-1}e^{-q_{nf_{k-1}}z_{k-1}}\right] \end{cases}$$

$$\Rightarrow \begin{pmatrix} B_k \\ C_k \end{pmatrix} = \left[T_{k-1}^\perp\right]^{-1} \begin{pmatrix} B_{k-1} \\ C_{k-1} \end{pmatrix} - \begin{pmatrix} e^{q_{nf_k}z_{k-1}} & e^{-q_{nf_k}z_{k-1}} \\ q_{nf_k}e^{q_{nf_k}z_{k-1}} & -q_{nf_k}e^{-q_{nf_k}z_{k-1}} \end{pmatrix}^{-1} \begin{pmatrix} 1 \\ q_{nf_k} \end{pmatrix} S_{nyy,f_k} e^{-q_{nf_k}z'} e^{q_{nf_k}z_{k-1}}$$

$$= \left[T_{k-1}^\perp\right]^{-1} \begin{pmatrix} B_{k-1} \\ C_{k-1} \end{pmatrix} - \frac{1}{2q_{nf_k}} \begin{pmatrix} q_{nf_k}e^{-q_{nf_k}z_{k-1}} & e^{-q_{nf_k}z_{k-1}} \\ q_{nf_k}e^{q_{nf_k}z_{k-1}} & -e^{q_{nf_k}z_{k-1}} \end{pmatrix} \begin{pmatrix} 1 \\ q_{nf_k} \end{pmatrix} S_{nyy,f_k} e^{-q_{nf_k}z'} e^{q_{nf_k}z_{k-1}}$$

$$= \left[T_{k-1}^\perp\right]^{-1} \begin{pmatrix} B_{k-1} \\ C_{k-1} \end{pmatrix} - \begin{pmatrix} 1 \\ 0 \end{pmatrix} S_{nyy,f_k} e^{-q_{nf_k}z'}$$

$$\Rightarrow \begin{pmatrix} B_{k-1} \\ C_{k-1} \end{pmatrix} = T_{k-1}^\perp \begin{pmatrix} B_k \\ C_k \end{pmatrix} + T_{k-1}^\perp \begin{pmatrix} 1 \\ 0 \end{pmatrix} S_{nyy,f_k} e^{-q_{nf_k}z'}$$

(317)

Combining Eqs. (316)-(317) with the transfer matrix method which is described in Eqs. (280)-(282), we obtain:

$$\begin{pmatrix} B_1 \\ C_1 \end{pmatrix} = T_1^\perp T_2^\perp \bullet\bullet T_{k-2}^\perp \begin{pmatrix} B_{k-1} \\ C_{k-1} \end{pmatrix} = T_1^\perp T_2^\perp \bullet\bullet T_{k-2}^\perp T_{k-1}^\perp \begin{pmatrix} B_k \\ C_k \end{pmatrix} + T_1^\perp T_2^\perp \bullet\bullet T_{k-2}^\perp T_{k-1}^\perp \begin{pmatrix} 1 \\ 0 \end{pmatrix} S_{nyy,f_k} e^{-q_{nf_k}z'}$$

$$= T_1^\perp T_2^\perp \bullet\bullet T_{k-2}^\perp T_{k-1}^\perp T_k^\perp \begin{pmatrix} B_{k+1} \\ C_{k+1} \end{pmatrix} - T_1^\perp T_2^\perp \bullet\bullet T_{k-2}^\perp T_{k-1}^\perp \begin{pmatrix} 0 \\ 1 \end{pmatrix} S_{nyy,f_k} e^{q_{nf_k}z'} + T_1^\perp T_2^\perp \bullet\bullet T_{k-2}^\perp T_{k-1}^\perp \begin{pmatrix} 1 \\ 0 \end{pmatrix} S_{nyy,f_k} e^{-q_{nf_k}z'}$$

$$\equiv \begin{pmatrix} M_{11}^\perp & M_{12}^\perp \\ M_{21}^\perp & M_{22}^\perp \end{pmatrix} \begin{pmatrix} B_N \\ C_N \end{pmatrix} - \begin{pmatrix} M_{11,k-1}^\perp & M_{12,k-1}^\perp \\ M_{21,k-1}^\perp & M_{22,k-1}^\perp \end{pmatrix} \begin{pmatrix} 0 \\ 1 \end{pmatrix} S_{nyy,f_k} e^{q_{nf_k}z'} + \begin{pmatrix} M_{11,k-1}^\perp & M_{12,k-1}^\perp \\ M_{21,k-1}^\perp & M_{22,k-1}^\perp \end{pmatrix} \begin{pmatrix} 1 \\ 0 \end{pmatrix} S_{nyy,f_k} e^{-q_{nf_k}z'}$$

$$= \begin{pmatrix} M_{11}^\perp B_N + M_{12}^\perp C_N \\ M_{21}^\perp B_N + M_{22}^\perp C_N \end{pmatrix} - \begin{pmatrix} M_{12,k-1}^\perp \\ M_{22,k-1}^\perp \end{pmatrix} S_{nyy,f_k} e^{q_{nf_k}z'} + \begin{pmatrix} M_{11,k-1}^\perp \\ M_{21,k-1}^\perp \end{pmatrix} S_{nyy,f_k} e^{-q_{nf_k}z'}$$

,

(318)

$$\begin{pmatrix} B_j \\ C_j \end{pmatrix} = T_j^\perp \bullet \bullet T_{k-2}^\perp \begin{pmatrix} B_{k-1} \\ C_{k-1} \end{pmatrix} = T_j^\perp \bullet \bullet T_{k-2}^\perp T_{k-1}^\perp \begin{pmatrix} B_k \\ C_k \end{pmatrix} + T_j^\perp \bullet \bullet T_{k-2}^\perp T_{k-1}^\perp \begin{pmatrix} 1 \\ 0 \end{pmatrix} S_{nyy,f_k} e^{-q_{nf_k} z'}$$

$$= T_j^\perp \bullet \bullet T_{k-2}^\perp T_{k-1}^\perp T_k^\perp \begin{pmatrix} B_{k+1} \\ C_{k+1} \end{pmatrix} - T_j^\perp \bullet \bullet T_{k-2}^\perp T_{k-1}^\perp \begin{pmatrix} 0 \\ 1 \end{pmatrix} S_{nyy,f_k} e^{q_{nf_k} z'} + T_j^\perp \bullet \bullet T_{k-2}^\perp T_{k-1}^\perp \begin{pmatrix} 1 \\ 0 \end{pmatrix} S_{nyy,f_k} e^{-q_{nf_k} z'}$$

$$\equiv \begin{pmatrix} M_{j,11}^\perp & M_{j,12}^\perp \\ M_{j,21}^\perp & M_{j,22}^\perp \end{pmatrix} \begin{pmatrix} B_N \\ C_N \end{pmatrix} - \begin{pmatrix} M_{j,11,k-1}^\perp & M_{j,12,k-1}^\perp \\ M_{j,21,k-1}^\perp & M_{j,22,k-1}^\perp \end{pmatrix} \begin{pmatrix} 0 \\ 1 \end{pmatrix} S_{nyy,f_k} e^{q_{nf_k} z'} + \begin{pmatrix} M_{j,11,k-1}^\perp & M_{j,12,k-1}^\perp \\ M_{j,21,k-1}^\perp & M_{j,22,k-1}^\perp \end{pmatrix} \begin{pmatrix} 1 \\ 0 \end{pmatrix} S_{nyy,f_k} e^{-q_{nf_k} z'}$$

$$= \begin{pmatrix} M_{j,11}^\perp B_N + M_{j,12}^\perp C_N \\ M_{j,21}^\perp B_N + M_{j,22}^\perp C_N \end{pmatrix} - \begin{pmatrix} M_{j,12,k-1}^\perp \\ M_{j,22,k-1}^\perp \end{pmatrix} S_{nyy,f_k} e^{q_{nf_k} z'} + \begin{pmatrix} M_{j,11,k-1}^\perp \\ M_{j,21,k-1}^\perp \end{pmatrix} S_{nyy,f_k} e^{-q_{nf_k} z'}, \; j = 2,3,\ldots,N-1, j \leq k-1$$

, (319)

$$\begin{pmatrix} B_k \\ C_k \end{pmatrix} = T_k^\perp \begin{pmatrix} B_{k+1} \\ C_{k+1} \end{pmatrix} - \begin{pmatrix} 0 \\ 1 \end{pmatrix} S_{nyy,f_k} e^{q_{nf_k} z'} = T_k^\perp \bullet \bullet T_{N-1}^\perp \begin{pmatrix} B_N \\ C_N \end{pmatrix} - \begin{pmatrix} 0 \\ 1 \end{pmatrix} S_{nyy,f_k} e^{q_{nf_k} z'}$$

$$\equiv \begin{pmatrix} M_{k,11}^\perp & M_{k,12}^\perp \\ M_{k,21}^\perp & M_{k,22}^\perp \end{pmatrix} \begin{pmatrix} B_N \\ C_N \end{pmatrix} - \begin{pmatrix} 0 \\ 1 \end{pmatrix} S_{nyy,f_k} e^{q_{nf_k} z'}, \quad (320)$$

$$= \begin{pmatrix} M_{k,11}^\perp B_N + M_{k,12}^\perp C_N \\ M_{k,21}^\perp B_N + M_{k,22}^\perp C_N \end{pmatrix} - \begin{pmatrix} 0 \\ 1 \end{pmatrix} S_{nyy,f_k} e^{q_{nf_k} z'}, \; j=2,3,\ldots,N-1, j=k$$

$$\begin{pmatrix} B_j \\ C_j \end{pmatrix} = T_j^\perp \begin{pmatrix} B_{j+1} \\ C_{j+1} \end{pmatrix} = T_j^\perp T_{j+1}^\perp \bullet \bullet \bullet T_{N-1}^\perp \begin{pmatrix} B_N \\ C_N \end{pmatrix} \equiv \begin{pmatrix} M_{j,11}^\perp & M_{j,12}^\perp \\ M_{j,21}^\perp & M_{j,22}^\perp \end{pmatrix} \begin{pmatrix} B_N \\ C_N \end{pmatrix},$$

$$= \begin{pmatrix} M_{j,11}^\perp B_N + M_{j,12}^\perp C_N \\ M_{j,21}^\perp B_N + M_{j,22}^\perp C_N \end{pmatrix}, \; j=2,3,\ldots,N-1, j \geq k+1 \quad (321)$$

and we obtain:

$$D = B_1 + C_1 = M_{11}^\perp R_{n\perp}^{f_N a} e^{-2q_{nf_N} z_N} C_N + M_{12}^\perp C_N - M_{12,k-1}^\perp S_{nyy,f_k} e^{q_{nf_k} z'} + M_{11,k-1}^\perp S_{nyy,f_k} e^{-q_{nf_k} z'}$$

$$+ M_{21}^\perp R_{n\perp}^{f_N a} e^{-2q_{nf_N} z_N} C_N + M_{22}^\perp C_N - M_{22,k-1}^\perp S_{nyy,f_k} e^{q_{nf_k} z'} + M_{21,k-1}^\perp S_{nyy,f_k} e^{-q_{nf_k} z'}$$

$$= \left( M_{11}^\perp R_{n\perp}^{f_N a} e^{-2q_{nf_N} z_N} + M_{12}^\perp + M_{21}^\perp R_{n\perp}^{f_N a} e^{-2q_{nf_N} z_N} + M_{22}^\perp \right) C_N - M_{12,k-1}^\perp S_{nyy,f_k} e^{q_{nf_k} z'} + M_{11,k-1}^\perp S_{nyy,f_k} e^{-q_{nf_k} z'}$$

$$- M_{22,k-1}^\perp S_{nyy,f_k} e^{q_{nf_k} z'} + M_{21,k-1}^\perp S_{nyy,f_k} e^{-q_{nf_k} z'}$$

, (322)

Furthermore, we get the following form to solve the coefficient $C_N$:

$$q_{nf_1}(B_1 - C_1) = q_{ns}(B_1 + C_1)$$

$$\Rightarrow (q_{ns} - q_{nf_1})B_1 = -(q_{ns} + q_{nf_1})C_1$$

$$\Rightarrow C_1 = -\frac{q_{ns} - q_{nf_1}}{q_{ns} + q_{nf_1}}B_1 = -R_{n\perp}^{sf_1}B_1$$

$$\Rightarrow M_{21}^{\perp}R_{n\perp}^{f_N a}C_N e^{-2q_{nf_N}z_N} + M_{22}^{\perp}C_N - M_{22,k-1}^{\perp}S_{nyy,f_k}e^{q_{nf_k}z'} + M_{21,k-1}^{\perp}S_{nyy,f_k}e^{-q_{nf_k}z'}$$

$$= -R_{n\perp}^{sf_1}\left(M_{11}^{\perp}R_{n\perp}^{f_N a}C_N e^{-2q_{nf_N}z_N} + M_{12}^{\perp}C_N - M_{12,k-1}^{\perp}S_{nyy,f_k}e^{q_{nf_k}z'} + M_{11,k-1}^{\perp}S_{nyy,f_k}e^{-q_{nf_k}z'}\right)$$

$$\Rightarrow \left(M_{21}^{\perp}R_{n\perp}^{f_N a}e^{-2q_{nf_N}z_N} + M_{22}^{\perp} + R_{n\perp}^{sf_1}M_{11}^{\perp}R_{n\perp}^{f_N a}e^{-2q_{nf_N}z_N} + R_{n\perp}^{sf_1}M_{12}^{\perp}\right)C_N =$$

$$M_{22,k-1}^{\perp}S_{nyy,f_k}e^{q_{nf_k}z'} - M_{21,k-1}^{\perp}S_{nyy,f_k}e^{-q_{nf_k}z'} + R_{n\perp}^{sf_1}M_{12,k-1}^{\perp}S_{nyy,f_k}e^{q_{nf_k}z'} - R_{n\perp}^{sf_1}M_{11,k-1}^{\perp}S_{nyy,f_k}e^{-q_{nf_k}z'}$$

$$\Rightarrow C_N = S_{nyy,f_k}\frac{M_{22,k-1}^{\perp}e^{q_{nf_k}z'} - M_{21,k-1}^{\perp}e^{-q_{nf_k}z'} + R_{n\perp}^{sf_1}M_{12,k-1}^{\perp}e^{q_{nf_k}z'} - R_{n\perp}^{sf_1}M_{11,k-1}^{\perp}e^{-q_{nf_k}z'}}{M_{21}^{\perp}R_{n\perp}^{f_N a}e^{-2q_{nf_N}z_N} + M_{22}^{\perp} + R_{n\perp}^{sf_1}M_{11}^{\perp}R_{n\perp}^{f_N a}e^{-2q_{nf_N}z_N} + R_{n\perp}^{sf_1}M_{12}^{\perp}}$$

. (323)

Hence we can get the other three coefficients as the following forms:

$$B_N = R_{n\perp}^{f_N a}C_N e^{-2q_{nf_N}z_N}$$

$$= R_{n\perp}^{f_N a}e^{-2q_{nf_N}z_N}S_{nyy,f_k}\frac{M_{22,k-1}^{\perp}e^{q_{nf_k}z'} - M_{21,k-1}^{\perp}e^{-q_{nf_k}z'} + R_{n\perp}^{sf_1}M_{12,k-1}^{\perp}e^{q_{nf_k}z'} - R_{n\perp}^{sf_1}M_{11,k-1}^{\perp}e^{-q_{nf_k}z'}}{M_{21}^{\perp}R_{n\perp}^{f_N a}e^{-2q_{nf_N}z_N} + M_{22}^{\perp} + R_{n\perp}^{sf_1}M_{11}^{\perp}R_{n\perp}^{f_N a}e^{-2q_{nf_N}z_N} + R_{n\perp}^{sf_1}M_{12}^{\perp}},$$

(324)

$$A = T_{n\perp}^{f_N a}e^{-q_{nf_N}z_N}e^{q_{na}z_N}C_N$$

$$= T_{n\perp}^{f_N a}e^{-q_{nf_N}z_N}e^{q_{na}z_N}S_{nyy,f_k}\frac{M_{22,k-1}^{\perp}e^{q_{nf_k}z'} - M_{21,k-1}^{\perp}e^{-q_{nf_k}z'} + R_{n\perp}^{sf_1}M_{12,k-1}^{\perp}e^{q_{nf_k}z'} - R_{n\perp}^{sf_1}M_{11,k-1}^{\perp}e^{-q_{nf_k}z'}}{M_{21}^{\perp}R_{n\perp}^{f_N a}e^{-2q_{nf_N}z_N} + M_{22}^{\perp} + R_{n\perp}^{sf_1}M_{11}^{\perp}R_{n\perp}^{f_N a}e^{-2q_{nf_N}z_N} + R_{n\perp}^{sf_1}M_{12}^{\perp}},$$

(325)

$$D = \left(M_{11}^{\perp}R_{n\perp}^{f_N a}e^{-2q_{nf_N}z_N} + M_{12}^{\perp} + M_{21}^{\perp}R_{n\perp}^{f_N a}e^{-2q_{nf_N}z_N} + M_{22}^{\perp}\right)C_N$$

$$-M_{12,k-1}^{\perp}S_{nyy,f_k}e^{q_{nf_k}z'} + M_{11,k-1}^{\perp}S_{nyy,f_k}e^{-q_{nf_k}z'} - M_{22,k-1}^{\perp}S_{nyy,f_k}e^{q_{nf_k}z'} + M_{21,k-1}^{\perp}S_{nyy,f_k}e^{-q_{nf_k}z'}$$

$$= \frac{S_{nyy,f_k}T_{n\perp}^{f_1 s}\left(M_{22,k-1}^{\perp}e^{q_{nf_k}z'} - M_{21,k-1}^{\perp}e^{-q_{nf_k}z'}\right)\left(M_{11}^{\perp}R_{n\perp}^{f_N a}e^{-2q_{nf_N}z_N} + M_{12}^{\perp}\right)}{M_{21}^{\perp}R_{n\perp}^{f_N a}e^{-2q_{nf_N}z_N} + M_{22}^{\perp} + R_{n\perp}^{sf_1}M_{11}^{\perp}R_{n\perp}^{f_N a}e^{-2q_{nf_N}z_N} + R_{n\perp}^{sf_1}M_{12}^{\perp}}$$

$$+ \frac{S_{nyy,f_k}T_{n\perp}^{f_1 s}\left(M_{11,k-1}^{\perp}e^{-q_{nf_k}z'} - M_{12,k-1}^{\perp}e^{q_{nf_k}z'}\right)\left(M_{21}^{\perp}R_{n\perp}^{f_N a}e^{-2q_{nf_N}z_N} + M_{22}^{\perp}\right)}{M_{21}^{\perp}R_{n\perp}^{f_N a}e^{-2q_{nf_N}z_N} + M_{22}^{\perp} + R_{n\perp}^{sf_1}M_{11}^{\perp}R_{n\perp}^{f_N a}e^{-2q_{nf_N}z_N} + R_{n\perp}^{sf_1}M_{12}^{\perp}}$$

. (326)

Hence we solve the coefficients in Eq. (299) completely.

<3> If $k = 1$

$$-q_{na}\left(B_N e^{q_{nf_N} z_N} + C_N e^{-q_{nf_N} z_N}\right) = q_{nf_N}\left(B_N e^{q_{nf_N} z_N} - C_N e^{-q_{nf_N} z_N}\right)$$

$$\Rightarrow \left(q_{nf_N} + q_{na}\right) B_N = \left(q_{nf_N} - q_{na}\right) C_N e^{-2q_{nf_N} z_N}$$

$$\Rightarrow B_N = \frac{q_{nf_N} - q_{na}}{q_{nf_N} + q_{na}} C_N e^{-2q_{nf_N} z_N} = R_{n\perp}^{f_N a} C_N e^{-2q_{nf_N} z_N} \quad , \quad (327)$$

$$\Rightarrow A = B_N e^{q_{nf_N} z_N} e^{q_{na} z_N} + C_N e^{-q_{nf_N} z_N} e^{q_{na} z_N} = R_{n\perp}^{f_N a} C_N e^{-q_{nf_N} z_N} e^{q_{na} z_N} + C_N e^{-q_{nf_N} z_N} e^{q_{na} z_N}$$

$$= T_{n\perp}^{f_N a} C_N e^{-q_{nf_N} z_N} e^{q_{na} z_N}$$

and

$$\begin{cases} B_2 e^{q_{nf_2} z_1} + C_2 e^{-q_{nf_2} z_1} = S_{nyy,f_1} e^{q_{nf_1} z'} e^{-q_{nf_1} z_1} + B_1 e^{q_{nf_1} z_1} + C_1 e^{-q_{nf_1} z_1} \\ q_{nf_2}\left[B_2 e^{q_{nf_2} z_1} - C_2 e^{-q_{nf_2} z_1}\right] = -q_{nf_1} S_{nyy,f_1} e^{q_{nf_1} z'} e^{-q_{nf_1} z_1} + q_{nf_1}\left[B_1 e^{q_{nf_1} z_1} - C_1 e^{-q_{nf_1} z_1}\right] \end{cases}$$

$$\Rightarrow \begin{pmatrix} B_1 \\ C_1 \end{pmatrix} = T_1^{\perp} \begin{pmatrix} B_2 \\ C_2 \end{pmatrix} + \begin{pmatrix} e^{q_{nf_1} z_1} & e^{-q_{nf_1} z_1} \\ q_{nf_1} e^{q_{nf_1} z_1} & -q_{nf_1} e^{-q_{nf_1} z_1} \end{pmatrix}^{-1} \begin{pmatrix} -1 \\ q_{nf_1} \end{pmatrix} S_{nyy,f_1} e^{q_{nf_1} z'} e^{-q_{nf_1} z_1}$$

$$= T_1^{\perp} \begin{pmatrix} B_2 \\ C_2 \end{pmatrix} + \frac{1}{2q_{nf_1}} \begin{pmatrix} q_{nf_1} e^{-q_{nf_1} z_1} & e^{-q_{nf_1} z_1} \\ q_{nf_1} e^{q_{nf_1} z_1} & -e^{q_{nf_1} z_1} \end{pmatrix} \begin{pmatrix} -1 \\ q_{nf_1} \end{pmatrix} S_{nyy,f_1} e^{q_{nf_1} z'} e^{-q_{nf_1} z_1}$$

$$= T_1^{\perp} \begin{pmatrix} B_2 \\ C_2 \end{pmatrix} - \begin{pmatrix} 0 \\ 1 \end{pmatrix} S_{nyy,f_1} e^{q_{nf_1} z'}$$

(328)

Combining Eq. (328) with the transfer matrix method which is described in Eqs. (280)-(282), we obtain:

$$\begin{pmatrix} B_1 \\ C_1 \end{pmatrix} = T_1^{\perp} \begin{pmatrix} B_2 \\ C_2 \end{pmatrix} - \begin{pmatrix} 0 \\ 1 \end{pmatrix} S_{nyy,f_1} e^{q_{nf_1} z'} = \begin{pmatrix} M_{11}^{\perp} & M_{12}^{\perp} \\ M_{21}^{\perp} & M_{22}^{\perp} \end{pmatrix} \begin{pmatrix} B_N \\ C_N \end{pmatrix} - \begin{pmatrix} 0 \\ 1 \end{pmatrix} S_{nyy,f_1} e^{q_{nf_1} z'}$$

$$= \begin{pmatrix} M_{11}^{\perp} B_N + M_{12}^{\perp} C_N \\ M_{21}^{\perp} B_N + M_{22}^{\perp} C_N \end{pmatrix} - \begin{pmatrix} 0 \\ 1 \end{pmatrix} S_{nyy,f_1} e^{q_{nf_1} z'} \quad , \quad (329)$$

$$\begin{pmatrix} B_j \\ C_j \end{pmatrix} = T_j^{\perp} \bullet \bullet \bullet T_{N-1}^{\perp} \begin{pmatrix} B_N \\ C_N \end{pmatrix} = \begin{pmatrix} M_{j,11}^{\perp} & M_{j,12}^{\perp} \\ M_{j,21}^{\perp} & M_{j,22}^{\perp} \end{pmatrix} \begin{pmatrix} B_N \\ C_N \end{pmatrix} = \begin{pmatrix} M_{j,11}^{\perp} B_N + M_{j,11}^{\perp} C_N \\ M_{j,21}^{\perp} B_N + M_{j,22}^{\perp} C_N \end{pmatrix}, j = 2,...,N-1$$

, (330)

and we obtain:

$$D = S_{nyy,f_1} e^{-q_{nf_1} z'} + B_1 + C_1 = M_{11}^{\perp} B_N + M_{12}^{\perp} C_N + M_{21}^{\perp} B_N + M_{22}^{\perp} C_N + S_{nyy,f_1} e^{-q_{nf_1} z'} - S_{nyy,f_1} e^{q_{nf_1} z'}$$

$$= \left(M_{11}^{\perp} R_{n\perp}^{f_N a} e^{-2q_{nf_N} z_N} + M_{12}^{\perp} + M_{21}^{\perp} R_{n\perp}^{f_N a} e^{-2q_{nf_N} z_N} + M_{22}^{\perp}\right) C_N + S_{nyy,f_1} e^{-q_{nf_1} z'} - S_{nyy,f_1} e^{q_{nf_1} z'}$$

. (331)

Furthermore, we get the following form to solve the coefficient $C_N$:

$$q_{nf_1} S_{nyy,f_1} e^{-q_{nf_1} z'} + q_{nf_1} (B_1 - C_1) = q_{ns} \left( S_{nyy,f_1} e^{-q_{nf_1} z'} + B_1 + C_1 \right)$$

$$\Rightarrow (q_{ns} + q_{nf_1}) C_1 = -(q_{ns} - q_{nf_1}) B_1 - (q_{ns} - q_{nf_1}) S_{nyy,f_1} e^{-q_{nf_1} z'}$$

$$\Rightarrow C_1 = -\frac{q_{ns} - q_{nf_1}}{q_{ns} + q_{nf_1}} B_1 - \frac{q_{ns} - q_{nf_1}}{q_{ns} + q_{nf_1}} S_{nyy,f_1} e^{-q_{nf_1} z'} = -R_{n\perp}^{sf_1} B_1 - R_{n\perp}^{sf_1} S_{nyy,f_1} e^{-q_{nf_1} z'}$$

$$\Rightarrow M_{21}^{\perp} R_{n\perp}^{f_N a} e^{-2 q_{nf_N} z_N} C_N + M_{22}^{\perp} C_N - S_{nyy,f_1} e^{q_{nf_1} z'} = -R_{n\perp}^{sf_1} \left( M_{11}^{\perp} R_{n\perp}^{f_N a} e^{-2 q_{nf_N} z_N} + M_{12}^{\perp} \right) C_N - R_{n\perp}^{sf_1} S_{nyy,f_1} e^{-q_{nf_1} z'}$$

$$\Rightarrow \left[ M_{21}^{\perp} R_{n\perp}^{f_N a} e^{-2 q_{nf_N} z_N} + M_{22}^{\perp} + R_{n\perp}^{sf_1} \left( M_{11}^{\perp} R_{n\perp}^{f_N a} e^{-2 q_{nf_N} z_N} + M_{12}^{\perp} \right) \right] C_N = S_{nyy,f_1} e^{q_{nf_1} z'} - R_{n\perp}^{sf_1} S_{nyy,f_1} e^{-q_{nf_1} z'}$$

$$\Rightarrow C_N = S_{nyy,f_1} \frac{e^{q_{nf_1} z'} - R_{n\perp}^{sf_1} e^{-q_{nf_1} z'}}{M_{21}^{\perp} R_{n\perp}^{f_N a} e^{-2 q_{nf_N} z_N} + M_{22}^{\perp} + R_{n\perp}^{sf_1} \left( M_{11}^{\perp} R_{n\perp}^{f_N a} e^{-2 q_{nf_N} z_N} + M_{12}^{\perp} \right)}$$

. (332)

Hence we can get the other three coefficients as the following forms:

$$B_N = R_{n\perp}^{f_N a} e^{-2 q_{nf_N} z_N} C_N = S_{nyy,f_1} \frac{R_{n\perp}^{f_N a} e^{-2 q_{nf_N} z_N} \left( e^{q_{nf_1} z'} - R_{n\perp}^{sf_1} e^{-q_{nf_1} z'} \right)}{M_{21}^{\perp} R_{n\perp}^{f_N a} e^{-2 q_{nf_N} z_N} + M_{22}^{\perp} + R_{n\perp}^{sf_1} \left( M_{11}^{\perp} R_{n\perp}^{f_N a} e^{-2 q_{nf_N} z_N} + M_{12}^{\perp} \right)}$$

, (333)

$$A = T_{n\perp}^{f_N a} e^{-q_{nf_N} z_N} e^{q_{na} z_N} C_N = S_{nyy,f_1} \frac{T_{n\perp}^{f_N a} e^{-q_{nf_N} z_N} e^{q_{na} z_N} \left( e^{q_{nf_1} z'} - R_{n\perp}^{sf_1} e^{-q_{nf_1} z'} \right)}{M_{21}^{\perp} R_{n\perp}^{f_N a} e^{-2 q_{nf_N} z_N} + M_{22}^{\perp} + R_{n\perp}^{sf_1} \left( M_{11}^{\perp} R_{n\perp}^{f_N a} e^{-2 q_{nf_N} z_N} + M_{12}^{\perp} \right)}$$

, (334)

$$D = \left( M_{11}^{\perp} R_{n\perp}^{f_N a} e^{-2 q_{nf_N} z_N} + M_{12}^{\perp} + M_{21}^{\perp} R_{n\perp}^{f_N a} e^{-2 q_{nf_N} z_N} + M_{22}^{\perp} \right) C_N + S_{nyy,f_1} e^{-q_{nf_1} z'} - S_{nyy,f_1} e^{q_{nf_1} z'}$$

$$= S_{nyy,f_1} \frac{\left( M_{11}^{\perp} R_{n\perp}^{f_N a} e^{-2 q_{nf_N} z_N} + M_{12}^{\perp} + M_{21}^{\perp} R_{n\perp}^{f_N a} e^{-2 q_{nf_N} z_N} + M_{22}^{\perp} \right) \left( e^{q_{nf_1} z'} - R_{n\perp}^{sf_1} e^{-q_{nf_1} z'} \right)}{M_{21}^{\perp} R_{n\perp}^{f_N a} e^{-2 q_{nf_N} z_N} + M_{22}^{\perp} + R_{n\perp}^{sf_1} \left( M_{11}^{\perp} R_{n\perp}^{f_N a} e^{-2 q_{nf_N} z_N} + M_{12}^{\perp} \right)}$$

$$+ S_{nyy,f_1} e^{-q_{nf_1} z'} - S_{nyy,f_1} e^{q_{nf_1} z'}$$

$$= S_{nyy,f_1} T_{n\perp}^{f_1 s} \frac{e^{q_{nf_1} z'} \left( M_{11}^{\perp} R_{n\perp}^{f_N a} e^{-2 q_{nf_N} z_N} + M_{12}^{\perp} \right) + e^{-q_{nf_1} z'} \left( M_{21}^{\perp} R_{n\perp}^{f_N a} e^{-2 q_{nf_N} z_N} + M_{22}^{\perp} \right)}{M_{21}^{\perp} R_{n\perp}^{f_N a} e^{-2 q_{nf_N} z_N} + M_{22}^{\perp} + R_{n\perp}^{sf_1} \left( M_{11}^{\perp} R_{n\perp}^{f_N a} e^{-2 q_{nf_N} z_N} + M_{12}^{\perp} \right)}$$

. (335)

Hence we solve the coefficients in Eq. (299) completely. Next we discuss the component $\tilde{g}_{nxx}$, starting from Eq. (145), we have:

$$\left( \partial_z^2 - q_n^2 \right) \tilde{g}_{nxx} = \frac{q_n^2}{k^2} \delta(z - z'). \tag{336}$$

Next we will consider the relative position between the field ($z$) and the source ($z'$). First we consider $z' < 0$ (without loss of generality, we may set $z_0 = 0$), the corresponding different equation in regions $z > z_N$, $z_{j-1} < z < z_j, j = N, N-1, ..., 2, 1$

and $z < 0$ are:

$$\begin{cases} \left(\partial_z^2 - q_{na}^2\right)\tilde{g}_{nxx} = 0, z > z_N \\ \left(\partial_z^2 - q_{nf_j}^2\right)\tilde{g}_{nxx} = 0, z_{j-1} < z < z_j, j = 1, 2, ..., N \\ \left(\partial_z^2 - q_{ns}^2\right)\tilde{g}_{nxx} = \frac{q_{ns}^2}{k_s^2}\delta(z - z'), z < 0 \end{cases} \quad (337)$$

Furthermore, the solutions of Eq. (337) are:

$$\tilde{g}_{nxx} = \begin{cases} Ae^{-q_{na}z}, z > z_N \\ B_j e^{q_{nf_j}z} + C_j e^{-q_{nf_j}z}, z_{j-1} < z < z_j, j = 1, 2, ..., N \\ S_{nxx,s} e^{-q_{ns}|z-z'|} + De^{q_{ns}z}, z < 0 \end{cases} \quad (338)$$

where $S_{nxx,s} \equiv -\dfrac{q_{ns}}{2k_s^2}$. In order to match the boundary conditions of $\tilde{g}_{nxx}$ as discuss in Eq. (148):

$$\begin{cases} \tilde{g}_{nxx}\big|_{z_{N+}} = \tilde{g}_{nxx}\big|_{z_{N-}}, \dfrac{\varepsilon_a}{q_{na}^2}\partial_z\tilde{g}_{nxx}\bigg|_{z_{N+}} = \dfrac{\varepsilon_{f_1}}{q_{nf_1}^2}\partial_z\tilde{g}_{nxx}\bigg|_{z_{N-}} \\ \tilde{g}_{nxx}\big|_{z_{j+}} = \tilde{g}_{nxx}\big|_{z_{j-}}, \dfrac{\varepsilon_{f_j}}{q_{nf_j}^2}\partial_z\tilde{g}_{nxx}\bigg|_{z_{j+}} = \dfrac{\varepsilon_{f_{j+1}}}{q_{nf_{j+1}}^2}\partial_z\tilde{g}_{nxx}\bigg|_{z_{j-}}, j = 1, 2, ..., N-1, \\ \tilde{g}_{nxx}\big|_{0+} = \tilde{g}_{nxx}\big|_{0-}, \dfrac{\varepsilon_{f_N}}{q_{nf_N}^2}\partial_z\tilde{g}_{nxx}\bigg|_{0+} = \dfrac{\varepsilon_s}{q_{ns}^2}\partial_z\tilde{g}_{nxx}\bigg|_{0-} \end{cases} \quad (339)$$

we have the following simulated equations:

$$\begin{cases} Ae^{-q_{na}z_N} = B_N e^{q_{nf_N}z_N} + C_N e^{-q_{nf_N}z_N}, z = z_N \\ B_j e^{q_{nf_j}z_j} + C_j e^{-q_{nf_j}z_j} = B_{j+1}e^{q_{nf_{j+1}}z_j} + C_{j+1}e^{-q_{nf_{j+1}}z_j}, z = z_j, j = 1, 2, ..., N-1, \\ B_1 + C_1 = S_{nxx,s}e^{q_{ns}z'} + D, z = 0 \end{cases} \quad (340)$$

and

$$\begin{cases} -\dfrac{\varepsilon_a}{q_{na}}Ae^{-q_{na}z_N} = \dfrac{\varepsilon_{f_N}}{q_{nf_N}}\left(B_N e^{q_{nf_N}z_N} - C_N e^{-q_{nf_N}z_N}\right), z = z_N \\ \dfrac{\varepsilon_{f_j}}{q_{nf_j}}\left[B_j e^{q_{nf_j}z_j} - C_j e^{-q_{nf_j}z_j}\right] = \dfrac{\varepsilon_{f_{j+1}}}{q_{nf_{j+1}}}\left[B_{j+1}e^{q_{nf_{j+1}}z_j} - C_{j+1}e^{-q_{nf_{j+1}}z_j}\right], z = z_j, j = 1, 2, ..., N-1. \\ \dfrac{\varepsilon_{f_1}}{q_{nf_1}}(B_1 - C_1) = -\dfrac{\varepsilon_s}{q_{ns}}S_{nxx,s}e^{q_{ns}z'} + \dfrac{\varepsilon_s}{q_{ns}}D, z = 0 \end{cases}$$

(341)

Furthermore, we obtain:

$$-\frac{\varepsilon_a}{q_{na}}\left(B_N e^{q_{nf_N} z_N} + C_N e^{-q_{nf_N} z_N}\right) = \frac{\varepsilon_{f_N}}{q_{nf_N}}\left(B_N e^{q_{nf_N} z_N} - C_N e^{-q_{nf_N} z_N}\right)$$

$$\Rightarrow \left(\frac{\varepsilon_{f_N}}{q_{nf_N}} + \frac{\varepsilon_a}{q_{na}}\right) B_N = \left(\frac{\varepsilon_{f_N}}{q_{nf_N}} - \frac{\varepsilon_a}{q_{na}}\right) C_N e^{-2q_{nf_N} z_N} \tag{342}$$

$$\Rightarrow B_N = \frac{\varepsilon_{f_N} q_{na} - \varepsilon_a q_{nf_N}}{\varepsilon_{f_N} q_{na} + \varepsilon_a q_{nf_N}} C_N e^{-2q_{nf_N} z_N} = R_{n//}^{f_N a} C_N e^{-2q_{nf_N} z_N}$$

$$\Rightarrow A = R_{n//}^{f_N a} C_N e^{-q_{nf_N} z_N} e^{q_{na} z_N} + C_N e^{-q_{nf_N} z_N} e^{q_{na} z_N} = T_{n//}^{f_N a} C_N e^{-q_{nf_N} z_N} e^{q_{na} z_N}$$

Combining Eq. (342) with the transfer matrix method which is described in Eqs. (217)-(219), we obtain:

$$\begin{pmatrix} B_1 \\ C_1 \end{pmatrix} = \mathrm{T}_1^{//}\begin{pmatrix} B_2 \\ C_2 \end{pmatrix} = \mathrm{T}_1^{//}\mathrm{T}_2^{//}\begin{pmatrix} B_3 \\ C_3 \end{pmatrix} = ... = \mathrm{T}_1^{//}\mathrm{T}_2^{//} \bullet\bullet\bullet \mathrm{T}_{N-1}^{//}\begin{pmatrix} B_N \\ C_N \end{pmatrix} \equiv \prod_{j=1}^{N-1} \mathrm{T}_j^{//}\begin{pmatrix} B_N \\ C_N \end{pmatrix}$$
$$\equiv \begin{pmatrix} M_{11}^{//} & M_{12}^{//} \\ M_{21}^{//} & M_{22}^{//} \end{pmatrix}\begin{pmatrix} B_N \\ C_N \end{pmatrix} = \begin{pmatrix} M_{11}^{//} B_N + M_{12}^{//} C_N \\ M_{21}^{//} B_N + M_{22}^{//} C_N \end{pmatrix}, \tag{343}$$

$$\begin{pmatrix} B_j \\ C_j \end{pmatrix} = \mathrm{T}_j^{//}\begin{pmatrix} B_{j+1} \\ C_{j+1} \end{pmatrix} = \mathrm{T}_j^{//}\mathrm{T}_{j+1}^{//}\begin{pmatrix} B_{j+2} \\ C_{j+2} \end{pmatrix} = ... = \mathrm{T}_j^{//}\mathrm{T}_{j+1}^{//} \bullet\bullet\bullet \mathrm{T}_{N-1}^{//}\begin{pmatrix} B_N \\ C_N \end{pmatrix}$$
$$\equiv \begin{pmatrix} M_{j,11}^{//} & M_{j,12}^{//} \\ M_{j,21}^{//} & M_{j,22}^{//} \end{pmatrix}\begin{pmatrix} B_N \\ C_N \end{pmatrix} = \begin{pmatrix} M_{j,11}^{//} B_N + M_{j,12}^{//} C_N \\ M_{j,21}^{//} B_N + M_{j,22}^{//} C_N \end{pmatrix}, \quad j = 2,3,...,N-1 \tag{344}$$

Furthermore, we obtain:

$$D = B_1 + C_1 - S_{nxx,s} e^{q_{ns} z'} = M_{11}^{//} B_N + M_{12}^{//} C_N + M_{21}^{//} B_N + M_{22}^{//} C_N - S_{nxx,s} e^{q_{ns} z'}$$
$$= M_{11}^{//} R_{n//}^{f_N a} e^{-2q_{nf_N} z_N} C_N + M_{12}^{//} C_N + M_{21}^{//} R_{n//}^{f_N a} e^{-2q_{nf_N} z_N} C_N + M_{22}^{//} C_N - S_{nxx,s} e^{q_{ns} z'}. \tag{345}$$

Then, we get the following form to solve the coefficient $C_N$:

$$\frac{\varepsilon_{f_1}}{q_{nf_1}}(B_1 - C_1) = -\frac{\varepsilon_s}{q_{ns}} S_{nxx,s} e^{q_{ns}z'} + \frac{\varepsilon_s}{q_{ns}}\left(B_1 + C_1 - S_{nxx,s} e^{q_{ns}z'}\right)$$

$$\Rightarrow \left(\frac{\varepsilon_s}{q_{ns}} - \frac{\varepsilon_{f_1}}{q_{nf_1}}\right) B_1 = -\left(\frac{\varepsilon_s}{q_{ns}} + \frac{\varepsilon_{f_1}}{q_{nf_1}}\right) C_1 + \frac{2\varepsilon_s}{q_{ns}} S_{nxx,s} e^{q_{ns}z'}$$

$$\Rightarrow B_1 = -\frac{\varepsilon_s q_{nf_1} + \varepsilon_{f_1} q_{ns}}{\varepsilon_s q_{nf_1} - \varepsilon_{f_1} q_{ns}} C_1 + \frac{2\varepsilon_s q_{nf_1}}{\varepsilon_s q_{nf_1} - \varepsilon_{f_1} q_{ns}} S_{nxx,s} e^{q_{ns}z'}$$

$$\Rightarrow C_1 = -\frac{\varepsilon_s q_{nf_1} - \varepsilon_{f_1} q_{ns}}{\varepsilon_s q_{nf_1} + \varepsilon_{f_1} q_{ns}} B_1 + \frac{2\varepsilon_s q_{nf_1}}{\varepsilon_s q_{nf_1} + \varepsilon_{f_1} q_{ns}} S_{nxx,s} e^{q_{ns}z'} = -R_{n//}^{sf_1} B_1 + T_{n//}^{sf_1} S_{nxx,s} e^{q_{ns}z'}$$

$$\Rightarrow M_{21}'' R_{n//}^{f_N a} e^{-2q_{nf_N} z_N} C_N + M_{22}'' C_N = -R_{n//}^{sf_1}\left(M_{11}'' R_{n//}^{f_N a} e^{-2q_{nf_N} z_N} + M_{12}''\right) C_N + T_{n//}^{sf_1} S_{nxx,s} e^{q_{ns}z'}$$

$$\Rightarrow \left[M_{21}'' R_{n//}^{f_N a} e^{-2q_{nf_N} z_N} + M_{22}'' + R_{n//}^{sf_1}\left(M_{11}'' R_{n//}^{f_N a} e^{-2q_{nf_N} z_N} + M_{12}''\right)\right] C_N = T_{n//}^{sf_1} S_{nxx,s} e^{q_{ns}z'}$$

$$\Rightarrow C_N = \frac{T_{n//}^{sf_1} S_{nxx,s} e^{q_{ns}z'}}{M_{21}'' R_{n//}^{f_N a} e^{-2q_{nf_N} z_N} + M_{22}'' + R_{n//}^{sf_1}\left(M_{11}'' R_{n//}^{f_N a} e^{-2q_{nf_N} z_N} + M_{12}''\right)}$$

. (346)

Hence we can get the other three coefficients as the following forms:

$$B_N = R_{n//}^{f_N a} C_N e^{-2q_{nf_N} z_N} = \frac{R_{n//}^{f_N a} T_{n//}^{sf_1} S_{nxx,s} e^{q_{ns}z'} e^{-2q_{nf_N} z_N}}{M_{21}'' R_{n//}^{f_N a} e^{-2q_{nf_N} z_N} + M_{22}'' + R_{n//}^{sf_1}\left(M_{11}'' R_{n//}^{f_N a} e^{-2q_{nf_N} z_N} + M_{12}''\right)}, \quad (347)$$

$$A = T_{n//}^{f_N a} e^{-q_{nf_N} z_N} e^{q_{na} z_N} C_N = \frac{T_{n//}^{f_N a} e^{-q_{nf_N} z_N} e^{q_{na} z_N} T_{n//}^{sf_1} S_{nxx,s} e^{q_{ns}z'}}{M_{21}'' R_{n//}^{f_N a} e^{-2q_{nf_N} z_N} + M_{22}'' + R_{n//}^{sf_1}\left(M_{11}'' R_{n//}^{f_N a} e^{-2q_{nf_N} z_N} + M_{12}''\right)}, \quad (348)$$

$$D = M_{11}'' R_{n//}^{f_N a} e^{-2q_{nf_N} z_N} C_N + M_{12}'' C_N + M_{21}'' R_{n//}^{f_N a} e^{-2q_{nf_N} z_N} C_N + M_{22}'' C_N - S_{nxx,s} e^{q_{ns}z'}$$

$$= \frac{T_{n//}^{sf_1} S_{nxx,s} e^{q_{ns}z'}\left(M_{11}'' R_{n//}^{f_N a} e^{-2q_{nf_N} z_N} + M_{12}'' + M_{21}'' R_{n//}^{f_N a} e^{-2q_{nf_N} z_N} + M_{22}''\right)}{M_{21}'' R_{n//}^{f_N a} e^{-2q_{nf_N} z_N} + M_{22}'' + R_{n//}^{sf_1}\left(M_{11}'' R_{n//}^{f_N a} e^{-2q_{nf_N} z_N} + M_{12}''\right)} - S_{nxx,s} e^{q_{ns}z'}. \quad (349)$$

$$= S_{nxx,s} e^{q_{ns}z'} \frac{M_{11}'' R_{n//}^{f_N a} e^{-2q_{nf_N} z_N} + M_{12}'' + R_{n//}^{sf_1}\left(M_{21}'' R_{n//}^{f_N a} e^{-2q_{nf_N} z_N} + M_{22}''\right)}{M_{21}'' R_{n//}^{f_N a} e^{-2q_{nf_N} z_N} + M_{22}'' + R_{n//}^{sf_1}\left(M_{11}'' R_{n//}^{f_N a} e^{-2q_{nf_N} z_N} + M_{12}''\right)}$$

Hence we solve the coefficients in Eq. (338) completely. Next we consider $z' > z_N$, the corresponding different equation in regions $z > z_N$, $z_{j-1} < z < z_j, j = N, N-1, ..., 2, 1$ and $z < 0$ are:

$$\begin{cases} \left(\partial_z^2 - q_{na}^2\right)\tilde{g}_{nxx} = \dfrac{q_{na}^2}{k_a^2}\delta(z-z'), z > z_N \\ \left(\partial_z^2 - q_{nf_j}^2\right)\tilde{g}_{nxx} = 0, z_{j-1} < z < z_j, j = 1,2,...,N \\ \left(\partial_z^2 - q_{ns}^2\right)\tilde{g}_{nxx} = 0, z < 0 \end{cases} \quad (350)$$

Furthermore, the solutions of Eq. (350) are:

$$\tilde{g}_{nxx} = \begin{cases} S_{nxx,a}e^{-q_{na}|z-z'|} + Ae^{-q_{na}z}, z > z_N \\ B_j e^{q_{nf_j}z} + C_j e^{-q_{nf_j}z}, z_{j-1} < z < z_j, j = 1,2,...,N \\ De^{q_{ns}z}, z < 0 \end{cases} \quad (351)$$

where $S_{nxx,a} \equiv -\dfrac{q_{na}}{2k_a^2}$. In order to match the boundary conditions of $\tilde{g}_{nxx}$ as discuss in Eq. (338), we have the following simulated equations:

$$\begin{cases} S_{nxx,a}e^{-q_{na}z'}e^{q_{na}z_N} + Ae^{-q_{na}z_N} = B_N e^{q_{nf_N}z_N} + C_N e^{-q_{nf_N}z_N}, z = z_N \\ B_j e^{q_{nf_j}z_j} + C_j e^{-q_{nf_j}z_j} = B_{j+1}e^{q_{nf_{j+1}}z_j} + C_{j+1}e^{-q_{nf_{j+1}}z_j}, z = z_j, j = 1,2,...,N-1 \\ B_1 + C_1 = D, z = 0 \end{cases} \quad (352)$$

and

$$\begin{cases} \dfrac{\varepsilon_a}{q_{na}}S_{nxx,a}e^{-q_{na}z'}e^{q_{na}z_N} - \dfrac{\varepsilon_a}{q_{na}}Ae^{-q_{na}z_N} = \dfrac{\varepsilon_{f_N}}{q_{nf_N}}\left(B_N e^{q_{nf_N}z_N} - C_N e^{-q_{nf_N}z_N}\right), z = z_N \\ \dfrac{\varepsilon_{f_j}}{q_{nf_j}}\left[B_j e^{q_{nf_j}z_j} - C_j e^{-q_{nf_j}z_j}\right] = \dfrac{\varepsilon_{f_{j+1}}}{q_{nf_{j+1}}}\left[B_{j+1}e^{q_{nf_{j+1}}z_j} - C_{j+1}e^{-q_{nf_{j+1}}z_j}\right], z = z_j, j = 1,2,...,N-1 \\ \dfrac{\varepsilon_{f_1}}{q_{nf_1}}(B_1 - C_1) = \dfrac{\varepsilon_s}{q_{ns}}D, z = 0 \end{cases} \quad (353)$$

Furthermore, we obtain:

$$\frac{\varepsilon_a}{q_{na}} S_{nxx,a} e^{-q_{na}z'} e^{q_{na}z_N} - \frac{\varepsilon_a}{q_{na}} \left( B_N e^{q_{nf_N} z_N} + C_N e^{-q_{nf_N} z_N} - S_{nxx,a} e^{-q_{na}z'} e^{q_{na}z_N} \right)$$

$$= \frac{\varepsilon_{f_N}}{q_{nf_N}} \left( B_N e^{q_{nf_N} z_N} - C_N e^{-q_{nf_N} z_N} \right)$$

$$\Rightarrow \left( \frac{\varepsilon_{f_N}}{q_{nf_N}} + \frac{\varepsilon_a}{q_{na}} \right) B_N = \left( \frac{\varepsilon_{f_N}}{q_{nf_N}} - \frac{\varepsilon_a}{q_{na}} \right) C_N e^{-2q_{nf_N} z_N} + \frac{2\varepsilon_a}{q_{na}} S_{nxx,a} e^{-q_{na}z'} e^{q_{na}z_N} e^{-q_{nf_N} z_N}$$

$$\Rightarrow B_N = \frac{\varepsilon_{f_N} q_{na} - \varepsilon_a q_{nf_N}}{\varepsilon_{f_N} q_{na} + \varepsilon_a q_{nf_N}} C_N e^{-2q_{nf_N} z_N} + \frac{2\varepsilon_a q_{nf_N}}{\varepsilon_{f_N} q_{na} + \varepsilon_a q_{nf_N}} S_{nxx,a} e^{-q_{na}z'} e^{q_{na}z_N} e^{-q_{nf_N} z_N}$$

$$= R_{n//}^{f_N a} C_N e^{-2q_{nf_N} z_N} + S_{nxx,a} T_{n//}^{af_N} e^{-q_{na}z'} e^{q_{na}z_N} e^{-q_{nf_N} z_N}$$

$$\Rightarrow A = B_N e^{q_{nf_N} z_N} e^{q_{na}z_N} + C_N e^{-q_{nf_N} z_N} e^{q_{na}z_N} - S_{nxx,a} e^{-q_{na}z'} e^{2q_{na}z_N}$$

$$= \left( R_{n//}^{f_N a} C_N e^{-2q_{nf_N} z_N} + S_{nxx,a} T_{n//}^{af_N} e^{-q_{na}z'} e^{q_{na}z_N} e^{-q_{nf_N} z_N} \right) e^{q_{nf_N} z_N} e^{q_{na}z_N} + C_N e^{-q_{nf_N} z_N} e^{q_{na}z_N} - S_{nxx,a} e^{-q_{na}z'} e^{2q_{na}z_N}$$

$$= T_{n//}^{f_N a} C_N e^{-q_{nf_N} z_N} e^{q_{na}z_N} + R_{n//}^{af_N} S_{nxx,a} e^{-q_{na}z'} e^{2q_{na}z_N}$$

, (354)

and

$$D = B_1 + C_1 = M_{11}^{//} B_N + M_{12}^{//} C_N + M_{21}^{//} B_N + M_{22}^{//} C_N$$

$$= M_{11}^{//} \left( R_{n//}^{f_N a} C_N e^{-2q_{nf_N} z_N} + S_{nxx,a} T_{n//}^{af_N} e^{-q_{na}z'} e^{q_{na}z_N} e^{-q_{nf_N} z_N} \right) + M_{12}^{//} C_N$$

$$+ M_{21}^{//} \left( R_{n//}^{f_N a} C_N e^{-2q_{nf_N} z_N} + S_{nxx,a} T_{n//}^{af_N} e^{-q_{na}z'} e^{q_{na}z_N} e^{-q_{nf_N} z_N} \right) + M_{22}^{//} C_N \quad . \quad (355)$$

$$= \left( M_{11}^{//} R_{n//}^{f_N a} e^{-2q_{nf_N} z_N} + M_{12}^{//} + M_{21}^{//} R_{n//}^{f_N a} e^{-2q_{nf_N} z_N} + M_{22}^{//} \right) C_N$$

$$+ M_{11}^{//} S_{nxx,a} T_{n//}^{af_N} e^{-q_{na}z'} e^{q_{na}z_N} e^{-q_{nf_N} z_N} + M_{21}^{//} S_{nxx,a} T_{n//}^{af_N} e^{-q_{na}z'} e^{q_{na}z_N} e^{-q_{nf_N} z_N}$$

Furthermore, we get the following form to solve the coefficient $C_N$:

$$\frac{\varepsilon_{f_1}}{q_{nf_1}}\left(M_{11}''R_{n//}^{f_N a}e^{-2q_{nf_N}z_N}+M_{12}''-M_{21}''R_{n//}^{f_N a}e^{-2q_{nf_N}z_N}-M_{22}''\right)C_N$$

$$+\frac{\varepsilon_{f_1}}{q_{nf_1}}S_{nxx,a}T_{n//}^{af_N}M_{11}''e^{-q_{na}z'}e^{q_{na}z_N}e^{-q_{nf_N}z_N}-\frac{\varepsilon_{f_1}}{q_{nf_1}}S_{nxx,a}T_{n//}^{af_N}M_{21}''e^{-q_{na}z'}e^{q_{na}z_N}e^{-q_{nf_N}z_N}$$

$$=\frac{\varepsilon_s}{q_{ns}}\left(M_{11}''R_{n//}^{f_N a}e^{-2q_{nf_N}z_N}+M_{12}''+M_{21}''R_{n//}^{f_N a}e^{-2q_{nf_N}z_N}+M_{22}''\right)C_N$$

$$+\frac{\varepsilon_s}{q_{ns}}M_{11}''S_{nxx,a}T_{n//}^{af_N}e^{-q_{na}z'}e^{q_{na}z_N}e^{-q_{nf_N}z_N}+\frac{\varepsilon_s}{q_{ns}}M_{21}''S_{nxx,a}T_{n//}^{af_N}e^{-q_{na}z'}e^{q_{na}z_N}e^{-q_{nf_N}z_N}$$

$$\Rightarrow\left[\left(\frac{\varepsilon_s}{q_{ns}}-\frac{\varepsilon_{f_1}}{q_{nf_1}}\right)\left(M_{11}''R_{n//}^{f_N a}e^{-2q_{nf_N}z_N}+M_{12}''\right)+\left(\frac{\varepsilon_s}{q_{ns}}+\frac{\varepsilon_{f_1}}{q_{nf_1}}\right)\left(M_{21}''R_{n//}^{f_N a}e^{-2q_{nf_N}z_N}+M_{22}''\right)\right]C_N$$

$$=-\left(\frac{\varepsilon_s}{q_{ns}}-\frac{\varepsilon_{f_1}}{q_{nf_1}}\right)S_{nxx,a}T_{n//}^{af_N}M_{11}''e^{-q_{na}z'}e^{q_{na}z_N}e^{-q_{nf_N}z_N}-\left(\frac{\varepsilon_s}{q_{ns}}+\frac{\varepsilon_{f_1}}{q_{nf_1}}\right)S_{nxx,a}T_{n//}^{af_N}M_{21}''e^{-q_{na}z'}e^{q_{na}z_N}e^{-q_{nf_N}z_N}$$

$$\Rightarrow C_N=-\frac{S_{nxx,a}T_{n//}^{af_N}\left(R_{n//}^{sf_1}M_{11}''+M_{21}''\right)e^{-q_{na}z'}e^{q_{na}z_N}e^{-q_{nf_N}z_N}}{R_{n//}^{sf_1}\left(M_{11}''R_{n//}^{f_N a}e^{-2q_{nf_N}z_N}+M_{12}''\right)+\left(M_{21}''R_{n//}^{f_N a}e^{-2q_{nf_N}z_N}+M_{22}''\right)}$$

. 
(356)

Hence we can get the other three coefficients as the following forms:

$$B_N=R_{n//}^{f_N a}C_N e^{-2q_{nf_N}z_N}+S_{nxx,a}T_{n//}^{af_N}e^{-q_{na}z'}e^{q_{na}z_N}e^{-q_{nf_N}z_N}$$

$$=-\frac{R_{n//}^{f_N a}S_{nxx,a}T_{n//}^{af_N}\left(R_{n//}^{sf_1}M_{11}''+M_{21}''\right)e^{-q_{na}z'}e^{q_{na}z_N}e^{-q_{nf_N}z_N}}{R_{n//}^{sf_1}\left(M_{11}''R_{n//}^{f_N a}e^{-2q_{nf_N}z_N}+M_{12}''\right)+\left(M_{21}''R_{n//}^{f_N a}e^{-2q_{nf_N}z_N}+M_{22}''\right)}e^{-2q_{nf_N}z_N}+S_{nxx,a}T_{n//}^{af_N}e^{-q_{na}z'}e^{q_{na}z_N}e^{-q_{nf_N}z_N}$$

$$=\frac{S_{nxx,a}e^{-q_{na}z'}e^{q_{na}z_N}e^{-q_{nf_N}z_N}T_{n//}^{af_N}\left(R_{n//}^{sf_1}M_{12}''+M_{22}''\right)}{R_{n//}^{sf_1}\left(M_{11}''R_{n//}^{f_N a}e^{-2q_{nf_N}z_N}+M_{12}''\right)+\left(M_{21}''R_{n//}^{f_N a}e^{-2q_{nf_N}z_N}+M_{22}''\right)}$$

, 
(357)

$$A=T_{n//}^{f_N a}C_N e^{-q_{nf_N}z_N}e^{q_{na}z_N}-R_{n//}^{f_N a}S_{nxx,a}e^{-q_{na}z'}e^{2q_{na}z_N}$$

$$=\frac{-T_{n//}^{f_N a}T_{n//}^{af_N}S_{nxx,a}\left(R_{n//}^{sf_1}M_{11}''+M_{21}''\right)e^{-q_{na}z'}e^{2q_{na}z_N}e^{-2q_{nf_N}z_N}}{R_{n//}^{sf_1}\left(M_{11}''R_{n//}^{f_N a}e^{-2q_{nf_N}z_N}+M_{12}''\right)+\left(M_{21}''R_{n//}^{f_N a}e^{-2q_{nf_N}z_N}+M_{22}''\right)}-R_{n//}^{f_N a}S_{nxx,a}e^{-q_{na}z'}e^{2q_{na}z_N},\quad(358)$$

$$=-S_{nxx,a}e^{-q_{na}z'}e^{2q_{na}z_N}\frac{R_{n//}^{sf_1}\left(M_{11}''+M_{21}''\right)e^{-2q_{nf_N}z_N}+R_{n//}^{f_N a}\left(R_{n//}^{sf_1}M_{12}''+M_{22}''\right)}{R_{n//}^{sf_1}\left(M_{11}''R_{n//}^{f_N a}e^{-2q_{nf_N}z_N}+M_{12}''\right)+\left(M_{21}''R_{n//}^{f_N a}e^{-2q_{nf_N}z_N}+M_{22}''\right)}$$

$$D = \left( M_{11}'' R_{n//}^{f_N a} e^{-2q_{nf_N} z_N} + M_{12}'' + M_{21}'' R_{n//}^{f_N a} e^{-2q_{nf_N} z_N} + M_{22}'' \right) C_N$$

$$+ M_{11}'' S_{nxx,a} T_{n//}^{af_N} e^{-q_{na} z'} e^{q_{na} z_N} e^{-q_{nf_N} z_N} + M_{21}'' S_{nxx,a} T_{n//}^{af_N} e^{-q_{na} z'} e^{q_{na} z_N} e^{-q_{nf_N} z_N}$$

$$= \frac{-S_{nxx,a} T_{n//}^{af_N} \left( M_{11}'' R_{n//}^{f_N a} e^{-2q_{nf_N} z_N} + M_{12}'' + M_{21}'' R_{n//}^{f_N a} e^{-2q_{nf_N} z_N} + M_{22}'' \right) \left( R_{n//}^{sf_1} M_{11}'' + M_{21}'' \right) e^{-q_{na} z'} e^{q_{na} z_N} e^{-q_{nf_N} z_N}}{R_{n//}^{sf_1} \left( M_{11}'' R_{n//}^{f_N a} e^{-2q_{nf_N} z_N} + M_{12}'' \right) + \left( M_{21}'' R_{n//}^{f_N a} e^{-2q_{nf_N} z_N} + M_{22}'' \right)}$$

$$+ M_{11}'' S_{nxx,a} T_{n//}^{af_N} e^{-q_{na} z'} e^{q_{na} z_N} e^{-q_{nf_N} z_N} + M_{21}'' S_{nxx,a} T_{n//}^{af_N} e^{-q_{na} z'} e^{q_{na} z_N} e^{-q_{nf_N} z_N}$$

$$= \frac{S_{nxx,a} T_{n//}^{af_N} T_{n//}^{f_1 s} \left( M_{22}'' M_{11}'' - M_{21}'' M_{12}'' \right) e^{-q_{na} z'} e^{q_{na} z_N} e^{-q_{nf_N} z_N}}{R_{n//}^{sf_1} \left( M_{11}'' R_{n//}^{f_N a} e^{-2q_{nf_N} z_N} + M_{12}'' \right) + \left( M_{21}'' R_{n//}^{f_N a} e^{-2q_{nf_N} z_N} + M_{22}'' \right)}$$

. (359)

According to Eq. (343) and Eq. (344), the coefficients $B_j$ and $C_j, j = 1, 2, ..., N-1$ are obtained. Hence we solve the coefficients in Eq. (351) completely. Next we consider the location of a point source is inside the film layer, that is $z_{k-1} < z' < z_k, k = 1, 2, ..., N$. Fixed the index $k$, the corresponding different equation in regions $z > z_N$, $z_{j-1} < z < z_j, j = N, N-1, ..., 2, 1$ and $z < 0$ are:

$$\begin{cases} \left( \partial_z^2 - q_{na}^2 \right) \tilde{g}_{nxx} = 0, z > z_N \\ \left( \partial_z^2 - q_{nf_j}^2 \right) \tilde{g}_{nxx} = \frac{q_{nf_k}^2}{k_{f_k}^2} \delta_{jk} \delta(z - z'), z_{j-1} < z < z_j, j = 1, 2, ..., N \\ \left( \partial_z^2 - q_{ns}^2 \right) \tilde{g}_{nxx} = 0, z < 0 \end{cases} \quad (360)$$

Furthermore, the solutions of Eq. (360) are:

$$\tilde{g}_{nxx} = \begin{cases} A e^{-q_{na} z}, z > z_N \\ S_{nxx, f_k} e^{-q_{nf_k} |z-z'|} \delta_{jk} + B_j e^{q_{nf_j} z} + C_j e^{-q_{nf_j} z}, z_{j-1} < z < z_j, j = 1, 2, ..., N, \\ D e^{q_{ns} z}, z < 0 \end{cases} \quad (361)$$

where $S_{nxx, f_k} \equiv -\dfrac{q_{nf_k}}{2k_{f_k}^2}$. In order to match the boundary conditions of $\tilde{g}_{nxx}$ as we discuss in Eq. (339), we have the following simulated equations:

<1> If $k = N$:

$$\begin{cases} A e^{-q_{na} z_N} = S_{nxx, f_N} e^{q_{nf_N} z'} e^{-q_{nf_N} z_N} + B_N e^{q_{nf_N} z_N} + C_N e^{-q_{nf_N} z_N} \\ S_{nxx, f_N} e^{-q_{nf_N} z'} e^{q_{nf_N} z_{N-1}} + B_N e^{q_{nf_N} z_{N-1}} + C_N e^{-q_{nf_N} z_{N-1}} = B_{N-1} e^{q_{nf_{N-1}} z_{N-1}} + C_{N-1} e^{-q_{nf_{N-1}} z_{N-1}} \\ B_j e^{q_{nf_j} z_j} + C_j e^{-q_{nf_j} z_j} = B_{j+1} e^{q_{nf_{j+1}} z_j} + C_{j+1} e^{-q_{nf_{j+1}} z_j}, j = 1, 2, ..., N-2 \\ B_1 + C_1 = D \end{cases}, \quad (362)$$

and

$$\begin{cases} -\dfrac{\varepsilon_a}{q_{na}} A e^{-q_{na} z_N} = -\dfrac{\varepsilon_{f_N}}{q_{nf_N}} S_{nxx,f_N} e^{q_{nf_N} z'} e^{-q_{nf_N} z_N} + \dfrac{\varepsilon_{f_N}}{q_{nf_N}} \left( B_N e^{q_{nf_N} z_N} - C_N e^{-q_{nf_N} z_N} \right) \\ \dfrac{\varepsilon_{f_N}}{q_{nf_N}} S_{nxx,f_N} e^{-q_{nf_N} z'} e^{q_{nf_N} z_{N-1}} + \dfrac{\varepsilon_{f_N}}{q_{nf_N}} \left[ B_N e^{q_{nf_N} z_{N-1}} - C_N e^{-q_{nf_N} z_{N-1}} \right] = \dfrac{\varepsilon_{f_{N-1}}}{q_{nf_{N-1}}} \left[ B_{N-1} e^{q_{nf_{N-1}} z_{N-1}} - C_{N-1} e^{-q_{nf_{N-1}} z_{N-1}} \right] \\ \dfrac{\varepsilon_{f_j}}{q_{nf_j}} \left[ B_j e^{q_{nf_j} z_j} - C_j e^{-q_{nf_j} z_j} \right] = \dfrac{\varepsilon_{f_{j+1}}}{q_{nf_{j+1}}} \left[ B_{j+1} e^{q_{nf_{j+1}} z_j} - C_{j+1} e^{-q_{nf_{j+1}} z_j} \right], j = 1, 2, ..., N-2 \\ \dfrac{\varepsilon_{f_1}}{q_{nf_1}} \left( B_1 - C_1 \right) = \dfrac{\varepsilon_{ns}}{q_{ns}} D \end{cases}$$

. (363)

<2> If $k = N-1, N-2, ..., 2$:

$$\begin{cases} A e^{-q_{na} z_N} = B_N e^{q_{nf_N} z_N} + C_N e^{-q_{nf_N} z_N} \\ B_{k+1} e^{q_{nf_{k+1}} z_k} + C_{k+1} e^{-q_{nf_{k+1}} z_k} = S_{nxx,f_k} e^{q_{nf_k} z'} e^{-q_{nf_k} z_k} + B_k e^{q_{nf_k} z_k} + C_k e^{-q_{nf_k} z_k} \\ S_{nxx,f_k} e^{-q_{nf_k} z'} e^{q_{nf_k} z_{k-1}} + B_k e^{q_{nf_k} z_{k-1}} + C_k e^{-q_{nf_k} z_{k-1}} = B_{k-1} e^{q_{nf_{k-1}} z_{k-1}} + C_{k-1} e^{-q_{nf_{k-1}} z_{k-1}} \\ B_j e^{q_{nf_j} z_j} + C_j e^{-q_{nf_j} z_j} = B_{j+1} e^{q_{nf_{j+1}} z_j} + C_{j+1} e^{-q_{nf_{j+1}} z_j}, j = 1, ..., N-1, j \neq k-1, k \\ B_1 + C_1 = D \end{cases}$$ (364)

and

$$\begin{cases} -\dfrac{\varepsilon_a}{q_{na}} A e^{-q_{na} z_N} = \dfrac{\varepsilon_{f_N}}{q_{nf_N}} \left( B_N e^{q_{nf_N} z_N} - C_N e^{-q_{nf_N} z_N} \right) \\ \dfrac{\varepsilon_{f_{k+1}}}{q_{nf_{k+1}}} \left[ B_{k+1} e^{q_{nf_{k+1}} z_k} - C_{k+1} e^{-q_{nf_{k+1}} z_k} \right] = -\dfrac{\varepsilon_{f_k}}{q_{nf_k}} S_{nxx,f_k} e^{q_{nf_k} z'} e^{-q_{nf_k} z_k} + \dfrac{\varepsilon_{f_k}}{q_{nf_k}} \left[ B_k e^{q_{nf_k} z_k} - C_k e^{-q_{nf_k} z_k} \right] \\ \dfrac{\varepsilon_{f_k}}{q_{nf_k}} S_{nxx,f_k} e^{-q_{nf_k} z'} e^{q_{nf_k} z_{k-1}} + \dfrac{\varepsilon_{f_k}}{q_{nf_k}} \left[ B_k e^{q_{nf_k} z_{k-1}} - C_k e^{-q_{nf_k} z_{k-1}} \right] = \dfrac{\varepsilon_{f_{k-1}}}{q_{nf_{k-1}}} \left[ B_{k-1} e^{q_{nf_{k-1}} z_{k-1}} - C_{k-1} e^{-q_{nf_{k-1}} z_{k-1}} \right] \\ \dfrac{\varepsilon_{f_j}}{q_{nf_j}} \left[ B_j e^{q_{nf_j} z_j} - C_j e^{-q_{nf_j} z_j} \right] = \dfrac{\varepsilon_{f_{j+1}}}{q_{nf_{j+1}}} \left[ B_{j+1} e^{q_{nf_{j+1}} z_j} - C_{j+1} e^{-q_{nf_{j+1}} z_j} \right], j = 1, ..., N-1, j \neq k-1, k \\ \dfrac{\varepsilon_{f_1}}{q_{nf_1}} \left( B_1 - C_1 \right) = \dfrac{\varepsilon_s}{q_{ns}} D \end{cases}$$

. (365)

<3> If $k = 1$:

$$\begin{cases} A e^{-q_{na} z_N} = B_N e^{q_{nf_N} z_N} + C_N e^{-q_{nf_N} z_N} \\ B_j e^{q_{nf_j} z_j} + C_j e^{-q_{nf_j} z_j} = B_{j+1} e^{q_{nf_{j+1}} z_j} + C_{j+1} e^{-q_{nf_{j+1}} z_j}, j = 2, ..., N-1 \\ B_2 e^{q_{nf_2} z_1} + C_2 e^{-q_{nf_2} z_1} = S_{nxx,f_1} e^{q_{nf_1} z'} e^{-q_{nf_1} z_1} + B_1 e^{q_{nf_1} z_1} + C_1 e^{-q_{nf_1} z_1} \\ S_{nxx,f_1} e^{-q_{nf_1} z'} + B_1 + C_1 = D \end{cases}$$, (366)

and

$$\begin{cases} -\dfrac{\varepsilon_a}{q_{na}} A e^{-q_{na} z_N} = \dfrac{\varepsilon_{f_N}}{q_{nf_N}} \left( B_N e^{q_{nf_N} z_N} - C_N e^{-q_{nf_N} z_N} \right) \\ \dfrac{\varepsilon_{f_j}}{q_{nf_j}} \left[ B_j e^{q_{nf_j} z_j} - C_j e^{-q_{nf_j} z_j} \right] = \dfrac{\varepsilon_{f_{j+1}}}{q_{nf_{j+1}}} \left[ B_{j+1} e^{q_{nf_{j+1}} z_j} - C_{j+1} e^{-q_{nf_{j+1}} z_j} \right], j = 2,\ldots, N-1 \\ \dfrac{\varepsilon_{f_2}}{q_{nf_2}} \left[ B_2 e^{q_{nf_2} z_1} - C_2 e^{-q_{nf_2} z_1} \right] = -\dfrac{\varepsilon_{f_1}}{q_{nf_1}} S_{nxx, f_1} e^{q_{nf_1} z'} e^{-q_{nf_1} z_1} + \dfrac{\varepsilon_{f_1}}{q_{nf_1}} \left[ B_1 e^{q_{nf_1} z_1} - C_1 e^{-q_{nf_1} z_1} \right] \\ \dfrac{\varepsilon_{f_1}}{q_{nf_1}} S_{nxx, f_1} e^{-q_{nf_1} z'} + \dfrac{\varepsilon_{f_1}}{q_{nf_1}} (B_1 - C_1) = \dfrac{\varepsilon_s}{q_{ns}} D \end{cases} \tag{367}$$

Furthermore, we simplify above these equations and show as the following forms:

<1> If $k = N$:

$$-\dfrac{\varepsilon_a}{q_{na}} \left( S_{nxx, f_N} e^{q_{nf_N} z'} e^{-q_{nf_N} z_N} + B_N e^{q_{nf_N} z_N} + C_N e^{-q_{nf_N} z_N} \right)$$

$$= -\dfrac{\varepsilon_{f_N}}{q_{nf_N}} S_{nxx, f_N} e^{q_{nf_N} z'} e^{-q_{nf_N} z_N} + \dfrac{\varepsilon_{f_N}}{q_{nf_N}} \left( B_N e^{q_{nf_N} z_N} - C_N e^{-q_{nf_N} z_N} \right)$$

$$\Rightarrow \left( \dfrac{\varepsilon_{f_N}}{q_{nf_N}} + \dfrac{\varepsilon_a}{q_{na}} \right) B_N = \left( \dfrac{\varepsilon_{f_N}}{q_{nf_N}} - \dfrac{\varepsilon_a}{q_{na}} \right) C_N e^{-2 q_{nf_N} z_N} + \left( \dfrac{\varepsilon_{f_N}}{q_{nf_N}} - \dfrac{\varepsilon_a}{q_{na}} \right) S_{nxx, f_N} e^{q_{nf_N} z'} e^{-2 q_{nf_N} z_N}$$

$$\Rightarrow B_N = \dfrac{\varepsilon_{f_N} q_{na} - \varepsilon_a q_{nf_N}}{\varepsilon_{f_N} q_{na} + \varepsilon_a q_{nf_N}} C_N e^{-2 q_{nf_N} z_N} + \dfrac{\varepsilon_{f_N} q_{na} - \varepsilon_a q_{nf_N}}{\varepsilon_{f_N} q_{na} + \varepsilon_a q_{nf_N}} S_{nxx, f_N} e^{q_{nf_N} z'} e^{-2 q_{nf_N} z_N} \tag{368}$$

$$= R_{n//}^{f_N a} C_N e^{-2 q_{nf_N} z_N} + R_{n//}^{f_N a} S_{nxx, f_N} e^{q_{nf_N} z'} e^{-2 q_{nf_N} z_N}$$

$$\Rightarrow A = S_{nxx, f_N} e^{q_{nf_N} z'} e^{-q_{nf_N} z_N} e^{q_{na} z_N} + B_N e^{q_{nf_N} z_N} e^{q_{na} z_N} + C_N e^{-q_{nf_N} z_N} e^{q_{na} z_N}$$

$$= S_{nxx, f_N} \left( 1 + R_{n//}^{f_N a} \right) e^{q_{nf_N} z'} e^{-q_{nf_N} z_N} e^{q_{na} z_N} + \left( 1 + R_{n//}^{f_N a} \right) C_N e^{-q_{nf_N} z_N} e^{q_{na} z_N}$$

$$= S_{nxx, f_N} T_{n//}^{f_N a} e^{q_{nf_N} z'} e^{-q_{nf_N} z_N} e^{q_{na} z_N} + T_{n//}^{f_N a} C_N e^{-q_{nf_N} z_N} e^{q_{na} z_N}$$

and

$$\begin{cases} S_{nxx,f_N} e^{-q_{nf_N} z'} e^{q_{nf_N} z_{N-1}} + B_N e^{q_{nf_N} z_{N-1}} + C_N e^{-q_{nf_N} z_{N-1}} = B_{N-1} e^{q_{nf_{N-1}} z_{N-1}} + C_{N-1} e^{-q_{nf_{N-1}} z_{N-1}} \\ \dfrac{\varepsilon_{f_N}}{q_{nf_N}} S_{nxx,f_N} e^{-q_{nf_N} z'} e^{q_{nf_N} z_{N-1}} + \dfrac{\varepsilon_{f_N}}{q_{nf_N}} \left[ B_N e^{q_{nf_N} z_{N-1}} - C_N e^{-q_{nf_N} z_{N-1}} \right] = \dfrac{\varepsilon_{f_{N-1}}}{q_{nf_{N-1}}} \left[ B_{N-1} e^{q_{nf_{N-1}} z_{N-1}} - C_{N-1} e^{-q_{nf_{N-1}} z_{N-1}} \right] \end{cases}$$

$$\Rightarrow \begin{pmatrix} B_N \\ C_N \end{pmatrix} = \left[ T_{N-1}^{//} \right]^{-1} \begin{pmatrix} B_{N-1} \\ C_{N-1} \end{pmatrix} - \begin{pmatrix} e^{q_{nf_N} z_{N-1}} & e^{-q_{nf_N} z_{N-1}} \\ \dfrac{\varepsilon_{f_N}}{q_{nf_N}} e^{q_{nf_N} z_{N-1}} & -\dfrac{\varepsilon_{f_N}}{q_{nf_N}} e^{-q_{nf_N} z_{N-1}} \end{pmatrix}^{-1} \begin{pmatrix} 1 \\ \dfrac{\varepsilon_{f_N}}{q_{nf_N}} \end{pmatrix} S_{nxx,f_N} e^{-q_{nf_N} z'} e^{q_{nf_N} z_{N-1}}$$

$$= \left[ T_{N-1}^{//} \right]^{-1} \begin{pmatrix} B_{N-1} \\ C_{N-1} \end{pmatrix} - \dfrac{q_{nf_N}}{2\varepsilon_{f_N}} \begin{pmatrix} \dfrac{\varepsilon_{f_N}}{q_{nf_N}} e^{-q_{nf_N} z_{N-1}} & e^{-q_{nf_N} z_{N-1}} \\ \dfrac{\varepsilon_{f_N}}{q_{nf_N}} e^{q_{nf_N} z_{N-1}} & -e^{q_{nf_N} z_{N-1}} \end{pmatrix} \begin{pmatrix} 1 \\ \dfrac{\varepsilon_{f_N}}{q_{nf_N}} \end{pmatrix} S_{nxx,f_N} e^{-q_{nf_N} z'} e^{q_{nf_N} z_{N-1}}$$

$$= \left[ T_{N-1}^{//} \right]^{-1} \begin{pmatrix} B_{N-1} \\ C_{N-1} \end{pmatrix} - \begin{pmatrix} 1 \\ 0 \end{pmatrix} S_{nxx,f_N} e^{-q_{nf_N} z'}$$

$$\Rightarrow \begin{pmatrix} B_{N-1} \\ C_{N-1} \end{pmatrix} = T_{N-1}^{//} \begin{pmatrix} B_N \\ C_N \end{pmatrix} + T_{N-1}^{//} \begin{pmatrix} 1 \\ 0 \end{pmatrix} S_{nxx,f_N} e^{-q_{nf_N} z'}$$

. (369)

Combining Eq. (369) with the transfer matrix method which is described in Eqs. (217)-(219), we obtain:

$$\begin{pmatrix} B_1 \\ C_1 \end{pmatrix} = T_1^{//} T_2^{//} \bullet\bullet\bullet T_{N-2}^{//} \begin{pmatrix} B_{N-1} \\ C_{N-1} \end{pmatrix} = T_1^{//} T_2^{//} \bullet\bullet\bullet T_{N-2}^{//} T_{N-1}^{//} \begin{pmatrix} B_N \\ C_N \end{pmatrix} + T_1^{//} T_2^{//} \bullet\bullet\bullet T_{N-2}^{//} T_{N-1}^{//} \begin{pmatrix} 1 \\ 0 \end{pmatrix} S_{nxx,f_N} e^{-q_{nf_N} z'}$$

$$\equiv \begin{pmatrix} M_{11}^{//} & M_{12}^{//} \\ M_{21}^{//} & M_{22}^{//} \end{pmatrix} \begin{pmatrix} B_N \\ C_N \end{pmatrix} + \begin{pmatrix} M_{11}^{//} & M_{12}^{//} \\ M_{21}^{//} & M_{22}^{//} \end{pmatrix} \begin{pmatrix} 1 \\ 0 \end{pmatrix} S_{nxx,f_N} e^{-q_{nf_N} z'}$$

$$= \begin{pmatrix} M_{11}^{//} B_N + M_{12}^{//} C_N \\ M_{21}^{//} B_N + M_{22}^{//} C_N \end{pmatrix} + \begin{pmatrix} M_{11}^{//} \\ M_{21}^{//} \end{pmatrix} S_{nxx,f_N} e^{-q_{nf_N} z'}$$

, (370)

$$\begin{pmatrix} B_j \\ C_j \end{pmatrix} = T_j^{//} \bullet\bullet\bullet T_{N-2}^{//} \begin{pmatrix} B_{N-1} \\ C_{N-1} \end{pmatrix} = T_j^{//} \bullet\bullet\bullet T_{N-2}^{//} T_{N-1}^{//} \begin{pmatrix} B_N \\ C_N \end{pmatrix} + T_j^{//} \bullet\bullet\bullet T_{N-2}^{//} T_{N-1}^{//} \begin{pmatrix} 1 \\ 0 \end{pmatrix} S_{nxx,f_N} e^{-q_{nf_N} z'}$$

$$\equiv \begin{pmatrix} M_{j,11}^{//} & M_{j,12}^{//} \\ M_{j,21}^{//} & M_{j,22}^{//} \end{pmatrix} \begin{pmatrix} B_N \\ C_N \end{pmatrix} + \begin{pmatrix} M_{j,11}^{//} & M_{j,12}^{//} \\ M_{j,21}^{//} & M_{j,22}^{//} \end{pmatrix} \begin{pmatrix} 1 \\ 0 \end{pmatrix} S_{nxx,f_N} e^{-q_{nf_N} z'}$$

$$= \begin{pmatrix} M_{j,11}^{//} B_N + M_{j,12}^{//} C_N \\ M_{j,21}^{//} B_N + M_{j,22}^{//} C_N \end{pmatrix} + \begin{pmatrix} M_{j,11}^{//} \\ M_{j,21}^{//} \end{pmatrix} S_{nxx,f_N} e^{-q_{nf_N} z'}, \quad j = 2,3,\ldots,N-2$$

, (371)

and we obtain:

$$D = B_1 + C_1 = M_{11}^{//} R_{n//}^{f_N a} C_N e^{-2q_{nf_N} z_N} + M_{11}^{//} R_{n//}^{f_N a} S_{nxx, f_N} e^{q_{nf_N} z'} e^{-2q_{nf_N} z_N} + M_{12}^{//} C_N + M_{11}^{//} S_{nxx, f_N} e^{-q_{nf_N} z'}$$

$$+ M_{21}^{//} R_{n//}^{f_N a} C_N e^{-2q_{nf_N} z_N} + M_{21}^{//} R_{n//}^{f_N a} S_{nxx, f_N} e^{q_{nf_N} z'} e^{-2q_{nf_N} z_N} + M_{22}^{//} C_N + M_{21}^{//} S_{nxx, f_N} e^{-q_{nf_N} z'}$$

$$= \left( M_{11}^{//} R_{n//}^{f_N a} e^{-2q_{nf_N} z_N} + M_{12}^{//} + M_{21}^{//} R_{n//}^{f_N a} e^{-2q_{nf_N} z_N} + M_{22}^{//} \right) C_N + M_{11}^{//} R_{n//}^{f_N a} S_{nxx, f_N} e^{q_{nf_N} z'} e^{-2q_{nf_N} z_N}$$

$$+ M_{11}^{//} S_{nxx, f_N} e^{-q_{nf_N} z'} + M_{21}^{//} R_{n//}^{f_N a} S_{nxx, f_N} e^{q_{nf_N} z'} e^{-2q_{nf_N} z_N} + M_{21}^{//} S_{nxx, f_N} e^{-q_{nf_N} z'}$$

. (372)

Furthermore, we get the following form to solve the coefficient $C_N$:

$$\frac{\varepsilon_{f_1}}{q_{nf_1}} (B_1 - C_1) = \frac{\varepsilon_{ns}}{q_{ns}} D = \frac{\varepsilon_{ns}}{q_{ns}} (B_1 + C_1)$$

$$\Rightarrow \left( \frac{\varepsilon_{ns}}{q_{ns}} - \frac{\varepsilon_{f_1}}{q_{nf_1}} \right) B_1 = -\left( \frac{\varepsilon_{ns}}{q_{ns}} + \frac{\varepsilon_{f_1}}{q_{nf_1}} \right) C_1 \Rightarrow C_1 = -\frac{\varepsilon_{ns} q_{nf_1} - \varepsilon_{f_1} q_{ns}}{\varepsilon_{ns} q_{nf_1} + \varepsilon_{f_1} q_{ns}} B_1 = R_{n//}^{sf_1} B_1$$

$$\Rightarrow M_{21}^{//} R_{n//}^{f_N a} C_N e^{-2q_{nf_N} z_N} + M_{21}^{//} R_{n//}^{f_N a} S_{nxx, f_N} e^{q_{nf_N} z'} e^{-2q_{nf_N} z_N} + M_{22}^{//} C_N + M_{21}^{//} S_{nxx, f_N} e^{-q_{nf_N} z'}$$

$$= R_{n//}^{sf_1} \left( M_{11}^{//} R_{n//}^{f_N a} C_N e^{-2q_{nf_N} z_N} + M_{11}^{//} R_{n//}^{f_N a} S_{nxx, f_N} e^{q_{nf_N} z'} e^{-2q_{nf_N} z_N} + M_{12}^{//} C_N + M_{11}^{//} S_{nxx, f_N} e^{-q_{nf_N} z'} \right)$$

$$\Rightarrow \left( M_{21}^{//} R_{n//}^{f_N a} e^{-2q_{nf_N} z_N} + M_{22}^{//} - R_{n//}^{sf_1} M_{11}^{//} R_{n//}^{f_N a} e^{-2q_{nf_N} z_N} - R_{n//}^{sf_1} M_{12}^{//} \right) C_N$$

$$= -M_{21}^{//} R_{n//}^{f_N a} S_{nxx, f_N} e^{q_{nf_N} z'} e^{-2q_{nf_N} z_N} - M_{21}^{//} S_{nxx, f_N} e^{-q_{nf_N} z'} + R_{n//}^{sf_1} M_{11}^{//} R_{n//}^{f_N a} S_{nxx, f_N} e^{q_{nf_N} z'} e^{-2q_{nf_N} z_N}$$

$$+ R_{n//}^{sf_1} M_{11}^{//} S_{nxx, f_N} e^{-q_{nf_N} z'}$$

$$\Rightarrow C_N = S_{nxx, f_N} \frac{R_{n//}^{sf_1} R_{n//}^{f_N a} M_{11}^{//} e^{q_{nf_N} z'} e^{-2q_{nf_N} z_N} - R_{n//}^{f_N a} M_{21}^{//} e^{q_{nf_N} z'} e^{-2q_{nf_N} z_N} - M_{21}^{//} e^{-q_{nf_N} z'} + R_{n//}^{sf_1} M_{11}^{//} e^{-q_{nf_N} z'}}{R_{n//}^{f_N a} M_{21}^{//} e^{-2q_{nf_N} z_N} + M_{22}^{//} - R_{n//}^{sf_1} R_{n//}^{f_N a} M_{11}^{//} e^{-2q_{nf_N} z_N} - R_{n//}^{sf_1} M_{12}^{//}}$$

$$= S_{nxx, f_N} \frac{\left( R_{n//}^{sf_1} M_{11}^{//} - M_{21}^{//} \right) \left( R_{n//}^{f_N a} e^{q_{nf_N} z'} e^{-2q_{nf_N} z_N} + e^{-q_{nf_N} z'} \right)}{R_{n//}^{f_N a} M_{21}^{//} e^{-2q_{nf_N} z_N} + M_{22}^{//} - R_{n//}^{sf_1} R_{n//}^{f_N a} M_{11}^{//} e^{-2q_{nf_N} z_N} - R_{n//}^{sf_1} M_{12}^{//}}$$

. (373)

Hence we can get the other three coefficients as the following forms:

$$B_N = \frac{R_{n//}^{f_N a} S_{nxx, f_N} \left( R_{n//}^{sf_1} M_{11}^{//} - M_{21}^{//} \right) \left( R_{n//}^{f_N a} e^{q_{nf_N} z'} e^{-2q_{nf_N} z_N} + e^{-q_{nf_N} z'} \right) e^{-2q_{nf_N} z_N}}{R_{n//}^{f_N a} M_{21}^{//} e^{-2q_{nf_N} z_N} + M_{22}^{//} - R_{n//}^{sf_1} R_{n//}^{f_N a} M_{11}^{//} e^{-2q_{nf_N} z_N} - R_{n//}^{sf_1} M_{12}^{//}} + R_{n//}^{f_N a} S_{nxx, f_N} e^{q_{nf_N} z'} e^{-2q_{nf_N} z_N}$$

$$= R_{n//}^{f_N a} S_{nxx, f_N} e^{-2q_{nf_N} z_N} \frac{-\left( M_{21}^{//} - R_{n//}^{sf_1} M_{11}^{//} \right) e^{-q_{nf_N} z'} + \left( M_{22}^{//} - R_{n//}^{sf_1} M_{12}^{//} \right) e^{q_{nf_N} z'}}{R_{n//}^{f_N a} \left( M_{21}^{//} - R_{n//}^{sf_1} M_{11}^{//} \right) e^{-2q_{nf_N} z_N} + M_{22}^{//} - R_{n//}^{sf_1} M_{12}^{//}}$$

, (374)

$$A = \frac{T_{n//}^{f_N a} S_{nxx, f_N} \left( R_{n//}^{sf_1} M_{11}^{//} - M_{21}^{//} \right) \left( R_{n//}^{f_N a} e^{q_{nf_N} z'} e^{-2q_{nf_N} z_N} + e^{-q_{nf_N} z'} \right) e^{-q_{nf_N} z_N} e^{q_{na} z_N}}{R_{n//}^{f_N a} M_{21}^{//} e^{-2q_{nf_N} z_N} + M_{22}^{//} - R_{n//}^{sf_1} R_{n//}^{f_N a} M_{11}^{//} e^{-2q_{nf_N} z_N} - R_{n//}^{sf_1} M_{12}^{//}} + S_{nxx, f_N} T_{n//}^{f_N a} e^{q_{nf_N} z'} e^{-q_{nf_N} z_N} e^{q_{na} z_N}$$

$$= T_{n//}^{f_N a} S_{nxx, f_N} e^{-q_{nf_N} z_N} e^{q_{na} z_N} \frac{-\left( M_{21}^{//} - R_{n//}^{sf_1} M_{11}^{//} \right) e^{-q_{nf_N} z'} + \left( M_{22}^{//} - R_{n//}^{sf_1} M_{12}^{//} \right) e^{q_{nf_N} z'}}{R_{n//}^{f_N a} \left( M_{21}^{//} - R_{n//}^{sf_1} M_{11}^{//} \right) e^{-2q_{nf_N} z_N} + M_{22}^{//} - R_{n//}^{sf_1} M_{12}^{//}}$$

, (375)

$$D = \left( M_{11}^{//} R_{n//}^{f_N a} e^{-2q_{nf_N} z_N} + M_{12}^{//} + M_{21}^{//} R_{n//}^{f_N a} e^{-2q_{nf_N} z_N} + M_{22}^{//} \right) C_N + M_{11}^{//} R_{n//}^{f_N a} S_{nxx,f_N} e^{q_{nf_N} z'} e^{-2q_{nf_N} z_N}$$

$$+ M_{11}^{//} S_{nxx,f_N} e^{-q_{nf_N} z'} + M_{21}^{//} R_{n//}^{f_N a} S_{nxx,f_N} e^{q_{nf_N} z'} e^{-2q_{nf_N} z_N} + M_{21}^{//} S_{nxx,f_N} e^{-q_{nf_N} z'}$$

$$= S_{nxx,f_N} \frac{\left( M_{11}^{//} R_{n//}^{f_N a} e^{-2q_{nf_N} z_N} + M_{12}^{//} + M_{21}^{//} R_{n//}^{f_N a} e^{-2q_{nf_N} z_N} + M_{22}^{//} \right) \left( R_{n//}^{sf_1} M_{11}^{//} - M_{21}^{//} \right) \left( R_{n//}^{f_N a} e^{q_{nf_N} z'} e^{-2q_{nf_N} z_N} + e^{-q_{nf_N} z'} \right)}{R_{n//}^{f_N a} M_{21}^{//} e^{-2q_{nf_N} z_N} + M_{22}^{//} - R_{n//}^{sf_1} R_{n//}^{f_N a} M_{11}^{//} e^{-2q_{nf_N} z_N} - R_{n//}^{sf_1} M_{12}^{//}}$$

$$+ M_{11}^{//} R_{n//}^{f_N a} S_{nxx,f_N} e^{q_{nf_N} z'} e^{-2q_{nf_N} z_N} + M_{11}^{//} S_{nxx,f_N} e^{-q_{nf_N} z'} + M_{21}^{//} R_{n//}^{f_N a} S_{nxx,f_N} e^{q_{nf_N} z'} e^{-2q_{nf_N} z_N} + M_{21}^{//} S_{nxx,f_N} e^{-q_{nf_N} z'}$$

$$= S_{nxx,f_N} T_{n//}^{sf_1} \frac{\left( M_{22}^{//} M_{11}^{//} - M_{12}^{//} M_{21}^{//} \right) R_{n//}^{f_N a} e^{q_{nf_N} z'} e^{-2q_{nf_N} z_N} + \left( M_{22}^{//} M_{11}^{//} - M_{12}^{//} M_{21}^{//} \right) e^{-q_{nf_N} z'}}{R_{n//}^{f_N a} M_{21}^{//} e^{-2q_{nf_N} z_N} + M_{22}^{//} - R_{n//}^{sf_1} R_{n//}^{f_N a} M_{11}^{//} e^{-2q_{nf_N} z_N} - R_{n//}^{sf_1} M_{12}^{//}}$$

. (376)

Then we solve the coefficients in Eq. (361) completely.

<2> If $k = 2, 3, ..., N-1$

$$-\frac{\varepsilon_a}{q_{na}} \left( B_N e^{q_{nf_N} z_N} + C_N e^{-q_{nf_N} z_N} \right) = \frac{\varepsilon_{f_N}}{q_{nf_N}} \left( B_N e^{q_{nf_N} z_N} - C_N e^{-q_{nf_N} z_N} \right)$$

$$\Rightarrow \left( \frac{\varepsilon_{f_N}}{q_{nf_N}} + \frac{\varepsilon_a}{q_{na}} \right) B_N = \left( \frac{\varepsilon_{f_N}}{q_{nf_N}} - \frac{\varepsilon_a}{q_{na}} \right) C_N e^{-2q_{nf_N} z_N} \quad , \quad (377)$$

$$\Rightarrow B_N = \frac{\varepsilon_{f_N} q_{na} - \varepsilon_a q_{nf_N}}{\varepsilon_{f_N} q_{na} + \varepsilon_a q_{nf_N}} C_N e^{-2q_{nf_N} z_N} = R_{n//}^{f_N a} C_N e^{-2q_{nf_N} z_N}$$

$$\Rightarrow A = R_{n//}^{f_N a} C_N e^{-q_{nf_N} z_N} e^{q_{na} z_N} + C_N e^{-q_{nf_N} z_N} e^{q_{na} z_N} = T_{n//}^{f_N a} C_N e^{-q_{nf_N} z_N} e^{q_{na} z_N}$$

which is the same as Eq. (216) and

$$\begin{cases} B_{k+1} e^{q_{nf_{k+1}} z_k} + C_{k+1} e^{-q_{nf_{k+1}} z_k} = S_{nxx,f_k} e^{q_{nf_k} z'} e^{-q_{nf_k} z_k} + B_k e^{q_{nf_k} z_k} + C_k e^{-q_{nf_k} z_k} \\ \frac{\varepsilon_{f_{k+1}}}{q_{nf_{k+1}}} \left[ B_{k+1} e^{q_{nf_{k+1}} z_k} - C_{k+1} e^{-q_{nf_{k+1}} z_k} \right] = -\frac{\varepsilon_{f_k}}{q_{nf_k}} S_{nxx,f_k} e^{q_{nf_k} z'} e^{-q_{nf_k} z_k} + \frac{\varepsilon_{f_k}}{q_{nf_k}} \left[ B_k e^{q_{nf_k} z_k} - C_k e^{-q_{nf_k} z_k} \right] \end{cases}$$

$$\Rightarrow \begin{pmatrix} B_k \\ C_k \end{pmatrix} = T_k^{//} \begin{pmatrix} B_{k+1} \\ C_{k+1} \end{pmatrix} + \begin{pmatrix} e^{q_{nf_k} z_k} & e^{-q_{nf_k} z_k} \\ \frac{\varepsilon_{f_k}}{q_{nf_k}} e^{q_{nf_k} z_k} & -\frac{\varepsilon_{f_k}}{q_{nf_k}} e^{-q_{nf_k} z_k} \end{pmatrix}^{-1} \begin{pmatrix} -1 \\ \frac{\varepsilon_{f_k}}{q_{nf_k}} \end{pmatrix} S_{nxx,f_k} e^{q_{nf_k} z'} e^{-q_{nf_k} z_k}$$

$$= T_k^{//} \begin{pmatrix} B_{k+1} \\ C_{k+1} \end{pmatrix} + \frac{q_{nf_k}}{2\varepsilon_{f_k}} \begin{pmatrix} \frac{\varepsilon_{f_k}}{q_{nf_k}} e^{-q_{nf_k} z_k} & e^{-q_{nf_k} z_k} \\ \frac{\varepsilon_{f_k}}{q_{nf_k}} e^{q_{nf_k} z_k} & -e^{q_{nf_k} z_k} \end{pmatrix} \begin{pmatrix} -1 \\ \frac{\varepsilon_{f_k}}{q_{nf_k}} \end{pmatrix} S_{nxx,f_k} e^{q_{nf_k} z'} e^{-q_{nf_k} z_k}$$

$$= T_k^{//} \begin{pmatrix} B_{k+1} \\ C_{k+1} \end{pmatrix} - \begin{pmatrix} 0 \\ 1 \end{pmatrix} S_{nxx,f_k} e^{q_{nf_k} z'}$$

, (378)

and

$$\begin{cases} S_{nxx,f_k} e^{-q_{nf_k} z'} e^{q_{nf_k} z_{k-1}} + B_k e^{q_{nf_k} z_{k-1}} + C_k e^{-q_{nf_k} z_{k-1}} = B_{k-1} e^{q_{nf_{k-1}} z_{k-1}} + C_{k-1} e^{-q_{nf_{k-1}} z_{k-1}} \\ \dfrac{\varepsilon_{f_k}}{q_{nf_k}} S_{nxx,f_k} e^{-q_{nf_k} z'} e^{q_{nf_k} z_{k-1}} + \dfrac{\varepsilon_{f_k}}{q_{nf_k}} \left[ B_k e^{q_{nf_k} z_{k-1}} - C_k e^{-q_{nf_k} z_{k-1}} \right] = \dfrac{\varepsilon_{f_{k-1}}}{q_{nf_{k-1}}} \left[ B_{k-1} e^{q_{nf_{k-1}} z_{k-1}} - C_{k-1} e^{-q_{nf_{k-1}} z_{k-1}} \right] \end{cases}$$

$$\begin{pmatrix} B_k \\ C_k \end{pmatrix} = \left[ \mathrm{T}_{k-1}^{//} \right]^{-1} \begin{pmatrix} B_{k-1} \\ C_{k-1} \end{pmatrix} - \begin{pmatrix} e^{q_{nf_k} z_{k-1}} & e^{-q_{nf} z_{k-1}} \\ \dfrac{\varepsilon_{f_k}}{q_{nf_k}} e^{q_{nf_k} z_{k-1}} & -\dfrac{\varepsilon_{f_k}}{q_{nf_k}} e^{-q_{nf_k} z_{k-1}} \end{pmatrix}^{-1} \begin{pmatrix} 1 \\ \dfrac{\varepsilon_{f_k}}{q_{nf_k}} \end{pmatrix} S_{nxx,f_k} e^{-q_{nf_k} z'} e^{q_{nf_k} z_{k-1}}$$

$$= \left[ \mathrm{T}_{k-1}^{//} \right]^{-1} \begin{pmatrix} B_{k-1} \\ C_{k-1} \end{pmatrix} - \dfrac{q_{nf_k}}{2\varepsilon_{f_k}} \begin{pmatrix} \dfrac{\varepsilon_{f_k}}{q_{nf_k}} e^{-q_{nf_k} z_{k-1}} & e^{-q_{nf} z_{k-1}} \\ \dfrac{\varepsilon_{f_k}}{q_{nf_k}} e^{q_{nf_k} z_{k-1}} & -e^{q_{nf_k} z_{k-1}} \end{pmatrix} \begin{pmatrix} 1 \\ \dfrac{\varepsilon_{f_k}}{q_{nf_k}} \end{pmatrix} S_{nxx,f_k} e^{-q_{nf_k} z'} e^{q_{nf_k} z_{k-1}}$$

$$= \left[ \mathrm{T}_{k-1}^{//} \right]^{-1} \begin{pmatrix} B_{k-1} \\ C_{k-1} \end{pmatrix} - \begin{pmatrix} 1 \\ 0 \end{pmatrix} S_{nxx,f_k} e^{-q_{nf_k} z'}$$

$$\Rightarrow \begin{pmatrix} B_{k-1} \\ C_{k-1} \end{pmatrix} = \mathrm{T}_{k-1}^{//} \begin{pmatrix} B_k \\ C_k \end{pmatrix} + \mathrm{T}_{k-1}^{//} \begin{pmatrix} 1 \\ 0 \end{pmatrix} S_{nxx,f_k} e^{-q_{nf_k} z'}$$

. 
(379)

Combining Eqs. (378)-(379) with the transfer matrix method which is described in Eqs. (217)-(219), we obtain:

$$\begin{pmatrix} B_1 \\ C_1 \end{pmatrix} = \mathrm{T}_1^{//} \mathrm{T}_2^{//} \bullet \bullet \mathrm{T}_{k-2}^{//} \begin{pmatrix} B_{k-1} \\ C_{k-1} \end{pmatrix} = \mathrm{T}_1^{//} \mathrm{T}_2^{//} \bullet \bullet \mathrm{T}_{k-2}^{//} \mathrm{T}_{k-1}^{//} \begin{pmatrix} B_k \\ C_k \end{pmatrix} + \mathrm{T}_1^{//} \mathrm{T}_2^{//} \bullet \bullet \mathrm{T}_{k-2}^{//} \mathrm{T}_{k-1}^{//} \begin{pmatrix} 1 \\ 0 \end{pmatrix} S_{nxx,f_k} e^{-q_{nf_k} z'}$$

$$= \mathrm{T}_1^{//} \mathrm{T}_2^{//} \bullet \bullet \mathrm{T}_{k-2}^{//} \mathrm{T}_{k-1}^{//} \mathrm{T}_k^{//} \begin{pmatrix} B_{k+1} \\ C_{k+1} \end{pmatrix} - \mathrm{T}_1^{//} \mathrm{T}_2^{//} \bullet \bullet \mathrm{T}_{k-2}^{//} \mathrm{T}_{k-1}^{//} \begin{pmatrix} 0 \\ 1 \end{pmatrix} S_{nxx,f_k} e^{q_{nf_k} z'} + \mathrm{T}_1^{//} \mathrm{T}_2^{//} \bullet \bullet \mathrm{T}_{k-2}^{//} \mathrm{T}_{k-1}^{//} \begin{pmatrix} 1 \\ 0 \end{pmatrix} S_{nxx,f_k} e^{-q_{nf_k} z'}$$

$$\equiv \begin{pmatrix} M_{11}^{//} & M_{12}^{//} \\ M_{21}^{//} & M_{22}^{//} \end{pmatrix} \begin{pmatrix} B_N \\ C_N \end{pmatrix} - \begin{pmatrix} M_{11,k-1}^{//} & M_{12,k-1}^{//} \\ M_{21,k-1}^{//} & M_{22,k-1}^{//} \end{pmatrix} \begin{pmatrix} 0 \\ 1 \end{pmatrix} S_{nxx,f_k} e^{q_{nf_k} z'} + \begin{pmatrix} M_{11,k-1}^{//} & M_{12,k-1}^{//} \\ M_{21,k-1}^{//} & M_{22,k-1}^{//} \end{pmatrix} \begin{pmatrix} 1 \\ 0 \end{pmatrix} S_{nxx,f_k} e^{-q_{nf_k} z'}$$

$$= \begin{pmatrix} M_{11}^{//} B_N + M_{12}^{//} C_N \\ M_{21}^{//} B_N + M_{22}^{//} C_N \end{pmatrix} - \begin{pmatrix} M_{12,k-1}^{//} \\ M_{22,k-1}^{//} \end{pmatrix} S_{nxx,f_k} e^{q_{nf_k} z'} + \begin{pmatrix} M_{11,k-1}^{//} \\ M_{21,k-1}^{//} \end{pmatrix} S_{nxx,f_k} e^{-q_{nf_k} z'}$$

,
(380)

$$\begin{pmatrix} B_j \\ C_j \end{pmatrix} = \mathrm{T}_j^{//} \bullet \bullet \mathrm{T}_{k-2}^{//} \begin{pmatrix} B_{k-1} \\ C_{k-1} \end{pmatrix} = \mathrm{T}_j^{//} \bullet \bullet \mathrm{T}_{k-2}^{//} \mathrm{T}_{k-1}^{//} \begin{pmatrix} B_k \\ C_k \end{pmatrix} + \mathrm{T}_j^{//} \bullet \bullet \mathrm{T}_{k-2}^{//} \mathrm{T}_{k-1}^{//} \begin{pmatrix} 1 \\ 0 \end{pmatrix} S_{nxx,f_k} e^{-q_{nf_k} z'}$$

$$= \mathrm{T}_j^{//} \bullet \bullet \mathrm{T}_{k-2}^{//} \mathrm{T}_{k-1}^{//} \mathrm{T}_k^{//} \begin{pmatrix} B_{k+1} \\ C_{k+1} \end{pmatrix} - \mathrm{T}_j^{//} \bullet \bullet \mathrm{T}_{k-2}^{//} \mathrm{T}_{k-1}^{//} \begin{pmatrix} 0 \\ 1 \end{pmatrix} S_{nxx,f_k} e^{q_{nf_k} z'} + \mathrm{T}_j^{//} \bullet \bullet \mathrm{T}_{k-2}^{//} \mathrm{T}_{k-1}^{//} \begin{pmatrix} 1 \\ 0 \end{pmatrix} S_{nxx,f_k} e^{-q_{nf_k} z'}$$

$$\equiv \begin{pmatrix} M_{j,11}^{//} & M_{j,12}^{//} \\ M_{j,21}^{//} & M_{j,22}^{//} \end{pmatrix} \begin{pmatrix} B_N \\ C_N \end{pmatrix} - \begin{pmatrix} M_{j,11,k-1}^{//} & M_{j,12,k-1}^{//} \\ M_{j,21,k-1}^{//} & M_{j,22,k-1}^{//} \end{pmatrix} \begin{pmatrix} 0 \\ 1 \end{pmatrix} S_{nxx,f_k} e^{q_{nf_k} z'} + \begin{pmatrix} M_{j,11,k-1}^{//} & M_{j,12,k-1}^{//} \\ M_{j,21,k-1}^{//} & M_{j,22,k-1}^{//} \end{pmatrix} \begin{pmatrix} 1 \\ 0 \end{pmatrix} S_{nxx,f_k} e^{-q_{nf_k} z'}$$

$$= \begin{pmatrix} M_{j,11}^{//} B_N + M_{j,12}^{//} C_N \\ M_{j,21}^{//} B_N + M_{j,22}^{//} C_N \end{pmatrix} - \begin{pmatrix} M_{j,12,k-1}^{//} \\ M_{j,22,k-1}^{//} \end{pmatrix} S_{nxx,f_k} e^{q_{nf_k} z'} + \begin{pmatrix} M_{j,11,k-1}^{//} \\ M_{j,21,k-1}^{//} \end{pmatrix} S_{nxx,f_k} e^{-q_{nf_k} z'}, \; j = 2, 3, ..., N-1, \; j \le k-1$$

$$\begin{pmatrix} B_k \\ C_k \end{pmatrix} = T_k^{//} \begin{pmatrix} B_{k+1} \\ C_{k+1} \end{pmatrix} - \begin{pmatrix} 0 \\ 1 \end{pmatrix} S_{nxx,f_k} e^{q_{nf_k} z'} = T_k^{//} \bullet \bullet T_{N-1}^{//} \begin{pmatrix} B_N \\ C_N \end{pmatrix} - \begin{pmatrix} 0 \\ 1 \end{pmatrix} S_{nxx,f_k} e^{q_{nf_k} z'} \qquad (381)$$

$$\equiv \begin{pmatrix} M_{k,11}^{//} & M_{k,12}^{//} \\ M_{k,21}^{//} & M_{k,22}^{//} \end{pmatrix} \begin{pmatrix} B_N \\ C_N \end{pmatrix} - \begin{pmatrix} 0 \\ 1 \end{pmatrix} S_{nxx,f_k} e^{q_{nf_k} z'}$$

$$= \begin{pmatrix} M_{k,11}^{//} B_N + M_{k,12}^{//} C_N \\ M_{k,21}^{//} B_N + M_{k,22}^{//} C_N \end{pmatrix} - \begin{pmatrix} 0 \\ 1 \end{pmatrix} S_{nxx,f_k} e^{q_{nf_k} z'}, \ j = 2,3,...,N-1, \ j = k \qquad (382)$$

$$\begin{pmatrix} B_j \\ C_j \end{pmatrix} = T_j^{//} \begin{pmatrix} B_{j+1} \\ C_{j+1} \end{pmatrix} = T_j^{//} T_{j+1}^{//} \bullet \bullet \bullet T_{N-1}^{//} \begin{pmatrix} B_N \\ C_N \end{pmatrix} \equiv \begin{pmatrix} M_{j,11}^{//} & M_{j,12}^{//} \\ M_{j,21}^{//} & M_{j,22}^{//} \end{pmatrix} \begin{pmatrix} B_N \\ C_N \end{pmatrix}$$

$$= \begin{pmatrix} M_{j,11}^{//} B_N + M_{j,12}^{//} C_N \\ M_{j,21}^{//} B_N + M_{j,22}^{//} C_N \end{pmatrix}, \ j = 2,3,...,N-1, \ j \geq k+1 \qquad (383)$$

Furthermore, we obtain:

$$D = B_1 + C_1 = M_{11}^{//} R_{n//}^{f_N a} C_N e^{-2q_{nf_N} z_N} + M_{12}^{//} C_N + M_{21}^{//} R_{n//}^{f_N a} C_N e^{-2q_{nf_N} z_N} + M_{22}^{//} C_N$$
$$- M_{12,k-1}^{//} S_{nxx,f_k} e^{q_{nf_k} z'} + M_{11,k-1}^{//} S_{nxx,f_k} e^{-q_{nf_k} z'} - M_{22,k-1}^{//} S_{nxx,f_k} e^{q_{nf_k} z'} + M_{21,k-1}^{//} S_{nxx,f_k} e^{-q_{nf_k} z'}. \qquad (384)$$

Furthermore, we get the following form to solve the coefficient $C_N$:

$$\frac{\varepsilon_{f_1}}{q_{nf_1}} \left( M_{11}^{//} R_{n//}^{f_N a} C_N e^{-2q_{nf_N} z_N} + M_{12}^{//} C_N - M_{12,k-1}^{//} S_{nxx,f_k} e^{q_{nf_k} z'} + M_{11,k-1}^{//} S_{nxx,f_k} e^{-q_{nf_k} z'} \right)$$

$$- \frac{\varepsilon_{f_1}}{q_{nf_1}} \left( M_{21}^{//} R_{n//}^{f_N a} C_N e^{-2q_{nf_N} z_N} + M_{22}^{//} C_N - M_{22,k-1}^{//} S_{nxx,f_k} e^{q_{nf_k} z'} + M_{21,k-1}^{//} S_{nxx,f_k} e^{-q_{nf_k} z'} \right)$$

$$= \frac{\varepsilon_s}{q_{ns}} \left( M_{11}^{//} R_{n//}^{f_N a} C_N e^{-2q_{nf_N} z_N} + M_{12}^{//} C_N + M_{21}^{//} R_{n//}^{f_N a} C_N e^{-2q_{nf_N} z_N} + M_{22}^{//} C_N \right)$$

$$- \frac{\varepsilon_s}{q_{ns}} \left( M_{12,k-1}^{//} S_{nxx,f_k} e^{q_{nf_k} z'} - M_{11,k-1}^{//} S_{nxx,f_k} e^{-q_{nf_k} z'} + M_{22,k-1}^{//} S_{nxx,f_k} e^{q_{nf_k} z'} - M_{21,k-1}^{//} S_{nxx,f_k} e^{-q_{nf_k} z'} \right)$$

$$\Rightarrow \left[ \left( \frac{\varepsilon_s}{q_{ns}} - \frac{\varepsilon_{f_1}}{q_{nf_1}} \right) \left( M_{11}^{//} R_{n//}^{f_N a} e^{-2q_{nf_N} z_N} + M_{12}^{//} \right) + \left( \frac{\varepsilon_s}{q_{ns}} + \frac{\varepsilon_{f_1}}{q_{nf_1}} \right) \left( M_{21}^{//} R_{n//}^{f_N a} e^{-2q_{nf_N} z_N} + M_{22}^{//} \right) \right] C_N$$

$$= \left[ \left( \frac{\varepsilon_s}{q_{ns}} - \frac{\varepsilon_{f_1}}{q_{nf_1}} \right) \left( M_{12,k-1}^{//} e^{q_{nf_k} z'} - M_{11,k-1}^{//} e^{-q_{nf_k} z'} \right) + \left( \frac{\varepsilon_s}{q_{ns}} + \frac{\varepsilon_{f_1}}{q_{nf_1}} \right) \left( M_{22,k-1}^{//} e^{q_{nf_k} z'} - M_{21,k-1}^{//} e^{-q_{nf_k} z'} \right) \right] S_{nxx,f_k}$$

$$\Rightarrow C_N = S_{nxx,f_k} \frac{\left( \frac{\varepsilon_s}{q_{ns}} - \frac{\varepsilon_{f_1}}{q_{nf_1}} \right) \left( M_{12,k-1}^{//} e^{q_{nf_k} z'} - M_{11,k-1}^{//} e^{-q_{nf_k} z'} \right) + \left( \frac{\varepsilon_s}{q_{ns}} + \frac{\varepsilon_{f_1}}{q_{nf_1}} \right) \left( M_{22,k-1}^{//} e^{q_{nf_k} z'} - M_{21,k-1}^{//} e^{-q_{nf_k} z'} \right)}{\left( \frac{\varepsilon_s}{q_{ns}} - \frac{\varepsilon_{f_1}}{q_{nf_1}} \right) \left( M_{11}^{//} R_{n//}^{f_N a} e^{-2q_{nf_N} z_N} + M_{12}^{//} \right) + \left( \frac{\varepsilon_s}{q_{ns}} + \frac{\varepsilon_{f_1}}{q_{nf_1}} \right) \left( M_{21}^{//} R_{n//}^{f_N a} e^{-2q_{nf_N} z_N} + M_{22}^{//} \right)}$$

$$= S_{nxx,f_k} \frac{R_{n//}^{sf_1} \left( M_{12,k-1}^{//} e^{q_{nf_k} z'} - M_{11,k-1}^{//} e^{-q_{nf_k} z'} \right) + \left( M_{22,k-1}^{//} e^{q_{nf_k} z'} - M_{21,k-1}^{//} e^{-q_{nf_k} z'} \right)}{R_{n//}^{sf_1} \left( M_{11}^{//} R_{n//}^{f_N a} e^{-2q_{nf_N} z_N} + M_{12}^{//} \right) + \left( M_{21}^{//} R_{n//}^{f_N a} e^{-2q_{nf_N} z_N} + M_{22}^{//} \right)}$$

. (385)

Hence we can get the other three coefficients as the following forms:

$$B_N = R_{n//}^{f_N a} S_{nxx,f_k} e^{-2q_{nf_N} z_N} \frac{R_{n//}^{sf_1}\left(M_{12,k-1}^{//} e^{q_{nf_k} z'} - M_{11,k-1}^{//} e^{-q_{nf_k} z'}\right) + \left(M_{22,k-1}^{//} e^{q_{nf_k} z'} - M_{21,k-1}^{//} e^{-q_{nf_k} z'}\right)}{R_{n//}^{sf_1}\left(M_{11}^{//} R_{n//}^{f_N a} e^{-2q_{nf_N} z_N} + M_{12}^{//}\right) + \left(M_{21}^{//} R_{n//}^{f_N a} e^{-2q_{nf_N} z_N} + M_{22}^{//}\right)},$$

(386)

$$A = T_{n//}^{f_N a} S_{nxx,f_k} e^{-q_{nf_N} z_N} e^{q_{na} z_N} \frac{R_{n//}^{sf_1}\left(M_{12,k-1}^{//} e^{q_{nf_k} z'} - M_{11,k-1}^{//} e^{-q_{nf_k} z'}\right) + \left(M_{22,k-1}^{//} e^{q_{nf_k} z'} - M_{21,k-1}^{//} e^{-q_{nf_k} z'}\right)}{R_{n//}^{sf_1}\left(M_{11}^{//} R_{n//}^{f_N a} e^{-2q_{nf_N} z_N} + M_{12}^{//}\right) + \left(M_{21}^{//} R_{n//}^{f_N a} e^{-2q_{nf_N} z_N} + M_{22}^{//}\right)}$$

, (387)

$$\begin{aligned}
D &= \left(M_{11}^{//} R_{n//}^{f_N a} e^{-2q_{nf_N} z_N} + M_{12}^{//} + M_{21}^{//} R_{n//}^{f_N a} e^{-2q_{nf_N} z_N} + M_{22}^{//}\right) C_N \\
&\quad - M_{12,k-1}^{//} S_{nxx,f_k} e^{q_{nf_k} z'} + M_{11,k-1}^{//} S_{nxx,f_k} e^{-q_{nf_k} z'} - M_{22,k-1}^{//} S_{nxx,f_k} e^{q_{nf_k} z'} + M_{21,k-1}^{//} S_{nxx,f_k} e^{-q_{nf_k} z'} \\
&= \frac{R_{n//}^{sf_1}\left(M_{12,k-1}^{//} e^{q_{nf_k} z'} - M_{11,k-1}^{//} e^{-q_{nf_k} z'}\right) + \left(M_{22,k-1}^{//} e^{q_{nf_k} z'} - M_{21,k-1}^{//} e^{-q_{nf_k} z'}\right)}{R_{n//}^{sf_1}\left(M_{11}^{//} R_{n//}^{f_N a} e^{-2q_{nf_N} z_N} + M_{12}^{//}\right) + \left(M_{21}^{//} R_{n//}^{f_N a} e^{-2q_{nf_N} z_N} + M_{22}^{//}\right)} M_{11}^{//} R_{n//}^{f_N a} e^{-2q_{nf_N} z_N} S_{nxx,f_k} \\
&\quad + \frac{R_{n//}^{sf_1}\left(M_{12,k-1}^{//} e^{q_{nf_k} z'} - M_{11,k-1}^{//} e^{-q_{nf_k} z'}\right) + \left(M_{22,k-1}^{//} e^{q_{nf_k} z'} - M_{21,k-1}^{//} e^{-q_{nf_k} z'}\right)}{R_{n//}^{sf_1}\left(M_{11}^{//} R_{n//}^{f_N a} e^{-2q_{nf_N} z_N} + M_{12}^{//}\right) + \left(M_{21}^{//} R_{n//}^{f_N a} e^{-2q_{nf_N} z_N} + M_{22}^{//}\right)} M_{12}^{//} S_{nxx,f_k} \\
&\quad + \frac{R_{n//}^{sf_1}\left(M_{12,k-1}^{//} e^{q_{nf_k} z'} - M_{11,k-1}^{//} e^{-q_{nf_k} z'}\right) + \left(M_{22,k-1}^{//} e^{q_{nf_k} z'} - M_{21,k-1}^{//} e^{-q_{nf_k} z'}\right)}{R_{n//}^{sf_1}\left(M_{11}^{//} R_{n//}^{f_N a} e^{-2q_{nf_N} z_N} + M_{12}^{//}\right) + \left(M_{21}^{//} R_{n//}^{f_N a} e^{-2q_{nf_N} z_N} + M_{22}^{//}\right)} M_{21}^{//} R_{n//}^{f_N a} e^{-2q_{nf_N} z_N} S_{nxx,f_k} \\
&\quad + \frac{R_{n//}^{sf_1}\left(M_{12,k-1}^{//} e^{q_{nf_k} z'} - M_{11,k-1}^{//} e^{-q_{nf_k} z'}\right) + \left(M_{22,k-1}^{//} e^{q_{nf_k} z'} - M_{21,k-1}^{//} e^{-q_{nf_k} z'}\right)}{R_{n//}^{sf_1}\left(M_{11}^{//} R_{n//}^{f_N a} e^{-2q_{nf_N} z_N} + M_{12}^{//}\right) + \left(M_{21}^{//} R_{n//}^{f_N a} e^{-2q_{nf_N} z_N} + M_{22}^{//}\right)} M_{22}^{//} S_{nxx,f_k} \\
&\quad - M_{12,k-1}^{//} S_{nxx,f_k} e^{q_{nf_k} z'} + M_{11,k-1}^{//} S_{nxx,f_k} e^{-q_{nf_k} z'} - M_{22,k-1}^{//} S_{nxx,f_k} e^{q_{nf_k} z'} + M_{21,k-1}^{//} S_{nxx,f_k} e^{-q_{nf_k} z'}.
\end{aligned}$$

(388)

Then we solve the coefficients in Eq. (361) completely.

<3> If $k = 1$

$$-\frac{\varepsilon_a}{q_{na}}\left(B_N e^{q_{nf_N} z_N} + C_N e^{-q_{nf_N} z_N}\right) = \frac{\varepsilon_{f_N}}{q_{nf_N}}\left(B_N e^{q_{nf_N} z_N} - C_N e^{-q_{nf_N} z_N}\right)$$

$$\Rightarrow \left(\frac{\varepsilon_{f_N}}{q_{nf_N}} + \frac{\varepsilon_a}{q_{na}}\right) B_N = \left(\frac{\varepsilon_{f_N}}{q_{nf_N}} - \frac{\varepsilon_a}{q_{na}}\right) C_N e^{-2q_{nf_N} z_N}$$, (389)

$$\Rightarrow B_N = \frac{\varepsilon_{f_N} q_{na} - \varepsilon_a q_{nf_N}}{\varepsilon_{f_N} q_{na} + \varepsilon_a q_{nf_N}} C_N e^{-2q_{nf_N} z_N} = R_{n//}^{f_N a} C_N e^{-2q_{nf_N} z_N}$$

$$\Rightarrow A = R_{n//}^{f_N a} C_N e^{-q_{nf_N} z_N} e^{q_{na} z_N} + C_N e^{-q_{nf_N} z_N} e^{q_{na} z_N} = T_{n//}^{f_N a} C_N e^{-q_{nf_N} z_N} e^{q_{na} z_N}$$

and

$$\begin{cases} B_2 e^{q_{nf_2} z_1} + C_2 e^{-q_{nf_2} z_1} = S_{nxx,f_1} e^{q_{nf_1} z'} e^{-q_{nf_1} z_1} + B_1 e^{q_{nf_1} z_1} + C_1 e^{-q_{nf_1} z_1} \\ \dfrac{\varepsilon_{f_2}}{q_{nf_2}} \left[ B_2 e^{q_{nf_2} z_1} - C_2 e^{-q_{nf_2} z_1} \right] = -\dfrac{\varepsilon_{f_1}}{q_{nf_1}} S_{nxx,f_1} e^{q_{nf_1} z'} e^{-q_{nf_1} z_1} + \dfrac{\varepsilon_{f_1}}{q_{nf_1}} \left[ B_1 e^{q_{nf_1} z_1} - C_1 e^{-q_{nf_1} z_1} \right] \end{cases}$$

$$\Rightarrow \begin{pmatrix} B_1 \\ C_1 \end{pmatrix} = T_1'' \begin{pmatrix} B_2 \\ C_2 \end{pmatrix} + \begin{pmatrix} e^{q_{nf_1} z_1} & e^{-q_{nf_1} z_1} \\ \dfrac{\varepsilon_{f_1}}{q_{nf_1}} e^{q_{nf_1} z_1} & -\dfrac{\varepsilon_{f_1}}{q_{nf_1}} e^{-q_{nf_1} z_1} \end{pmatrix}^{-1} \begin{pmatrix} -1 \\ \dfrac{\varepsilon_{f_1}}{q_{nf_1}} \end{pmatrix} S_{nxx,f_1} e^{q_{nf_1} z'} e^{-q_{nf_1} z_1}$$

(390)

$$= T_1'' \begin{pmatrix} B_2 \\ C_2 \end{pmatrix} + \dfrac{q_{nf_1}}{2\varepsilon_{f_1}} \begin{pmatrix} \dfrac{\varepsilon_{f_1}}{q_{nf_1}} e^{-q_{nf_1} z_1} & e^{-q_{nf_1} z_1} \\ \dfrac{\varepsilon_{f_1}}{q_{nf_1}} e^{q_{nf_1} z_1} & -e^{q_{nf_1} z_1} \end{pmatrix} \begin{pmatrix} -1 \\ \dfrac{\varepsilon_{f_1}}{q_{nf_1}} \end{pmatrix} S_{nxx,f_1} e^{q_{nf_1} z'} e^{-q_{nf_1} z_1}$$

$$= T_1'' \begin{pmatrix} B_2 \\ C_2 \end{pmatrix} - \begin{pmatrix} 0 \\ 1 \end{pmatrix} S_{nxx,f_1} e^{q_{nf_1} z'}$$

Combining Eq. (390) with the transfer matrix method which is described in Eqs. (217)-(219), we obtain:

$$\begin{pmatrix} B_1 \\ C_1 \end{pmatrix} = T_1'' \begin{pmatrix} B_2 \\ C_2 \end{pmatrix} - \begin{pmatrix} 0 \\ 1 \end{pmatrix} S_{nxx,f_1} e^{q_{nf_1} z'} = \begin{pmatrix} M_{11}'' & M_{12}'' \\ M_{21}'' & M_{22}'' \end{pmatrix} \begin{pmatrix} B_N \\ C_N \end{pmatrix} - \begin{pmatrix} 0 \\ 1 \end{pmatrix} S_{nxx,f_1} e^{q_{nf_1} z'}$$

$$= \begin{pmatrix} M_{11}'' B_N + M_{12}'' C_N \\ M_{21}'' B_N + M_{22}'' C_N \end{pmatrix} - \begin{pmatrix} 0 \\ 1 \end{pmatrix} S_{nxx,f_1} e^{q_{nf_1} z'}$$

(391)

$$\begin{pmatrix} B_j \\ C_j \end{pmatrix} = T_j'' \bullet \bullet \bullet T_{N-1}'' \begin{pmatrix} B_N \\ C_N \end{pmatrix} = \begin{pmatrix} M_{j,11}'' & M_{j,12}'' \\ M_{j,21}'' & M_{j,22}'' \end{pmatrix} \begin{pmatrix} B_N \\ C_N \end{pmatrix} = \begin{pmatrix} M_{j,11}'' B_N + M_{j,11}'' C_N \\ M_{j,21}'' B_N + M_{j,22}'' C_N \end{pmatrix}, \; j=2,...,N-1$$

(392)

Furthermore, we obtain:

$$D = S_{nxx,f_1} e^{-q_{nf_1} z'} + B_1 + C_1 = S_{nxx,f_1} e^{-q_{nf_1} z'} + M_{11}'' R_{n//}^{f_N a} C_N e^{-2 q_{nf_N} z_N} + M_{12}'' C_N$$
$$+ M_{21}'' R_{n//}^{f_N a} C_N e^{-2 q_{nf_N} z_N} + M_{22}'' C_N - S_{nxx,f_1} e^{q_{nf_1} z'}$$
$$= \left( M_{11}'' R_{n//}^{f_N a} e^{-2 q_{nf_N} z_N} + M_{12}'' + M_{21}'' R_{n//}^{f_N a} e^{-2 q_{nf_N} z_N} + M_{22}'' \right) C_N + S_{nxx,f_1} e^{-q_{nf_1} z'} - S_{nxx,f_1} e^{q_{nf_1} z'}$$

(393)

Furthermore, we get the following form to solve the coefficient $C_N$:

$$\frac{\varepsilon_{f_1}}{q_{nf_1}} S_{nxx,f_1} e^{-q_{nf_1} z'} + \frac{\varepsilon_{f_1}}{q_{nf_1}} (B_1 - C_1) = \frac{\varepsilon_s}{q_{ns}} \left( S_{nxx,f_1} e^{-q_{nf_1} z'} + B_1 + C_1 \right)$$

$$\Rightarrow \left( \frac{\varepsilon_s}{q_{ns}} - \frac{\varepsilon_{f_1}}{q_{nf_1}} \right) B_1 = -\left( \frac{\varepsilon_s}{q_{ns}} + \frac{\varepsilon_{f_1}}{q_{nf_1}} \right) C_1 - \left( \frac{\varepsilon_s}{q_{ns}} - \frac{\varepsilon_{f_1}}{q_{nf_1}} \right) S_{nxx,f_1} e^{-q_{nf_1} z'}$$

$$\Rightarrow M_{11}^{//} R_{n//}^{f_N a} C_N e^{-2q_{nf_N} z_N} + M_{12}^{//} C_N = -\frac{1}{R_{n//}^{sf_1}} \left( M_{21}^{//} R_{n//}^{f_N a} C_N e^{-2q_{nf_N} z_N} + M_{22}^{//} C_N - S_{nxx,f_1} e^{q_{nf_1} z'} \right) - S_{nxx,f_1} e^{-q_{nf_1} z'}$$

$$\Rightarrow \left( M_{11}^{//} R_{n//}^{sf_1} R_{n//}^{f_N a} e^{-2q_{nf_N} z_N} + M_{12}^{//} R_{n//}^{sf_1} + M_{21}^{//} R_{n//}^{f_N a} e^{-2q_{nf_N} z_N} + M_{22}^{//} \right) C_N = S_{nxx,f_1} e^{q_{nf_1} z'} - S_{nxx,f_1} R_{n//}^{sf_1} e^{-q_{nf_1} z'}$$

$$\Rightarrow C_N = S_{nxx,f_1} \frac{e^{q_{nf_1} z'} - R_{n//}^{sf_1} e^{-q_{nf_1} z'}}{M_{11}^{//} R_{n//}^{sf_1} R_{n//}^{f_N a} e^{-2q_{nf_N} z_N} + M_{12}^{//} R_{n//}^{sf_1} + M_{21}^{//} R_{n//}^{f_N a} e^{-2q_{nf_N} z_N} + M_{22}^{//}}$$

(394)

Hence we can get the other three coefficients as the following forms:

$$B_N = R_{n//}^{f_N a} S_{nxx,f_1} \frac{e^{q_{nf_1} z'} - R_{n//}^{sf_1} e^{-q_{nf_1} z'}}{M_{11}^{//} R_{n//}^{sf_1} R_{n//}^{f_N a} e^{-2q_{nf_N} z_N} + M_{12}^{//} R_{n//}^{sf_1} + M_{21}^{//} R_{n//}^{f_N a} e^{-2q_{nf_N} z_N} + M_{22}^{//}} e^{-2q_{nf_N} z_N}, \quad (395)$$

$$A = T_{n//}^{f_N a} S_{nxx,f_1} \frac{e^{q_{nf_1} z'} - R_{n//}^{sf_1} e^{-q_{nf_1} z'}}{M_{11}^{//} R_{n//}^{sf_1} R_{n//}^{f_N a} e^{-2q_{nf_N} z_N} + M_{12}^{//} R_{n//}^{sf_1} + M_{21}^{//} R_{n//}^{f_N a} e^{-2q_{nf_N} z_N} + M_{22}^{//}} e^{-q_{nf_N} z_N} e^{q_{na} z_N},$$

(396)

$$D = \left( M_{11}^{//} R_{n//}^{f_N a} e^{-2q_{nf_N} z_N} + M_{12}^{//} + M_{21}^{//} R_{n//}^{f_N a} e^{-2q_{nf_N} z_N} + M_{22}^{//} \right) C_N + S_{nxx,f_1} e^{-q_{nf_1} z'} - S_{nxx,f_1} e^{q_{nf_1} z'}$$

$$= S_{nxx,f_1} \frac{\left( M_{11}^{//} R_{n//}^{f_N a} e^{-2q_{nf_N} z_N} + M_{12}^{//} + M_{21}^{//} R_{n//}^{f_N a} e^{-2q_{nf_N} z_N} + M_{22}^{//} \right) \left( e^{q_{nf_1} z'} - R_{n//}^{sf_1} e^{-q_{nf_1} z'} \right)}{M_{11}^{//} R_{n//}^{sf_1} R_{n//}^{f_N a} e^{-2q_{nf_N} z_N} + M_{12}^{//} R_{n//}^{sf_1} + M_{21}^{//} R_{n//}^{f_N a} e^{-2q_{nf_N} z_N} + M_{22}^{//}} + S_{nxx,f_1} e^{-q_{nf_1} z'} - S_{nxx,f_1} e^{q_{nf_1} z'}$$

$$= S_{nxx,f_1} T_{n//}^{f_1 s} \frac{e^{q_{nf_1} z'} \left( M_{11}^{//} R_{n//}^{f_N a} e^{-2q_{nf_N} z_N} + M_{12}^{//} \right) + e^{-q_{nf_1} z'} \left( M_{21}^{//} R_{n//}^{f_N a} e^{-2q_{nf_N} z_N} + M_{22}^{//} \right)}{M_{11}^{//} R_{n//}^{sf_1} R_{n//}^{f_N a} e^{-2q_{nf_N} z_N} + M_{12}^{//} R_{n//}^{sf_1} + M_{21}^{//} R_{n//}^{f_N a} e^{-2q_{nf_N} z_N} + M_{22}^{//}}$$

(397)

Then we solve the coefficients in Eq. (361) completely. Next we discuss the component $\tilde{g}_{nzx}$. According to Eq. (177), we have:

$$\tilde{g}_{nzx} = -\frac{ik_n}{q_n^2} \partial_z \tilde{g}_{nxx}. \tag{398}$$

Next we will consider the relative position between the field ($z$) and the source ($z'$). First we consider $z' < 0$ (without loss of generality, we may set $z_0 = 0$), the corresponding different equation in regions $z > z_N$, $z_{j-1} < z < z_j, j = N, N-1, ..., 2, 1$ and $z < 0$ are:

$$\tilde{g}_{nzx} = \begin{cases} -\dfrac{ik_n}{q_{na}^2} A \partial_z e^{-q_{na}z}, & z > z_N \\ -\dfrac{ik_n}{q_{nf_j}^2} B_j \partial_z e^{q_{nf_j}z} - \dfrac{ik_n}{q_{nf_j}^2} C_j \partial_z e^{-q_{nf_j}z}, & z_{j-1} < z < z_j, j=1,2,\ldots,N \\ -\dfrac{ik_n}{q_{ns}^2} S_{nxx,s} \partial_z e^{-q_{ns}|z-z'|} - \dfrac{ik_n}{q_{ns}^2} D \partial_z e^{q_{ns}z}, & z < 0 \end{cases}, \quad (399)$$

$$= \begin{cases} A \dfrac{ik_n}{q_{na}} e^{-q_{na}z}, & z > z_N \\ -B_j \dfrac{ik_n}{q_{nf_j}} e^{q_{nf_j}z} + C_j \dfrac{ik_n}{q_{nf_j}} e^{-q_{nf_j}z}, & z_{j-1} < z < z_j, j=1,2,\ldots,N \\ S_{nxx,s} \dfrac{ik_n}{q_{ns}} \mathrm{sgn}(z-z') e^{-q_{ns}|z-z'|} - D \dfrac{ik_n}{q_{ns}} e^{q_{ns}z}, & z < 0 \end{cases}$$

where $S_{nxx,s} \equiv -\dfrac{q_{ns}}{2k_a^2}$ and the basic four coefficients $A$, $B_N$, $C_N$, $D$, $B_j$ and $C_j$ ($j=1,2,\ldots,N-1$) have already defined in Eq. (348), (347), (346), (349), (343), (344). These explicit forms are:

$$A = \frac{T_{n//}^{f_N a} e^{-q_{nf_N} z_N} e^{q_{na} z_N} T_{n//}^{sf_1} S_{nxx,s} e^{q_{ns} z'}}{M_{21}^{//} R_{n//}^{f_N a} e^{-2q_{nf_N} z_N} + M_{22}^{//} + R_{n//}^{sf_1}\left(M_{11}^{//} R_{n//}^{f_N a} e^{-2q_{nf_N} z_N} + M_{12}^{//}\right)}, \quad (400)$$

$$B_N = \frac{R_{n//}^{f_N a} T_{n//}^{sf_1} S_{nxx,s} e^{q_{ns} z'} e^{-2q_{nf_N} z_N}}{M_{21}^{//} R_{n//}^{f_N a} e^{-2q_{nf_N} z_N} + M_{22}^{//} + R_{n//}^{sf_1}\left(M_{11}^{//} R_{n//}^{f_N a} e^{-2q_{nf_N} z_N} + M_{12}^{//}\right)}, \quad (401)$$

$$C_N = \frac{T_{n//}^{sf_1} S_{nxx,s} e^{q_{ns} z'}}{M_{21}^{//} R_{n//}^{f_N a} e^{-2q_{nf_N} z_N} + M_{22}^{//} + R_{n//}^{sf_1}\left(M_{11}^{//} R_{n//}^{f_N a} e^{-2q_{nf_N} z_N} + M_{12}^{//}\right)}, \quad (402)$$

$$D = S_{nxx,s} e^{q_{ns} z'} \frac{M_{11}^{//} R_{n//}^{f_N a} e^{-2q_{nf_N} z_N} + M_{12}^{//} + R_{n//}^{sf_1}\left(M_{21}^{//} R_{n//}^{f_N a} e^{-2q_{nf_N} z_N} + M_{22}^{//}\right)}{M_{21}^{//} R_{n//}^{f_N a} e^{-2q_{nf_N} z_N} + M_{22}^{//} + R_{n//}^{sf_1}\left(M_{11}^{//} R_{n//}^{f_N a} e^{-2q_{nf_N} z_N} + M_{12}^{//}\right)}, \quad (403)$$

$$\begin{pmatrix} B_1 \\ C_1 \end{pmatrix} = \mathrm{T}_1^{//} \mathrm{T}_2^{//} \bullet\bullet\bullet \mathrm{T}_{N-1}^{//} \begin{pmatrix} B_N \\ C_N \end{pmatrix} \equiv \begin{pmatrix} M_{11}^{//} & M_{12}^{//} \\ M_{21}^{//} & M_{22}^{//} \end{pmatrix}\begin{pmatrix} B_N \\ C_N \end{pmatrix} = \begin{pmatrix} M_{11}^{//} B_N + M_{12}^{//} C_N \\ M_{21}^{//} B_N + M_{22}^{//} C_N \end{pmatrix}, \quad (404)$$

$$\begin{pmatrix} B_j \\ C_j \end{pmatrix} = \mathrm{T}_j^{//} \mathrm{T}_{j+1}^{//} \bullet\bullet\bullet \mathrm{T}_{N-1}^{//} \begin{pmatrix} B_N \\ C_N \end{pmatrix} \equiv \begin{pmatrix} M_{j,11}^{//} & M_{j,12}^{//} \\ M_{j,21}^{//} & M_{j,22}^{//} \end{pmatrix}\begin{pmatrix} B_N \\ C_N \end{pmatrix} = \begin{pmatrix} M_{j,11}^{//} B_N + M_{j,12}^{//} C_N \\ M_{j,21}^{//} B_N + M_{j,22}^{//} C_N \end{pmatrix}, j=2,3,\ldots,N-1$$

. (405)

Next we consider $z' > z_N$, the corresponding different equation in regions $z > z_N$, $z_{j-1} < z < z_j, j=N, N-1,\ldots,2,1$ and $z < 0$ are:

$$\tilde{g}_{nzx} = \begin{cases} -\dfrac{ik_n}{q_{na}^2} S_{nxx,a} \partial_z e^{-q_{na}|z-z'|} - \dfrac{ik_n}{q_{na}^2} A \partial_z e^{-q_{na}z}, z > z_N \\ -\dfrac{ik_n}{q_{nf_j}^2} B_j \partial_z e^{q_{nf_j}z} - \dfrac{ik_n}{q_{nf_j}^2} C_j \partial_z e^{-q_{nf_j}z}, z_{j-1} < z < z_j, j=1,2,\ldots,N \\ -\dfrac{ik_n}{q_{ns}^2} D \partial_z e^{q_{ns}z}, z < 0 \end{cases}$$

$$= \begin{cases} S_{nxx,a} \dfrac{ik_n}{q_{na}} \operatorname{sgn}(z-z') e^{-q_{na}|z-z'|} + A \dfrac{ik_n}{q_{na}} e^{-q_{na}z}, z > z_N \\ -B_j \dfrac{ik_n}{q_{nf_j}} e^{q_{nf_j}z} + C_j \dfrac{ik_n}{q_{nf_j}} e^{-q_{nf_j}z}, z_{j-1} < z < z_j, j=1,2,\ldots,N \\ -D \dfrac{ik_n}{q_{ns}} e^{q_{ns}z}, z < 0 \end{cases}, \quad (406)$$

where $S_{nxx,a} \equiv -\dfrac{q_{na}}{2k_a^2}$ and the basic four coefficients $A$, $B_N$, $C_N$ and $D$ have already defined in Eq. (358), (357), (356), (359). These explicit forms are:

$$A = -S_{nxx,a} e^{-q_{na}z'} e^{2q_{na}z_N} \frac{R_{n//}^{sf_1}\left(M_{11}^{//} + M_{21}^{//}\right) e^{-2q_{nf_N}z_N} + R_{n//}^{f_N a}\left(R_{n//}^{sf_1} M_{12}^{//} + M_{22}^{//}\right)}{R_{n//}^{sf_1}\left(M_{11}^{//} R_{n//}^{f_N a} e^{-2q_{nf_N}z_N} + M_{12}^{//}\right) + \left(M_{21}^{//} R_{n//}^{f_N a} e^{-2q_{nf_N}z_N} + M_{22}^{//}\right)}, \quad (407)$$

$$B_N = \frac{S_{nxx,a} e^{-q_{na}z'} e^{q_{na}z_N} e^{-q_{nf_N}z_N} T_{n//}^{af_N}\left(R_{n//}^{sf_1} M_{12}^{//} + M_{22}^{//}\right)}{R_{n//}^{sf_1}\left(M_{11}^{//} R_{n//}^{f_N a} e^{-2q_{nf_N}z_N} + M_{12}^{//}\right) + \left(M_{21}^{//} R_{n//}^{f_N a} e^{-2q_{nf_N}z_N} + M_{22}^{//}\right)}, \quad (408)$$

$$C_N = -\frac{S_{nxx,a} T_{n//}^{af_N}\left(R_{n//}^{sf_1} M_{11}^{//} + M_{21}^{//}\right) e^{-q_{na}z'} e^{q_{na}z_N} e^{-q_{nf_N}z_N}}{R_{n//}^{sf_1}\left(M_{11}^{//} R_{n//}^{f_N a} e^{-2q_{nf_N}z_N} + M_{12}^{//}\right) + \left(M_{21}^{//} R_{n//}^{f_N a} e^{-2q_{nf_N}z_N} + M_{22}^{//}\right)}. \quad (409)$$

$$D = \frac{S_{nxx,a} T_{n//}^{af_N} T_{n//}^{f_1 s}\left(M_{22}^{//} M_{11}^{//} - M_{21}^{//} M_{12}^{//}\right) e^{-q_{na}z'} e^{q_{na}z_N} e^{-q_{nf_N}z_N}}{R_{n//}^{sf_1}\left(M_{11}^{//} R_{n//}^{f_N a} e^{-2q_{nf_N}z_N} + M_{12}^{//}\right) + \left(M_{21}^{//} R_{n//}^{f_N a} e^{-2q_{nf_N}z_N} + M_{22}^{//}\right)}. \quad (410)$$

The coefficients $B_j$ and $C_j$ ($j=1,2,\ldots,N-1$) are shown in Eq. (404) and Eq. (405). Next we consider the location of a point source is inside the film layer, that is $z_{k-1} < z' < z_k, k=1,2,\ldots,N$. Fixed the index $k$, the corresponding different equation in regions $z > z_N$, $z_{j-1} < z < z_j, j = N, N-1, \ldots, 2, 1$ and $z < 0$ are:

$$\tilde{g}_{nzx} = \begin{cases} -\dfrac{ik_n}{q_{na}^2} A \partial_z e^{-q_{na}z}, \; z > z_N \\ -\dfrac{ik_n}{q_{nf_k}^2} S_{nxx,f_k} \partial_z e^{-q_{nf_k}|z-z'|} \delta_{jk} - \dfrac{ik_n}{q_{nf_j}^2} B_j \partial_z e^{q_{nf_j}z} - \dfrac{ik_n}{q_{nf_j}^2} C_j \partial_z e^{-q_{nf_j}z}, \; z_{j-1} < z < z_j, \; j=1,2,\ldots,N \\ -\dfrac{ik_n}{q_{ns}^2} D \partial_z e^{q_{ns}z}, \; z < 0 \end{cases}$$

$$= \begin{cases} \dfrac{ik_n}{q_{na}} A e^{-q_{na}z}, \; z > z_N \\ S_{nxx,f_k} \dfrac{ik_n}{q_{nf_k}} \mathrm{sgn}(z-z') e^{-q_{nf_k}|z-z'|} \delta_{jk} - \dfrac{ik_n}{q_{nf_j}} B_j e^{q_{nf_j}z} + \dfrac{ik_n}{q_{nf_j}} C_j e^{-q_{nf_j}z}, \; z_{j-1} < z < z_j, \; j=1,2,\ldots,N \\ -\dfrac{ik_n}{q_{ns}} D e^{q_{ns}z}, \; z < 0 \end{cases}$$

, (411)

where $S_{nxx,f_k} \equiv -\dfrac{q_{nf_k}}{2k_{f_k}^2}$. The coefficients $A$, $B_N$, $C_N$, $D$, $B_j$ and $C_j$

($j = 1, 2, \ldots, N-1$) in Eq. (411) have already defined in Eq. (375), (374), (373), (376), (370), (371), (369) for $k = N$, Eq. (387), (386), (385), (388), (380), (381), (382), (383) for $k = 2, 3, \ldots, N-1$ and Eq. (396), (395), (394), (397), (391), (392) for $k = 1$, respectively. These explicit forms are:

<1> If $k = N$:

$$A = T_{n//}^{f_N a} S_{nxx,f_N} e^{-q_{nf_N} z_N} e^{q_{na} z_N} \frac{-\left(M_{21}^{//} - R_{n//}^{sf_1} M_{11}^{//}\right) e^{-q_{nf_N} z'} + \left(M_{22}^{//} - R_{n//}^{sf_1} M_{12}^{//}\right) e^{q_{nf_N} z'}}{R_{n//}^{f_N a} \left(M_{21}^{//} - R_{n//}^{sf_1} M_{11}^{//}\right) e^{-2q_{nf_N} z_N} + M_{22}^{//} - R_{n//}^{sf_1} M_{12}^{//}}, \quad (412)$$

$$B_N = R_{n//}^{f_N a} S_{nxx,f_N} e^{-2q_{nf_N} z_N} \frac{-\left(M_{21}^{//} - R_{n//}^{sf_1} M_{11}^{//}\right) e^{-q_{nf_N} z'} + \left(M_{22}^{//} - R_{n//}^{sf_1} M_{12}^{//}\right) e^{q_{nf_N} z'}}{R_{n//}^{f_N a} \left(M_{21}^{//} - R_{n//}^{sf_1} M_{11}^{//}\right) e^{-2q_{nf_N} z_N} + M_{22}^{//} - R_{n//}^{sf_1} M_{12}^{//}}, \quad (413)$$

$$C_N = S_{nxx,f_N} \frac{\left(R_{n//}^{sf_1} M_{11}^{//} - M_{21}^{//}\right)\left(R_{n//}^{f_N a} e^{q_{nf_N} z'} e^{-2q_{nf_N} z_N} + e^{-q_{nf_N} z'}\right)}{R_{n//}^{f_N a} M_{21}^{//} e^{-2q_{nf_N} z_N} + M_{22}^{//} - R_{n//}^{sf_1} R_{n//}^{f_N a} M_{11}^{//} e^{-2q_{nf_N} z_N} - R_{n//}^{sf_1} M_{12}^{//}}, \quad (414)$$

$$D = S_{nxx,f_N} T_{n//}^{sf_1} \frac{\left(M_{22}^{//} M_{11}^{//} - M_{12}^{//} M_{21}^{//}\right) R_{n//}^{f_N a} e^{q_{nf_N} z'} e^{-2q_{nf_N} z_N} + \left(M_{22}^{//} M_{11}^{//} - M_{12}^{//} M_{21}^{//}\right) e^{-q_{nf_N} z'}}{R_{n//}^{f_N a} M_{21}^{//} e^{-2q_{nf_N} z_N} + M_{22}^{//} - R_{n//}^{sf_1} R_{n//}^{f_N a} M_{11}^{//} e^{-2q_{nf_N} z_N} - R_{n//}^{sf_1} M_{12}^{//}}, (415)$$

$$\begin{pmatrix} B_1 \\ C_1 \end{pmatrix} = T_1^{//} T_2^{//} \bullet\bullet\bullet T_{N-2}^{//} \begin{pmatrix} B_{N-1} \\ C_{N-1} \end{pmatrix} = T_1^{//} T_2^{//} \bullet\bullet\bullet T_{N-2}^{//} T_{N-1}^{//} \begin{pmatrix} B_N \\ C_N \end{pmatrix} + T_1^{//} T_2^{//} \bullet\bullet\bullet T_{N-2}^{//} T_{N-1}^{//} \begin{pmatrix} 1 \\ 0 \end{pmatrix} S_{nxx,f_N} e^{-q_{nf_N} z'}$$

$$\equiv \begin{pmatrix} M_{11}^{//} & M_{12}^{//} \\ M_{21}^{//} & M_{22}^{//} \end{pmatrix} \begin{pmatrix} B_N \\ C_N \end{pmatrix} + \begin{pmatrix} M_{11}^{//} & M_{12}^{//} \\ M_{21}^{//} & M_{22}^{//} \end{pmatrix} \begin{pmatrix} 1 \\ 0 \end{pmatrix} S_{nxx,f_N} e^{-q_{nf_N} z'}$$

$$= \begin{pmatrix} M_{11}^{//} B_N + M_{12}^{//} C_N \\ M_{21}^{//} B_N + M_{22}^{//} C_N \end{pmatrix} + \begin{pmatrix} M_{11}^{//} \\ M_{21}^{//} \end{pmatrix} S_{nxx,f_N} e^{-q_{nf_N} z'}$$

, (416)

$$\begin{pmatrix} B_j \\ C_j \end{pmatrix} = T_j^{//} \bullet\bullet\bullet T_{N-2}^{//} T_{N-1}^{//} \begin{pmatrix} B_N \\ C_N \end{pmatrix} + T_j^{//} \bullet\bullet\bullet T_{N-2}^{//} T_{N-1}^{//} \begin{pmatrix} 1 \\ 0 \end{pmatrix} S_{nxx,f_N} e^{-q_{nf_N} z'}$$

$$\equiv \begin{pmatrix} M_{j,11}^{//} & M_{j,12}^{//} \\ M_{j,21}^{//} & M_{j,22}^{//} \end{pmatrix} \begin{pmatrix} B_N \\ C_N \end{pmatrix} + \begin{pmatrix} M_{j,11}^{//} & M_{j,12}^{//} \\ M_{j,21}^{//} & M_{j,22}^{//} \end{pmatrix} \begin{pmatrix} 1 \\ 0 \end{pmatrix} S_{nxx,f_N} e^{-q_{nf_N} z'} \quad , \quad (417)$$

$$= \begin{pmatrix} M_{j,11}^{//} B_N + M_{j,12}^{//} C_N \\ M_{j,21}^{//} B_N + M_{j,22}^{//} C_N \end{pmatrix} + \begin{pmatrix} M_{j,11}^{//} \\ M_{j,21}^{//} \end{pmatrix} S_{nxx,f_N} e^{-q_{nf_N} z'}, \ j = 2,3,...,N-2$$

$$\begin{pmatrix} B_{N-1} \\ C_{N-1} \end{pmatrix} = T_{N-1}^{//} \begin{pmatrix} B_N \\ C_N \end{pmatrix} + T_{N-1}^{//} \begin{pmatrix} 1 \\ 0 \end{pmatrix} S_{nxx,f_N} e^{-q_{nf_N} z'}. \quad (418)$$

<2> If $k = 2,3,...,N-1$:

$$A = T_{n//}^{f_N a} S_{nxx,f_k} e^{-q_{nf_N} z_N} e^{q_{na} z_N} \frac{R_{n//}^{sf_1} \left( M_{12,k-1}^{//} e^{q_{nf_k} z'} - M_{11,k-1}^{//} e^{-q_{nf_k} z'} \right) + \left( M_{22,k-1}^{//} e^{q_{nf_k} z'} - M_{21,k-1}^{//} e^{-q_{nf_k} z'} \right)}{R_{n//}^{sf_1} \left( M_{11}^{//} R_{n//}^{f_N a} e^{-2q_{nf_N} z_N} + M_{12}^{//} \right) + \left( M_{21}^{//} R_{n//}^{f_N a} e^{-2q_{nf_N} z_N} + M_{22}^{//} \right)}$$

(419)

$$B_N = R_{n//}^{f_N a} S_{nxx,f_k} e^{-2q_{nf_N} z_N} \frac{R_{n//}^{sf_1} \left( M_{12,k-1}^{//} e^{q_{nf_k} z'} - M_{11,k-1}^{//} e^{-q_{nf_k} z'} \right) + \left( M_{22,k-1}^{//} e^{q_{nf_k} z'} - M_{21,k-1}^{//} e^{-q_{nf_k} z'} \right)}{R_{n//}^{sf_1} \left( M_{11}^{//} R_{n//}^{f_N a} e^{-2q_{nf_N} z_N} + M_{12}^{//} \right) + \left( M_{21}^{//} R_{n//}^{f_N a} e^{-2q_{nf_N} z_N} + M_{22}^{//} \right)},$$

(420)

$$C_N = S_{nxx,f_k} \frac{R_{n//}^{sf_1} \left( M_{12,k-1}^{//} e^{q_{nf_k} z'} - M_{11,k-1}^{//} e^{-q_{nf_k} z'} \right) + \left( M_{22,k-1}^{//} e^{q_{nf_k} z'} - M_{21,k-1}^{//} e^{-q_{nf_k} z'} \right)}{R_{n//}^{sf_1} \left( M_{11}^{//} R_{n//}^{f_N a} e^{-2q_{nf_N} z_N} + M_{12}^{//} \right) + \left( M_{21}^{//} R_{n//}^{f_N a} e^{-2q_{nf_N} z_N} + M_{22}^{//} \right)}, \quad (421)$$

$$D = \frac{R_{n//}^{sf_1}\left(M_{12,k-1}^{//}e^{q_{nf_k}z'} - M_{11,k-1}^{//}e^{-q_{nf_k}z'}\right) + \left(M_{22,k-1}^{//}e^{q_{nf_k}z'} - M_{21,k-1}^{//}e^{-q_{nf_k}z'}\right)}{R_{n//}^{sf_1}\left(M_{11}^{//}R_{n//}^{f_N a}e^{-2q_{nf_N}z_N} + M_{12}^{//}\right) + \left(M_{21}^{//}R_{n//}^{f_N a}e^{-2q_{nf_N}z_N} + M_{22}^{//}\right)}M_{11}^{//}R_{n//}^{f_N a}e^{-2q_{nf_N}z_N}S_{nxx,f_k}$$

$$+ \frac{R_{n//}^{sf_1}\left(M_{12,k-1}^{//}e^{q_{nf_k}z'} - M_{11,k-1}^{//}e^{-q_{nf_k}z'}\right) + \left(M_{22,k-1}^{//}e^{q_{nf_k}z'} - M_{21,k-1}^{//}e^{-q_{nf_k}z'}\right)}{R_{n//}^{sf_1}\left(M_{11}^{//}R_{n//}^{f_N a}e^{-2q_{nf_N}z_N} + M_{12}^{//}\right) + \left(M_{21}^{//}R_{n//}^{f_N a}e^{-2q_{nf_N}z_N} + M_{22}^{//}\right)}M_{12}^{//}S_{nxx,f_k}$$

$$+ \frac{R_{n//}^{sf_1}\left(M_{12,k-1}^{//}e^{q_{nf_k}z'} - M_{11,k-1}^{//}e^{-q_{nf_k}z'}\right) + \left(M_{22,k-1}^{//}e^{q_{nf_k}z'} - M_{21,k-1}^{//}e^{-q_{nf_k}z'}\right)}{R_{n//}^{sf_1}\left(M_{11}^{//}R_{n//}^{f_N a}e^{-2q_{nf_N}z_N} + M_{12}^{//}\right) + \left(M_{21}^{//}R_{n//}^{f_N a}e^{-2q_{nf_N}z_N} + M_{22}^{//}\right)}M_{21}^{//}R_{n//}^{f_N a}e^{-2q_{nf_N}z_N}S_{nxx,f_k}$$

$$+ \frac{R_{n//}^{sf_1}\left(M_{12,k-1}^{//}e^{q_{nf_k}z'} - M_{11,k-1}^{//}e^{-q_{nf_k}z'}\right) + \left(M_{22,k-1}^{//}e^{q_{nf_k}z'} - M_{21,k-1}^{//}e^{-q_{nf_k}z'}\right)}{R_{n//}^{sf_1}\left(M_{11}^{//}R_{n//}^{f_N a}e^{-2q_{nf_N}z_N} + M_{12}^{//}\right) + \left(M_{21}^{//}R_{n//}^{f_N a}e^{-2q_{nf_N}z_N} + M_{22}^{//}\right)}M_{22}^{//}S_{nxx,f_k}$$

$$- M_{12,k-1}^{//}S_{nxx,f_k}e^{q_{nf_k}z'} + M_{11,k-1}^{//}S_{nxx,f_k}e^{-q_{nf_k}z'} - M_{22,k-1}^{//}S_{nxx,f_k}e^{q_{nf_k}z'} + M_{21,k-1}^{//}S_{nxx,f_k}e^{-q_{nf_k}z'}$$

, (422)

$$\begin{pmatrix}B_1\\C_1\end{pmatrix} = T_1^{//}T_2^{//}\bullet\bullet T_{k-2}^{//}\begin{pmatrix}B_{k-1}\\C_{k-1}\end{pmatrix} = T_1^{//}T_2^{//}\bullet\bullet T_{k-2}^{//}T_{k-1}^{//}\begin{pmatrix}B_k\\C_k\end{pmatrix} + T_1^{//}T_2^{//}\bullet\bullet T_{k-2}^{//}T_{k-1}^{//}\begin{pmatrix}1\\0\end{pmatrix}S_{nxx,f_k}e^{-q_{nf_k}z'}$$

$$= T_1^{//}T_2^{//}\bullet\bullet T_{k-2}^{//}T_{k-1}^{//}T_k^{//}\begin{pmatrix}B_{k+1}\\C_{k+1}\end{pmatrix} - T_1^{//}T_2^{//}\bullet\bullet T_{k-2}^{//}T_{k-1}^{//}\begin{pmatrix}0\\1\end{pmatrix}S_{nxx,f_k}e^{q_{nf_k}z'} + T_1^{//}T_2^{//}\bullet\bullet T_{k-2}^{//}T_{k-1}^{//}\begin{pmatrix}1\\0\end{pmatrix}S_{nxx,f_k}e^{-q_{nf_k}z'}$$

$$\equiv \begin{pmatrix}M_{11}^{//} & M_{12}^{//}\\M_{21}^{//} & M_{22}^{//}\end{pmatrix}\begin{pmatrix}B_N\\C_N\end{pmatrix} - \begin{pmatrix}M_{11,k-1}^{//} & M_{12,k-1}^{//}\\M_{21,k-1}^{//} & M_{22,k-1}^{//}\end{pmatrix}\begin{pmatrix}0\\1\end{pmatrix}S_{nxx,f_k}e^{q_{nf_k}z'} + \begin{pmatrix}M_{11,k-1}^{//} & M_{12,k-1}^{//}\\M_{21,k-1}^{//} & M_{22,k-1}^{//}\end{pmatrix}\begin{pmatrix}1\\0\end{pmatrix}S_{nxx,f_k}e^{-q_{nf_k}z'}$$

$$= \begin{pmatrix}M_{11}^{//}B_N + M_{12}^{//}C_N\\M_{21}^{//}B_N + M_{22}^{//}C_N\end{pmatrix} - \begin{pmatrix}M_{12,k-1}^{//}\\M_{22,k-1}^{//}\end{pmatrix}S_{nxx,f_k}e^{q_{nf_k}z'} + \begin{pmatrix}M_{11,k-1}^{//}\\M_{21,k-1}^{//}\end{pmatrix}S_{nxx,f_k}e^{-q_{nf_k}z'}$$

, (423)

$$\begin{pmatrix}B_j\\C_j\end{pmatrix} = T_j^{//}\bullet\bullet T_{k-2}^{//}\begin{pmatrix}B_{k-1}\\C_{k-1}\end{pmatrix} = T_j^{//}\bullet\bullet T_{k-2}^{//}T_{k-1}^{//}\begin{pmatrix}B_k\\C_k\end{pmatrix} + T_j^{//}\bullet\bullet T_{k-2}^{//}T_{k-1}^{//}\begin{pmatrix}1\\0\end{pmatrix}S_{nxx,f_k}e^{-q_{nf_k}z'}$$

$$= T_j^{//}\bullet\bullet T_{k-2}^{//}T_{k-1}^{//}T_k^{//}\begin{pmatrix}B_{k+1}\\C_{k+1}\end{pmatrix} - T_j^{//}\bullet\bullet T_{k-2}^{//}T_{k-1}^{//}\begin{pmatrix}0\\1\end{pmatrix}S_{nxx,f_k}e^{q_{nf_k}z'} + T_j^{//}\bullet\bullet T_{k-2}^{//}T_{k-1}^{//}\begin{pmatrix}1\\0\end{pmatrix}S_{nxx,f_k}e^{-q_{nf_k}z'}$$

$$\equiv \begin{pmatrix}M_{j,11}^{//} & M_{j,12}^{//}\\M_{j,21}^{//} & M_{j,22}^{//}\end{pmatrix}\begin{pmatrix}B_N\\C_N\end{pmatrix} - \begin{pmatrix}M_{j,11,k-1}^{//} & M_{j,12,k-1}^{//}\\M_{j,21,k-1}^{//} & M_{j,22,k-1}^{//}\end{pmatrix}\begin{pmatrix}0\\1\end{pmatrix}S_{nxx,f_k}e^{q_{nf_k}z'} + \begin{pmatrix}M_{j,11,k-1}^{//} & M_{j,12,k-1}^{//}\\M_{j,21,k-1}^{//} & M_{j,22,k-1}^{//}\end{pmatrix}\begin{pmatrix}1\\0\end{pmatrix}S_{nxx,f_k}e^{-q_{nf_k}z'}$$

$$= \begin{pmatrix}M_{j,11}^{//}B_N + M_{j,12}^{//}C_N\\M_{j,21}^{//}B_N + M_{j,22}^{//}C_N\end{pmatrix} - \begin{pmatrix}M_{j,12,k-1}^{//}\\M_{j,22,k-1}^{//}\end{pmatrix}S_{nxx,f_k}e^{q_{nf_k}z'} + \begin{pmatrix}M_{j,11,k-1}^{//}\\M_{j,21,k-1}^{//}\end{pmatrix}S_{nxx,f_k}e^{-q_{nf_k}z'}, \quad j = 2,3,...,N-1, j \leq k-1$$

, (424)

$$\begin{pmatrix} B_k \\ C_k \end{pmatrix} = T_k^{//} \begin{pmatrix} B_{k+1} \\ C_{k+1} \end{pmatrix} - \begin{pmatrix} 0 \\ 1 \end{pmatrix} S_{nxx,f_k} e^{q_{nf_k} z'} = T_k^{//} \bullet \bullet T_{N-1}^{//} \begin{pmatrix} B_N \\ C_N \end{pmatrix} - \begin{pmatrix} 0 \\ 1 \end{pmatrix} S_{nxx,f_k} e^{q_{nf_k} z'}$$
$$\equiv \begin{pmatrix} M_{k,11}^{//} & M_{k,12}^{//} \\ M_{k,21}^{//} & M_{k,22}^{//} \end{pmatrix} \begin{pmatrix} B_N \\ C_N \end{pmatrix} - \begin{pmatrix} 0 \\ 1 \end{pmatrix} S_{nxx,f_k} e^{q_{nf_k} z'}, \quad (425)$$
$$= \begin{pmatrix} M_{k,11}^{//} B_N + M_{k,12}^{//} C_N \\ M_{k,21}^{//} B_N + M_{k,22}^{//} C_N \end{pmatrix} - \begin{pmatrix} 0 \\ 1 \end{pmatrix} S_{nxx,f_k} e^{q_{nf_k} z'}, \quad j = 2,3,...,N-1, \ j = k$$

$$\begin{pmatrix} B_j \\ C_j \end{pmatrix} = T_j^{//} \begin{pmatrix} B_{j+1} \\ C_{j+1} \end{pmatrix} = T_j^{//} T_{j+1}^{//} \bullet \bullet \bullet T_{N-1}^{//} \begin{pmatrix} B_N \\ C_N \end{pmatrix} \equiv \begin{pmatrix} M_{j,11}^{//} & M_{j,12}^{//} \\ M_{j,21}^{//} & M_{j,22}^{//} \end{pmatrix} \begin{pmatrix} B_N \\ C_N \end{pmatrix}$$
$$= \begin{pmatrix} M_{j,11}^{//} B_N + M_{j,12}^{//} C_N \\ M_{j,21}^{//} B_N + M_{j,22}^{//} C_N \end{pmatrix}, \quad j = 2,3,...,N-1, \ j \geq k+1 \quad (426)$$

<3> If $k = 1$:

$$A = T_{n//}^{f_N a} S_{nxx,f_1} \frac{e^{q_{nf_1} z'} - R_{n//}^{sf_1} e^{-q_{nf_1} z'}}{M_{11}^{//} R_{n//}^{sf_1} R_{n//}^{f_N a} e^{-2q_{nf_N} z_N} + M_{12}^{//} R_{n//}^{sf_1} + M_{21}^{//} R_{n//}^{f_N a} e^{-2q_{nf_N} z_N} + M_{22}^{//}} e^{-q_{nf_N} z_N} e^{q_{na} z_N}, \quad (427)$$

$$B_N = R_{n//}^{f_N a} S_{nxx,f_1} \frac{e^{q_{nf_1} z'} - R_{n//}^{sf_1} e^{-q_{nf_1} z'}}{M_{11}^{//} R_{n//}^{sf_1} R_{n//}^{f_N a} e^{-2q_{nf_N} z_N} + M_{12}^{//} R_{n//}^{sf_1} + M_{21}^{//} R_{n//}^{f_N a} e^{-2q_{nf_N} z_N} + M_{22}^{//}} e^{-2q_{nf_N} z_N}, \quad (428)$$

$$C_N = S_{nxx,f_1} \frac{e^{q_{nf_1} z'} - R_{n//}^{sf_1} e^{-q_{nf_1} z'}}{M_{11}^{//} R_{n//}^{sf_1} R_{n//}^{f_N a} e^{-2q_{nf_N} z_N} + M_{12}^{//} R_{n//}^{sf_1} + M_{21}^{//} R_{n//}^{f_N a} e^{-2q_{nf_N} z_N} + M_{22}^{//}}, \quad (429)$$

$$D = S_{nxx,f_1} T_{n//}^{f_1 s} \frac{e^{q_{nf_1} z'} \left( M_{11}^{//} R_{n//}^{f_N a} e^{-2q_{nf_N} z_N} + M_{12}^{//} \right) + e^{-q_{nf_1} z'} \left( M_{21}^{//} R_{n//}^{f_N a} e^{-2q_{nf_N} z_N} + M_{22}^{//} \right)}{M_{11}^{//} R_{n//}^{sf_1} R_{n//}^{f_N a} e^{-2q_{nf_N} z_N} + M_{12}^{//} R_{n//}^{sf_1} + M_{21}^{//} R_{n//}^{f_N a} e^{-2q_{nf_N} z_N} + M_{22}^{//}}, \quad (430)$$

$$\begin{pmatrix} B_1 \\ C_1 \end{pmatrix} = T_1^{//} \begin{pmatrix} B_2 \\ C_2 \end{pmatrix} - \begin{pmatrix} 0 \\ 1 \end{pmatrix} S_{nxx,f_1} e^{q_{nf_1} z'} = \begin{pmatrix} M_{11}^{//} & M_{12}^{//} \\ M_{21}^{//} & M_{22}^{//} \end{pmatrix} \begin{pmatrix} B_N \\ C_N \end{pmatrix} - \begin{pmatrix} 0 \\ 1 \end{pmatrix} S_{nxx,f_1} e^{q_{nf_1} z'}$$
$$= \begin{pmatrix} M_{11}^{//} B_N + M_{12}^{//} C_N \\ M_{21}^{//} B_N + M_{22}^{//} C_N \end{pmatrix} - \begin{pmatrix} 0 \\ 1 \end{pmatrix} S_{nxx,f_1} e^{q_{nf_1} z'}, \quad (431)$$

$$\begin{pmatrix} B_j \\ C_j \end{pmatrix} = T_j^{//} \bullet \bullet \bullet T_{N-1}^{//} \begin{pmatrix} B_N \\ C_N \end{pmatrix} = \begin{pmatrix} M_{j,11}^{//} & M_{j,12}^{//} \\ M_{j,21}^{//} & M_{j,22}^{//} \end{pmatrix} \begin{pmatrix} B_N \\ C_N \end{pmatrix} = \begin{pmatrix} M_{j,11}^{//} B_N + M_{j,11}^{//} C_N \\ M_{j,21}^{//} B_N + M_{j,22}^{//} C_N \end{pmatrix}, \quad j = 2,...,N-1$$
. (432)

Finally we discuss the component $\tilde{g}_{nzz}$. According to Eq. (193), we have:

$$\tilde{g}_{nzz} = \frac{1}{q_n^2} \delta(z - z') - \frac{ik_n}{q_n^2} \partial_z \tilde{g}_{nxz}. \quad (433)$$

Next we will consider the relative position between the field ($z$) and the source ($z'$). First we consider $z' < 0$ (without loss of generality, we may set $z_0 = 0$), the corresponding different equation in regions $z > z_N$, $z_{j-1} < z < z_j, j = N, N-1, \ldots, 2, 1$ and $z < 0$ are:

$$\tilde{g}_{nzz} = \begin{cases} -\dfrac{ik_n}{q_{na}^2} A \partial_z e^{-q_{na}z}, z > z_N \\ -\dfrac{ik_n}{q_{nf_j}^2} B_j \partial_z e^{q_{nf_j}z} - \dfrac{ik_n}{q_{nf_j}^2} C_j \partial_z e^{-q_{nf_j}z}, z_{j-1} < z < z_j, j = 1, 2, \ldots, N \\ \dfrac{1}{q_{ns}^2} \delta(z-z') - \dfrac{ik_n}{q_{ns}^2} S_{nxz,s} \partial_z \left[ \text{sgn}(z-z') e^{-q_{ns}|z-z'|} \right] - \dfrac{ik_n}{q_{ns}^2} D \partial_z e^{q_{ns}z}, z < 0 \end{cases}$$

$$= \begin{cases} \dfrac{ik_n}{q_{na}} A e^{-q_{na}z}, z > z_N \\ -\dfrac{ik_n}{q_{nf_j}} B_j e^{q_{nf}z} + \dfrac{ik_n}{q_{nf_j}} C_j e^{-q_{nf}z}, z_{j-1} < z < z_j, j = 1, 2, \ldots, N \\ \dfrac{1}{q_{ns}^2} \delta(z-z') - \dfrac{ik_n}{q_{ns}^2} S_{nxz,s} \left[ 2\delta(z-z') e^{-q_{ns}|z-z'|} - q_{ns} e^{-q_{ns}|z-z'|} \right] - \dfrac{ik_n}{q_{ns}} D e^{q_{ns}z}, z < 0 \end{cases}$$

$$= \begin{cases} \dfrac{ik_n}{q_{na}} A e^{-q_{na}z}, z > z_N \\ -\dfrac{ik_n}{q_{nf_j}} B_j e^{q_{nf}z} + \dfrac{ik_n}{q_{nf_j}} C_j e^{-q_{nf}z}, z_{j-1} < z < z_j, j = 1, 2, \ldots, N \\ -\dfrac{1}{k_s^2} \delta(z-z') + \dfrac{ik_n}{q_{ns}} S_{nxz,s} e^{-q_{ns}|z-z'|} - \dfrac{ik_n}{q_{ns}} D e^{q_{ns}z}, z < 0 \end{cases} \tag{434}$$

where $S_{nxz,s} = -\dfrac{ik_n}{2k_s^2}$ and the basic four coefficients $A$, $B_N$, $C_N$ and $D$ have already defined in Eq. (223), (222), (221), (224). These following explicit forms are:

$$A = \dfrac{-T_{n//}^{f_N a} T_{n//}^{sf_1} S_{nxz,s} e^{q_{ns}z'} e^{-q_{nf_N} z_N} e^{q_{na} z_N}}{R_{n//}^{sf_1} \left( M_{11}^{//} R_{n//}^{f_N a} e^{-2q_{nf_N} z_N} + M_{12}^{//} \right) + \left( M_{21}^{//} R_{n//}^{f_N a} e^{-2q_{nf_N} z_N} + M_{22}^{//} \right)}, \tag{435}$$

$$B_N = \dfrac{-R_{n//}^{f_N a} T_{n//}^{sf_1} S_{nxz,s} e^{q_{ns}z'} e^{-2q_{nf_N} z_N}}{R_{n//}^{sf_1} \left( M_{11}^{//} R_{n//}^{f_N a} e^{-2q_{nf_N} z_N} + M_{12}^{//} \right) + \left( M_{21}^{//} R_{n//}^{f_N a} e^{-2q_{nf_N} z_N} + M_{22}^{//} \right)}, \tag{436}$$

$$C_N = \dfrac{-T_{n//}^{sf_1} S_{nxz,s} e^{q_{ns}z'}}{R_{n//}^{sf_1} \left( M_{11}^{//} R_{n//}^{f_N a} e^{-2q_{nf_N} z_N} + M_{12}^{//} \right) + \left( M_{21}^{//} R_{n//}^{f_N a} e^{-2q_{nf_N} z_N} + M_{22}^{//} \right)}, \tag{437}$$

$$D = -\frac{\left[\left(T_{n//}^{sf_1} + R_{n//}^{sf_1}\right)M_{11}^{//}R_{n//}^{f_Na} + \left(T_{n//}^{sf_1} + 1\right)M_{21}^{//}R_{n//}^{f_Na}\right]e^{-2q_{nf_N}z_N} + \left(T_{n//}^{sf_1} + R_{n//}^{sf_1}\right)M_{12}^{//} + \left(T_{n//}^{sf_1} + 1\right)M_{22}^{//}}{R_{n//}^{sf_1}\left(M_{11}^{//}R_{n//}^{f_Na}e^{-2q_{nf_N}z_N} + M_{12}^{//}\right) + \left(M_{21}^{//}R_{n//}^{f_Na}e^{-2q_{nf_N}z_N} + M_{22}^{//}\right)} S_{nxz,s} e^{q_{ns}z'}$$

(438)

The coefficients $B_j$ and $C_j$ ($j = 1, 2, ..., N-1$) are easily obtained via the transfer matrix method which is shown in Eq. (217) and described in Eq. (404) and Eq. (405). Next we consider $z' > z_N$, the corresponding different equation in regions $z > z_N$, $z_{j-1} < z < z_j, j = N, N-1, ..., 2, 1$ and $z < 0$ are:

$$\tilde{g}_{nzz} = \begin{cases} \dfrac{1}{q_{na}^2}\delta(z-z') - \dfrac{ik_n}{q_{na}^2}S_{nxz,a}\partial_z\left[\mathrm{sgn}(z-z')e^{-q_{na}|z-z'|}\right] - \dfrac{ik_n}{q_{na}^2}A\partial_z e^{-q_{na}z}, z > z_N \\ -\dfrac{ik_n}{q_{nf_j}^2}B_j\partial_z e^{q_{nf_j}z} - \dfrac{ik_n}{q_{nf_j}^2}C_j\partial_z e^{-q_{nf_j}z}, z_{j-1} < z < z_j, j = 1, 2, ..., N \\ -\dfrac{ik_n}{q_{ns}^2}D\partial_z e^{q_{ns}z}, z < 0 \end{cases}$$

$$= \begin{cases} \dfrac{1}{q_{na}^2}\delta(z-z') - \dfrac{ik_n}{q_{na}^2}S_{nxz,a}\left[2\delta(z-z')e^{-q_{na}|z-z'|} - q_{na}e^{-q_{na}|z-z'|}\right] + \dfrac{ik_n}{q_{na}}Ae^{-q_{na}z}, z > z_N \\ -\dfrac{ik_n}{q_{nf_j}}B_j e^{q_{nf_j}z} + \dfrac{ik_n}{q_{nf_j}}C_j e^{-q_{nf_j}z}, z_{j-1} < z < z_j, j = 1, 2, ..., N \\ -\dfrac{ik_n}{q_{ns}}De^{q_{ns}z}, z < 0 \end{cases}$$

$$= \begin{cases} -\dfrac{1}{k_a^2}\delta(z-z') + \dfrac{ik_n}{q_{na}}S_{nxz,a}e^{-q_{na}|z-z'|} + \dfrac{ik_n}{q_{na}}Ae^{-q_{na}z}, z > z_N \\ -\dfrac{ik_n}{q_{nf_j}}B_j e^{q_{nf_j}z} + \dfrac{ik_n}{q_{nf_j}}C_j e^{-q_{nf_j}z}, z_{j-1} < z < z_j, j = 1, 2, ..., N \\ -\dfrac{ik_n}{q_{ns}}De^{q_{ns}z}, z < 0 \end{cases},$$

(439)

where $S_{nxz,a} = -\dfrac{ik_n}{2k_a^2}$ and the basic four coefficients $A$, $B_N$, $C_N$ and $D$ have already defined in Eq. (233), (232), (231), (234). These following explicit forms are:

$$A = S_{nxz,a} e^{-q_{na}z'} e^{2q_{na}z_N} \frac{R_{n//}^{sf_1}\left(M_{11}^{//} + M_{21}^{//}\right)e^{-2q_{nf_N}z_N} + R_{n//}^{f_Na}\left(R_{n//}^{sf_1}M_{12}^{//} + M_{22}^{//}\right)}{R_{n//}^{sf_1}\left(M_{11}^{//}R_{n//}^{f_Na}e^{-2q_{nf_N}z_N} + M_{12}^{//}\right) + \left(M_{21}^{//}R_{n//}^{f_Na}e^{-2q_{nf_N}z_N} + M_{22}^{//}\right)},$$ (440)

$$B_N = \frac{-S_{nxz,a} e^{-q_{na}z'} e^{q_{na}z_N} e^{-q_{nf_N}z_N} T_{n//}^{af_N} \left( R_{n//}^{sf_1} M_{12}^{//} + M_{22}^{//} \right)}{R_{n//}^{sf_1} \left( M_{11}^{//} R_{n//}^{f_N a} e^{-2q_{nf_N}z_N} + M_{12}^{//} \right) + \left( M_{21}^{//} R_{n//}^{f_N a} e^{-2q_{nf_N}z_N} + M_{22}^{//} \right)}, \tag{441}$$

$$C_N = \frac{S_{nxz,a} T_{n//}^{af_N} \left( R_{n//}^{sf_1} M_{11}^{//} + M_{21}^{//} \right) e^{-q_{na}z'} e^{q_{na}z_N} e^{-q_{nf_N}z_N}}{R_{n//}^{sf_1} \left( M_{11}^{//} R_{n//}^{f_N a} e^{-2q_{nf_N}z_N} + M_{12}^{//} \right) + \left( M_{21}^{//} R_{n//}^{f_N a} e^{-2q_{nf_N}z_N} + M_{22}^{//} \right)}. \tag{442}$$

$$D = \frac{-S_{nxz,a} T_{n//}^{af_N} T_{n//}^{f_1 s} \left( M_{22}^{//} M_{11}^{//} - M_{21}^{//} M_{12}^{//} \right) e^{-q_{na}z'} e^{q_{na}z_N} e^{-q_{nf_N}z_N}}{R_{n//}^{sf_1} \left( M_{11}^{//} R_{n//}^{f_N a} e^{-2q_{nf_N}z_N} + M_{12}^{//} \right) + \left( M_{21}^{//} R_{n//}^{f_N a} e^{-2q_{nf_N}z_N} + M_{22}^{//} \right)}. \tag{443}$$

The coefficients $B_j$ and $C_j$ ($j = 1, 2, ..., N-1$) are easily obtained via the transfer matrix method which is shown in Eq. (217) and described in Eq. (404) and Eq. (405). Next we consider the location of a point source is inside the film layer, that is $z_{k-1} < z' < z_k, k = 1, 2, ..., N$. Fixed the index $k$, the corresponding different equation in regions $z > z_N$, $z_{j-1} < z < z_j, j = N, N-1, ..., 2, 1$ and $z < 0$ are:

$$\tilde{g}_{nzz} = \begin{cases} -\dfrac{ik_n}{q_{na}^2} A \partial_z e^{-q_{na}z}, z > z_N \\[6pt] \dfrac{1}{q_{nf_j}^2} \delta(z-z') - \dfrac{ik_n}{q_{nf_j}^2} S_{nxz,f_j} \partial_z \left[ \text{sgn}(z-z') e^{-q_{nf_j}|z-z'|} \right] - \dfrac{ik_n}{q_{nf_j}^2} B_j \partial_z e^{q_{nf_j}z} - \dfrac{ik_n}{q_{nf_j}^2} C_j \partial_z e^{-q_{nf_j}z} \\ , z_{j-1} < z < z_j, j = 1, 2, ..., N \\[6pt] -\dfrac{ik_n}{q_{ns}^2} D \partial_z e^{q_{ns}z}, z < 0 \end{cases}$$

$$= \begin{cases} \dfrac{ik_n}{q_{na}} A e^{-q_{na}z}, z > z_N \\[6pt] \dfrac{1}{q_{nf_j}^2} \delta(z-z') - \dfrac{ik_n}{q_{nf_j}^2} S_{nxz,f_j} \left[ 2\delta(z-z') e^{-q_{nf_j}|z-z'|} - q_{nf_j} e^{-q_{nf_j}|z-z'|} \right] - \dfrac{ik_n}{q_{nf_j}} B_j e^{q_{nf_j}z} + \dfrac{ik_n}{q_{nf_j}} C_j e^{-q_{nf_j}z} \\ , z_{j-1} < z < z_j, j = 1, 2, ..., N \\[6pt] -\dfrac{ik_n}{q_{ns}} D e^{q_{ns}z}, z < 0 \end{cases}$$

$$= \begin{cases} \dfrac{ik_n}{q_{na}} A e^{-q_{na}z}, z > z_N \\[6pt] -\dfrac{1}{k_{f_j}^2} \delta(z-z') + \dfrac{ik_n}{q_{nf_j}} S_{nxz,f_j} e^{-q_{nf_j}|z-z'|} - \dfrac{ik_n}{q_{nf_j}} B_j e^{q_{nf_j}z} + \dfrac{ik_n}{q_{nf_j}} C_j e^{-q_{nf_j}z}, z_{j-1} < z < z_j, j = 1, 2, ..., N \\[6pt] -\dfrac{ik_n}{q_{ns}} D e^{q_{ns}z}, z < 0 \end{cases}$$

,
(444)

where $S_{nxz,f_k} = -\dfrac{ik_n}{2k_{f_k}^2}$. The coefficients $A$, $B_N$, $C_N$, $D$, $B_j$ and $C_j$ ($j=1,2,...,N-1$) in Eq. (444) have already defined in Eq. (250), (249), (248), (251), (245), (246), (244) for $k=N$, Eq. (262), (261), (260), (263), (255), (256), (257), (258) for $k=2,3,...,N-1$ and Eq. (271), (270), (269), (272), (266), (267) for $k=1$, respectively. These following explicit forms are:

<1> If $k=N$:

$$A = T_{n//}^{f_N a} S_{nxz,f_N} e^{-q_{nf_N} z_N} e^{q_{na} z_N} \frac{\left(M_{21}^{//} - R_{n//}^{sf_1} M_{11}^{//}\right)e^{-q_{nf_N} z'} + \left(M_{22}^{//} - R_{n//}^{sf_1} M_{12}^{//}\right)e^{q_{nf_N} z'}}{R_{n//}^{f_N a}\left(M_{21}^{//} - R_{n//}^{sf_1} M_{11}^{//}\right)e^{-2q_{nf_N} z_N} + M_{22}^{//} - R_{n//}^{sf_1} M_{12}^{//}}, \qquad (445)$$

$$B_N = R_{n//}^{f_N a} S_{nxz,f_N} e^{-2q_{nf_N} z_N} \frac{\left(M_{21}^{//} - R_{n//}^{sf_1} M_{11}^{//}\right)e^{-q_{nf_N} z'} + \left(M_{22}^{//} - R_{n//}^{sf_1} M_{12}^{//}\right)e^{q_{nf_N} z'}}{R_{n//}^{f_N a}\left(M_{21}^{//} - R_{n//}^{sf_1} M_{11}^{//}\right)e^{-2q_{nf_N} z_N} + M_{22}^{//} - R_{n//}^{sf_1} M_{12}^{//}}, \qquad (446)$$

$$C_N = S_{nxz,f_N} \frac{\left(R_{n//}^{sf_1} M_{11}^{//} - M_{21}^{//}\right)\left(R_{n//}^{f_N a} e^{q_{nf_N} z'} e^{-2q_{nf_N} z_N} - e^{-q_{nf_N} z'}\right)}{R_{n//}^{f_N a} M_{21}^{//} e^{-2q_{nf_N} z_N} + M_{22}^{//} - R_{n//}^{sf_1} R_{n//}^{f_N a} M_{11}^{//} e^{-2q_{nf_N} z_N} - R_{n//}^{sf_1} M_{12}^{//}}, \qquad (447)$$

$$D = S_{nxz,f_N} T_{n//}^{sf_1} \frac{\left(M_{22}^{//} M_{11}^{//} - M_{12}^{//} M_{21}^{//}\right) R_{n//}^{f_N a} e^{q_{nf_N} z'} e^{-2q_{nf_N} z_N} - \left(M_{22}^{//} M_{11}^{//} - M_{12}^{//} M_{21}^{//}\right) e^{-q_{nf_N} z'}}{R_{n//}^{f_N a} M_{21}^{//} e^{-2q_{nf_N} z_N} + M_{22}^{//} - R_{n//}^{sf_1} R_{n//}^{f_N a} M_{11}^{//} e^{-2q_{nf_N} z_N} - R_{n//}^{sf_1} M_{12}^{//}},$$

(448)

$$\begin{pmatrix} B_1 \\ C_1 \end{pmatrix} = T_1^{//} T_2^{//} \bullet\bullet\bullet T_{N-2}^{//} \begin{pmatrix} B_{N-1} \\ C_{N-1} \end{pmatrix} = T_1^{//} T_2^{//} \bullet\bullet\bullet T_{N-2}^{//} T_{N-1}^{//} \begin{pmatrix} B_N \\ C_N \end{pmatrix} - T_1^{//} T_2^{//} \bullet\bullet\bullet T_{N-2}^{//} T_{N-1}^{//} \begin{pmatrix} 1 \\ 0 \end{pmatrix} S_{nxz,f_N} e^{-q_{nf_N} z'}$$

$$\equiv \begin{pmatrix} M_{11}^{//} & M_{12}^{//} \\ M_{21}^{//} & M_{22}^{//} \end{pmatrix} \begin{pmatrix} B_N \\ C_N \end{pmatrix} - \begin{pmatrix} M_{11}^{//} & M_{12}^{//} \\ M_{21}^{//} & M_{22}^{//} \end{pmatrix} \begin{pmatrix} 1 \\ 0 \end{pmatrix} S_{nxz,f_N} e^{-q_{nf_N} z'}$$

$$= \begin{pmatrix} M_{11}^{//} B_N + M_{12}^{//} C_N \\ M_{21}^{//} B_N + M_{22}^{//} C_N \end{pmatrix} - \begin{pmatrix} M_{11}^{//} \\ M_{21}^{//} \end{pmatrix} S_{nxz,f_N} e^{-q_{nf_N} z'}$$

,

(449)

$$\begin{pmatrix} B_j \\ C_j \end{pmatrix} = T_j^{//} \bullet\bullet\bullet T_{N-2}^{//} \begin{pmatrix} B_{N-1} \\ C_{N-1} \end{pmatrix} = T_j^{//} \bullet\bullet\bullet T_{N-2}^{//} T_{N-1}^{//} \begin{pmatrix} B_N \\ C_N \end{pmatrix} - T_j^{//} \bullet\bullet\bullet T_{N-2}^{//} T_{N-1}^{//} \begin{pmatrix} 1 \\ 0 \end{pmatrix} S_{nxz,f_N} e^{-q_{nf_N} z'}$$

$$\equiv \begin{pmatrix} M_{j,11}^{//} & M_{j,12}^{//} \\ M_{j,21}^{//} & M_{j,22}^{//} \end{pmatrix} \begin{pmatrix} B_N \\ C_N \end{pmatrix} - \begin{pmatrix} M_{j,11}^{//} & M_{j,12}^{//} \\ M_{j,21}^{//} & M_{j,22}^{//} \end{pmatrix} \begin{pmatrix} 1 \\ 0 \end{pmatrix} S_{nxz,f_N} e^{-q_{nf_N} z'}$$

$$= \begin{pmatrix} M_{j,11}^{//} B_N + M_{j,12}^{//} C_N \\ M_{j,21}^{//} B_N + M_{j,22}^{//} C_N \end{pmatrix} - \begin{pmatrix} M_{j,11}^{//} \\ M_{j,21}^{//} \end{pmatrix} S_{nxz,f_N} e^{-q_{nf_N} z'}, \quad j=2,3,...,N-2$$

,

(450)

$$\begin{pmatrix} B_{N-1} \\ C_{N-1} \end{pmatrix} = T_{N-1}^{//} \begin{pmatrix} B_N \\ C_N \end{pmatrix} - T_{N-1}^{//} \begin{pmatrix} 1 \\ 0 \end{pmatrix} S_{nxz,f_N} e^{-q_{nf_N} z'}. \qquad (451)$$

<2> If $k = 2, 3, ..., N-1$:

$$A = T_{n//}^{f_N a} S_{nxz,f_k} e^{-q_{nf_N} z_N} e^{q_{na} z_N} \frac{R_{n//}^{sf_1}\left(M_{12,k-1}^{//} e^{q_{nf_k} z'} + M_{11,k-1}^{//} e^{-q_{nf_k} z'}\right) + \left(M_{22,k-1}^{//} e^{q_{nf_k} z'} + M_{21,k-1}^{//} e^{-q_{nf_k} z'}\right)}{R_{n//}^{sf_1}\left(M_{11}^{//} R_{n//}^{f_N a} e^{-2q_{nf_N} z_N} + M_{12}^{//}\right) + \left(M_{21}^{//} R_{n//}^{f_N a} e^{-2q_{nf_N} z_N} + M_{22}^{//}\right)},$$

(452)

$$B_N = R_{n//}^{f_N a} S_{nxz,f_k} e^{-2q_{nf_N} z_N} \frac{R_{n//}^{sf_1}\left(M_{12,k-1}^{//} e^{q_{nf_k} z'} + M_{11,k-1}^{//} e^{-q_{nf_k} z'}\right) + \left(M_{22,k-1}^{//} e^{q_{nf_k} z'} + M_{21,k-1}^{//} e^{-q_{nf_k} z'}\right)}{R_{n//}^{sf_1}\left(M_{11}^{//} R_{n//}^{f_N a} e^{-2q_{nf_N} z_N} + M_{12}^{//}\right) + \left(M_{21}^{//} R_{n//}^{f_N a} e^{-2q_{nf_N} z_N} + M_{22}^{//}\right)},$$

(453)

$$C_N = S_{nxz,f_k} \frac{R_{n//}^{sf_1}\left(M_{12,k-1}^{//} e^{q_{nf_k} z'} + M_{11,k-1}^{//} e^{-q_{nf_k} z'}\right) + \left(M_{22,k-1}^{//} e^{q_{nf_k} z'} + M_{21,k-1}^{//} e^{-q_{nf_k} z'}\right)}{R_{n//}^{sf_1}\left(M_{11}^{//} R_{n//}^{f_N a} e^{-2q_{nf_N} z_N} + M_{12}^{//}\right) + \left(M_{21}^{//} R_{n//}^{f_N a} e^{-2q_{nf_N} z_N} + M_{22}^{//}\right)},$$ (454)

$$D = \frac{R_{n//}^{sf_1}\left(M_{12,k-1}^{//} e^{q_{nf_k} z'} + M_{11,k-1}^{//} e^{-q_{nf_k} z'}\right) + \left(M_{22,k-1}^{//} e^{q_{nf_k} z'} + M_{21,k-1}^{//} e^{-q_{nf_k} z'}\right)}{R_{n//}^{sf_1}\left(M_{11}^{//} R_{n//}^{f_N a} e^{-2q_{nf_N} z_N} + M_{12}^{//}\right) + \left(M_{21}^{//} R_{n//}^{f_N a} e^{-2q_{nf_N} z_N} + M_{22}^{//}\right)} M_{11}^{//} R_{n//}^{f_N a} e^{-2q_{nf_N} z_N} S_{nxz,f_k}$$

$$+ \frac{R_{n//}^{sf_1}\left(M_{12,k-1}^{//} e^{q_{nf_k} z'} + M_{11,k-1}^{//} e^{-q_{nf_k} z'}\right) + \left(M_{22,k-1}^{//} e^{q_{nf_k} z'} + M_{21,k-1}^{//} e^{-q_{nf_k} z'}\right)}{R_{n//}^{sf_1}\left(M_{11}^{//} R_{n//}^{f_N a} e^{-2q_{nf_N} z_N} + M_{12}^{//}\right) + \left(M_{21}^{//} R_{n//}^{f_N a} e^{-2q_{nf_N} z_N} + M_{22}^{//}\right)} M_{12}^{//} S_{nxz,f_k}$$

$$+ \frac{R_{n//}^{sf_1}\left(M_{12,k-1}^{//} e^{q_{nf_k} z'} + M_{11,k-1}^{//} e^{-q_{nf_k} z'}\right) + \left(M_{22,k-1}^{//} e^{q_{nf_k} z'} + M_{21,k-1}^{//} e^{-q_{nf_k} z'}\right)}{R_{n//}^{sf_1}\left(M_{11}^{//} R_{n//}^{f_N a} e^{-2q_{nf_N} z_N} + M_{12}^{//}\right) + \left(M_{21}^{//} R_{n//}^{f_N a} e^{-2q_{nf_N} z_N} + M_{22}^{//}\right)} M_{21}^{//} R_{n//}^{f_N a} e^{-2q_{nf_N} z_N} S_{nxz,f_k}$$

$$+ \frac{R_{n//}^{sf_1}\left(M_{12,k-1}^{//} e^{q_{nf_k} z'} + M_{11,k-1}^{//} e^{-q_{nf_k} z'}\right) + \left(M_{22,k-1}^{//} e^{q_{nf_k} z'} + M_{21,k-1}^{//} e^{-q_{nf_k} z'}\right)}{R_{n//}^{sf_1}\left(M_{11}^{//} R_{n//}^{f_N a} e^{-2q_{nf_N} z_N} + M_{12}^{//}\right) + \left(M_{21}^{//} R_{n//}^{f_N a} e^{-2q_{nf_N} z_N} + M_{22}^{//}\right)} M_{22}^{//} S_{nxz,f_k}$$

$$- M_{12,k-1}^{//} S_{nxz,f_k} e^{q_{nf_k} z'} - M_{11,k-1}^{//} S_{nxz,f_k} e^{-q_{nf_k} z'} - M_{22,k-1}^{//} S_{nxz,f_k} e^{q_{nf_k} z'} - M_{21,k-1}^{//} S_{nxz,f_k} e^{-q_{nf_k} z'}$$

, (455)

$$\begin{pmatrix} B_1 \\ C_1 \end{pmatrix} = T_1^{//} T_2^{//} \bullet \bullet T_{k-2}^{//} \begin{pmatrix} B_{k-1} \\ C_{k-1} \end{pmatrix} = T_1^{//} T_2^{//} \bullet \bullet T_{k-2}^{//} T_{k-1}^{//} \begin{pmatrix} B_k \\ C_k \end{pmatrix} - T_1^{//} T_2^{//} \bullet \bullet T_{k-2}^{//} T_{k-1}^{//} \begin{pmatrix} 1 \\ 0 \end{pmatrix} S_{nxz,f_k} e^{-q_{nf_k} z'}$$

$$= T_1^{//} T_2^{//} \bullet \bullet T_{k-2}^{//} T_{k-1}^{//} T_k^{//} \begin{pmatrix} B_{k+1} \\ C_{k+1} \end{pmatrix} - T_1^{//} T_2^{//} \bullet \bullet T_{k-2}^{//} T_{k-1}^{//} \begin{pmatrix} 0 \\ 1 \end{pmatrix} S_{nxz,f_k} e^{q_{nf_k} z'} - T_1^{//} T_2^{//} \bullet \bullet T_{k-2}^{//} T_{k-1}^{//} \begin{pmatrix} 1 \\ 0 \end{pmatrix} S_{nxz,f_k} e^{-q_{nf_k} z'}$$

$$\equiv \begin{pmatrix} M_{11}^{//} & M_{12}^{//} \\ M_{21}^{//} & M_{22}^{//} \end{pmatrix} \begin{pmatrix} B_N \\ C_N \end{pmatrix} - \begin{pmatrix} M_{11,k-1}^{//} & M_{12,k-1}^{//} \\ M_{21,k-1}^{//} & M_{22,k-1}^{//} \end{pmatrix} \begin{pmatrix} 0 \\ 1 \end{pmatrix} S_{nxz,f_k} e^{q_{nf_k} z'} - \begin{pmatrix} M_{11,k-1}^{//} & M_{12,k-1}^{//} \\ M_{21,k-1}^{//} & M_{22,k-1}^{//} \end{pmatrix} \begin{pmatrix} 1 \\ 0 \end{pmatrix} S_{nxz,f_k} e^{-q_{nf_k} z'}$$

$$= \begin{pmatrix} M_{11}^{//} B_N + M_{12}^{//} C_N \\ M_{21}^{//} B_N + M_{22}^{//} C_N \end{pmatrix} - \begin{pmatrix} M_{12,k-1}^{//} \\ M_{22,k-1}^{//} \end{pmatrix} S_{nxz,f_k} e^{q_{nf_k} z'} - \begin{pmatrix} M_{11,k-1}^{//} \\ M_{21,k-1}^{//} \end{pmatrix} S_{nxz,f_k} e^{-q_{nf_k} z'}$$

, (456)

$$\begin{pmatrix} B_j \\ C_j \end{pmatrix} = \mathrm{T}_j^{//} \bullet \bullet \mathrm{T}_{k-2}^{//} \begin{pmatrix} B_{k-1} \\ C_{k-1} \end{pmatrix} = \mathrm{T}_j^{//} \bullet \bullet \mathrm{T}_{k-2}^{//} \mathrm{T}_{k-1}^{//} \begin{pmatrix} B_k \\ C_k \end{pmatrix} - \mathrm{T}_j^{//} \bullet \bullet \mathrm{T}_{k-2}^{//} \mathrm{T}_{k-1}^{//} \begin{pmatrix} 1 \\ 0 \end{pmatrix} S_{nxz,f_k} e^{-q_{nf_k} z'}$$

$$= \mathrm{T}_j^{//} \bullet \bullet \mathrm{T}_{k-2}^{//} \mathrm{T}_{k-1}^{//} \mathrm{T}_k^{//} \begin{pmatrix} B_{k+1} \\ C_{k+1} \end{pmatrix} - \mathrm{T}_j^{//} \bullet \bullet \mathrm{T}_{k-2}^{//} \mathrm{T}_{k-1}^{//} \begin{pmatrix} 0 \\ 1 \end{pmatrix} S_{nxz,f_k} e^{q_{nf_k} z'} - \mathrm{T}_j^{//} \bullet \bullet \mathrm{T}_{k-2}^{//} \mathrm{T}_{k-1}^{//} \begin{pmatrix} 1 \\ 0 \end{pmatrix} S_{nxz,f_k} e^{-q_{nf_k} z'}$$

$$\equiv \begin{pmatrix} M_{j,11}^{//} & M_{j,12}^{//} \\ M_{j,21}^{//} & M_{j,22}^{//} \end{pmatrix} \begin{pmatrix} B_N \\ C_N \end{pmatrix} - \begin{pmatrix} M_{j,11,k-1}^{//} & M_{j,12,k-1}^{//} \\ M_{j,21,k-1}^{//} & M_{j,22,k-1}^{//} \end{pmatrix} \begin{pmatrix} 0 \\ 1 \end{pmatrix} S_{nxz,f_k} e^{q_{nf_k} z'} - \begin{pmatrix} M_{j,11,k-1}^{//} & M_{j,12,k-1}^{//} \\ M_{j,21,k-1}^{//} & M_{j,22,k-1}^{//} \end{pmatrix} \begin{pmatrix} 1 \\ 0 \end{pmatrix} S_{nxz,f_k} e^{-q_{nf_k} z'}$$

$$= \begin{pmatrix} M_{j,11}^{//} B_N + M_{j,12}^{//} C_N \\ M_{j,21}^{//} B_N + M_{j,22}^{//} C_N \end{pmatrix} - \begin{pmatrix} M_{j,12,k-1}^{//} \\ M_{j,22,k-1}^{//} \end{pmatrix} S_{nxz,f_k} e^{q_{nf_k} z'} - \begin{pmatrix} M_{j,11,k-1}^{//} \\ M_{j,21,k-1}^{//} \end{pmatrix} S_{nxz,f_k} e^{-q_{nf_k} z'}, \quad j = 2,3,\ldots,N-1, j \leq k-1$$

, (457)

$$\begin{pmatrix} B_k \\ C_k \end{pmatrix} = \mathrm{T}_k^{//} \begin{pmatrix} B_{k+1} \\ C_{k+1} \end{pmatrix} - \begin{pmatrix} 0 \\ 1 \end{pmatrix} S_{nxz,f_k} e^{q_{nf_k} z'} = \mathrm{T}_k^{//} \bullet \bullet \mathrm{T}_{N-1}^{//} \begin{pmatrix} B_N \\ C_N \end{pmatrix} - \begin{pmatrix} 0 \\ 1 \end{pmatrix} S_{nxz,f_k} e^{q_{nf_k} z'}$$

$$\equiv \begin{pmatrix} M_{k,11}^{//} & M_{k,12}^{//} \\ M_{k,21}^{//} & M_{k,22}^{//} \end{pmatrix} \begin{pmatrix} B_N \\ C_N \end{pmatrix} - \begin{pmatrix} 0 \\ 1 \end{pmatrix} S_{nxz,f_k} e^{q_{nf_k} z'}, \quad (458)$$

$$= \begin{pmatrix} M_{k,11}^{//} B_N + M_{k,12}^{//} C_N \\ M_{k,21}^{//} B_N + M_{k,22}^{//} C_N \end{pmatrix} - \begin{pmatrix} 0 \\ 1 \end{pmatrix} S_{nxz,f_k} e^{q_{nf_k} z'}, \quad j = 2,3,\ldots,N-1, j = k$$

$$\begin{pmatrix} B_j \\ C_j \end{pmatrix} = \mathrm{T}_j^{//} \begin{pmatrix} B_{j+1} \\ C_{j+1} \end{pmatrix} = \mathrm{T}_j^{//} \mathrm{T}_{j+1}^{//} \bullet \bullet \bullet \mathrm{T}_{N-1}^{//} \begin{pmatrix} B_N \\ C_N \end{pmatrix} \equiv \begin{pmatrix} M_{j,11}^{//} & M_{j,12}^{//} \\ M_{j,21}^{//} & M_{j,22}^{//} \end{pmatrix} \begin{pmatrix} B_N \\ C_N \end{pmatrix}$$

$$= \begin{pmatrix} M_{j,11}^{//} B_N + M_{j,12}^{//} C_N \\ M_{j,21}^{//} B_N + M_{j,22}^{//} C_N \end{pmatrix}, j = 2,3,\ldots,N-1, j \geq k+1 \quad . \quad (459)$$

<3> If $k = 1$:

$$A = T_{n//}^{f_N a} S_{nxz,f_1} \frac{e^{q_{nf_1} z'} + R_{n//}^{sf_1} e^{-q_{nf_1} z'}}{M_{11}^{//} R_{n//}^{sf_1} R_{n//}^{f_N a} e^{-2q_{nf_N} z_N} + M_{12}^{//} R_{n//}^{sf_1} + M_{21}^{//} R_{n//}^{f_N a} e^{-2q_{nf_N} z_N} + M_{22}^{//}} e^{-q_{nf_N} z_N} e^{q_{na} z_N},$$

(460)

$$B_N = R_{n//}^{f_N a} S_{nxz,f_1} \frac{e^{q_{nf_1} z'} + R_{n//}^{sf_1} e^{-q_{nf_1} z'}}{M_{11}^{//} R_{n//}^{sf_1} R_{n//}^{f_N a} e^{-2q_{nf_N} z_N} + M_{12}^{//} R_{n//}^{sf_1} + M_{21}^{//} R_{n//}^{f_N a} e^{-2q_{nf_N} z_N} + M_{22}^{//}} e^{-2q_{nf_N} z_N}, \quad (461)$$

$$C_N = S_{nxz,f_1} \frac{e^{q_{nf_1} z'} + R_{n//}^{sf_1} e^{-q_{nf_1} z'}}{M_{11}^{//} R_{n//}^{sf_1} R_{n//}^{f_N a} e^{-2q_{nf_N} z_N} + M_{12}^{//} R_{n//}^{sf_1} + M_{21}^{//} R_{n//}^{f_N a} e^{-2q_{nf_N} z_N} + M_{22}^{//}}, \quad (462)$$

$$D = S_{nxz,f_1} T_{n//}^{f_1 s} \frac{e^{q_{nf_1} z'} \left( M_{11}^{//} R_{n//}^{f_N a} e^{-2q_{nf_N} z_N} + M_{12}^{//} \right) - e^{-q_{nf_1} z'} \left( M_{21}^{//} R_{n//}^{f_N a} e^{-2q_{nf_N} z_N} + M_{22}^{//} \right)}{M_{11}^{//} R_{n//}^{sf_1} R_{n//}^{f_N a} e^{-2q_{nf_N} z_N} + M_{12}^{//} R_{n//}^{sf_1} + M_{21}^{//} R_{n//}^{f_N a} e^{-2q_{nf_N} z_N} + M_{22}^{//}}, \quad (463)$$

$$\begin{pmatrix} B_1 \\ C_1 \end{pmatrix} = T_1^{//} \begin{pmatrix} B_2 \\ C_2 \end{pmatrix} - \begin{pmatrix} 0 \\ 1 \end{pmatrix} S_{nxz,f_1} e^{q_{nf_1} z'} = \begin{pmatrix} M_{11}^{//} & M_{12}^{//} \\ M_{21}^{//} & M_{22}^{//} \end{pmatrix} \begin{pmatrix} B_N \\ C_N \end{pmatrix} - \begin{pmatrix} 0 \\ 1 \end{pmatrix} S_{nxz,f_1} e^{q_{nf_1} z'}$$
$$= \begin{pmatrix} M_{11}^{//} B_N + M_{12}^{//} C_N \\ M_{21}^{//} B_N + M_{22}^{//} C_N \end{pmatrix} - \begin{pmatrix} 0 \\ 1 \end{pmatrix} S_{nxz,f_1} e^{q_{nf_1} z'}$$
, (464)

$$\begin{pmatrix} B_j \\ C_j \end{pmatrix} = T_j^{//} \bullet \bullet \bullet T_{N-1}^{//} \begin{pmatrix} B_N \\ C_N \end{pmatrix} = \begin{pmatrix} M_{j,11}^{//} & M_{j,12}^{//} \\ M_{j,21}^{//} & M_{j,22}^{//} \end{pmatrix} \begin{pmatrix} B_N \\ C_N \end{pmatrix} = \begin{pmatrix} M_{j,11}^{//} B_N + M_{j,11}^{//} C_N \\ M_{j,21}^{//} B_N + M_{j,22}^{//} C_N \end{pmatrix}, j = 2,...,N-1$$
. (465)

## 5. Validity of the derived multi-planar dyadic Green's functions

In order to ensure the validity of the above results obtained for multi-planar dyadic Green's functions by Fourier expansion method, we need to compare this with the literatures [6-7]. To this end, we consider a single dipole source with two different orientations (vertical and horizontal) located inside the film, as shown in Fig. 4.

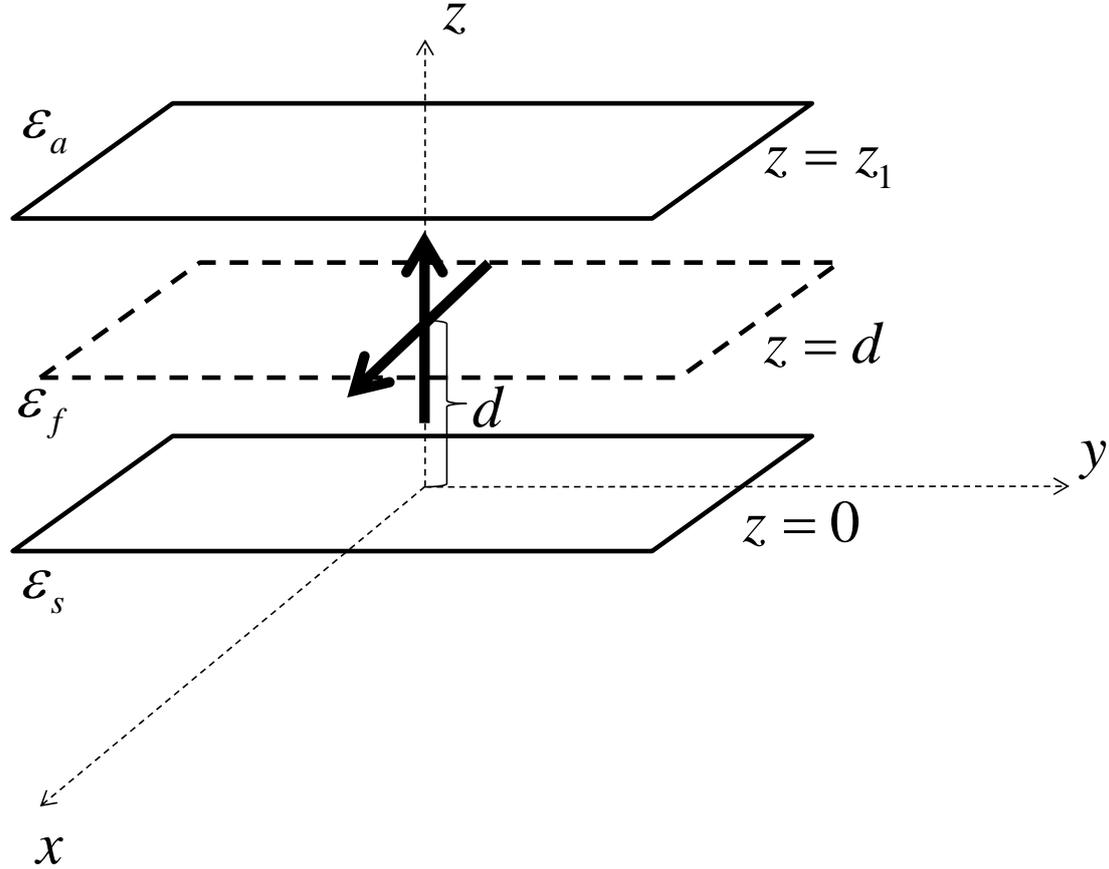

**Fig. 4** Geometry of the problem that a dipole with both vertical and horizontal direction inside the film

### 5-1. A vertically oriented dipole

First, we consider a vertically oriented dipole with moment $p_0 e^{-i\omega t} \mathbf{e}_z$, which locates at the position $z = d$. The current density is [7]:

$$\mathbf{J}(\mathbf{r}') = -i\omega p_0 e^{-i\omega t} \delta(\mathbf{r}' - d\hat{z})\mathbf{e}_z. \tag{466}$$

Substitute the above identity into Eq. (15) to calculate the reflected electric field at the dipole position and then obtain:

$$\begin{aligned} E_z(d\mathbf{e}_z) &= \omega^2 \mu_0 p_0 e^{-i\omega t} \mathbf{e}_z \bullet \mathbf{G}(d\mathbf{e}_z, d\mathbf{e}_z) \bullet \mathbf{e}_z = \omega^2 \mu_0 p_0 e^{-i\omega t} G_{zz}(d\mathbf{e}_z, d\mathbf{e}_z) \\ &= \frac{1}{(2\pi)^2} \omega^2 \mu_0 p_0 e^{-i\omega t} \lim_{\substack{\boldsymbol{\rho} \to 0 \\ \boldsymbol{\rho}' \to 0}} \int d^2 \mathbf{k}_n e^{i\mathbf{k}_n \bullet (\boldsymbol{\rho} - \boldsymbol{\rho}')} g_{nzz}(d,d) \\ &= \frac{1}{(2\pi)^2} \omega^2 \mu_0 p_0 e^{-i\omega t} \lim_{\substack{\boldsymbol{\rho} \to 0 \\ \boldsymbol{\rho}' \to 0}} \int d^2 \mathbf{k}_n e^{i\mathbf{k}_n \bullet (\boldsymbol{\rho} - \boldsymbol{\rho}')} \tilde{g}_{nzz}(d,d) \end{aligned} \tag{467}$$

where the explicit form of integration $\lim_{\substack{\boldsymbol{\rho} \to 0 \\ \boldsymbol{\rho}' \to 0}} \int d^2 \mathbf{k}_n e^{i\mathbf{k}_n \bullet (\boldsymbol{\rho} - \boldsymbol{\rho}')} \tilde{g}_{nzz}(d,d)$ is:

$$\begin{aligned} &\lim_{\substack{\boldsymbol{\rho} \to 0 \\ \boldsymbol{\rho}' \to 0}} \int d^2 \mathbf{k}_n e^{i\mathbf{k}_n \bullet (\boldsymbol{\rho} - \boldsymbol{\rho}')} \tilde{g}_{nzz}(d,d) = \int d^2 \mathbf{k}_n \left( -\frac{ik_n}{q_{nf}} B_{zz} e^{q_{nf} d} + \frac{ik_n}{q_{nf}} C_{zz} e^{-q_{nf} d} \right) \\ &= -i \int d^2 \mathbf{k}_n S_{nxz,f} \frac{k_n}{q_{nf}} \left( \frac{e^{q_{nf} d} - R_{n//}^{fs} e^{-q_{nf} d}}{1 - R_{n//}^{fs} R_{n//}^{fa} e^{-2q_{nf} z_1}} R_{n//}^{fa} e^{-2q_{nf} z_1} e^{q_{nf} d} - \frac{R_{n//}^{fa} e^{q_{nf} d} e^{-2q_{nf} z_1} - e^{-q_{nf} d}}{1 - R_{n//}^{fs} R_{n//}^{fa} e^{-2q_{nf} z_1}} R_{n//}^{fs} e^{-q_{nf} d} \right) \\ &= -\int_0^\infty dk_n \frac{k_n^2}{2k_f^2} \int_0^{2\pi} d\varphi_n \frac{k_n}{q_{nf}} \left( \frac{e^{q_{nf} d} - R_{n//}^{fs} e^{-q_{nf} d}}{1 - R_{n//}^{fs} R_{n//}^{fa} e^{-2q_{nf} z_1}} R_{n//}^{fa} e^{-2q_{nf} z_1} e^{q_{nf} d} - \frac{R_{n//}^{fa} e^{q_{nf} d} e^{-2q_{nf} z_1} - e^{-q_{nf} d}}{1 - R_{n//}^{fs} R_{n//}^{fa} e^{-2q_{nf} z_1}} R_{n//}^{fs} e^{-q_{nf} d} \right) \\ &= -\frac{\pi}{k_f^2} \int_0^\infty dk_n \frac{k_n^3}{q_{nf}} \left( \frac{e^{q_{nf} d} - R_{n//}^{fs} e^{-q_{nf} d}}{1 - R_{n//}^{fs} R_{n//}^{fa} e^{-2q_{nf} z_1}} R_{n//}^{fa} e^{-2q_{nf} z_1} e^{q_{nf} d} - \frac{R_{n//}^{fa} e^{q_{nf} d} e^{-2q_{nf} z_1} - e^{-q_{nf} d}}{1 - R_{n//}^{fs} R_{n//}^{fa} e^{-2q_{nf} z_1}} R_{n//}^{fs} e^{-q_{nf} d} \right) \\ &= -\frac{\pi}{k_f^2} \int_0^\infty dk_n \frac{k_n^3}{q_{nf}} \frac{R_{n//}^{fa} e^{2q_{nf} d} e^{-2q_{nf} z_1} - 2R_{n//}^{fs} R_{n//}^{fa} e^{-2q_{nf} z_1} + R_{n//}^{fs} e^{-2q_{nf} d}}{1 - R_{n//}^{fs} R_{n//}^{fa} e^{-2q_{nf} z_1}} \end{aligned}$$

, (468)

where the coefficients $B_{zz}$ and $C_{zz}$ have been presented in Eqs. (202) and (203), with $z' = d$ as follows:

$$B_{zz} = S_{nxz,f} R_{n//}^{fa} \frac{e^{q_{nf} d} - R_{n//}^{fs} e^{-q_{nf} d}}{1 - R_{n//}^{fs} R_{n//}^{fa} e^{-2q_{nf} z_1}} e^{-2q_{nf} z_1}, \tag{469}$$

$$C_{zz} = \frac{R_{n//}^{fa} e^{q_{nf} d} e^{-2q_{nf} z_1} - e^{-q_{nf} d}}{1 - R_{n//}^{fs} R_{n//}^{fa} e^{-2q_{nf} z_1}} R_{n//}^{fs} S_{nxz,f}. \tag{470}$$

Then Eq. (467) becomes:

$$E_z(d\mathbf{e}_z) = \frac{1}{(2\pi)^2} \lim_{\substack{\rho \to 0 \\ \rho' \to 0}} \int d^2\mathbf{k}_n e^{i\mathbf{k}_n \cdot (\rho-\rho')} \tilde{g}_{nzz}(d,d)$$

$$= -\frac{\omega^2 \mu_0 p_0 e^{-i\omega t}}{4\pi k_f^2} \int_0^\infty dk_n \frac{k_n^3}{q_{nf}} \frac{R_{n//}^{fa} e^{2q_{nf}d} e^{-2q_{nf}z_1} - 2R_{n//}^{fs} R_{n//}^{fa} e^{-2q_{nf}z_1} + R_{n//}^{fs} e^{-2q_{nf}d}}{1 - R_{n//}^{fs} R_{n//}^{fa} e^{-2q_{nf}z_1}}$$

$$= -\frac{\omega^2 \mu_0 p_0 e^{-i\omega t}}{4\pi k_f^2} \int_0^\infty dk_n \frac{k_n^3}{-ih_{nf}} \frac{R_{n//}^{fa} e^{2q_{nf}d} e^{-2q_{nf}z_1} - 2R_{n//}^{fs} R_{n//}^{fa} e^{-2q_{nf}z_1} + R_{n//}^{fs} e^{-2q_{nf}d}}{1 - R_{n//}^{fs} R_{n//}^{fa} e^{-2q_{nf}z_1}} \quad (471)$$

$$= \frac{i\omega^2 \mu_0 p_0 e^{-i\omega t}}{4\pi k_f^2} \int_0^\infty dk_n \frac{k_n^3}{h_{nf}} \frac{2R_{n//}^{fs} R_{n//}^{fa} e^{-2q_{nf}z_1} - R_{n//}^{fa} e^{2q_{nf}d} e^{-2q_{nf}z_1} - R_{n//}^{fs} e^{-2q_{nf}d}}{1 - R_{n//}^{fs} R_{n//}^{fa} e^{-2q_{nf}z_1}}$$

which is the same as Eq. (27) of the previous work [7].

### 5-2. A horizontally oriented dipole

Second, we consider a horizontally oriented dipole with moment $p_0 e^{-i\omega t}\mathbf{e}_x$, which locates at the position $z = d$. The current density is [7]:

$$\mathbf{J}(\mathbf{r}') = -i\omega p_0 e^{-i\omega t} \delta(\mathbf{r}' - d\hat{z})\mathbf{e}_x. \quad (472)$$

Similar to the case of vertical dipole as discussed above, substituting the above identity into Eq. (15) to calculate the reflected electric field at the dipole position and then obtain:

$$E_x(d\mathbf{e}_z) = \omega^2 \mu_0 p_0 e^{-i\omega t} \mathbf{e}_x \cdot \mathbf{G}(d\mathbf{e}_z, d\mathbf{e}_z) \cdot \mathbf{e}_x = \omega^2 \mu_0 p_0 e^{-i\omega t} G_{xx}(d\mathbf{e}_z, d\mathbf{e}_z)$$

$$= \omega^2 \mu_0 p_0 e^{-i\omega t} \frac{1}{(2\pi)^2} \lim_{\substack{\rho \to 0 \\ \rho' \to 0}} \int d^2\mathbf{k}_n e^{i\mathbf{k}_n \cdot (\rho-\rho')} g_{nxx}(d,d)$$

$$= \frac{\omega^2 \mu_0 p_0 e^{-i\omega t}}{(2\pi)^2} \lim_{\substack{\rho \to 0 \\ \rho' \to 0}} \int d^2\mathbf{k}_n e^{i\mathbf{k}_n \cdot (\rho-\rho')} \left[ \tilde{g}_{nxx}(d,d) \cos^2\varphi_n + \tilde{g}_{nyy}(d,d) \sin^2\varphi_n \right], \quad (473)$$

$$= \frac{\omega^2 \mu_0 p_0 e^{-i\omega t}}{(2\pi)^2} \int d^2\mathbf{k}_n \left[ \tilde{g}_{nxx}(d,d) \cos^2\varphi_n + \tilde{g}_{nyy}(d,d) \sin^2\varphi_n \right]$$

$$= \frac{\omega^2 \mu_0 p_0 e^{-i\omega t}}{4\pi} \int_0^\infty dk_n k_n \left[ \tilde{g}_{nxx}(d,d) + \tilde{g}_{nyy}(d,d) \right]$$

where the explicit form of integration $\int_0^\infty dk_n k_n \left[ \tilde{g}_{nxx}(d,d) + \tilde{g}_{nyy}(d,d) \right]$ is:

$$\int_0^\infty dk_n k_n \left[ \tilde{g}_{nxx}(d,d) + \tilde{g}_{nyy}(d,d) \right]$$
$$= \int_0^\infty dk_n k_n \left[ B_{xx} e^{q_{nf}d} + C_{xx} e^{-q_{nf}d} \right] + \int_0^\infty dk_n k_n \left[ B_{yy} e^{q_{nf}d} + C_{yy} e^{-q_{nf}d} \right], \quad (474)$$

where the coefficients $B_{xx}$, $C_{xx}$, $B_{yy}$ and $C_{yy}$ have been presented in Eqs. (164), (163), (132) and (131), with $z' = d$ as follows:

$$B_{xx} = \frac{R_{n//}^{fa} R_{n//}^{fs} S_{nxx,f} e^{-q_{nf} d} e^{-2q_{nf} z_1} + R_{n//}^{fa} S_{nxx,f} e^{q_{nf} d} e^{-2q_{nf} z_1}}{1 - R_{n//}^{fs} R_{n//}^{fa} e^{-2q_{nf} z_1}}, \tag{475}$$

$$C_{xx} = \frac{R_{n//}^{fs} S_{nxx,f} e^{-q_{nf} d} + R_{n//}^{fs} R_{n//}^{fa} S_{nxx,f} e^{q_{nf} d} e^{-2q_{nf} z_1}}{1 - R_{n//}^{fs} R_{n//}^{fa} e^{-2q_{nf} z_1}}, \tag{476}$$

$$B_{yy} = S_{nyy,f} e^{-2q_{nf} z_1} R_{n\perp}^{fa} \frac{R_{n\perp}^{fs} e^{-q_{nf} d} + e^{q_{nf} d}}{1 - R_{n\perp}^{fs} R_{n\perp}^{fa} e^{-2q_{nf} z_1}}, \tag{477}$$

$$C_{yy} = R_{n\perp}^{fs} \frac{e^{-q_{nf} d} + R_{n\perp}^{fa} e^{q_{nf} d} e^{-2q_{nf} z_1}}{1 - R_{n\perp}^{fs} R_{n\perp}^{fa} e^{-2q_{nf} z_1}} S_{nyy,f}. \tag{478}$$

Then Eq. (474) becomes:

$$\int_0^\infty dk_n k_n \left[ B_{xx} e^{q_{nf} d} + C_{xx} e^{-q_{nf} d} \right] + \int_0^\infty dk_n k_n \left[ B_{yy} e^{q_{nf} d} + C_{yy} e^{-q_{nf} d} \right]$$

$$= \int_0^\infty dk_n k_n \frac{R_{n//}^{fa} R_{n//}^{fs} S_{nxx,f} e^{-2q_{nf} z_1} + R_{n//}^{fa} S_{nxx,f} e^{2q_{nf} d} e^{-2q_{nf} z_1} + R_{n//}^{fs} S_{nxx,f} e^{-2q_{nf} d} + R_{n//}^{fs} R_{n//}^{fa} S_{nxx,f} e^{-2q_{nf} z_1}}{1 - R_{n//}^{fs} R_{n//}^{fa} e^{-2q_{nf} z_1}}$$

$$+ \int_0^\infty dk_n k_n \frac{S_{nyy,f} e^{-2q_{nf} z_1} R_{n\perp}^{fa} R_{n\perp}^{fs} + S_{nyy,f} e^{-2q_{nf} z_1} R_{n\perp}^{fa} e^{2q_{nf} d} + S_{nyy,f} R_{n\perp}^{fs} e^{-2q_{nf} d} + S_{nyy,f} R_{n\perp}^{fs} R_{n\perp}^{fa} e^{-2q_{nf} z_1}}{1 - R_{n\perp}^{fs} R_{n\perp}^{fa} e^{-2q_{nf} z_1}}$$

$$= -\frac{1}{2k_f^2} \int_0^\infty dk_n q_{nf} k_n \frac{2 R_{n//}^{fa} R_{n//}^{fs} e^{-2q_{nf} z_1} + R_{n//}^{fa} e^{2q_{nf} d} e^{-2q_{nf} z_1} + R_{n//}^{fs} e^{-2q_{nf} d}}{1 - R_{n//}^{fs} R_{n//}^{fa} e^{-2q_{nf} z_1}}$$

$$+ \frac{1}{2} \int_0^\infty dk_n \frac{k_n}{q_{nf}} \frac{e^{-2q_{nf} z_1} R_{n\perp}^{fa} e^{2q_{nf} d} + R_{n\perp}^{fs} e^{-2q_{nf} d} + 2 R_{n\perp}^{fs} R_{n\perp}^{fa} e^{-2q_{nf} z_1}}{1 - R_{n\perp}^{fs} R_{n\perp}^{fa} e^{-2q_{nf} z_1}}$$

. \tag{479}

Hence Eq. (473) becomes:

$$\begin{aligned}
E_x(d\mathbf{e}_z) &= \frac{\omega^2 \mu_0 p_0 e^{-i\omega t}}{4\pi} \int_0^\infty dk_n k_n \left[ \tilde{g}_{nxx}(d,d) + \tilde{g}_{nyy}(d,d) \right] \\
&= -\frac{\omega^2 \mu_0 p_0 e^{-i\omega t}}{8\pi k_f^2} \int_0^\infty dk_n q_{nf} k_n \frac{2R_{n//}^{fa} R_{n//}^{fs} e^{-2q_{nf} z_1} + R_{n//}^{fa} e^{2q_{nf} d} e^{-2q_{nf} z_1} + R_{n//}^{fs} e^{-2q_{nf} d}}{1 - R_{n//}^{fs} R_{n//}^{fa} e^{-2q_{nf} z_1}} \\
&\quad + \frac{\omega^2 \mu_0 p_0 e^{-i\omega t}}{8\pi} \int_0^\infty dk_n \frac{k_n}{q_{nf}} \frac{e^{-2q_{nf} z_1} R_{n\perp}^{fa} e^{2q_{nf} d} + R_{n\perp}^{fs} e^{-2q_{nf} d} + 2R_{n\perp}^{fs} R_{n\perp}^{fa} e^{-2q_{nf} z_1}}{1 - R_{n\perp}^{fs} R_{n\perp}^{fa} e^{-2q_{nf} z_1}} \\
&= -\frac{\omega^2 \mu_0 p_0 e^{-i\omega t}}{8\pi k_f^2} \int_0^\infty dk_n (-ih_{nf}) k_n \frac{2R_{n//}^{fa} R_{n//}^{fs} e^{-2q_{nf} z_1} + R_{n//}^{fa} e^{2q_{nf} d} e^{-2q_{nf} z_1} + R_{n//}^{fs} e^{-2q_{nf} d}}{1 - R_{n//}^{fs} R_{n//}^{fa} e^{-2q_{nf} z_1}} \\
&\quad + \frac{\omega^2 \mu_0 p_0 e^{-i\omega t}}{8\pi \omega^2 \varepsilon_0 \varepsilon_1 \mu_0} \int_0^\infty dk_n k_f^2 \frac{k_n}{-ih_{nf}} \frac{e^{-2q_{nf} z_1} R_{n\perp}^{fa} e^{2q_{nf} d} + R_{n\perp}^{fs} e^{-2q_{nf} d} + 2R_{n\perp}^{fs} R_{n\perp}^{fa} e^{-2q_{nf} z_1}}{1 - R_{n\perp}^{fs} R_{n\perp}^{fa} e^{-2q_{nf} z_1}} \\
&= \frac{ip_0 e^{-i\omega t}}{8\pi \varepsilon_0 \varepsilon_1} \int_0^\infty dk_n h_{nf} k_n \frac{2R_{n//}^{fa} R_{n//}^{fs} e^{-2q_{nf} z_1} + R_{n//}^{fa} e^{2q_{nf} d} e^{-2q_{nf} z_1} + R_{n//}^{fs} e^{-2q_{nf} d}}{1 - R_{n//}^{fs} R_{n//}^{fa} e^{-2q_{nf} z_1}} \\
&\quad + \frac{ip_0 e^{-i\omega t}}{8\pi \varepsilon_0 \varepsilon_1} \int_0^\infty dk_n k_f^2 \frac{k_n}{h_{nf}} \frac{e^{-2q_{nf} z_1} R_{n\perp}^{fa} e^{2q_{nf} d} + R_{n\perp}^{fs} e^{-2q_{nf} d} + 2R_{n\perp}^{fs} R_{n\perp}^{fa} e^{-2q_{nf} z_1}}{1 - R_{n\perp}^{fs} R_{n\perp}^{fa} e^{-2q_{nf} z_1}} \\
&= \frac{ip_0 e^{-i\omega t}}{8\pi \varepsilon_0 \varepsilon_1} \int_0^\infty dk_n \frac{k_n}{h_{nf}} \left( h_{nf}^2 \frac{2R_{n//}^{fa} R_{n//}^{fs} e^{-2q_{nf} z_1} + R_{n//}^{fa} e^{2q_{nf} d} e^{-2q_{nf} z_1} + R_{n//}^{fs} e^{-2q_{nf} d}}{1 - R_{n//}^{fs} R_{n//}^{fa} e^{-2q_{nf} z_1}} \right. \\
&\quad \left. + k_f^2 \frac{R_{n\perp}^{fa} e^{-2q_{nf} z_1} e^{2q_{nf} d} + R_{n\perp}^{fs} e^{-2q_{nf} d} + 2R_{n\perp}^{fs} R_{n\perp}^{fa} e^{-2q_{nf} z_1}}{1 - R_{n\perp}^{fs} R_{n\perp}^{fa} e^{-2q_{nf} z_1}} \right)
\end{aligned} \qquad (480)$$

which is the same as Eq. (32) of the previous work [7].

## 6. Application to a dipole source with arbitrary orientation

In the previous section, we have ensured the validity of our derived multi-planar Green's functions. Now we apply this to calculate reflected electric field from a dipole source with arbitrary orientation. To simplify this problem, without loss of generality, we merely consider one film structure shown in Fig. 5. An illustration to arbitrary orientation of a dipole is shown in Fig. 5(b). In "localized" spherical coordinate, a dipole moment has a magnitude $p_0$, a polar angle $\theta_d$ and an azimuth angle $\varphi_d$. According to Fig. 5 (a), we consider a dipole located at the position $z = d$ with moment $p_0 e^{-i\omega t} \mathbf{e}_0$ where $\mathbf{e}_0$ denotes a unit vector with the same direction of $\mathbf{p_0}$. Hence the current density is:

$$\mathbf{J}(\mathbf{r}') = -i\omega p_0 e^{-i\omega t} \delta(\mathbf{r}' - d\hat{z}) \mathbf{e}_0. \qquad (481)$$

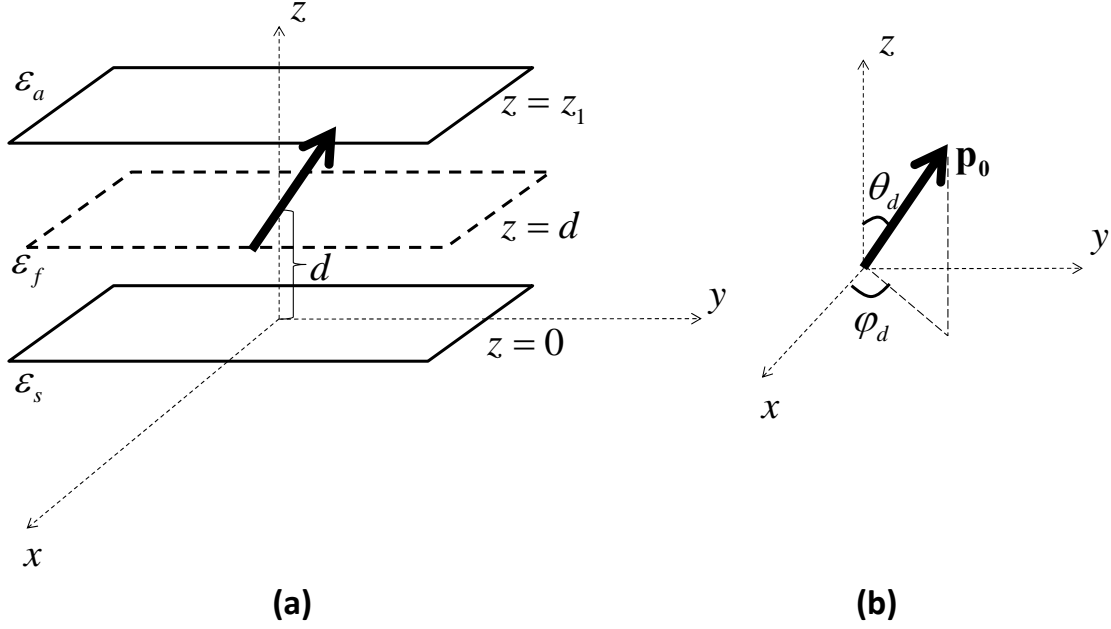

**Fig. 5** (a) 3D Geometry of the problem that a dipole with arbitrary orientation inside the film (b) Illustration of a dipole with arbitrary orientation

The reflected electric field at the dipole position is:

$$\begin{aligned} E_0(d\mathbf{e}_z) &= \mathbf{e}_0 \bullet \mathbf{E}(d\mathbf{e}_z) = \omega^2 \mu_0 p_0 e^{-i\omega t} \mathbf{e}_0 \bullet \int d^3\mathbf{r}' \mathbf{G}(d\mathbf{e}_z, \mathbf{r}') \bullet \delta(\mathbf{r}' - d\mathbf{e}_z)\mathbf{e}_0 \\ &= \omega^2 \mu_0 p_0 e^{-i\omega t} \mathbf{e}_0 \bullet \mathbf{G}(d\mathbf{e}_z, d\mathbf{e}_z) \bullet \mathbf{e}_0 \end{aligned} \qquad (482)$$

where $\mathbf{e}_0 = (\sin\theta_d \cos\varphi_d, \sin\theta_d \sin\varphi_d, \cos\theta_d)$ and $\mathbf{e}_0 \bullet \mathbf{G}(d\mathbf{e}_z, d\mathbf{e}_z) \bullet \mathbf{e}_0$ becomes:

$$\begin{aligned} \mathbf{e}_0 \bullet \mathbf{G}(d\mathbf{e}_z, d\mathbf{e}_z) \bullet \mathbf{e}_0 &= \sin^2\theta_d \cos^2\varphi_d G_{xx}(d\mathbf{e}_z, d\mathbf{e}_z) \\ &+ \sin^2\theta_d \sin^2\varphi_d G_{yy}(d\mathbf{e}_z, d\mathbf{e}_z) + \cos^2\theta_d G_{zz}(d\mathbf{e}_z, d\mathbf{e}_z) \\ &+ \sin^2\theta_d \cos\varphi_d \sin\varphi_d G_{xy}(d\mathbf{e}_z, d\mathbf{e}_z) + \sin\theta_d \cos\theta_d \cos\varphi_d G_{xz}(d\mathbf{e}_z, d\mathbf{e}_z) \\ &+ \sin\theta_d \cos\theta_d \sin\varphi_d G_{yz}(d\mathbf{e}_z, d\mathbf{e}_z) + \sin^2\theta_d \cos\varphi_d \sin\varphi_d G_{yx}(d\mathbf{e}_z, d\mathbf{e}_z) \\ &+ \sin\theta_d \cos\theta_d \cos\varphi_d G_{zx}(d\mathbf{e}_z, d\mathbf{e}_z) + \sin\theta_d \cos\theta_d \sin\varphi_d G_{zy}(d\mathbf{e}_z, d\mathbf{e}_z) \\ &\equiv \sum_{i,j=x,y,z} \alpha_{ij} G_{ij}(d\mathbf{e}_z, d\mathbf{e}_z) \end{aligned} \qquad (483)$$

Hence Eq. (482) becomes:

$$E_0(d\mathbf{e}_z) = \omega^2 \mu_0 p_0 e^{-i\omega t} \sum_{i,j=x,y,z} \alpha_{ij} G_{ij}(d\mathbf{e}_z, d\mathbf{e}_z)$$

$$= \omega^2 \mu_0 p_0 e^{-i\omega t} \frac{1}{(2\pi)^2} \lim_{\substack{\rho \to 0 \\ \rho' \to 0}} \int d^2\mathbf{k}_n e^{i\mathbf{k}_n \bullet (\rho-\rho')} \sum_{i,j=x,y,z} \alpha_{ij} g_{nij}(d,d)$$

$$= \frac{\omega^2 \mu_0 p_0 e^{-i\omega t}}{(2\pi)^2} \int_0^\infty dk_n k_n \int_0^{2\pi} d\varphi_n \sum_{i,j=x,y,z} \alpha_{ij} g_{nij}(d,d)$$

$$= \frac{\omega^2 \mu_0 p_0 e^{-i\omega t}}{4\pi} \int_0^\infty dk_n k_n \left[ (\alpha_{xx} + \alpha_{yy}) \tilde{g}_{nxx} + (\alpha_{xx} + \alpha_{yy}) \tilde{g}_{nyy} + 2\alpha_{zz} \tilde{g}_{nzz} \right]$$

$$= \frac{\omega^2 \mu_0 p_0 e^{-i\omega t}}{4\pi} \sin^2 \theta_d \int_0^\infty dk_n k_n \tilde{g}_{nxx} + \frac{\omega^2 \mu_0 p_0 e^{-i\omega t}}{4\pi} \sin^2 \theta_d \int_0^\infty dk_n k_n \tilde{g}_{nyy}$$

$$+ \frac{\omega^2 \mu_0 p_0 e^{-i\omega t}}{2\pi} \cos^2 \theta_d \int_0^\infty dk_n k_n \tilde{g}_{nzz}$$

(484)

where three integrals $\int_0^\infty dk_n k_n \tilde{g}_{nxx}$, $\int_0^\infty dk_n k_n \tilde{g}_{nyy}$ and $\int_0^\infty dk_n k_n \tilde{g}_{nzz}$ are:

$$\int_0^\infty dk_n k_n \tilde{g}_{nxx}$$

$$= \int_0^\infty dk_n k_n \frac{\left( R_{n//}^{fa} R_{n//}^{fs} S_{nxx,f} e^{-2q_{nf} z_1} + R_{n//}^{fa} S_{nxx,f} e^{2q_{nf}d} e^{-2q_{nf} z_1} \right) + \left( R_{n//}^{fs} S_{nxx,f} e^{-2q_{nf}d} + R_{n//}^{fs} R_{n//}^{fa} S_{nxx,f} e^{-2q_{nf} z_1} \right)}{1 - R_{n//}^{fs} R_{n//}^{fa} e^{-2q_{nf} z_1}}$$

$$= \int_0^\infty dk_n k_n \frac{2 R_{n//}^{fa} R_{n//}^{fs} S_{nxx,f} e^{-2q_{nf} z_1} + R_{n//}^{fa} S_{nxx,f} e^{2q_{nf}d} e^{-2q_{nf} z_1} + R_{n//}^{fs} S_{nxx,f} e^{-2q_{nf}d}}{1 - R_{n//}^{fs} R_{n//}^{fa} e^{-2q_{nf} z_1}}$$

, (485)

$$\int_0^\infty dk_n k_n \tilde{g}_{nyy}$$

$$= \int_0^\infty dk_n k_n \frac{S_{nyy,f} e^{-2q_{nf} z_1} R_{n\perp}^{fa} R_{n\perp}^{fs} + S_{nyy,f} e^{-2q_{nf} z_1} R_{n\perp}^{fa} e^{2q_{nf}d} + S_{nyy,f} R_{n\perp}^{fs} e^{-2q_{nf}d} + S_{nyy,f} R_{n\perp}^{fs} R_{n\perp}^{fa} e^{-2q_{nf} z_1}}{1 - R_{n\perp}^{fs} R_{n\perp}^{fa} e^{-2q_{nf} z_1}}$$

$$= \int_0^\infty dk_n k_n \frac{2 S_{nyy,f} R_{n\perp}^{fa} R_{n\perp}^{fs} e^{-2q_{nf} z_1} + S_{nyy,f} R_{n\perp}^{fa} e^{-2q_{nf} z_1} e^{2q_{nf}d} + S_{nyy,f} R_{n\perp}^{fs} e^{-2q_{nf}d}}{1 - R_{n\perp}^{fs} R_{n\perp}^{fa} e^{-2q_{nf} z_1}}$$

and (486)

$$\int_0^\infty dk_n k_n \tilde{g}_{nzz}$$

$$= \int_0^\infty dk_n k_n \frac{ik_n}{q_{nf}} \frac{-S_{nxz,f} R_{n//}^{fa} e^{-2q_{nf} z_1} \left( e^{q_{nf}d} - R_{n//}^{fs} e^{-q_{nf}d} \right) e^{q_{nf}d} + R_{n//}^{fs} S_{nxz,f} \left( R_{n//}^{fa} e^{q_{nf}d} e^{-2q_{nf} z_1} - e^{-q_{nf}d} \right) e^{-q_{nf}d}}{1 - R_{n//}^{fs} R_{n//}^{fa} e^{-2q_{nf} z_1}}$$

$$= \int_0^\infty dk_n \frac{ik_n^2}{q_{nf}} \frac{2 S_{nxz,f} R_{n//}^{fa} e^{-2q_{nf} z_1} R_{n//}^{fs} - S_{nxz,f} R_{n//}^{fa} e^{-2q_{nf} z_1} e^{2q_{nf}d} - R_{n//}^{fs} S_{nxz,f} e^{-2q_{nf}d}}{1 - R_{n//}^{fs} R_{n//}^{fa} e^{-2q_{nf} z_1}}$$

. (487)

To check if our formulation is correct, we go back to consider a single dipole source with vertical and horizontal orientations as shown in Fig. 4. In this case, for vertical dipole, we put $\theta_d = \varphi_d = 0$ and then obtain:

$$E_0(d\mathbf{e}_z) = \omega^2 \mu_0 p_0 e^{-i\omega t} \sum_{i,j=x,y,z} \alpha_{ij} G_{ij}(d\mathbf{e}_z, d\mathbf{e}_z) = \frac{\omega^2 \mu_0 p_0 e^{-i\omega t}}{2\pi} \int_0^\infty dk_n k_n \tilde{g}_{nzz}$$

$$= \frac{\omega^2 \mu_0 p_0 e^{-i\omega t}}{2\pi} \int_0^\infty dk_n \frac{ik_n^2}{q_{nf}} \frac{2 S_{nxz,f} R_{n//}^{fs} R_{n//}^{fa} e^{-2q_{nf}z_1} - S_{nxz,f} R_{n//}^{fa} e^{-2q_{nf}z_1} e^{2q_{nf}d} - R_{n//}^{fs} S_{nxz,f} e^{-2q_{nf}d}}{1 - R_{n//}^{fs} R_{n//}^{fa} e^{-2q_{nf}z_1}},$$

$$= \frac{i\omega^2 \mu_0 p_0 e^{-i\omega t}}{4\pi k_f^2} \int_0^\infty dk_n \frac{k_n^3}{h_{nf}} \frac{2 R_{n//}^{fs} R_{n//}^{fa} e^{-2q_{nf}z_1} - R_{n//}^{fa} e^{-2q_{nf}z_1} e^{2q_{nf}d} - R_{n//}^{fs} e^{-2q_{nf}d}}{1 - R_{n//}^{fs} R_{n//}^{fa} e^{-2q_{nf}z_1}}$$

(488)

which is the same as Eq. (471). For horizontally case, we put $\theta_d = \frac{\pi}{2}, \varphi_d = 0$ and then obtain:

$$E_0(d\mathbf{e}_z) = \omega^2 \mu_0 p_0 e^{-i\omega t} \sum_{i,j=x,y,z} \alpha_{ij} G_{ij}(d\mathbf{e}_z, d\mathbf{e}_z)$$

$$= \frac{\omega^2 \mu_0 p_0 e^{-i\omega t}}{4\pi} \int_0^\infty dk_n k_n \frac{2 R_{n//}^{fa} R_{n//}^{fs} S_{nxx,f} e^{-2q_{nf}z_1} + R_{n//}^{fa} S_{nxx,f} e^{2q_{nf}d} e^{-2q_{nf}z_1} + R_{n//}^{fs} S_{nxx,f} e^{-2q_{nf}d}}{1 - R_{n//}^{fs} R_{n//}^{fa} e^{-2q_{nf}z_1}}$$

$$+ \frac{\omega^2 \mu_0 p_0 e^{-i\omega t}}{4\pi} \int_0^\infty dk_n k_n \frac{2 S_{nyy,f} R_{n\perp}^{fa} R_{n\perp}^{fs} e^{-2q_{nf}z_1} + S_{nyy,f} R_{n\perp}^{fa} e^{-2q_{nf}z_1} e^{2q_{nf}d} + S_{nyy,f} R_{n\perp}^{fs} e^{-2q_{nf}d}}{1 - R_{n\perp}^{fs} R_{n\perp}^{fa} e^{-2q_{nf}z_1}},$$

$$= -\frac{\omega^2 \mu_0 p_0 e^{-i\omega t}}{8\pi k_f^2} \int_0^\infty dk_n k_n q_{nf} \frac{2 R_{n//}^{fa} R_{n//}^{fs} e^{-2q_{nf}z_1} + R_{n//}^{fa} e^{2q_{nf}d} e^{-2q_{nf}z_1} + R_{n//}^{fs} e^{-2q_{nf}d}}{1 - R_{n//}^{fs} R_{n//}^{fa} e^{-2q_{nf}z_1}}$$

$$+ \frac{\omega^2 \mu_0 p_0 e^{-i\omega t}}{8\pi} \int_0^\infty dk_n \frac{k_n}{q_{nf}} \frac{2 R_{n\perp}^{fa} R_{n\perp}^{fs} e^{-2q_{nf}z_1} + R_{n\perp}^{fa} e^{-2q_{nf}z_1} e^{2q_{nf}d} + R_{n\perp}^{fs} e^{-2q_{nf}d}}{1 - R_{n\perp}^{fs} R_{n\perp}^{fa} e^{-2q_{nf}z_1}}$$

(489)

which is the same as Eq. (480).

## 7. Conclusion and further prospect

In this paper, we have presented a method for the construction of the multi-planar Green's functions by Fourier expansion in detail. Under the two-dimensional reciprocal space with suitable coordinates chosen, we have converted highly-complicated three-dimensional wave equation into one-variable simple differential equation. Furthermore, in order to ensure its validity, without loss of generality, we consider a vertically (horizontally) oriented dipole located inside a film, calculating the reflected electric field at its position and obtain the same results in the literature [6-7]. In addition, we have also extended our calculation for the reflected electric field of a dipole source with arbitrary orientation, which leads back to the correct results in special cases (both vertically and horizontally). The results will be useful for the modeling of fluorescing molecules (as emitting dipoles) in the vicinity of the multi-planner structure. For case with more than one molecule, we may investigate the Förster fluorescence energy transfer (FRET) between a donor and an acceptor near the multi-planner structure. Furthermore, if considering more realistic structures that experimentalists concern like metallic nanoparticles on or embedded planner substrate, in which localized surface plasmon of metallic nanoparticles can achieve strong enhancement of electric fields as well as surface enhanced Raman spectroscope, we may apply the Lippmann-Schwinger equation (or say Dyson's equation) to obtain greater insight into the phenomena.

We are in the process of building a FORTRAN code for our derived multi-planar Green's functions. Based on its formulation, we will try to extend it to include anisotropic effects of the medium. Nonlocal effect will be also considered in our future study. Also, we will try to find generalized dyadic Green's function which can accommodate for boundaries with arbitrary shapes. In our ultimate goal, we wish to control light theoretically by constructing the powerful packages which are more efficient and accurate for the application to other physical processes besides molecular fluorescence modeling.

## 8. Acknowledgment

I would like to thank my family for supporting this work in my spare time and also thank Prof. P. T. Leung for fruitful suggestions.